\documentclass[12pt,centertags,dvips,twoside]{report}
\usepackage{a4}
\usepackage{fancyheadings}
\newcommand{\clearemptydoublepage}{%
    \newpage{\pagestyle{empty}\cleardoublepage}}
\usepackage{amssymb}
\usepackage{amsfonts}
\newlength{\diracchlen}

\newlength{\vecchlen}
\newlength{\vecchhgt}
\newcommand{\lvec}[1]{\ensuremath{\settowidth{\vecchlen}{$#1$}%
  \settoheight{\vecchhgt}{$#1$}\addtolength{\vecchhgt}{1pt}
  #1\hspace{-0.5\vecchlen}%
  \makebox[0pt]{\raisebox{\vecchhgt}{$\scriptscriptstyle\leftarrow$}}%
  \hspace{0.5\vecchlen}%
}}

\usepackage{subfigure}
\usepackage{array}
\usepackage{epic}
\usepackage{eepic}
\usepackage{epsfig}
\usepackage{afterpage}
\usepackage{axodraw}
\makeatletter
\renewcommand{\@makecaption}[2]{%
  \vskip\abovecaptionskip
  \sbox\@tempboxa{#1: \emph{#2}}%
  \ifdim \wd\@tempboxa >\hsize
    #1: \emph{#2}\par
  \else
    \global \@minipagefalse
    \hb@xt@\hsize{\hfil\box\@tempboxa\hfil}%
  \fi
  \vskip\belowcaptionskip}
\makeatother
\usepackage{german}
\usepackage{dsfont}
\usepackage{axodraw}
\usepackage{subfigure}
\usepackage{rotating}
\usepackage{varioref}
\usepackage{cite}
\usepackage{citesort}
\usepackage{amsmath}
\newlength{\eqoff}
\newlength{\eqofftwenty}
\newlength{\eqoffsixty}
\newcommand{\col}{~,}
\newcommand{\pnt}{~.}
\newcommand{\AdS}{\text{AdS}}
\newcommand{\AdSS}{\AdS_{d+1}\times \text{S}^{d'+1}}
\newcommand{\twob}{{\text{II}\,\text{B}}}
\newcommand{\YM}{\text{YM}}
\newcommand{\GF}{\text{GF}}
\newcommand{\FP}{\text{FP}}
\newcommand{\GeV}{\,\text{GeV}}
\newcommand{\arsh}{\operatorname{arsh}}
\newcommand{\arccot}{\operatorname{arccot}}
\newcommand{\parderiv}[2]{\frac{\partial #1}{\partial #2}}
\newcommand{\preparderiv}[1]{\frac{\partial}{\partial #1}}
\newcommand{\funcderiv}[2]{\frac{\delta #1}{\delta #2}}
\newcommand{\prefuncderiv}[1]{\frac{\delta}{\delta #1}}
\newcommand{\prefuncderivi}[1]{\frac{\delta}{i\delta #1}}

\newcommand{\tprefuncderivi}[1]{\tfrac{\delta}{i\delta #1}}
\newcommand{\doublepreparderiv}[1]{\frac{\partial^2}{\partial {#1}^2}}
\newcommand{\fuud}[3]{f^{#1#2}_{\phantom{#1#2}#3}}
\newcommand{\duud}[3]{d^{#1#2}_{\phantom{#1#2}#3}}
\newcommand{\fddu}[3]{f_{#1#2}^{\phantom{#1#2}#3}}
\newcommand{\dddu}[3]{d_{#1#2}^{\phantom{#1#2}#3}}
\newcommand{\hypergeometric}[4]{F\big(#1,#2;#3;#4\big)}
\newcommand{\dummylen}{\phantom{x}}
\newcommand{\comm}[2]{\left[#1\smash[b]{\mathbin{,}}#2\right]}
\newcommand{\acomm}[2]{\left\{#1\smash[b]{\mathbin{,}}#2\right\}}
\newcommand{\astcomm}[2]{\left[#1\smash[b]{\mathbin{\overset{\ast}{,}}}#2\right]}
\newcommand{\astacomm}[2]{\left\{\smash[b]{#1\mathbin{\overset{\ast}{,}}}#2\right\}}

\newcommand{\astcos}{\cos\big(\lvec\partial_\rho\tfrac{1}{2}\theta^{\rho\sigma}\vec\partial_\sigma\big)}
\newcommand{\astsin}{\sin\big(\lvec\partial_\rho\tfrac{1}{2}\theta^{\rho\sigma}\vec\partial_\sigma\big)}
\newcommand{\astwedgecos}{\cos\big(\lvec\partial\wedge\vec\partial\big)}
\newcommand{\astwedgesin}{\sin\big(\lvec\partial\wedge\vec\partial\big)}

\newcommand{\astposwedgecos}[2]{\cos\big((\partial)_{#1}\wedge(\partial)_{#2}\big)}
\newcommand{\astposwedgesin}[2]{\sin\big((\partial)_{#1}\wedge(\partial)_{#2}\big)}

\newcommand{\astmomwedgecos}[2]{\cos(#1\wedge#2)}
\newcommand{\astmomwedgesin}[2]{\sin(#1\wedge#2)}
\newcommand{\no}{\mathord{:}}
\newcommand{\de}{\operatorname{d}\!}
\newcommand{\Ruddd}[4]{\mathcal{R}^{#1}_{\phantom{#1}#2#3#4}}
\newcommand{\Gudd}[3]{\Gamma^{#1}_{\phantom{#1}#2#3}}
\newcommand{\gudd}[3]{\gamma^{#1}_{\phantom{#1}#2#3}}
\newcommand{\Cuddd}[4]{\mathcal{C}^{#1}_{\phantom{#1}#2#3#4}}
\newcommand{\e}{\operatorname{e}}

\newlength{\diameter}
\newcommand{\olett}[1]{  \ensuremath{ \text{$\settowidth{\diameter}{$\bigcirc$}%
  \bigcirc%
  \hspace{-0.5\diameter}%
  \makebox[0pt][c]{$\scriptstyle{#1}$}%
  \hspace{0.5\diameter}$}}}    
\newlength{\neglength}
\newcommand{\negphantom}[1]{\text{\settowidth{\neglength}{$#1$}$
                             \hspace{-\neglength}$}}         
\makeatletter
\newcommand{\redefinelabel}[1]{
  \def\@currentlabel{#1}}
\makeatother
\DeclareMathOperator{\tr}{tr}
\DeclareMathOperator{\sgn}{sgn}

\DeclareMathOperator{\T}{T}
\DeclareMathOperator{\wop}{\hat{\mathcal{W}}}
\DeclareMathOperator{\dop}{\hat{\Delta}}
\DeclareMathOperator{\vol}{vol}
\allowdisplaybreaks

\begin{document}
\selectlanguage{english}
\setlength{\baselineskip}{0.65cm}
\setlength{\parskip}{1ex}
\renewcommand{\arraystretch}{1.3}  

\begin{titlepage}
September 2004 \\
\begin{flushright}
HU Berlin-EP-04/49\\ 
\end{flushright}
\mbox{ }  \hfill hep-th/0409053
\vspace{5ex}
\large
\begin{center}
PhD Thesis
\end{center}
\Large
\begin {center}     
{\bf Aspects of Noncommutativity and Holography in Field Theory and String
Theory}
\end {center}
\large
\vspace{1ex}
\begin{center}
Christoph Sieg \footnote{csieg@physik.hu-berlin.de}
\end{center}
\begin{center}
Humboldt--Universit\"at zu Berlin, Institut f\"ur Physik\\
Newtonstra\ss{}e 15, D-12489 Berlin\\[2mm]  
\end{center}
\vspace{4ex}
\rm
\begin{center}
{\bf Abstract}
\end{center}
\normalsize 
This thesis addresses two topics: 
noncommutative Yang-Mills theories and the $\AdS/\text{CFT}$ correspondence. 

Noncommutative Yang-Mills theories with arbitrary gauge groups have been
defined as a power expansion in the noncommutativity parameter $\theta$, 
circumventing the a priori restriction to $U(N)$ groups.
We study a partial summation of this $\theta$-expanded perturbation theory. 
We review the Seiberg-Witten map as an essential ingredient for the 
formulation and then summarize different proposals of how to extend the gauge
group to $G\subseteq U(N)$. 
The partial summation of the $\theta$-expanded perturbation theory allows us to
study issues of possible $\theta$-exact Feynman rules for these generalized 
theories. On diagrammatic level  a summation 
procedure is established, which in the $U(N)$ case delivers the full
$\ast$-product induced rules. Thereby we uncover a cancellation mechanism 
between certain diagrams, which is crucial in the $U(N)$ case, but set
out of work if $G\subset U(N)$, $G\neq U(M)$, $M<N$.
In addition, an explicit proof is given that there is no
partial summation of the $\theta$-expanded rules resulting in new
Feynman rules using the $U(N)$ star-product vertices and  
besides suitable modified propagators at most a \emph{finite} 
number of additional
building blocks. Finally, we disprove certain $SO(N)$
Feynman rules conjectured in the literature. 

Within the second topic we study quantities which are important
for the realization of the holographic principle in the
$\AdS/\text{CFT}$ correspondence: boundaries, geodesics and propagators of 
scalar fields. They should play a role in the holographic setup in the 
BMN limit as well. 
We first give a review of the relevant backgrounds and required limits that
have to be taken. We then describe in brief the $\AdS/\text{CFT}$
correspondence and its BMN limit and summarize different 
proposals of how holography
could be understood in this limit. The full $\AdS_5\times\text{S}^5$ background is transformed via the Penrose limit to the $10$-dimensional plane wave. 
By projecting on a suitable subset of coordinates we observe the degeneration
of the conformal $\AdS_5\times\text{S}^5$ boundary in this limit.
We construct all $\AdS_5\times\text{S}^5$ and plane wave geodesics in their 
integrated form. Performing the Penrose limit, the approach of null geodesics
reaching the conformal boundary of $\AdS_5\times\text{S}^5$ to that of the
plane wave is studied in detail.
Based on a relation between the scalar bulk-to-boundary and 
bulk-to-bulk propagators in the $\AdS/\text{CFT}$ correspondence we argue that 
bulk-to-bulk propagators should be useful for identifying the holographic
setup in the plane wave limit. 
With this motivation 
the propagator of a scalar field in generic $\AdSS$ backgrounds is 
discussed in detail. In special cases we find powerlike solutions 
in terms of the total chordal distance in $\AdSS$. For these solutions we
discuss the possible $\delta$-sources. For
$\AdS_5\times\text{S}^5$ we relate the propagator to the expression in the  
$10$-dimensional plane wave and find a geometric interpretation of the
variables occurring in the known explicit construction on the plane wave.
As a byproduct of comparing different techniques, including the
KK mode summation,  a theorem for summing certain products of
Legendre and Gegenbauer functions is derived.

Detailed calculations, introductions to some required formalisms and a
collection of useful formulae are delegated to several Appendices.


\vfill
\end{titlepage} 
\clearemptydoublepage
\pagenumbering{roman} 
\pagestyle{empty}
\vspace*{95mm}
\begin{center}
To my parents
\end{center}
\pagestyle{fancyplain}
\tableofcontents
\clearemptydoublepage

\selectlanguage{english}
\clearemptydoublepage
\pagenumbering{arabic} 
\part{Introduction}
\clearemptydoublepage
\chapter{General Introduction}
\label{intro}
In the energy range accessible with present experiments, 
three of the four fundamental interactions are successfully described by 
quantum field theories. The latter provide a theoretical description of 
the strong, weak and electromagnetic interactions which are 
collectively denoted as the Standard Model (of particle physics). 
The remaining interaction is gravity and 
its classical description is given by the theory of General 
Relativity.

The quantum field theories in the Standard Model are all examples of gauge 
theories, where spin one particles are responsible for transmitting the
interaction. Gauge
theories contain more degrees of freedom than necessary for the description of
the physical (measurable) quantities. The gauge
transformations relate physically equivalent field
configurations, and thus form a
group such that a sequential application of two
gauge transformations is itself a gauge transformation.
For example, the gauge group of the Standard Model is given by 
$SU(3)\times SU(2)\times U(1)$, where the first factor refers to the
theory of the strong 
interaction which is called quantum chromodynamics (QCD), 
and the second and third factors describe the electroweak interaction that
includes the weak and the electromagnetic parts. 
In addition to the gauge fields, a gauge theory may contain additional fields
the gauge fields interact with. In the case of the Standard Model these
are the fermionic quark and lepton fields and the scalar Higgs field.

Naively, one would expect that a direct observation of the particles
that are associated with the Standard Model fields is possible in their
elementary form. Depending on the interaction, one furthermore should find
several compound systems 
of the elementary particles, bound together by the fundamental forces.
But this naive expectation is not quite true. For instance, one does
not observe free quarks.  
The hadrons, which are bound states of the the strong interaction,  behave
differently compared to electromagnetic bound states.
Consider for instance positronium which is a state of an
electron and a positron bound via the electromagnetic force.
If one separates the electron from 
the positron, the force of electromagnetism decreases, and it is possible to
break up a positronium into its elementary constituents. 
In contrast to this,
trying to separate two quarks inside the hadron leads to an effectively 
constant force, corresponding to an effective potential which is linearly 
increasing with the distance. Hence at a certain point, where the potential 
energy is high enough to create a quark anti-quark pair, the hadron breaks up
into two hadrons, such that one does not find free quarks. 
A necessary (but not sufficient) ingredient for this behaviour is the
interaction of the gauge bosons with each other. This is one of the main
differences between QCD and quantum electrodynamics (QED) and is due to the
gauge group of QCD being non-Abelian, whereas QED has an Abelian gauge group.

The absence of unbound quarks in nature makes it difficult to show their
existence. Hence, it becomes understandable that before the formulation
of QCD, much work was spent in trying to explain the spectrum of
hadrons without taking into account that they might be non-fundamental
composite objects. One of the surprising issues of the hadronic
spectrum is that the hadrons can be sorted into groups in such a way
that, within every group, one finds a linear relation between the mass
squares $m^2$ of the hadrons and their spins $J$. In the linear relation 
$\alpha'm^2=\alpha_0+J$ the slope $\alpha'\sim 1\GeV^{-2}$ is 
universal for all hadrons, only
the intercept $\alpha_0$ is different for each group of hadrons. 
If one plots $m^2$ versus $J$ one obtains the famous Regge trajectories
\cite{Regge:1959mz,Regge:1960zc}.
They were realized within dual  
models\footnote{For an introduction with a detailed summary of the historical
  developments see \cite{Scherk:1975jj}.}  
\cite{Veneziano:1968yb,Virasoro:1969me,Shapiro:1970gy} which successfully
describe the behaviour of hadron scattering amplitudes in the so called  
Regge regime. 
For example, a $2\rightarrow 2$ scattering process in four dimensions
is specified by the three Mandelstam variables $s$, $t$ and $u$ as kinematical
invariants. Only two of them, choose for instance $s$ and $t$ being
related to the center of mass energy and the scattering
angle in the process, are independent.
In the Regge regime, which is given by $s\to\infty$ with
$t=\text{fixed}$, the scaling behaviour of the scattering amplitude
proportional to $s^{\alpha(t)}$ with 
$\alpha(t)=\alpha_0+\alpha't$ is successfully reproduced by the dual models.
It was discovered \cite{Nambu:1970,Nielsen:1970,Susskind:1970xm} that the dual
models describe the dynamics of relativistic strings and thus we will
henceforth denote them as (hadronic) string theories. 

Besides the above outlined successes these string theories
contain some issues that are unwanted
in a theory of hadrons. They predict the existence of a variety
of massless particles not detected in the hadronic spectrum. 
Depending on the concrete model, different values for the intercept
$\alpha_0$ are theoretically favoured\footnote{An enhancement of symmetries 
\cite{Virasoro:1970zu} and the
   decoupling of negative norm states \cite{Brower:1972wj,Goddard:1972iy} is
   found.}  
that are in disagreement with the phenomenologically preferred value 
$\alpha_0=\frac{1}{2}$.
Furthermore, the behaviour of the scattering amplitudes in the hard fixed angle
scattering regime, where $s$ and $t$ are large with fixed ratio $\frac{s}{t}$,
is too soft compared to the experimental results. Here the string
theories predict an exponential falloff, 
while experimentally one finds a powerlike behaviour. 
Experimental data indicate that the probed structure in this
regime is not an extended object but instead a pointlike particle. 
This means that the probe particles no longer interact with the 
hadron as a whole but with pointlike constituents from which the latter 
is built. 

Hence, a theory of hadrons is not a theory of fundamental objects. It
turned out that QCD is the appropriate description of the constituents from
which hadrons are built.
But the strong interactions are far from being understood completely. 
Perturbative QCD is very successful in describing the phenomenology of strong
interactions at high energies but the effects at low energy, where the coupling
constant is large and hence perturbation theory is not applicable, are still
hard to analyze. 
The absence of free quarks described above is an effect of this  
property called confinement, that until now cannot be successfully explained.
The hadronic string theories can be interpreted as effective descriptions 
of confinement where the open string collects the effects of the flux tubes 
transmitting the forces between the quarks that are situated at the 
endpoints of this QCD string (see e.\ g.\
\cite{Schwarz:2000yy,Polyakov:1997tj,Polyakov:1996nc,Polchinski:1992vg,Brower:2003cd}
and references therein). 
To be able to analyze quantum field theories non-perturbatively and thereby
understand the mechanism of confinement beyond such an effective description 
is a challenge for theoretical physics. 

Besides this fact there is another lack of understanding. General Relativity
is the relevant theory of gravity at length scales that are large compared to
the fundamental length of gravity, the Planck length
$l_\text{P}=M_\text{P}^{-1}$ ($M_\text{P}=1.22\times 10^{19}\GeV$ is the
Planck mass). But a description of gravity  
at very short distances comparable to the Planck length requires a quantum
theory of gravity. The naive attempt to quantize general relativity fails. 
To be more precise, quantum field
theories suffer from divergencies that have to be regularized and absorbed
into the physical parameters of the theory. This procedure is known as
renormalization. In contrast to QCD and the electroweak interaction which
are renormalizable quantum field theories\footnote{See \cite{'tHooft:1971fh,'tHooft:1971rn,'tHooft:1972fi} and further references
  given in \cite{Collins:1984xc,Muta:1998vi}.}, naive quantum gravity is
non-renormalizable.
The fields with highest spin for which a
quantum field theoretical description is known are the spin one gauge fields,
but the graviton (the quantum of gravitation) carries spin two.

One fundamental difference between gravity and gauge theories is that
spacetime itself is affected by gravity and hence is dynamical. 
A good way to understand this is to think about what happens if one wants to
observe smaller and smaller structures in spacetime.  
The probe wavelength has to be comparable or smaller than the minimum length
one wants to resolve. This means one has to put higher and higher energy 
into the system. Energy is a source of gravitation and hence the spacetime is
deformed in the measurement process.
This is a good reason why the spacetime at small scales of the order of
magnitude of the Planck length would look differently from 
what one observes at large scales. 
In particular, if one increases the energy above a critical value, the
gravitational collapse of the region would be inevitable and the 
desired information would be absorbed by the black hole which would form. 
The density of quantum states 
of a black hole turns out to depend on the area of the horizon and it suggests 
that only one bit of information can be stored in a surface element of the
order of the squared Planck length 
\cite{'tHooft:1993gx,Susskind:1995vu,'tHooft:1999bw}.
Hence, a formulation of quantum gravity should include a mechanism 
that makes it impossible to resolve structures that 
are smaller than the Planck length. This
clearly introduces an upper bound on momenta and hence an ultraviolet cutoff
in the quantum theory.  

One way to realize such a cutoff is to introduce extended objects to replace
the pointlike particles as fundamental objects. Heuristically speaking, to
smear out interactions and make them non-local prohibits one from  
probing the spacetime at arbitrarily small scales. 
The string theories, that originally appeared as a proposal to describe the
hadrons, are theories of $1$-dimensional extended objects. If they are 
formulated as a
theory of gravity \cite{Scherk:1974ca} where now $\alpha'\sim
M_\text{P}^{-2}$, the disadvantage that  
occurred in the form of the exponential falloff of the amplitudes then turns
into an advantage to guarantee a nice high energy behaviour. Moreover, the
appearance of the unwanted massless modes is important. These modes contain
the desired graviton. 
Quantization leads to constraints on the dimension and 
the shape of the spacetime in which the string theories live. 
The so called
critical dimension which is necessary for a consistent
quantum theory of strings is much higher than four, 
in particular it is given by
$D=26$ and $D=10$ for the bosonic and the supersymmetric string theories,
respectively.\footnote{It is not strictly necessary to work in the critical
  dimension. The price to pay is that the worldsheet metric becomes dynamical
  even in conformal gauge and introduces one new degree of freedom. 
Another possibility is to 
work with a non-constant dilaton such that the critical dimension is modified.
This clearly breaks Poincar\'e invariance in the target space.}  
The additional dimensions do not rule out string theories as theories of
gravity. Since we do not know how spacetime looks like at short distances
comparable to the Planck length, there is the possibility that additional
compact dimensions exist. They should simply be highly curved and thus so tiny
that it is impossible to detect them at energy scales that are accessible
today. Much better, the additional  
dimensions naturally lead to a unified description of gauge theories and
gravity in our four-dimensional perspective.

Besides fixing the dimension, the consistency of the quantum
theory further restricts 
the choice of a classical background on which the string theories can be
formulated. By classical background we mean that one specifies a spacetime
plus values for the fields of the theory. A consistent classical background 
has to fulfill certain differential equations that at the leading order 
turn out to be the Einstein equations for the metric and the Yang-Mills field
equations for gauge fields. This means that at energies small
compared to the Planck scale one finds string theories encompass general 
relativity \cite{Yoneya:1973ca,Yoneya:1974jg,Scherk:1974ca} and gauge theories
\cite{Neveu:1972mu}.

Putting together all the above given observations, string theories 
are promising candidates for a unified formulation of the 
four fundamental interactions \cite{Scherk:1974ca}. In the following will not 
review string theories further and refer the reader to the literature
\cite{Green:1987sp,Green:1987mn,Lust:1989tj,Polchinski:1998rq,Polchinski:1998rr}. 

The next two Sections contain a short introduction and motivation of the two
topics we are dealing with in this thesis. A short summary of the structure of
the corresponding part will be given at the end of each of these sections.

\section{Noncommutative Yang-Mills theories}
\label{NCintro}
We have seen that the understanding of quantum field theories and of quantum
gravity is highly relevant for a successful 
description of all fundamental interactions. Quantum field theories and in
particular Yang-Mills (YM) theories are far from being understood
completely. One way to learn more about them is to analyze modified YM
theories which do not necessarily play a direct role in the description of
nature.  

For instance, one can deform the spacetime on which YM theories are formulated.
The case of a noncommutative spacetime is of particular interest.
In the canonical case that will be of importance here, 
the commutator of two spacetime coordinates $x^\mu$ and $x^\nu$ is given by
\begin{equation}
\comm{x^\mu}{x^\nu}=i\theta^{\mu\nu}\col
\end{equation}
where $\theta^{\mu\nu}$ is a constant tensor that necessarily has length
dimension two. In this way one has introduced a new parameter
$\sqrt{\Vert\theta\Vert}$ into the theory that can be used as an expansion
parameter for a perturbative 
analysis.\footnote{$\Vert\theta\Vert$ denotes the
  maximum of the absolute values of all entries of $\theta^{\mu\nu}$ in its
  canonical skew-diagonal form, see \cite{Szabo:2001kg}.} 

Noncommutativity gives rise to a topological
classification of Feynman diagrams
\cite{'tHooft:1974jz,Bessis:1980ss}.
One replaces each line of a graph by a double line such that one obtains the
so called ribbon graphs. The genus $h$ of a Feynman diagram is then defined as 
the minimal genus of all surfaces on which its ribbon graph can be drawn
without the crossing of lines.
It then turns out that planar diagrams in the noncommutative 
theory are given by essentially the same expressions that one finds in the
ordinary (commutative) theory. The only difference is that an overall phase 
factor multiplies each planar graph, being the same for each graph with fixed
external momenta \cite{Filk:1996dm}.

The situation is different for non-planar 
diagrams. Inside the loop integrals of the corresponding expressions, phase
factors occur in the noncommutative case that depend on the loop momentum $k$.
In the limit of larger and larger $\theta^{\mu\nu}$ 
these phase factors oscillate with smaller and smaller period and hence they 
increasingly suppress the non-planar contributions.  
Particularly, in the perturbation expansion for maximal
noncommutativity, i.\ e.\ $\theta^{\mu\nu}\to\infty$ or 
alternatively all momenta 
being large at fixed $\theta^{\mu\nu}$, only planar diagrams survive.
Hence, $\frac{1}{\Vert\theta\Vert}$ plays the role of a topological expansion
parameter, 
i.\ e.\ $\Vert\theta\Vert$ is the analog of the rank $N$ of the gauge 
group in the large $N$ genus expansion \cite{'tHooft:1974jz} that will be
described in Section \ref{dualityintro} below. 

Making spacetime coordinates noncommutative is not only useful for learning
more about gauge theories.  
In addition it enables one to study some aspects of gravity without 
using gravity itself. The noncommutativity of the coordinates gives rise to 
the non-locality of interactions. Hence, one can study the influence of
non-locality on renormalization properties in the framework of gauge theories
and one avoids to work with theories of gravity. 
Noncommutativity and hence non-locality of the interactions 
do not improve the ultraviolet behaviour of planar diagrams since only a
multiplicative phase factor arises. Non-planar diagrams, however, behave
differently. The oscillatoric behaviour of the phase factors inside the loop
integrals renders all one-loop non-planar diagrams finite\footnote{This does
  not hold for all non-planar diagrams at higher loops.}.   
This gives rise to the remarkable issue of UV/IR mixing
\cite{Minwalla:1999px}. The effective UV cutoff depends on the external
momenta $p$ in such a way that the UV divergencies reappear whenever
$p_\mu\theta^{\mu\nu}\to0$. This means that at small momenta the
noncommutative phase factors inside the loop integrals are irrelevant and
hence turning on noncommutativity replaces the standard ultraviolet
divergencies with a singular infrared behaviour.

As we have seen before, non-locality and improved high energy behaviour of the
amplitudes should be important ingredients in a theory of quantum gravity.
The facts that noncommutative YM theories capture some aspects of
gravity and that gravity naturally occurs in string theories, 
leads to the question of how deeply gauge and string theories are related. 
Could it be possible that 
some gauge and string theories describe the same physics? 
Before describing examples of this kind in section \ref{dualityintro}, we 
motivate and summarize the analysis in Part \ref{NC} of this thesis. 

During investigating noncommutative field theories one 
could have the idea of simplifying the problem by studying a truncated
version of the
full noncommutative theory. One expands in powers of $\theta^{\mu\nu}$ and
only considers some of the leading terms. However, in such an expansion
effects like UV/IR mixing are lost. They require the full
$\theta^{\mu\nu}$-dependence. 
On the other hand, a frequently addressed task in the context of
noncommutative gauge theories is their consistent formulation for 
gauge groups different from $U(N)$. In contrast to the case of ordinary gauge 
theories this question is highly non-trivial since a priori noncommutative 
theories appear to be consistent only for $U(N)$ gauge groups.
In some approaches that deal with such a modification, an explicit
expansion in powers of $\theta^{\mu\nu}$ is required and this prevents one 
from studying effects like UV/IR mixing for these theories.

Part \ref{NC} deals with the question of how compatible the extension of
noncommutative YM theories to arbitrary gauge groups is with keeping the 
exact dependence on the noncommutativity parameter $\theta^{\mu\nu}$.

\begin{itemize}
\item
In Chapter \ref{chap:NCfromstrings} we start with a review of how
noncommutative YM theories arise from string theories. This connection 
implies the existence of a map between the noncommutative and the ordinary 
description. The map is of particular importance in a formulation of
noncommutative YM theories with arbitrary gauge groups.  
\item 
In Chapter \ref{chap:NCYM} we introduce noncommutative YM theories.
We review the effects of noncommutativity on the choice of the gauge
group. Then we present the Faddeev-Popov gauge fixing procedure and
use it to derive the map between the noncommutative and ordinary ghost fields.
We will first extract 
the well-known Feynman rules for noncommutative $U(N)$ YM theories. 
We summarize the aforementioned approaches of how to implement other gauge 
groups in noncommutative gauge theories. 
Based on our work \cite{Dorn:2002ah}, we will then discuss
what happens in a construction of $\theta^{\mu\nu}$ exact 
Feynman rules if the gauge group is a subgroup of $U(N)$. 
\item
A short introduction to the used formalism as well as some detailed 
calculations that are useful for fixing notations and support 
the analysis of Part \ref{NC} are collected in Appendix \ref{app:NC}.
\end{itemize}

\section{Dualities of gauge and string theories}
\label{dualityintro}
In 1974 'tHooft \cite{'tHooft:1974jz} 
found a property of $SU(N)$ gauge theories that suggests a
relation to string theory. He observed that the perturbation expansion can be
regarded as an expansion in powers of $\lambda=g^2N$ instead of an expansion
in powers of the YM coupling constant $g$. In addition to the expansion in
$\lambda$, one can classify Feynman graphs in powers of $N^{-2}$.
A perturbative expansion in $N^{-2}$ is possible in the case of large $N$
with $\lambda$ kept fixed. It can be interpreted with the help of the double 
line notation for Feynman graphs \cite{'tHooft:1974jz,Bessis:1980ss}, 
that was already introduced in section \ref{NCintro}. Here,
each line now represents one fundamental
index of the $N\times N$ representation matrices of the gauge group. The
large $N$ expansion then turns out to be a topological expansion in which a
Feynman diagram with genus $h$ is of order $N^{-2h}$. 
If one now takes the so called 'tHooft limit, where $N\to\infty$ and $\lambda$
is kept fixed, only the planar graphs which can be drawn on a sphere ($h=0$)
survive. 
The genus expansion of Feynman diagrams in a gauge theory resembles the 
genus expansion of string theory, such that a relation between these two 
different types of theories was conjectured.

\subsection{The $\AdS/\text{CFT}$ correspondence}
The first concrete example of a gauge/string duality was proposed by Maldacena
in 1997 \cite{Maldacena:1998re} and is known as $\AdS/\text{CFT}$
correspondence in the literature. 
One case of particular importance is the conjecture that 
the $4$-dimensional supersymmetric ($\mathcal{N}=4$) YM theory 
should describe the same physics as type $\twob$ string theory on an 
$\AdS_5\times\text{S}^5$ background with some additional Ramond-Ramond flux 
switched on.

Up to now the conjecture could not be analyzed in full generality,
but it has passed several non-trivial checks. 
One regime in which concrete computations can be done on the string theory side
is where the string theory can be replaced by type $\twob$ supergravity. 
On the gauge theory side this regime corresponds to first taking $N\to\infty$
at fixed $\lambda$ and then taking $\lambda\to\infty$. It is therefore 
perfectly inaccessible with perturbation theory.
But turning the argument around, if one assumes that the $\AdS/\text{CFT}$
correspondence is correct, then there is a nice tool to analyze this 
concrete gauge theory in the large $N$ limit at strong 'tHooft coupling
$\lambda$  by working in the dual supergravity description. 

How can it happen that a lower dimensional gauge theory contains the same
amount of information as the higher dimensional string theory? 
The physical picture behind this is that the $\AdS/\text{CFT}$ correspondence 
is holographic. Since this issue is one of the main motivations for
Part \ref{BMN} of this thesis, let us explain it in more detail. 

Originally if we were to talk about holography we refer to 
a particular technique in photography. 
The use of a coherent light source like a laser
enables one to store not only brightness and colour information of a three 
dimensional object on the two dimensional film. One splits the laser beam
into two parts such that one hits the film directly and the other hits the 
object on the side that is directed towards the film. The direct part
of the beam and the light that is reflected from the object generate an
interference pattern that is stored on the film. 
The result is that one has encoded information about the varying distance
between the surface of the object and the film.
Viewing a holographic film, the object then appears three dimensional and the
part that was directed towards the film could in principle be 
completely reproduced without information loss, at least if the resolution
of the film were arbitrarily high.  
This is in sharp contrast
to the ordinary photography which is simply a (orthogonal) projection, such
that all the information about length scales perpendicular to the film is
lost. 
In the real case of a finite resolution of the film there is an
information loss in both cases, both images appear blurred at sufficiently
small distances. But the essential difference between the 
holographic and the ordinary image, that one does or does not have any
information about the third dimension, remains.  

In quantum gravity, holography
\cite{'tHooft:1993gx,Susskind:1995vu,'tHooft:1999bw} appears in connection
with a black hole which information is stored on its horizon. 
The holographic screen is the horizon and a single grain of the film 
corresponds to an area element of Planck size. It is the minimum area on which
information can be stored. One could now argue \cite{'tHooft:1999bw} that  
the emergence of holography in this context is only a special case of 
a holographic principle that is of universal validity in theories of 
gravity.  
Indeed, the holographic principle plays an essential role in connection with 
the $\AdS/\text{CFT}$ correspondence.

Let us now draw the analogy to the $\AdS/\text{CFT}$ correspondence.
The complete three dimensional space in the photography example becomes the 
$5$-dimensional anti-de Sitter spacetime $\AdS_5$.\footnote{The additional
  $5$-dimensional sphere to form a $10$-dimensional spacetime simply produces
  the Kaluza-Klein tower of states.} The lower dimensional holographic screen
which corresponds to the two dimensional film in the above example is given by
the $4$-dimensional boundary of $\AdS_5$. The easiest way to understand this
is to represent $\AdS_5$ as a full cylinder. The boundary of the cylinder then
is the holographic screen and the radial coordinate 
is the coordinate perpendicular to the film. It is called the 
holographic direction. 
Strings live in the full cylinder, and the four dimensional theory lives on its
boundary. The interference pattern that encodes the position perpendicular to
the screen in holographic photography corresponds to the energy scale in the 
boundary theory. Like in the case of the black hole, the Planck length
corresponds to the finite resolution of the film \cite{Susskind:1995vu}.
  
Although one finds many similarities between holographic photography and 
the $\AdS/\text{CFT}$ correspondence one should not drive the analogy too
far. In contrast to photography where it should in principle not matter where
the screen is put, in the $\AdS/\text{CFT}$
correspondence this choice can have non-trivial effects. 
The reason is that the geometry is not only given by $\AdS_5$ but instead 
by the $10$-dimensional product space $\AdS_5\times\text{S}^5$.
For instance, the choice to take the boundary of $\AdS_5$ as a holographic
screen implies that it is four dimensional, because there the $\text{S}^5$ 
is shrunk to a point. Any slice 
at a constant holographic coordinate value somewhere in 
the interior of $\AdS_5\times\text{S}^5$ is $9$-dimensional because there the
$\text{S}^5$ has finite size. Hence, such a slice would define a
$9$-dimensional holographic screen. 
Indeed, this difference is one main reason why in a
certain limit of the $\AdS/\text{CFT}$ correspondence a unique holographic 
description has not yet been found. 

Before we describe this limit it is worth to recapitulate what happened
with string theories until now.  
They were originally introduced to describe the strong interactions.
Then they were replaced by QCD that has many advantages but is little under
control beyond perturbation theory. String theories were proposed as the 
theories of gravity. But they reentered the
regime of gauge theories as  
possible dual descriptions that might be a key tool to 
study non-perturbative effects in gauge theories. 

This seems to explain why in
their original formulation as theories of hadrons string theories 
covered some aspects of hadron physics. In particular it becomes more
plausible why strings are an effective description of the QCD flux tubes.
Furthermore, it sheds new light on 
the unwanted issues of string theories in this direct formulation describing 
hadrons. 
They could be regarded as a hint that one was 
working in the wrong setup (one should work in the spirit of the 
$\AdS/\text{CFT}$ correspondence instead of trying to formulate them directly
as theories of hadrons) including the wrong choice of scales (one should use  
$\alpha'\sim M_\text{P}^{-2}$ instead of $\alpha'\sim 1\GeV^{-2}$).
For instance, it was indeed found that the too soft behaviour of string
amplitudes in the hard fixed angle regime can be avoided in the framework of
the $\AdS/\text{CFT}$ correspondence \cite{Polchinski:2001tt}.

\subsection{The BMN limit of the $\AdS/\text{CFT}$ correspondence}
At the present time, a proof of the $\AdS/\text{CFT}$ correspondence is
still out of reach. Even an analysis of string theory on
$\AdS_5\times\text{S}^5$ beyond the supergravity approximation
lies outside present capabilities. Berenstein, Maldacena and Nastase
\cite{Berenstein:2002jq} 
formulated a new limit of the $\AdS/\text{CFT}$ correspondence that 
goes beyond the supergravity approximation. The proposal is based on the
observation \cite{Blau:2002mw} 
that the $\AdS_5\times\text{S}^5$ background can be transformed 
with the so called Penrose-G\"uven limit \cite{Penrose:1976,Gueven:2000ru} 
to a plane wave background \cite{Blau:2001ne} on which string
theory is quantizable \cite{Metsaev:2001bj,Metsaev:2002re}. Berenstein, 
Maldacena and Nastase translated the limit to 
the gauge theory side and proposed that a certain subsector of operators 
in the gauge theory should then be related to type $\twob$ string theory on
the plane wave. This limit reveals further aspects of the presumably deep
connection between gauge and string theories. A very nice picture is that 
the operators which are compound objects made from the
fields of the gauge theory can be interpreted as discretized 
strings where each field operator represents a single string bit.

Later it was found \cite{Gubser:2002tv,Frolov:2002av} that the BMN limit
can be embedded into the more
general  framework of expanding the string theory about a classical string
solution on $\AdS_5\times\text{S}^5$ and that explicit checks for several
classical solutions can be performed because the relevant gauge theory 
parts are integrable.   
 
But what is still missing is a satisfactory description of how holography in
the $\AdS/\text{CFT}$ correspondence appears in its BMN limit.
Connected with this, it is not clear what the dual gauge theory is and where it
lives. Is it a one dimensional theory on the one dimensional boundary of the
plane wave spacetime, or does it live on a screen in the interior that 
could have any dimension between one and nine?

In Part \ref{BMN} of this thesis we discuss the behaviour of some
geometrical and field theoretical quantities in the limiting process from 
the $\AdS/\text{CFT}$ correspondence to its BMN limit. The idea is that 
one should observe how quantities that are relevant in the holographic
description are transformed in the limit and hence learn more about the fate 
of holography. 

\begin{itemize}
\item
In Chapter \ref{chap:BMNAdSpw} we review the ingredients that are
essential for an understanding of the backgrounds in the $\AdS/\text{CFT}$
correspondence and in its BMN limit. Furthermore, we will review
the limiting processes with which some of the backgrounds can be related and 
which play an important role in understanding the idea of the 
$\AdS/\text{CFT}$ correspondence and how its BMN limit arises.
\item
In Chapter \ref{chap:BMNholo} we review how holography is understood in the 
$\AdS/\text{CFT}$ correspondence in more detail. We then turn our attention 
to its plane wave limit and give a brief summary of some work to define a
holographic setup in the BMN limit. This gives an impression that the 
picture is less unique than in the $\AdS/\text{CFT}$ correspondence itself. 
\item
In Chapter \ref{chap:boundgeo} 
we analyze how the boundary structure behaves in
the limiting process. Furthermore, to get some information on the causal
structure, we derive and classify all geodesics in the
original background and discuss their fate in the limit. This Chapter is
completely based on our analysis in \cite{Dorn:2003ct}.  
\item
In Chapter \ref{chap:propaga} we come back to the arguments given in Chapter
\ref{chap:BMNholo}, that the propagators are ingredients of
particular importance in a holographic setup.
We focus on the bulk-to-bulk
propagator in generic $\AdSS$ backgrounds. We compute it in
particular cases and analyze some of its properties in detail. In particular,
we show how the well-known result in the $10$-dimensional plane wave spacetime
is obtained by taking the Penrose limit. This chapter is mainly based on 
our analysis in \cite{Dorn:2003au}.
\item
Some detailed calculations and useful formulas that refer to Part \ref{BMN}
are collected in Appendix \ref{BMNapp}.
\end{itemize}


\clearemptydoublepage
\part{Noncommutative geometry}
\label{NC}
\clearemptydoublepage
\chapter{Noncommutative geometry from string theory}
\label{chap:NCfromstrings}
The intention of this Chapter is to review how noncommutative gauge theories
arise from open strings in the presence of a constant $B$-field. There exist 
several approaches 
\cite{Abouelsaood:1987gd,Douglas:1998fm,Cheung:1998nr,Chu:1998qz,Chu:1999gi,Schomerus:1999ug,Ardalan:1998ks,Ardalan:1998ce,Ardalan:1999av} 
to this problem but we will mainly follow the argumentation of
\cite{Seiberg:1999vs}.\footnote{See also the introduction in
  \cite{Torrielli:2003sf}.} 
First we will discuss the $\sigma$-model description and
then later on we will focus on the low energy effective action. 
Besides giving insights into the mechanism of how field theories can be 
derived from string theories, the discussion leads us to an 
essential ingredient for our own analysis:
the Seiberg-Witten (SW) map, which provides a translation between ordinary 
and noncommutative gauge theory quantities.

\section{The low energy limit of string theory with constant background
  $B$-field}
\subsection{$\mathbf{\sigma}$-model description}
The action of a bosonic string which propagates in a background consisting of 
a spacetime metric $g_{MN}$, an antisymmetric
field $B_{MN}$ and a dilaton $\phi$  can be written as
\begin{equation}\label{genstringaction}
S=\frac{1}{4\pi\alpha'}\int_\Sigma\de^2\sigma\,\Big[\big(\delta^{ab}g_{MN}(X)-2\pi
i\alpha'\epsilon^{ab}B_{MN}(X)\big)\partial_aX^M\partial_bX^N+\alpha'\mathcal{R}\phi(X)\Big]\pnt
\end{equation}   
Here the string sweeps out a worldsheet $\Sigma$ with 
scalar curvature $\mathcal{R}$
and we have chosen conformal gauge and work with Euclidean signature such that
the worldsheet metric is 
$\delta_{ab}$. Furthermore, we assume a $(D=26)$-dimensional target space
which  
leads to a critical string theory where the worldsheet metric remains 
non dynamical in conformal gauge after 
quantization.\footnote{We of course choose 
string backgrounds which preserve the conformal invariance.} 
If the theory describes open strings we may add additional terms such as
\begin{equation}
-i\int_{\partial\Sigma}\de t\,A_M(X)\partial_\parallel
 X^M+\frac{1}{2\pi}\int_{\partial\Sigma}\de t\,k\,\phi(X)
\end{equation}  
to the action \eqref{genstringaction}, which couple a background gauge 
field $A_M$ and the dilaton $\phi$ to
the boundary $\partial\Sigma$ of the worldsheet $\Sigma$.
Here $\partial_\parallel$ denotes the tangential derivative along the
worldsheet boundary and $k$ is the geodesic curvature of the boundary. 
If the $2$-form $B$ fulfills $\de B=0$, which includes the case of
a constant $B_{MN}$, and if the dilaton $\phi$ is constant, the action for 
open strings can be cast into the form
\begin{equation}\label{ostringaction}
S=\frac{1}{4\pi\alpha'}\int_\Sigma\de^2\sigma\,
g_{MN}\partial_aX^M\partial^aX^N-\frac{i}{2}\int_{\partial\Sigma}\de
t\,\Big(B_{MN}X^M+2A_N\Big)\partial_\parallel X^N+\phi\chi\col
\end{equation} 
where $\chi$ denotes the Euler number of the worldsheet $\Sigma$. 
The second integral shows that one can alternatively describe the constant 
$B$-field by a gauge field $A_M=-\frac{1}{2}B_{MN}X^N$. The field 
strength derived from it is $F_{MN}=B_{MN}$ with the usual definition
\begin{equation}\label{U1fieldstrength}
F_{MN}=\partial_MA_N-\partial_NA_M\pnt
\end{equation} 

Up to now the endpoints of the open strings can move unconstrained in the
spacetime, the string obeys Neumann boundary conditions. 
In the framework of $\text{D}p$-branes this setup corresponds
to the case of a spacetime filling $\text{D}(D-1)$-brane. 
If instead we impose Neumann boundary conditions in $p+1$ dimensions
and Dirichlet boundary conditions in the remaining $D-p-1$ dimensions, this
defines the string endpoints to lie on a
$\text{D}p$-brane. This means that they can move freely in $p+1$ spacetime
directions and are stuck at fixed positions in the remaining $D-p-1$ spatial
dimensions.
We split the coordinate indices in the following way
\begin{equation}\label{coordsplit}
M,N=0,\dots,D\col\qquad\mu,\nu=0,\dots,p\col\qquad m,n=p+1,\dots,D\col
\end{equation}
such that capital Latin indices $M,N,\dots$ run over all spacetime directions
whereas lower case Greek ($\mu,\nu,\dots$) and Latin ($m,n,\dots$) indices 
denote directions respectively parallel and perpendicular to the
$\text{D}p$-brane. The boundary of the string worldsheet lies 
completely on
the brane. This means that the coordinates $X^m|_{\partial\Sigma}$ are constant and thus $\partial_\parallel X^m|_{\partial\Sigma}=0$. 
Then only the components
$B_{\mu\nu}$ and $A_\nu$ along the brane contribute to the boundary term of
\eqref{ostringaction} and we can set all other  
components to zero without loss of generality. 
To simplify the 
analysis we will now in addition assume that the field $B_{\mu\nu}$ on the
$\text{D}p$-brane has maximum rank and that the metric $g_{MN}$ 
is independent of $X^M$ and of block-diagonal 
form with respect to the coordinate split \eqref{coordsplit}. The background
fields then read
\begin{equation}\label{NCbackground}
g_{MN}=\begin{pmatrix} g_{\mu\nu} & 0 \\ 0 & g_{mn} \end{pmatrix}\col\qquad
B_{MN}=\begin{pmatrix} B_{\mu\nu} & 0 \\ 0 & 0 \end{pmatrix}\pnt
\end{equation}
In this setup the boundary terms in \eqref{ostringaction} (with $A_N=0$) 
modify the boundary conditions of the open strings. One finds
\begin{equation}
\begin{aligned}\label{BCgeneral}
X^m\big|_{\partial\Sigma}&=0\col\\
\big(g_{\mu\nu}\partial_\perp+2\pi i\alpha'B_{\mu\nu}\partial_\parallel \big)X^\nu\big|_{\partial\Sigma}&=0
\end{aligned}
\end{equation}
for the directions perpendicular and parallel to the $\text{D}p$-brane 
respectively. $\partial_\perp$ denotes the derivative normal to the boundary 
of the worldsheet. It is important to mention that the equations of motion
are not affected by the addition of the boundary terms in
\eqref{ostringaction}.
To analyze the theory with the boundary conditions \eqref{BCgeneral} it is
advantageous to choose coordinates $(\tau,\sigma)$ in which the boundary
of the open string worldsheet is parameterized by $\tau$ and is located at 
constant $\sigma=0,\pi$. 
The string worldsheet then describes a strip in the complex plane of 
$w=\sigma-i\tau$ which can be conformally mapped to the upper half plane 
via the holomorphic transformation $z=\e^{iw}=\e^{\tau+i\sigma}$. The 
origin of the half plane corresponds to $\tau\to-\infty$ and 
half circles around the origin refer to constant $\tau$, see
Fig.\ \ref{fig:complexzplane}. %
\newlength{\arlength}
\newlength{\arheight}
\newlength{\ardepth}
\newlength{\shift}
\newcommand{\rad}{|z|=\e^\tau}
{\setlength{\fboxsep}{0pt}
\setlength{\fboxrule}{0pt}
\settowidth{\arlength}{\fbox{$\rad$}}
\settoheight{\arheight}{\fbox{$\rad$}}
\settodepth{\ardepth}{\fbox{$\rad$}}
\addtolength{\arheight}{\ardepth}
\setlength{\shift}{0.5\arheight}
\addtolength{\shift}{0.2ex}
\addtolength{\shift}{-\ardepth}
\begin{figure}[t]
\begin{center}
\begin{picture}(300,160)(0,0)
\SetWidth{1}
\SetOffset(150,25)
\SetScale{1}
\LongArrow(-110,0)(110,0)\Text(115,0)[lc]{$\Re z$}
\Text(0,-10)[ct]{$0$}
\LongArrow(0,-5)(0,110)\Text(0,115)[cb]{$\Im z$}
\CArc(0,0)(100,0,180)
\LongArrow(0,0)(70,70)\Text(30,40)[]{\begin{rotate}{45}\raisebox{-0.5\shift}[\arheight][\arlength]{\hspace{-0.5\arlength}\fbox{$\rad$}}\end{rotate}}
\ArrowArc(0,0)(20,0,45)\Text(11,5)[]{$\sigma$}
\end{picture}
\caption{Description of the string worldsheet with the complex coordinate
  $z$. The worldsheet of the open string is given by the upper half plane with
  $\Im z\ge 0$ and its boundary is the real axis. 
A slice of constant worldsheet `time' $\tau$ is given by a half
  circle with radius $|z|=\e^\tau$. The infinite past and future are given by
  the origin and the circle with infinite radius respectively.}
\label{fig:complexzplane}
\end{center}
\end{figure}
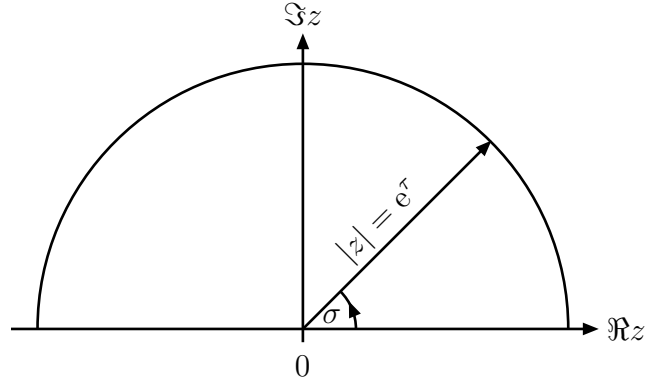%

In these coordinates the boundary of the worldsheet is the real axis where
$z=\bar z$ and the derivatives $\partial_\tau$ and $\partial_\sigma$
are given by
\begin{equation}
\partial_\tau=z\partial+\bar
z\bar\partial\col\qquad\partial_\sigma=iz\partial-i\bar
z\bar\partial\col\qquad\partial=\preparderiv{z}\col\qquad\bar\partial=\preparderiv{\bar
  z}\pnt
\end{equation}   
Using \eqref{NCbackground}, the boundary conditions \eqref{BCgeneral} then 
read in $(z,\bar z)$ coordinates
\begin{equation}
\begin{aligned}\label{BCzbarz}
X^m\big|_{z=\bar z}&=0\col\\
\Big(g_{\mu\nu}(\partial-\bar\partial)+2\pi\alpha'B_{\mu\nu}(\partial+\bar\partial)\Big)X^\mu\big|_{z=\bar
  z}&=0\pnt
\end{aligned}
\end{equation}
Since the boundary terms in the action \eqref{ostringaction} do
not modify the equations of motion, the propagator is standard for
the directions perpendicular to the $\text{D}p$-brane. The complete solution
with the boundary conditions \eqref{BCzbarz} in the
constant background \eqref{NCbackground} is given by
\cite{Abouelsaood:1987gd,Seiberg:1999vs} 
\begin{equation}\label{ostringprop}
\begin{aligned}
\big\langle
X^\mu(z)X^\nu(z')\big\rangle&=-\alpha'\Big[g^{\mu\nu}\ln\frac{|z-z'|}{|z-\bar
z'|}+G^{\mu\nu}\ln|z-\bar z'|^2+\frac{1}{2\pi\alpha'}\theta^{\mu\nu}\ln\frac{z-\bar
  z'}{\bar z-z'}+D^{\mu\nu} \Big]\col \\
\big\langle
X^\mu(z)X^n(z')\big\rangle&=0\col \\
\big\langle
X^m(z)X^n(z')\big\rangle&=-\alpha'g^{mn}\ln|z-z'|\col 
\end{aligned}
\end{equation} 
where $D^{\mu\nu}$ is a constant and we have used the abbreviations
\begin{equation}\label{effgtheta}
\begin{aligned}
G_{MN}&=g_{MN}-(2\pi\alpha')^2\big(B\frac{1}{g}B\big)_{MN}\col\\
G^{MN}&=\Big(\frac{1}{g+2\pi\alpha'B}\Big)^{MN}_\text{S}=\Big(\frac{1}{g+2\pi\alpha'B}g\frac{1}{g-2\pi\alpha'B}\Big)^{MN}\col\\
\theta^{MN}&=2\pi\alpha'\Big(\frac{1}{g+2\pi\alpha'B}\Big)^{MN}_\text{A}=-(2\pi\alpha')^2\Big(\frac{1}{g+2\pi\alpha'B}B\frac{1}{g-2\pi\alpha'B}\Big)^{MN}\pnt
\end{aligned}
\end{equation}
They are related via 
\begin{equation}\label{effgthetarel}
G^{MN}+\frac{\theta^{MN}}{2\pi\alpha'}=\Big(\frac{1}{g+2\pi\alpha'B}\Big)^{MN}\col
\end{equation}
which is trivially fulfilled for the directions
$(M,N)=\{(m,\nu),(\mu,n),(m,n)\}$, where $B_{MN}=0$, $\theta^{MN}=0$. 

Due to the third term in the first line of \eqref{ostringprop}, the propagator
is single-valued if the branch cut of the logarithm lies in the lower half
plane. As a consistency check for $(M,N)=(\mu,\nu)$ one recovers the
propagator on the disk with Neumann boundary conditions
\cite{Polchinski:1998rq} for
$B_{\mu\nu}=0$.

The quantities given in \eqref{ostringprop} and \eqref{effgtheta} have a
simple interpretation. 
In conformal field theories there exists a map between 
asymptotic states of incoming
and outgoing fields and operators which are called vertex operators. 
If one wants to
compute string theoretical scattering amplitudes 
one has to insert the vertex operators into 
the two dimensional surface. The topology of the latter determines
the order in the string coupling constant and the type of vertex operators 
that can couple. Closed string vertex operators couple to all
surfaces, inserting them  at points in the interior. However, 
open string vertex require that the surface possesses a boundary where they 
have to be inserted.
The short distance singularity of two vertex operators that approach each 
other can either be read off from the propagator or from their operator product
expansion. The anomalous dimensions of the vertex operators determine
the short distance behaviour in the operator product expansion.
Hence, one concludes that the singularity of the propagator \eqref{ostringprop}
if two interior points coincide determines the anomalous dimensions of closed
string vertex operators.
From \eqref{ostringprop} one finds in this case that for $(M,N)=(\mu,\nu)$ the
only singular term is the numerator of the first logarithmic term which is
similar to the term for $(M,N)=(m,n)$. 
The short-distance behaviour is
\begin{equation}
\big\langle X^M(z)X^N(z')\big\rangle\sim-\alpha'g^{MN}\ln|z-z'|
\end{equation}
and its coefficient enters the expressions for the anomalous 
dimensions of the closed string vertex operators. Thus, $g_{MN}$ is the metric
seen by closed strings.

On the other hand since open strings couple to the disk by inserting the
corresponding vertex operators into the boundary of the disk, their anomalous
dimensions are determined by the short distance singularity of 
\eqref{ostringprop}
for both points at the boundary, where $z=\bar z=s$. One finds
\begin{equation}\label{ostringboundprop}
\big\langle
X^\mu(s)X^\nu(s')\big\rangle=-\alpha'G^{\mu\nu}\ln(s-s')^2+\frac{i}{2}\theta^{\mu\nu}\epsilon(s-s')\col
\end{equation}
with an appropriately chosen $D^{\mu\nu}=-\frac{i}{2\alpha'}\theta^{\mu\nu}$ and with
\begin{equation}\label{signfuncdef}
\epsilon(s)=\begin{cases}-1 & s<0 \\ 1 & s>0\end{cases}\pnt
\end{equation}
Open string vertex operators see the metric $G_{\mu\nu}$. We will therefore 
denote $G_{\mu\nu}$ as the open string metric. 

We want to show that 
the sign function $\epsilon(s)$ in \eqref{ostringboundprop} is responsible for
the  non vanishing of the commutator of the two fields $X^\mu$, $X^\nu$ at the
same boundary point. Remember first that according to the 
previous discussion around fig.\ \ref{fig:complexzplane}, the radial coordinate
of the half plane refers to the worldsheet `time'. `Time' ordering therefore
translates to radial ordering on the complex $z$ plane.
The equal time commutator of two fields is defined as the difference of two
limits of the time ordered product of these fields. The first [second] limit
is to let the time coordinate of the second field approach the time coordinate 
of the first one from below [above]. The translation to radial ordering
(denoted by $\operatorname{R}$) is
obvious and one obtains 
using \eqref{ostringboundprop}
\begin{equation}\label{XXcomm}
\big\langle\big[X^\mu(s),X^\nu(s)\big]\big\rangle=\big\langle\lim_{\varepsilon\to 0}\operatorname{R}\big(X^\mu(s)X^\nu(s-\varepsilon)-X^\mu(s)X^\nu(s+\varepsilon)\big)\big\rangle=\frac{i}{2}\big(\theta^{\mu\nu}-\theta^{\nu\mu}\big)=i\theta^{\mu\nu}\pnt
\end{equation}
The parameter $\theta^{\mu\nu}$ can be interpreted as the noncommutativity
parameter in a space where the embedding coordinates on the $\text{D}p$-brane
describe the noncommutative coordinates. 

We will now determine the effect of the additional terms in
\eqref{ostringaction} on string scattering amplitudes. As we have mentioned
above, an element of the string 
S-matrix is a correlator of vertex operators that describe the 
asymptotic states. The correlators (at fixed order in the string coupling) 
are defined as a path integral over all fields $X^M$ and metrics 
of the $2$-dimensional surface of particular topology with vertex operators 
inserted and with the action \eqref{ostringaction} (with $A_N=0$). 
An appropriate gauge
fixing procedure is also understood. 
Consider first the simplest vertex operator for an open string tachyon with
momentum $p$ 
which is given by $\no\e^{ip_\mu X^\mu}\no$. 
Here $:$ denotes normal ordering and indices are raised and lowered with 
the metric in \eqref{effgtheta}.
Using \eqref{ostringboundprop}, the operator
product of two open string tachyon vertex operators for $s>s'$ is given by
\begin{equation}\label{tVOprod}
\no\e^{ip\cdot X}(s)\no\no\e^{iq\cdot X}(s')\no\sim(s-s')^{2\alpha'p\cdot
  q}\e^{-\frac{i}{2}\theta^{\mu\nu}p_\mu q_\nu}\no\e^{i(p+q)\cdot X}(s')\no\col
\end{equation} 
where `$\sim$' denotes that we have only kept the most singular terms and we
have defined $p\cdot q=G^{\mu\nu}p_\mu q_\nu$. 
One can capture the complete $\theta$-dependence on the R.\ H.\ S.\ 
with the Moyal-Weyl $\ast$-product \cite{Moyal:1949sk} which is defined as 
\begin{equation}\label{astproddef}
f(x)\ast g(x)=\e^{\frac{i}{2}\theta^{\mu\nu}\preparderiv{\xi^\mu}\preparderiv{\eta^\nu}}f(x+\xi)g(x+\eta)\Big|_{\xi=\eta=0}
\end{equation}
and which is a special example of a noncommutative $\ast$-product, see
Appendix \eqref{app:NCastprod} for some details.
One especially finds for this product
\begin{equation}
x^\mu\ast x^\nu-x^\nu\ast x^\mu=i\theta^{\mu\nu}
\end{equation}
in accord with \eqref{XXcomm}.
The product of two tachyon vertex operators \eqref{tVOprod} can then be
written as
\begin{equation}\label{tVOprodast}
\no\e^{ip\cdot X}(s)\no\ast\no\e^{iq\cdot X}(s')\no\Big|_{\left\langle
X^\mu(s)X^\nu(s')\right\rangle_{\theta=0}}=\no\e^{ip\cdot X}(s)\no\no
\e^{iq\cdot X}(s')\no\pnt  
\end{equation}
 The above result means
that the normal ordering of two tachyon vertex
operators can either be performed by using \eqref{ostringboundprop} for
contractions or alternatively by replacing the 
ordinary product with the $\ast$-product and
contracting with the two point function \eqref{ostringboundprop} without 
the $\theta$-dependent term. 
Both procedures capture the entire $\theta$-dependence. 

This discussion can be generalized to products of arbitrary open string vertex
operators.  A generic open string vertex operator is given by 
a polynomial $P$ that depends on derivatives of the $X$ and an exponential
function to ensure the right behaviour under translations
\begin{equation}
V(p,s)=\no P[\partial X,\partial^2X,\dots]\e^{ip\cdot X}(s)\no\pnt 
\end{equation}
If one now normal orders products of these vertex operators, one finds that
only the exponential factors generate a dependence on $\theta^{\mu\nu}$ 
in contrast to the contractions which include at least one field of the
polynomial prefactors. The reason for
this is that in the two point function \eqref{ostringboundprop} the
$\theta$-dependent term can be disregarded if derivatives are taken and 
an appropriate regularization is used (like point splitting regularization in 
Subsection \ref{subsec:LEEA}).  
Therefore, one obtains the same $\theta$-dependent exponential factor 
as if one had used tachyon vertex operators. 
It is now simple to see the $\theta$-dependence of string amplitudes.
One has to insert the vertex operators into the boundary of the
string worldsheet, and one has to 
integrate over the insertion points.\footnote{Note that the gauge fixing
  procedure does not fix the worldsheet metric completely. Depending on the
  topology of the worldsheet, a remnant, the conformal Killing group (CKG), 
  has to be fixed by inserting some vertex operators at fixed points without
  performing an integration. 
  For the disk, three vertex operator positions have to be fixed.}
In case of an $n$-point amplitude with vertex operators $V_k$, $k=1,\dots,n$
one obtains 
\begin{equation}\label{VOcorrthetadependence}
\Big\langle\prod_{k=1}^nV_k(p_k,s_k)\Big\rangle_{G,\theta}=\exp\Big\{-\frac{i}{2}\sum_{k>l}\theta^{\mu\nu}(p_k)_\mu(p_l)_\nu\epsilon(s_k-s_l)\Big\}\Big\langle\prod_{k=1}^nV_k(p_k,s_k)\Big\rangle_{G,\theta=0}\pnt
\end{equation}
The subscript of a correlator denotes which parameters the correlator 
has to be computed with.
The complete $\theta$-dependence is thus described by the exponential
prefactor if the theory is formulated in terms of the open string metric
$G_{MN}$. Due to momentum conservation the prefactor is invariant under 
cyclic permutations of the $p_k$.
Performing the integrations of the above expression over 
(some of) the insertion points $s_k$ then produces contributions with
different phase factors  
if the vertex operators exchange their positions in a non cyclic way.  
The appearance of the prefactor can be exactly described by the
$\ast$-product \eqref{astproddef} of the corresponding $n$ fields, and one
then rewrites \eqref{VOcorrthetadependence} as 
\begin{equation}\label{VOcorr}
\Big\langle V_1(p_1,s_1)\cdots
V_n(p_n,s_n)\Big\rangle_{G,\theta}=\Big\langle
V_1(p_1,s_1)\ast\cdots\ast V_n(p_n,s_n)\Big\rangle_{G,\theta=0}\pnt
\end{equation}
The above expression is an important result for the
discussion of the low energy effective theory. 

\subsection{The Seiberg-Witten limit}
In order to find the low energy description of the theory with action
\eqref{ostringaction} one has to get rid of stringy effects in
the correlation functions \eqref{VOcorr}. 
It is clear that in an appropriate limit one should send to zero the
parameter $\alpha'$, which is proportional to the square of the string length. 
In this case, where one wants to keep the effects of a constant
$B$-field, one cannot simply keep constant the other parameters. 
As can be seen from \eqref{effgtheta}, the $B$-dependence and
especially $\theta^{\mu\nu}$ vanishes and one finds the same theory without
initial $B$-field, if one keeps constant the other
parameters in the limit.
One should instead keep the open sting metric $G_{\mu\nu}$ and
$\theta^{\mu\nu}$ finite and different from zero. This is natural
because the correlators \eqref{VOcorr} which will be discussed in the limit 
depend on these quantities and on $\alpha'$.  
In the Seiberg-Witten limit some components of the closed string metric
$g_{\mu\nu}$ and $\alpha'$ are sent to zero at constant $B$-field in the following way
\begin{equation}\label{SWlimit}
\alpha'\sim\sqrt{\epsilon}\to 0\col\qquad g_{\mu\nu}\sim\epsilon\to 0\col\qquad
g_{mn}=\text{const.}\col\qquad B_{\mu\nu}=\text{const.}
\end{equation}
This then ensures that $G_{MN}$ and $\theta^{MN}$ have reasonable
limits. The expressions \eqref{effgtheta} become 
\begin{equation}
\begin{aligned}
G_{MN}&=\begin{cases}-(2\pi\alpha')^2\big(Bg^{-1}B\big)_{\mu\nu} & \\ g_{mn} & \end{cases}\col\\
G^{MN}&=\begin{cases}-\frac{1}{(2\pi\alpha')^2}\big(B^{-1}gB^{-1}\big)^{\mu\nu}
  & \\ g^{mn} & \end{cases}\col\\
\theta^{MN}&=\begin{cases}\big(B^{-1}\big)^{\mu\nu} & \\ 0 & \end{cases}\col
\end{aligned}
\end{equation}
The propagator \eqref{ostringboundprop} at the boundary reads
\begin{equation}\label{SWlimostringprop}
\big\langle
X^\mu(s)X^\nu(s')\big\rangle=\frac{i}{2}\theta^{\mu\nu}\epsilon(s-s')
\end{equation}
in the Seiberg-Witten limit. Thus, from \eqref{tVOprod} and \eqref{tVOprodast}
it can be seen that the only remnant of the $\theta$-dependence is  
the exponential prefactor that is described by the $\ast$-product 
\eqref{astproddef}. 
Instead of taking the Seiberg-Witten limit one can equivalently  
send $B_{\mu\nu}\to\infty$ without scaling the 
metric $g_{\mu\nu}$ and taking the $\alpha\to0$ limit. 

The interpretation of the Seiberg-Witten limit is
as follows. Sending $\alpha'\to 0$ is an infinite tension limit. However,
instead of leading to a point particle, the rescaling of the metric
$g_{\mu\nu}$ leads to the improvement 
of the resolution such that a remnant of the 
1-dimensionality of the string survives. The different positions of the two 
endpoints of the string
remain observable. Simultaneously, as the tension runs to infinity, the 
string is made rigid (excitations in the form of massive oscillator modes are
removed from the spectrum). It can be seen as a rigid rod, possessing two 
distinct endpoints but no further internal structure. In
\cite{Sheikh-Jabbari:1999vm} it was shown that the two endpoints of the
string on the $\text{D}p$-brane are separated if a $B$-field is
present. They can be described by a dipole
\cite{Sheikh-Jabbari:1999vm,Susskind:2001fb}. In the Seiberg-Witten limit, the 
boundary degrees of freedom (the endpoints of the dipole) are governed by the 
action
\begin{equation}\label{Baction}
-\frac{i}{2}\int_{\partial\Sigma}\de
t\, B_{\mu\nu}X^\mu\partial_\parallel X^\nu\pnt
\end{equation}  
This can be regarded as the action of charged particles with 
the world lines $X^i$ which move in a strong magnetic field $B_{\mu\nu}$
\cite{Magro:2003bs}. 

\subsection{The low energy effective action}
\label{subsec:LEEA}
The next step is to determine the effective low energy description of 
the open strings moving in the constant $B$-field.

At first let us neglect that there is a $B$-field present and discuss
the general procedure. The part of the 
low energy effective action (LEEA) for the open strings is obtained in the
following way: compute the correlation functions of the vertex operators 
for the massless open string states and expand in powers of $\alpha'$.
The coefficient at a certain order in $\alpha'$ of the one-particle
irreducible piece of one of the amplitudes
then enters the effective action at this particular
order in $\alpha'$. It describes the coupling of the fields that 
correspond to the vertex operators in this string amplitude.
In this way one finds that the leading non constant term in the effective
action describes a $U(1)$ gauge theory.  
Furthermore, it turns out \cite{Fradkin:1985qd} 
that the LEEA with full dependence 
on $\alpha'$ is
the Dirac-Born-Infeld (DBI) action as long as the field strength
\eqref{U1fieldstrength} fulfills
$\sqrt{2\pi\alpha'}\big|\frac{\partial F}{F}\big|\ll1$.
Its Lagrangian in presence of a background $B$-field with $B_{\mu n}=B_{mn}=0$
and with a gauge field $a_\mu$ on the $\text{D}p$-brane is given by
\begin{equation}\label{DBIlag1}
\mathcal{L}[g,B,a]=T_p\sqrt{\det\big[g^\text{pb}_{\mu\nu}+2\pi\alpha'(B_{\mu\nu}+f_{\mu\nu})\big]}\col\qquad
f_{\mu\nu}=\partial_\mu a_\nu-\partial_\nu a_\mu\col
\end{equation}
where we have used lower case notation for the gauge field and the field
strength for reasons that will become clear later. $T_p$ denotes the 
$\text{D}p$-brane tension at zero
$B$-field 
\begin{equation}\label{stringtension}
T_p=\frac{1}{g_\text{s}(2\pi)^p(\alpha')^\frac{p+1}{2}}\col
\end{equation}
where $g_\text{s}$ denotes the closed string coupling constant.
The pullback metric is defined in terms of the $D-p-1$ scalars
$2\pi\alpha'\phi^m=X^m$ as
\begin{equation}\label{pbmetric}
g^\text{pb}_{\mu\nu}=g_{MN}\partial_\mu X^M\partial_\nu X^N=g_{\mu\nu}+(2\pi\alpha')^2g_{mn}\partial_\mu\phi^m\partial_\nu\phi^m
\end{equation}
and the last equality follows in static gauge where $X^\mu=\xi^\mu$ and for a
block diagonal metric, see \eqref{NCbackground}. For our purpose the scalars
play no role and so we will neglect them in the following.  

Although we have started the discussion without $B$-field on the 
$\text{D}p$-brane, we have 
introduced it in \eqref{DBIlag1}. That this is a consistent generalization 
can be seen from the symmetries 
of the underlying $\sigma$-model. Since derivatives of $B_{\mu\nu}$ and
$f_{\mu\nu}$ have been neglected in the DBI action, we can regard them as 
constant and thus the DBI action should possess the same symmetries as the
underlying $\sigma$-model description \eqref{ostringaction}. The latter is
invariant under the transformation 
\begin{equation}\label{Batrafo}
B\to B'=B+\de\omega\col\qquad a\to a'=a-\omega\col\qquad \omega\in\Omega^1\col
\end{equation}
where $\omega$ is a one-form. This symmetry is respected by
$(B+f)_{\mu\nu}=(B'+f')_{\mu\nu}$ where $f'_{\mu\nu}$ has to be computed with $a'_\mu$.
We have therefore found one low energy description of the $\sigma$-model 
\eqref{ostringaction} with constant $B$-field. 
However, one has to be careful if one wants to expand \eqref{DBIlag1}
in powers of $\alpha'$ to get the Seiberg-Witten limit
\eqref{SWlimit} because the open string metric is not
kept constant.
As we will now see, there exists a second description in terms of the open
string metric and different fields where it is easier to take the
Seiberg-Witten limit. 

We remember that the correlation functions \eqref{VOcorr} which are formulated
with the open string metric $G_{\mu\nu}$ have a simple 
dependence on $\theta^{\mu\nu}$ and there is no explicit 
$B$-dependence left.
The Seiberg-Witten limit \eqref{SWlimit} in
this formulation is the ordinary $\alpha'\to0$ limit.  
For $\theta^{\mu\nu}=0$ the DBI Lagrangian would then simply be given
by \eqref{DBIlag1} but with $g_{MN}$ replaced by $G_{MN}$ and with
$B_{\mu\nu}=0$. In addition one should rename the gauge
field
to stress that they are different from the ones in the other formulation. 
Let us first give a more intuitive argument of how a nontrivial 
$\theta^{\mu\nu}$ modifies the description.  
We have already seen that the product between the vertex operators is simply
replaced by the $\ast$-product \eqref{astproddef} to describe the dependence
on $\theta^{\mu\nu}$. This dependence has to
be reproduced by the LEEA. Moreover one can check that the field theory limit
of the correlator of three gauge 
fields\footnote{See e.\ g.\ \cite{Polchinski:1998rq}.} 
does not vanish anymore
if $\ast$-products occur between them. 
We find self interactions of the gauge
fields and therefore the $U(1)$ gauge theory becomes non-Abelian. 
This motivates that the recipe to construct the
DBI Lagrangian for non vanishing $\theta^{\mu\nu}$ is to replace the field strength 
$f_{\mu\nu}$ by a non-Abelian version and to replace all products of fields by
the $\ast$-product.
The DBI Lagrangian should thus be given by\footnote{In principle one should
  evaluate the determinant using $\ast$-products. Effectively, however, this
  does not make a difference here because in the DBI action terms of order
  $\mathcal{O}(\partial F)$ are neglected. The difference caused by
  choosing the ordinary product instead of the $\ast$-product is exactly of
  this form. Of course one must not neglect the $\ast$-product between the 
 gauge fields.} 
\begin{equation}\label{DBIlag2}
\mathcal{L}[G,B,A]=\frac{g_\text{s}}{G_\text{o}^2}T_p\sqrt{\det\big[G^\text{pb}_{\mu\nu}+2\pi\alpha'F_{\mu\nu}\big]}\col\qquad
F_{\mu\nu}=\partial_\mu A_\nu-\partial_\nu A_\mu-iA_\mu\ast A_\nu+iA_\nu\ast A_\mu\col
\end{equation}
where the first fraction replaces the string coupling $g_\text{s}$ inside the 
tension by $G_\text{o}$, the open string coupling%
, which has to be determined later and which
should be appropriately rescaled in the $\alpha'\to0$ limit. In the infinite 
tension limit $\alpha'\to0$ one finds with 
the relation \eqref{detexpansion} that 
\eqref{DBIlag2} describes a noncommutative $U(1)$ gauge theory with
action\footnote{Here we have written the $\ast$-product between the $F_{\mu\nu}$.
The expression is exactly the same as if we had taken the ordinary product
because of the property \eqref{optracebilin}. But even if both actions were 
not exactly the same they would
only differ by terms $\mathcal{O}(\partial F)$ that have been neglected
anyway.}
\begin{equation}\label{NCYMlag}
\frac{(\alpha')^{\frac{3-p}{2}}}{4(2\pi)^{p-2}G_\text{o}^2}\int\de^{p+1}\xi\,\sqrt{G}G^{\mu\mu'}G^{\nu\nu'}F_{\mu\nu}\ast
  F_{\mu'\nu'}\pnt
\end{equation}
The above result has been verified by a direct computation of scattering
amplitudes of massless open strings \cite{Sheikh-Jabbari:1998ac}.
A complete derivation of the noncommutative DBI action \eqref{DBIlag2} 
along the lines 
of \cite{Fradkin:1985qd} can be found in \cite{Lee:1999cc}. 

Here we will give another argument based on the $\sigma$-model picture that
motivates the appearance of the $\ast$-product in \eqref{DBIlag2} and thus in
\eqref{NCYMlag}.
In the $\sigma$-model action \eqref{ostringaction} we already included a
background gauge field and it is easy to see that a constant $B_{\mu\nu}$ on
the $\text{D}p$-brane  
can alternatively be described by switching on a background gauge field
$A_\nu=\frac{1}{2}B_{\mu\nu}X^\mu$ with field strength
$F_{\mu\nu}=B_{\mu\nu}$. This is a consequence of the symmetry under
\eqref{Batrafo}. Absorbing the $B$-field completely into the gauge
field $A_\mu$, the part of the action that survives the Seiberg-Witten limit
\eqref{Baction} therefore becomes 
\begin{equation}
S_b=-i\int_{\partial\Sigma}\de t\,A_\mu\partial_\parallel X^\mu=-i\int\de
s\,A_\mu\partial_s X^\mu\pnt 
\end{equation}
Naively this action is invariant under the gauge transformation
\begin{equation}\label{gaugetraf}
A_\mu\to A'_\mu=A_\mu+\delta_\lambda A_\mu\col\qquad \delta_\lambda A_\mu=\partial_\mu\lambda\pnt
\end{equation} 
However,this is no longer true if we quantize the fields $X^\mu$. In this case
the product of operators has to be regularized and this can
change the naive gauge invariance.
Under the gauge transformation \eqref{gaugetraf} the integrand of the path
integral transforms as follows
\begin{equation}\label{pintintegrand}
\delta\e^{-S_b}=\e^{-S_b}i\int\de s\,\no\partial_\mu\lambda\partial_sX^\mu\no=
\sum_{n=1}^\infty\frac{i^{n+1}}{n!}\Big(\int\de s\,\no A_\mu\partial_sX^\mu\no\Big)^n\int\de s\,\no\partial_s\lambda\no\pnt
\end{equation}   
So for instance the $n=1$ term in the above sum is the product
\begin{equation}\label{n=1var}
-\int_{\partial\Sigma}\de s\,\no A_\mu\partial_sX^\mu\no\int\de s'\,\no\partial_{s'}\lambda\no
\end{equation}
which has to be regularized because the integrations along the boundary lead
to divergencies at $s=s'$. Using point
splitting regularization, we cut out the region $|s-s'|<\varepsilon$ and
we obtain
\begin{equation}
\begin{aligned}
\int_{\partial\Sigma}\de s\,\no A_\mu\partial_sX^\mu\no\int\de s'\,\no\partial_{s'}\lambda\no
&=\int_{\partial\Sigma}\de
s\,\no A_\mu\partial_sX^\mu\no\Big(\int_{-\infty}^{s-\varepsilon}\de
s'+\int_{s+\varepsilon}^{\infty}\de
s'\Big)\no\partial_{s'}\lambda\no\\
&=\int_{\partial\Sigma}\de s\,\no A_\mu\partial_sX^\mu\no\no\big(\lambda(X(s+\varepsilon))-\lambda(X(s-\varepsilon))\big)\no
\end{aligned}
\end{equation}
for $\lambda$ sufficiently fast vanishing at $s\to\pm\infty$. If we now take 
the limit $\varepsilon\to 0$ and use the propagator \eqref{SWlimostringprop}
for contractions and the definition \eqref{astproddef} for the
$\ast$-product we find
\begin{equation}
\int\de s\,\no\big(A_\mu\ast\lambda-\lambda\ast A_\mu\big)\partial_sX^\mu\no\pnt
\end{equation}
Applying the point splitting regularization method, one therefore 
discovers that the gauge transformation has to be modified to remain a
symmetry of the quantized theory.
The new gauge transformation is
\begin{equation}\label{NCU1gaugetraf}
\hat\delta_\lambda A_\mu=\partial_\mu\lambda+i\lambda\ast A_\mu-iA_\mu\ast\lambda\pnt
\end{equation}
This expression is complete, even if the other terms in
\eqref{pintintegrand} with $n>1$ are considered \cite{Seiberg:1999vs}.

The consequences for the LEEA are now as follows. A background gauge field 
can be seen as a coherent state of open massless strings. The LEEA
should therefore be invariant under the gauge transformations which transform
the background fields. The gauge transformation \eqref{NCU1gaugetraf} 
is a non-Abelian one and the DBI Lagrangian \eqref{DBIlag2} respects this
symmetry with the non-Abelian field strength $F_{\mu\nu}$.  The same gauge
invariance is obviously present in the noncommutative $U(1)$ gauge theory 
\eqref{NCYMlag} as it is a limit of the DBI action \eqref{DBIlag2}.

In \cite{Seiberg:1999vs} the
above change in the description of the theory by either keeping the explicit 
dependence on $B_{\mu\nu}$, or by interpreting it as a boundary condition 
and thus absorbing it into the definition of $G_{MN}$ and
$\theta^{\mu\nu}$, is denoted as background independence. In fact one can
choose arbitrary steps in between and absorb only parts of the $B_{\mu\nu}$
dependence. This leads to different values for $G_{MN}$ and
$\theta^{\mu\nu}$, but the low energy descriptions are equivalent.

Besides discussing this issue 
one can ask the question in which sense the above 
procedure depends on the chosen regularization scheme. Assume in the following
that the entire $B$-dependence is dealt with in the boundary
conditions for the CFT and therefore it modifies the propagator like in
\eqref{ostringboundprop}. If we had chosen another method (like
e.\ g.\ Pauli-Villars regularization) to 
regularize the operator products in in \eqref{pintintegrand} then the standard
Abelian $U(1)$ gauge transformation \eqref{gaugetraf} would have been
preserved and this would have to be respected by the LEEA. 
We already know how the 
corresponding LEEA with all required symmetries looks like. It is the DBI
Lagrangian given in \eqref{DBIlag1}. 
In \cite{Seiberg:1999vs} the authors 
formulate a smooth interpolation between the two descriptions by introducing a 
two-form $\Phi$ such that the relation
\eqref{effgthetarel} is replaced by 
\begin{equation}\label{geneffgthetarel}
\Big(\frac{1}{G+2\pi\alpha'\Phi}\Big)^{MN}+\frac{\theta^{MN}}{2\pi\alpha'}=\Big(\frac{1}{g+2\pi\alpha'B}\Big)^{MN}
\end{equation}
and the corresponding DBI Lagrangian with the parameters that fulfill this 
relation reads 
\begin{equation}\label{DBIlaginterpol}
\mathcal{L}[G,\Phi,A^\Phi]=\frac{g_\text{s}}{(G^\Phi_o)^2}T_p\sqrt{\det\big[G^\text{pb}_{\mu\nu}+2\pi\alpha'(\Phi_{\mu\nu}+F_{\mu\nu})\big]}\pnt
\end{equation}
In particular, the $\ast$-product above has to be evaluated with
$\theta^{\mu\nu}$ that obeys \eqref{geneffgthetarel} and
the open string coupling $G^\Phi_o$ carries an index
$\Phi$ to indicate that its value depends on $\Phi$.
For the same reason the field strength 
\begin{equation}
F_{\mu\nu}=\partial_\mu A^\Phi_\nu-\partial_\nu A^\Phi_\mu-iA^\Phi_\mu\ast A^\Phi_\nu+iA^\Phi_\nu\ast A^\Phi_\mu
\end{equation}
is constructed with gauge fields $A^\Phi_\mu$ to indicate that they are 
different quantities for different $\Phi$.
The two LEEAs \eqref{DBIlag1} and \eqref{DBIlag2} follow from the above 
generalized description evaluated at $\Phi_{\mu\nu}=B_{\mu\nu}$ and $\Phi_{\mu\nu}=0$
with the identification $A^B_\mu=a_\mu$, $A^0_\mu=A_\mu$ respectively.
Remember that different values for $\Phi$ describe the
same theory but regularized in different ways. 
The value of $\Phi$ is 
connected to the parameters of a
regularization scheme that interpolates smoothly between Pauli-Villars
($\Phi_{\mu\nu}=B_{\mu\nu}$) and the
point splitting regularization ($\Phi_{\mu\nu}=0$). Schemes that show the 
existence of $\Phi$ were found in \cite{Andreev:1999pv}.
There should thus exist a map that relates the gauge fields $A^\Phi_\mu$ for
different $\Phi$. In particular for the extreme cases ($\Phi_{\mu\nu}=0,B_{\mu\nu}$) one should find
\begin{equation}
a_\mu\to A_\mu[a]=a_\mu+A'_\mu[a]
\end{equation}
that relates the field $a_\mu$ in the first ordinary 
effective description to the field $A_\mu$ in the second noncommutative one. 
This map is essential for our analysis and we will deal with it in the 
next Section.
It is very important to stress here that the relation between the two 
descriptions is essentially different from the aforementioned background field
transformations. There, the regularization scheme was always 
point splitting and the low energy description depends on the separation of 
$B_{\mu\nu}$ into background part and contribution to the boundary conditions 
of the CFT. Here, however, we have chosen a fixed separation and varied the
regularization of the underlying CFT. For further investigations concerning
the relation between the ordinary and the noncommutative descriptions see
\cite{Cornalba:1999ah,Ishibashi:1999vi}. 

At the end of this Section the open string coupling constant $G^\Phi_o$ in 
\eqref{DBIlaginterpol} will be related to the closed string
coupling constant $g_\text{s}$ of \eqref{stringtension} by using the
equivalence 
\begin{equation}\label{DBIlagequiv}
\mathcal{L}[g,B,a]=\mathcal{L}[G,\Phi,A^\Phi]+\mathcal{O}(\partial
F)+\text{total derivatives}\col 
\end{equation}
which is the statement that the effective descriptions \eqref{DBIlaginterpol}
are the same for all $\Phi_{\mu\nu}$ and are in particular related to $\Phi_{\mu\nu}=B_{\mu\nu}$. The additional total derivatives arise in the 
exact equality because the derivation of the effective actions is insensitive
to them. The corrections by derivative are caused by the fact that 
in the approximation by the DBI actions they are neglected.  
We determine the open string coupling constant $G^\Phi_o$ from this equality by
setting all dynamical fields to zero. One then finds
\begin{equation}\label{genostrincouplingrel}
(G^\Phi_o)^2=g_\text{s}\sqrt{\frac{\det(G+2\pi\alpha'\Phi)}{\det(g+2\pi\alpha'B)}}\pnt
\end{equation}
For $\Phi_{\mu\nu}=0$ this gives $G_\text{o}$, see \eqref{DBIlag2} and
\eqref{NCYMlag} 
\begin{equation}\label{ostringcouplingrel}
G_\text{o}^2=g_\text{s}\sqrt{\frac{\det G}{\det(g+2\pi\alpha'B)}}=\sqrt{\frac{\det(g-2\pi\alpha'B)}{\det g}}\col
\end{equation}
where the last equality follows with the help of \eqref{effgtheta}.
In the Seiberg-Witten limit it becomes 
\begin{equation}
G_\text{o}^2=g_\text{s}\sqrt{\det\Big(2\pi\alpha'B\frac{1}{g}\Big)}\pnt
\end{equation}
The effective Yang-Mills coupling is given by the prefactor of the $F^2$ term 
in the expansion of \eqref{DBIlag2} which is \eqref{NCYMlag} and it reads
\begin{equation}\label{NCYMcouplingrel}
\frac{1}{g^2}=\frac{(\alpha')^\frac{3-p}{2}}{(2\pi)^{p-2}G_\text{o}^2}=\frac{(\alpha')^\frac{3-p}{2}}{(2\pi)^{p-2}g_\text{s}}\sqrt{\frac{\det(g-2\pi\alpha'B)}{\det g}}\pnt
\end{equation}
As already mentioned in the discussion of \eqref{NCYMlag}, to keep $g^2$ finite in the limit $\alpha'\to 0$ we should scale the string coupling constants like
\begin{equation}
G_\text{o}\sim\epsilon^\frac{3-p}{4}\col\qquad g_\text{s}\sim\epsilon^2\pnt
\end{equation}
The corresponding expression for arbitrary rank of $B_{\mu\nu}$ can be found
in \cite{Seiberg:1999vs}.

The above presented analysis can be generalized in the 
presence of $N$ coincident
$\text{D}p$-branes. The gauge symmetry is then enhanced and again, depending 
on the chosen regularization, one finds an ordinary or a noncommutative
Yang-Mills theory in the $\alpha'\to0$ limit. The DBI action for $N$ parallel
$\text{D}p$-branes \cite{Myers:1999ps}, however, is more complicated and
only confirmed to coincide with the direct computation of amplitudes up to
fourth order in $\alpha'$.\footnote{I am grateful to S.\ Stieberger for this
   comment.} For a discussion of the non-Abelian DBI action in connection with
noncommutativity see \cite {Cornalba:2000ua}.   
  
\section{Construction of the Seiberg-Witten map}
In Subsection \ref{subsec:LEEA} we have seen that, depending on the chosen
regularization, the Seiberg-Witten limit of string theory in the presence 
of a constant background $B$-field leads to equivalent effective description
on the worldvolume of the $\text{D}p$-brane. 
In particular one finds an ordinary and a noncommutative formulation.
The unique origin of these two descriptions predicts the existence of a map
between the gauge fields $A_\mu$ and $a_\mu$ of the 
noncommutative and the ordinary YM theories.
In this Section we will now analyze this mapping in more detail. All fields
and gauge parameters will refer to the YM case.  
The first naive assumption is that the map could have the form
$A_\mu=A_\mu[a,\partial a,\partial\partial a,\dots]$,
$\Lambda=\Lambda[\lambda,\partial\lambda,\partial\partial\lambda,\dots]$.
However, it is easy to see that this cannot
be true. Remember that the ordinary and the noncommutative gauge
transformation for a $U(1)$ gauge field are given by
\begin{equation}\label{U1gaugetrafs}
\delta_\lambda
a_\mu=\partial_\mu\lambda\col\qquad\hat\delta_\Lambda A_\mu=\partial\Lambda_\mu+i\astcomm{\Lambda}{A_\mu}\col
\end{equation}
where we define the $\ast$-commutator and $\ast$-anticommutator as follows
\begin{equation}\label{astcommacommdef}
\astcomm{A}{B}=A\ast B-B\ast A\col\qquad\astacomm{A}{B}=B\ast A+B\ast A\pnt
\end{equation} 
In the case of a $U(1)$ gauge group, the $\ast$-product is simply given by the 
expression in
\eqref{astproddef}. In the non-Abelian case we will denote the tensor product
of \eqref{astproddef} with the matrix multiplication as $\ast$-product.
The ordinary gauge transformation in \eqref{U1gaugetrafs} is Abelian whereas the noncommutative gauge
transformation is a non-Abelian one. A simple redefinition of the gauge
parameter $\lambda\to\Lambda=\Lambda[\lambda,\partial
\lambda,\partial\partial\lambda,\dots]$ can never change the Abelian 
to a non-Abelian gauge group. The only requirement on the map is that gauge 
equivalent configurations in the ordinary theory are mapped to 
gauge equivalent configurations in the noncommutative theory. 
This led Seiberg and Witten to the ansatz that the 
gauge parameter of one theory depends on both, the gauge parameter and 
the gauge field of the other theory \cite{Seiberg:1999vs}. In short we will
write 
\begin{equation}\label{SWmapansatz}
A=A[a,\partial
a,\partial\partial a,\dots]=A[a]\col\qquad\Lambda=\Lambda[\lambda,\partial
\lambda,\partial\partial\lambda,\dots,a,\partial
a,\partial\partial a,\dots]=\Lambda[\lambda,a]\pnt
\end{equation}
The requirement that gauge orbits are mapped to gauge orbits reads for 
infinitesimal transformations
\begin{equation}\label{gaugeorbitmap}
A_\mu[a]+\hat\delta_{\Lambda}A_\mu[a]=A_\mu[a+\delta_\lambda a]\pnt
\end{equation}  
We will now evaluate this expression to first order in the noncommutativity
parameter $\theta^{\mu\nu}$.
One inserts the definitions for infinitesimal 
ordinary and noncommutative gauge transformations of the non-Abelian 
gauge fields
\begin{equation}\label{YMgaugetrafs}
\delta a_\mu=\partial\lambda+i\comm{\lambda}{a_\mu}\col\qquad\hat\delta_{\Lambda}A_\mu=\partial_\mu\Lambda+i\astcomm{\Lambda}{A_\mu}
\end{equation}
into \eqref{gaugeorbitmap}. Then one expands the $\ast$-product
\eqref{astproddef} and 
the maps in powers of $\theta^{\mu\nu}$
\begin{equation}
A[a]=a+A'[a]\col\qquad\Lambda[\lambda,a]=\lambda+\Lambda'[\lambda,a]\col
\end{equation}
where $A'[a]$ and $\Lambda'[\lambda,a]$ depend linearly on 
$\theta^{\mu\nu}$. In this way one finds from 
\eqref{gaugeorbitmap} the
expression
\begin{equation}\label{gaugeorbitmaplintheta}
A'_\mu[a+\delta_\lambda
a]-A'_\mu[a]-i\comm{\lambda}{A'_\mu[a]}-i\comm{\Lambda'[\lambda,a]}{a_\mu}=-\frac{1}{2}\theta^{\alpha\beta}\big(\partial_\alpha\lambda\partial_\beta
a_\mu+\partial_\alpha a_\mu\partial_\beta\lambda\big)\col
\end{equation}
where $\comm{\dummylen}{\dummylen}$ denotes the commutator with
ordinary matrix product.
The above equation is solved by
\begin{equation}
A'_\mu(a)=-\frac{1}{4}\theta^{\alpha\beta}\acomm{a_\alpha}{\partial_\beta a_\mu+f_{\beta\mu}}\col\qquad\Lambda'(\lambda,a)=\frac{1}{4}\theta^{\alpha\beta}\acomm{\partial_\beta\lambda}{a_\alpha}
\end{equation}
and hence the Seiberg-Witten map at order $\mathcal{O}(\theta)$ reads
\cite{Seiberg:1999vs}
\begin{equation}\label{SWmaplin}
\begin{aligned}
A_\mu(a)&=a_\mu-\frac{1}{4}\theta^{\alpha\beta}\acomm{a_\alpha}{\partial_\beta
  a_\mu+f_{\beta\mu}}\col\\
\Lambda(\lambda,a)&=\lambda+\frac{1}{4}\theta^{\alpha\beta}\acomm{\partial_\beta\lambda}{a_\alpha}\col
\end{aligned}
\end{equation}
where $\acomm{\dummylen}{\dummylen}$ denotes the ordinary matrix
anticommutator and $f_{\mu\nu}$ the ordinary YM field strength
\begin{equation}\label{ordinaryYMfieldstrength}
f_{\mu\nu}=\partial_\mu a_\nu-\partial_\nu a_\mu-i\comm{a_\mu}{a_\nu}\pnt
\end{equation} 

The result \eqref{SWmaplin} is of central importance for our later analysis of
noncommutative YM theories with gauge groups which are not $U(N)$.
It is easy to derive  
from the above transformation formula for the gauge field the mapping of the
field strength
\begin{equation}
F_{\mu\nu}=\partial_\mu A_\nu-\partial_\nu A_\mu-i\astcomm{A_\mu}{A_\nu}\col
\end{equation}
for which one finds
\begin{equation}
F_{\mu\nu}=f_{\mu\nu}-\frac{1}{4}\theta^{\alpha\beta}\big(2\acomm{f_{\mu\alpha}}{f_{\beta\nu}}-\acomm{a_\alpha}{D_\beta
  f_{\mu\nu}+\partial_\beta f_{\mu\nu}}\big)\col\qquad D_\beta=\partial_\beta+i\comm{\dummylen}{a_\beta}\pnt
\end{equation}

In \cite{Seiberg:1999vs} the authors derive differential equations which 
determine the Seiberg-Witten map. One can evaluate \eqref{SWmaplin} for an 
infinitesimal variation of $\theta^{\mu\nu}$.
The differential equations then read
\begin{equation}\label{SWdiffeq}
\begin{aligned}
\delta A_\mu&=\delta\theta^{\alpha\beta}\preparderiv{\theta^{\alpha\beta}}
A_\mu=-\frac{1}{4}\delta\theta^{\alpha\beta}\astacomm{A_\alpha}{\partial_\beta
  A_\mu+F_{\beta\mu}}\col\\
\delta\Lambda&=\delta\theta^{\alpha\beta}\preparderiv{\theta^{\alpha\beta}}
\Lambda=\frac{1}{4}\delta\theta^{\alpha\beta}\astacomm{\partial_\alpha\Lambda}{A_\beta}\col\\
\delta F_{\mu\nu}&=\delta\theta^{\alpha\beta}\preparderiv{\theta^{\alpha\beta}}
F_{\mu\nu}
=\frac{1}{4}\delta\theta^{\alpha\beta}\big(2\astacomm{F_{\mu\alpha}}{F_{\nu\beta}}-\astacomm{A_\alpha}{D_\beta
  F_{\mu\nu}+\partial_\beta
  F_{\mu\nu}}\big)\pnt
\end{aligned}
\end{equation}
In Appendix \ref{DBIlagequivcheck} we present a check that these differential
equations lead to the  
invariance \eqref{DBIlagequiv} of the DBI action \eqref{DBIlaginterpol} under
variations of $\Phi_{\mu\nu}$. 
The invariance under finite changes of $\Phi_{\mu\nu}$ is proven in
\cite{Liu:2000mj}. There the author uses 
an expression for the (inverse) Seiberg-Witten map in the $U(1)$ case
that is exact in $\theta^{\mu\nu}$, the validity of which is proven  
in \cite{
Okawa:2001mv,Mukhi:2001vx,Liu:2001pk}. 
The differential equation \eqref{SWdiffeq} is solved exactly in
$\theta^{\mu\nu}$ but as an expansion in powers of the noncommutative 
$U(1)$ gauge field $A_\mu$ in \cite{Mehen:2000vs}.
An investigation of the Seiberg-Witten map order by order in $\theta^{\mu\nu}$
can be found in \cite{Fidanza:2001qm} and up to quadratic order in
\cite{Jurco:2001rq}.  

At the end of this Section we note that the Seiberg-Witten
map is by far not unique but it possesses enormous freedom. 
This has been observed by \cite{Asakawa:1999cu}, see also
\cite{Bichl:2001cq,Jurco:2001rq}.
For our analysis, it will be sufficient to work with the expression
\eqref{SWmaplin}, because in linear order in $\theta^{\mu\nu}$ the 
ambiguity of $A_\mu$ has the form of a gauge transformation
\cite{Asakawa:1999cu}.


\clearemptydoublepage
\clearemptydoublepage
\chapter[Noncommutative Yang-Mills (NCYM) theories]{Noncommutative Yang-Mills 
theories}
\label{chap:NCYM}
In this Chapter we will deal with noncommutative YM theories. They are the 
generalizations of the noncommutative $U(1)$ gauge theory, that appeared 
in Chapter \ref{chap:NCfromstrings}, to different gauge groups.   
The noncommutative YM theory with gauge group $U(N)$ is the
generalization of the $U(1)$ gauge theory (see \eqref{NCYMlag} with the YM
coupling constant given in \eqref{NCYMcouplingrel}), 
if one considers open strings  
on a stack of $N$ $\text{D}p$-branes instead of a single
$\text{D}p$-brane with a constant $B$-field on the branes. 
The noncommutative $U(N)$ YM theory then is the effective 
description of open strings in this background with the 
boundary conditions \eqref{BCgeneral} 
in the Seiberg-Witten limit \eqref{SWlimit} 
if one uses point splitting regularization.  
A $d$-dimensional gauge theory corresponds to the choice of
$\text{D}(d-1)$-branes which world volume coordinates we denote with
$x^\mu$. For simplicity we assume
that the open string metric introduced in Chapter \ref{chap:NCfromstrings}
obeys $G=|\det G|=1$. Worldvolume indices are understood to 
be lowered and raised with respectively $G_{\mu\nu}$ and $G^{\mu\nu}$, which
signature we will fix to `mostly minus' in this Chapter.
The noncommutative YM action with gauge group $U(N)$ then reads
\begin{equation}\label{NCYMaction}
S_\text{YM}=-\frac{1}{2g^2}\int\de^dx\,\tr\big(F_{\mu\nu}\ast
F^{\mu\nu}\big)=-\frac{1}{2g^2}\int\de^dx\,\tr\big(F_{\mu\nu}F^{\mu\nu}\big)\pnt
\end{equation}
Here `$\tr$' denotes the trace w.\ r. t.\ the gauge group. The $\ast$-product 
in the second equality has been removed because of the following reason: 
the spacetime integration corresponds to the operator trace in the space of 
Weyl operators which can be used to describe the noncommutativity of the 
coordinates (see Appendices \ref{app:NCWeylop},\eqref{app:NCastprod} and \ref{app:NCYM} for a short introduction of this
formalism). Here we will work in ordinary spacetime where noncommutativity is
described by the $\ast$-product and the operator trace becomes a spacetime
integral. The cyclicity of the operator trace of two Weyl operators then
translates into the rule that one can remove the $\ast$-product between two 
functions under a spacetime integral, see \eqref{optracebilin}. 
The noncommutative field strength is defined as
\begin{equation}\label{NCYMfieldstrength}
F_{\mu\nu}=\partial_\mu A_\nu-\partial_\nu A_\mu-i\astcomm{A_\mu}{A_\nu}\col
\end{equation}
where the $\ast$-product in the $\ast$-commutator above is
defined as the tensor product of $\ast$-multiplication of functions with
ordinary matrix multiplication of the representation matrices of the gauge Lie
algebra (see \eqref{astcommacommdef}). 

In the next Section we will analyze the $\ast$-commutator in more detail and 
discuss which influence noncommutativity exerts on the choice of the gauge
group. The Faddeev-Popov gauge fixing procedure will be applied in
Section \ref{sec:NCYMgf} to noncommutative YM theories. This leads to a
relation between the noncommutative and ordinary sets of ghost fields. 
In Section \ref{sec:FrulesNCYM} the Feynman rules for noncommutative $U(N)$ YM
theories will be worked out. Two proposals that allow one to choose other gauge
groups than $U(N)$ for noncommutative YM theories 
will be summarized in Section \ref{sec:NCYMggG}. 
In Section \ref{FrulesggG} we will then analyze for these theories which kind
of Feynman rules can be defined.

\section[The gauge groups in NCYM theories]{The gauge groups in noncommutative Yang-Mills theories}
\label{sec:NCYMgaugegroups}
Ordinary YM theories can be realized with different gauge groups and for 
some of them string theory setups are known. This is different for 
noncommutative YM theories. 
This Section briefly reviews the complications which arise in the definition
of noncommutative gauge theories with non-Abelian gauge groups in general and 
shows that $U(N)$ groups play a special role. 
The arguments are based on \cite{Madore:2000en}. 

Let $T^A$, $A=1,\dots,N^2$ denote the $N\times N$ matrix representation of the
generators of the $U(N)$ Lie algebra $\mathfrak{u}(N)$ which
fulfill\footnote{In contrast to the usual definition in most textbooks we have 
included the trace part into $\duud{A}{B}{C}$. The reason is to keep
expressions simple.}
\begin{equation}\label{Liebase}
\comm{T^A}{T^B}=i\fuud{A}{B}{C}T^C\col\qquad\acomm{T^A}{T^B}=\duud{A}{B}{C}T^C\col\qquad\tr(T^AT^B)=\frac{1}{2}\delta^{AB}\pnt
\end{equation}  
Here $\fuud{A}{B}{C}$ denote the antisymmetric structure constants and $\duud{A}{B}{C}$ are 
symmetric in all indices. 
Obviously, the anticommutation relation does not hold in the
Lie algebra $\mathfrak{u}(N)$, but in its $N\times N$ matrix-representation. 
That it closes on the representation matrices is a specialty
of $\mathfrak{u}(N)$ because it respects the Hermiticity condition on the
elements. It does, however, not respect further conditions (i.\ e.\
tracelessness) which may enter the definition of subalgebras
of $\mathfrak{u}(N)$.

Let $\mathfrak{t}^A$, $A=1,\dots,N^2$ be the generators of $\mathfrak{u}(N)$ and let
 $\{\mathfrak{t}^a\}\subset\{\mathfrak{t}^A\}$, $\mathfrak{t}^{a'}\in\{\mathfrak{t}^A\}\setminus\{\mathfrak{t}^a\}$, $ \{\mathfrak{t}^A\}=\{\mathfrak{t}^a\}\cup\{\mathfrak{t}^{a'}\}$.
A subalgebra $\mathfrak{g}$ of $\mathfrak{u}(N)$ is defined exactly as a
subset of generators $\mathfrak{g}=\{\mathfrak{t}^a\}$
that closes under the commutator, i.\ e.\ 
$\comm{\mathfrak{t}^a}{\mathfrak{t}^b}=\fuud{a}{b}{c}\mathfrak{t}^c$ 
and thus $\fuud{a}{b}{c'}=0$. 
For a corresponding matrix representation one therefore has
\begin{equation}\label{Liesubalgebrarel}
\comm{T^a}{T^b}=i\fuud{a}{b}{c}T^c\col\qquad\acomm{T^a}{T^b}=\duud{a}{b}{C}T^C=\duud{a}{b}{c}T^c+\duud{a}{b}{c'}T^{c'}\pnt
\end{equation} 
As is obvious from this expression, the anticommutator does not necessarily
close on the representation of the Lie subalgebra. It is easy to check 
that the tracelessness $\tr T^a=0$ is not preserved by the anticommutator and 
that therefore it does not close onto the representation of
the $\mathfrak{su}$, $\mathfrak{so}$ and $\mathfrak{sp}$ subalgebras of 
$\mathfrak{u}(N)$. 

The importance of the previous discussion becomes evident if one computes the 
$\ast$-commutator (as defined in \eqref{astcommacommdef}) of two Lie algebra
valued functions $f(x)=f_a(x)\mathfrak{t}^a$, $g(x)=g_a(x)\mathfrak{t}^a$. 
It can be decomposed as
\begin{equation}\label{astcommdecomp}
\begin{aligned}
\astcomm{f}{g}&=\frac{1}{2}(f_a\ast g_b+g_b\ast
f_a)\comm{\mathfrak{t}^a}{\mathfrak{t}^b}+\frac{1}{2}(f_a\ast g_b-g_b\ast
f_a)\acomm{\mathfrak{t}^a}{\mathfrak{t}^b}\\
&=
\frac{1}{2}\astacomm{f_a}{g_b}\comm{\mathfrak{t}^a}{\mathfrak{t}^b}+\frac{1}{2}\astcomm{f_a}{g_b}\acomm{\mathfrak{t}^a}{\mathfrak{t}^b}\\
&=
f_a\astcos g_b\comm{\mathfrak{t}^a}{\mathfrak{t}^b}+if_a\astsin g_b\acomm{\mathfrak{t}^a}{\mathfrak{t}^b}\\
\end{aligned}
\end{equation}
and it is immediately clear that the 
noncommutative gauge transformations defined in
\eqref{NCgaugetrafoinfinit} are not Lie algebra  
valued because they exactly contain the $\ast$-commutator which depends on the 
anticommutator of two Lie algebra generators. 
One is therefore in general lead to define the noncommutative gauge theories
to take values in the enveloping algebra which has as basis vectors all
symmetric products of the original Lie algebra generators and which is
therefore infinite dimensional. The generators read \cite{Jurco:2001rq}
\begin{equation}\label{envalggen}
t^{A_1}\cdots t^{A_n}=\frac{1}{n!}\sum_{\pi\in S_n}\mathfrak{t}^{A_{\pi(1)}}\cdots\mathfrak{t}^{A_{\pi(n)}}\col
\end{equation}
where the sum runs over all $n!$ permutations of $1,\dots,n$. The Lie algebra 
generators are simply the ones with $n=1$. 
The infinite tower of basis elements requires 
an infinite number of coefficients, given by infinitely many field components.
This seems to prevent one from formulating a
reasonable theory. In Section \ref{sec:NCYMggG} we will discuss proposals for 
a solution of this problem that works for generic gauge groups.

For a $U(N)$ gauge group, however, the problem solves itself. As we already
mentioned, the corresponding $\mathfrak{u}(N)$ algebra is special because the
Hermiticity condition on the representation matrices is respected by the
anticommutator. 
This means that even the anticommutator of two representation matrices 
closes and can be expressed as a linear combination of the matrices, see 
\eqref{Liebase}. 
The symmetric products of generators defined in \eqref{envalggen} therefore
collapse in a representation (they become linear combinations of the $n=1$
elements), and 
one finds for the components of the $\ast$-anticommutator with
$f(x)=f_A(x)T^A$, $g(x)=g_A(x)T^A$, 
\begin{equation}\label{repastcommdecomp}
\begin{aligned}
2\tr\big(\astcomm{f}{g}T_C\big)&=\frac{i}{2}(f_A\ast g_B+g_B\ast
f_A)\fuud{A}{B}{C}+\frac{1}{2}(f_A\ast g_B-g_B\ast
f_A)\duud{A}{B}{C}\\
&=
\frac{i}{2}\astacomm{f_A}{g_B}\fuud{A}{B}{C}+\frac{1}{2}\astcomm{f_A}{g_B}\duud{A}{B}{C}\\
&=if_A\astcos g_B\fuud{A}{B}{C}+if_A\astsin g_B\duud{A}{B}{C}\pnt\\
\end{aligned}
\end{equation}
The above given discussion shows that 
YM theories with gauge groups $U(N)$ are special. Their
noncommutative counterparts can directly be formulated. More effort is needed 
if one wants to construct noncommutative YM theories with different gauge 
groups. We will summarize two different approaches in Section 
\ref{sec:NCYMggG}.

\section[Gauge fixing in NCYM theories]{Gauge fixing in noncommutative Yang-Mills theories}
\label{sec:NCYMgf}
In the following we will review the Faddeev-Popov gauge fixing procedure
\cite{Faddeev:1967fc} for YM theories\footnote{See e.\ g.\
  \cite{Itzykson:1980rh,Ryder:1985wq,Muta:1998vi} for more details.}. 
After a general description we will apply the 
formalism to the case of noncommutative YM
theories and discuss its behaviour under the field redefinition defined by
the Seiberg-Witten map.  
In the general description we always use capital variables to describe the
gauge quantities. Most of the formalism applies similarly to the
ordinary and the noncommutative YM theories. We will explicitly point out 
where differences occur.
 
Due to gauge invariance, path integration in YM theory should run over gauge 
inequivalent configurations only. Alternatively one can integrate over all
configurations and divide by the volume $V_\text{g}$
of the gauge orbits such that the partition function is given by 
\begin{equation}\label{pathintgtansatz}
Z[0]=\int\frac{\mathcal{D}A}{V_\text{g}}\e^{iS[A]}
\end{equation} 
with the integration measure
\begin{equation}
\mathcal{D}A=\prod_{a,\mu}\mathcal{D}A^a_\mu\pnt
\end{equation}
Here the indices $a$ and $\mu$ run over all components w.\ r.\ t.\ the gauge
group and spacetime respectively.   
The Faddeev-Popov trick allows one to split the integration over all
configurations into gauge inequivalent (gauge fixed) ones and integrations
along the gauge orbits. This then enables one to cancel the volume factor
$V_\text{g}$ and to formulate a gauge fixed version of
\eqref{pathintgtansatz}. Consider a gauge field configuration $A^a_{0\mu}$.
The gauge orbit that includes $A^a_{0\mu}$ is
spanned by acting with the gauge transformation \eqref{ordinarygaugetrafo} or \eqref{NCgaugetrafoWop} with all
possible choices of the gauge parameter on the gauge field configuration
$A^a_{0\mu}$. 
A configuration that is gauge equivalent to
$A^a_{0\mu}$ can be seen as a functional $A^a_\mu[A_0,\Lambda]$.

We now single out the configuration $A^a_{0\mu}$ by imposing a condition 
that is fulfilled by exactly one element\footnote{Strictly speaking this is
  only true in a perturbative analysis.} $A^a_{0\mu}$ 
\begin{equation}\label{gaugefixingfunc}
\mathcal{F}_a[A_0]=0\col
\end{equation}
where $\mathcal{F}_a$ is a gauge fixing functional for each component 
w.\ r.\ t.\ the gauge group. 
This then allows for a separation between $A^a_{0\mu}$ and the gauge equivalent configurations.
The trick is to manipulate the measure in the path integral 
\eqref{pathintgt} by introducing an identity
\begin{equation}\label{pathintid}
1=\int\mathcal{D}\Lambda\,\delta[\Lambda]=\int\mathcal{D}\Lambda\,\det\funcderiv{\mathcal{F}\big[A[A_0,\Lambda]\big]}{\Lambda}\,\delta\big[\mathcal{F}\big[A[A_0,\Lambda]\big]\big]\pnt
\end{equation}
Here $\delta[\dummylen]$ is the $\delta$-functional which can be seen 
as an infinite product of $\delta$-functions at each point in spacetime and
the second equality is based on the functional analog of the well known
identity
\begin{equation}
\delta\big(f(x)\big)=\frac{1}{|\det\partial_if_j|}\delta(x)
\end{equation} 
for the ordinary $\delta$-function depending on a vector valued function $f_i$
of arguments $x_i$, $i=1,\dots,d$ with only $f_i(0)=0$.
The determinant in \eqref{pathintid} refers to group and spacetime `indices' 
which means more
explicitly that one has to see its argument as a matrix carrying two 
independent pairs of `indices' $(a,b)$ and $(x,y)$
\begin{equation}
\funcderiv{\mathcal{F}_a\big[A[A_0(x),\Lambda(x)]\big]}{\Lambda_b(y)}\pnt
\end{equation}
The gauge fixing \eqref{gaugefixingfunc} and the obvious identity
$A^a_\mu[A_0,0]=A^a_{0\mu}$ now allows one to write
\begin{equation}\label{FPdet}
1=\Delta[A_0]\int\mathcal{D}\Lambda\,\delta\big[\mathcal{F}\big[A[A_0,\Lambda]\big]\big]
\col\qquad\Delta[A_0]=\det\funcderiv{\mathcal{F}\big[A[A_0,\Lambda]\big]}{\Lambda}\bigg|_{\Lambda=0}\pnt
\end{equation}
This is an obvious consequence of the functional form of the identity 
$f(x)\delta(x)=f(0)\delta(x)$.
It is important to notice that $\Delta[A_0]$ is invariant under gauge 
transformations. Let $A'=A[A_0,\Lambda']$ then one finds 
\begin{equation}\label{Deltainv}
1=\Delta[A']\int\mathcal{D}\Lambda\,\delta\big[\mathcal{F}\big[A[A[A_0,\Lambda'],\Lambda]\big]\big]=\Delta[A']\int\mathcal{D}\tilde\Lambda\,\delta\big[\mathcal{F}\big[A[A_0,\tilde\Lambda]\big]\big]=\frac{\Delta[A']}{\Delta[A_0]}\col
\end{equation}
where one has to use that two sequential gauge transformations are again
a gauge transformation and that the measure fulfills
$\mathcal{D}\Lambda=\mathcal{D}(\Lambda\Lambda')=\mathcal{D}(\tilde\Lambda)$.
In particular $u(\Lambda)u(\Lambda')=u(\tilde\Lambda)$ or
$U(\Lambda)\ast U(\Lambda')=U(\tilde\Lambda)$ enter the ordinary and
noncommutative gauge transformations \eqref{ordinarygaugetrafo} and \eqref{NCgaugetrafoWop} respectively.
One now inserts \eqref{FPdet} into the path integral \eqref{pathintgtansatz} to
obtain 
\begin{equation}\label{pathintgt}
Z[0]=\int\frac{\mathcal{D}
  A'\,\mathcal{D}\Lambda}{V_\text{g}}\Delta[A_0]\,\delta\big[\mathcal{F}\big[A[A_0,\Lambda]\big]\big]\e^{iS[A']}=\int\frac{\mathcal{D}
  A'\,\mathcal{D}\Lambda}{V_\text{g}}\Delta[A']\,\delta\big[\mathcal{F}\big[A[A',\Lambda]\big]\big]\e^{iS[A']}\col
\end{equation} 
where we have renamed the integration measure and the argument of the action
and then used the invariance \eqref{Deltainv}.
Due to the invariance of $\Delta[A_0]$, of the action
and of the path integral measure under gauge transformations we can now
perform a gauge transformation that transforms $A'$ to $A$. The integrand,
especially the argument 
of $\mathcal{F}$,  then becomes independent of $\Lambda$ such that the
integration can be performed to cancel precisely $V_\text{g}$ in the
denominator. One thus obtains for the gauge fixed path integral
\begin{equation}\label{pathintgtintermediate}
Z[0]=\int\mathcal{D}A\,\Delta[A]\,\delta\big[\mathcal{F}[A]\big]\e^{iS[A]}\pnt
\end{equation} 
The determinant $\Delta$ as defined in \eqref{FPdet} can
be formulated as a fermionic Gaussian path integral 
\begin{equation}\label{FPfunc}
\Delta[A]=\int\mathcal{D}C\,\mathcal{D}\bar C\,
\exp\bigg\{-i\int\de^dx\de^dy\,\bar C^a(x)\mathcal{M}_{ab}(x,y)C^b(y)\bigg\}\col
\end{equation}
with the integration measures $\mathcal{D}C=\prod_a\mathcal{D}C^a$,
$\mathcal{D}\bar C=\prod_a\mathcal{D}\bar C^a$ 
and the definition
\begin{equation}\label{FPmatrix}
\mathcal{M}_{ab}(x,y)=\funcderiv{\mathcal{F}_a\big[A[A_0(x),\Lambda(x)]\big]}{\Lambda_b(y)}\bigg|_{\Lambda=0}
\end{equation}
Finally, the $\delta$-functional in 
\eqref{pathintgtintermediate} can be removed by replacing its argument with 
$\mathcal{F}_a[A]-f_a$, where $f_a$ are functions of
spacetime and compute the average of all $f_a$ with a Gaussian
weight.
This means integrate the functional with
\begin{equation}
\int\mathcal{D}f\,\e^{-\frac{i}{2\kappa}\int\de^dxf_af^a}\col
\end{equation}
where $\kappa$ is a real parameter.
The $\delta$ functional then effectively cancels against the integration and
it replaces $f_a$ in the exponential by  $\mathcal{F}_a[A]$. The final form of
the gauge fixed functional therefore is (up to an unimportant constant $N'$ in 
front)
\begin{equation}\label{gaugefixedfunctional}
Z[0]=N'\int\mathcal{D}A\,\mathcal{D}C\,\mathcal{D}\bar C\,\exp\bigg\{iS[A]-i\int\de^dx\Big(\frac{1}{2\kappa}\mathcal{F}_a[A]\mathcal{F}^a[A]+\bar C^aM_{ab}C^b\Big)\bigg\}\col
\end{equation}
where $M_{ab}$ is an operator that acts on $C_b$. It is defined as an operator
that has to act on the $\delta$-function to give the matrix
$\mathcal{M}_{ab}(x,y)$ which is given in \eqref{FPmatrix} 
\begin{equation}\label{FPmatrixdelta}
M_{ab}\delta(x-y)=\mathcal{M}_{ab}(x,y)\pnt
\end{equation}

It is not difficult to evaluate \eqref{FPmatrix}, since ($\Lambda$ is set to
zero at the end) it is sufficient to replace
$A[A_0,\Lambda]$ by the infinitesimal version
\eqref{ordinarygaugetrafoinfinit}  or \eqref{NCgaugetrafoinfinit} of the
respectively ordinary or noncommutative gauge transformation. 
Furthermore, we take the gauge fixing functional $\mathcal{F}$ to be linear
in its argument such that it can be described by 
\begin{equation}
\mathcal{F}_a[A]=\mathcal{O}_{ab}^\mu A^b_\mu\col
\end{equation}
where an operator $\mathcal{O}_{ab}^\mu$ acts on $A^a_\mu$.
One then finds for $M_{ab}$ 
\begin{equation}\label{FPmatrixexplicit}
\begin{aligned}
M_{ab}\delta(x-y)&=\mathcal{O}_{ac}^\mu\Big(\delta^c_b\partial_\mu\delta(x-y) 
-\frac{1}{2}(\delta(x-y)\bullet
A^d_\mu+A^d_\mu\bullet \delta(x-y))\fddu{b}{d}{c}\\
&\phantom{{}=\mathcal{O}_{ac}^\mu\Big(\delta^c_b\partial_\mu\delta(x-y)}
+\frac{i}{2}(\delta(x-y)\bullet A^d_\mu-A^d_\mu\bullet
\delta(x-y))\dddu{b}{d}{c}\Big)\pnt
\end{aligned}
\end{equation}
Here the $\bullet$-product either denotes the
$\ast$-product or the ordinary one.  

At the end of this general discussion we compare gauge fixing in
the ordinary and in the noncommutative case and specifically work out how the 
ghost fields are related in both cases.
Here we will now explicitly use lower case and capital 
letters for variables of the ordinary and noncommutative case respectively.
We begin with the gauge fixing functional and use the Seiberg-Witten map
\eqref{SWmapansatz} to translate this condition between the ordinary and
noncommutative case
\begin{equation}
\mathcal{F}[A]=\mathcal{F}[A[a]]
\end{equation}
One can then write the expressions \eqref{FPmatrix} in both cases as
\begin{equation}\label{NCandordFPmatrix}
\begin{aligned}
\mathcal{M}^\text{NC}_{ab}(x,y)&=\funcderiv{\mathcal{F}_a[A(x)]}{A}\cdot\funcderiv{A[A_0,\Lambda]}{\Lambda_b(y)}\bigg|_{\Lambda=0}\col\\
\mathcal{M}^\text{ord}_{ab}(x,y)&=\funcderiv{\mathcal{F}_a[A(x)]}{A}\cdot\funcderiv{A[a]}{a}\cdot\funcderiv{a[a_0,\lambda]}{\lambda_b(y)}\bigg|_{\lambda=0}\col
\end{aligned}
\end{equation}
where $\cdot$ abbreviates the summation and integration over all omitted
indices and spacetime coordinates. A relation between $\mathcal{M}^\text{NC}$
and $\mathcal{M}^\text{ord}$ follows from a comparison of the R.\ H. S.\ in
the above given expressions. It reads
\begin{equation}\label{Mrel}
\mathcal{M}^\text{NC}\cdot\funcderiv{A[a]}{a}\cdot\funcderiv{a[a_0,\lambda]}{\lambda_b}\bigg|_{\lambda=0}=\mathcal{M}^\text{ord}\cdot\funcderiv{A[A_0,\Lambda]}{\Lambda_b}\bigg|_{\Lambda=0}\pnt
\end{equation}
Since $\Delta=\det\mathcal{M}$,
taking the determinant on both sides then relates
$\Delta^\text{NC}$ and $\Delta^\text{ord}$. 
Both can be expressed as in \eqref{FPfunc}
using $\mathcal{M}^\text{NC}$ and the ghost fields $\bar C$, $C$ in the
noncommutative 
and $\mathcal{M}^\text{ord}$, $\bar c$, $c$ in the ordinary formulation. 
According to \eqref{FPfunc}, the exponents are then given by
\begin{equation}\label{exponents}
\bar C^a\cdot\mathcal{M}^\text{NC}_{ab}\cdot C^b\col\qquad\bar
c^a\cdot\mathcal{M}^\text{ord}_{ab}\cdot c^b
\end{equation}
in the noncommutative and the ordinary formulation. 
A variable transformation between the two 
sets of ghost fields in the path integrals should produce the required
determinants that are necessary to fulfill \eqref{Mrel}.
If both exponents given in \eqref{exponents} are set equal, the transformation
properties of fermionic path integrals guarantee that the required determinant
is generated.\footnote{It is important to remember that the
fermionic path integral measure transforms with the reciprocal determinant,
see e.\ g.\ \cite{Ryder:1985wq}.}
Using the relation \eqref{Mrel} the equality of both expressions in 
\eqref{exponents} can be achieved with the choice
\begin{equation}\label{ghostmap}
\bar C=\bar c\col\qquad \funcderiv{A[A_0,\Lambda]}{\Lambda}\bigg|_{\Lambda=0}C=\funcderiv{A[a]}{a}\cdot\funcderiv{a[a_0,\lambda]}{\lambda}\bigg|_{\lambda=0}c\col
\end{equation}
where again indices and spacetime dependence have been omitted.
The solution to the above equation is already known. It is precisely the 
Seiberg-Witten map for the gauge parameter that maps the ghosts.
The basic relation from which the map has been derived is
\eqref{gaugeorbitmap}
which can be cast into the form
\begin{equation}
A[a]+\hat\delta_{\Lambda}A[a]=A[a+\delta_\lambda a]=A[a]+\funcderiv{A[a]}{a}\delta_\lambda a
\end{equation}  
for infinitesimal gauge transformations. For these one has the relation
\begin{equation}
\hat\delta_{\Lambda}A[a]=\funcderiv{A[A_0,\Lambda]}{\Lambda}\bigg|_{\Lambda=0}\Lambda\col\qquad\delta_\lambda a=\funcderiv{a[a_0,\lambda]}{\lambda}\bigg|_{\lambda=0}\lambda\col
\end{equation}
such that the relation for the gauge parameters read
\begin{equation}
\funcderiv{A[A_0,\Lambda]}{\Lambda}\bigg|_{\Lambda=0}\Lambda=\funcderiv{A[a]}{a}\cdot\funcderiv{a[a_0,\lambda]}{\lambda}\bigg|_{\lambda=0}\lambda\pnt
\end{equation}
This is exactly the same as \eqref{ghostmap} for the ghosts. Therefore, the
solution for $C$ in \eqref{ghostmap} is simply given by the expression for the 
gauge parameter $\Lambda=\Lambda[\lambda,a]$ with $\lambda$ replaced by $c$.
At order $\mathcal{O}(\theta)$ one thus finds from \eqref{SWmaplin}
\begin{equation}\label{SWmapghostlin}
C[c,a]=c+\frac{1}{4}\theta^{\alpha\beta}\acomm{\partial_\beta c}{a_\alpha}\pnt
\end{equation}
This relation is found in \cite{Bichl:2001gu} for the Abelian case via a
discussion of the Becchi-Rouet-Stora-Tyutin (BRST) transformation 
\cite{Becchi:1976nq}. 
The above
connection between the transformation of the gauge parameter and the ghosts is
not surprising if one remembers that the BRST transformation of the gauge
field is given by a gauge transformation with the anticommuting ghost as a
parameter. 

\section[Feynman rules for NCYM theories with gauge groups $U(N)$ ]{Feynman rules for noncommutative Yang-Mills theories with
  gauge groups $U(N)$}
\label{sec:FrulesNCYM}
In this Section we will show how to extract the known Feynman rules \cite{Bonora:2000ga,Armoni:2000xr,Szabo:2001kg} for noncommutative YM
theories with $U(N)$ gauge groups, which can be consistently formulated as 
discussed in Section \ref{sec:NCYMgaugegroups}. We will extract them from the
path integral formulation that was presented in Appendix 
\ref{subsec:Frulespint}.
The action is given by \eqref{NCYMaction} and we fix the gauge with the 
Lorentz condition
\begin{equation}\label{lorenzgaugefix}
\mathcal{F}_a[A]=\mathcal{O}_{ab}^\mu A^b_\mu=\frac{1}{g}\delta_{ab}\partial^\mu A^b_\mu=0\pnt
\end{equation}
One can now write down the complete action that enters 
the gauge fixed functional \eqref{gaugefixedfunctional} with $M_{ab}$ given in 
\eqref{FPmatrixexplicit} for the 
above given $\mathcal{O}_{ab}^\mu$ and with $\ast$-products. 
The complete Lagrangian is
\begin{equation}
\mathcal{L}=\mathcal{L}_\YM+\mathcal{L}_\GF+\mathcal{L}_\FP\pnt
\end{equation}
With the noncommutative field strength \eqref{NCYMfieldstrength}, each piece
in the above expression reads
\begin{equation}\label{NCLYMLGFLFPG}
\begin{aligned}
\mathcal{L}_\YM&=-\frac{1}{2g^2}\tr\big(F_{\mu\nu}F^{\mu\nu}\big)\col\\
\mathcal{L}_\GF&=-\frac{1}{2g^2\kappa}\tr\big((\partial^\mu A_\mu)^2\big)\col\\
\mathcal{L}_\FP&=-2\tr\big(\bar C\partial^\mu(\partial_\mu C+i\astcomm{C}{A_\mu})\big)\col
\end{aligned}
\end{equation}
or after evaluating the trace over the gauge group
\begin{equation}
\begin{aligned}
\mathcal{L}_\YM&=-\frac{1}{4g^2}F^A_{\mu\nu}F_A^{\mu\nu}\col\\
\mathcal{L}_\GF&=-\frac{1}{4g^2\kappa}(\partial^\mu A^A_\mu)(\partial_\nu
A_A^\nu)\col\\ 
\mathcal{L}_\FP&=-\bar
C^A\partial^\mu\Big(\partial_\mu C_A-\frac{1}{2}\astacomm{C^B}{A^C_\mu}f_{BCA}+\frac{i}{2}\astcomm{C^B}{A^C_\mu}d_{BCA}\Big)\col\\
&=-\bar C^A\partial^\mu\Big(\partial_\mu C_A-
C^B\astcos A^C_\mu f_{BCA}- C^B\astsin A^C_\mu d_{BCA}\Big)\pnt
\end{aligned}
\end{equation}
To find the above results we have used \eqref{Liebase},
\eqref{astcommdecomp} and \eqref{astsincos}. To make the expressions more
readable we will use the abbreviation
\begin{equation}
\mathcal{O}_\rho\tfrac{1}{2}\theta^{\rho\sigma}\mathcal{P}_\sigma=\mathcal{O}\wedge\mathcal{P}\col
\end{equation}
where $\mathcal{O}$ and $\mathcal{P}$ denote any operators that carry one
spacetime index. The components of the field strength tensor are then
given by 
\begin{equation}
\begin{aligned}
F^A_{\mu\nu}&=\partial_\mu A^A_\nu-\partial_\nu
A^A_\mu+\frac{1}{2}\astacomm{A^B_\mu}{A^C_\nu}\fddu{B}{C}{A}-\frac{i}{2}\astcomm{A^B_\mu}{A^C_\nu}\dddu{B}{C}{A}\\ 
&=\partial_\mu A^A_\nu-\partial_\nu A^A_\mu+A^B_\mu\astwedgecos
A^C_\nu\fddu{B}{C}{A}+A^B_\mu\astwedgesin A^C_\nu\dddu{B}{C}{A}\pnt
\end{aligned}
\end{equation}
One finds for $\mathcal{L}_\YM+\mathcal{L}_\GF$ after adding appropriate total
derivative terms 
\begin{equation}\label{LYMLGFexpand}
\begin{aligned}
\mathcal{L}_\YM+\mathcal{L}_\GF&\cong-\frac{1}{g^2}\tr\Big(-A_\mu\big(G^{\mu\nu}\Box
-(1-\kappa^{-1})\partial^\mu\partial^\nu\big)A_\nu\\
&\phantom{-\cong\frac{1}{g^2}\tr\Big({}}-2i\partial_\mu A_\nu\astcomm{\smash[t]{A^\mu}}{\smash[t]{A^\nu}}-\frac{1}{2}\astcomm{A_\mu}{A_\nu}\astcomm{\smash[t]{A^\mu}}{\smash[t]{A^\nu}}\Big)\\
&\negphantom{\qquad}=\frac{1}{2g^2}A^A_\alpha\delta_{AB}\big(G^{\alpha\beta}\Box
-(1-\kappa^{-1})\partial^\alpha\partial^\beta\big)A^B_\beta\\
&\negphantom{\qquad}\phantom{{}=}-\frac{1}{g^2}G^{\mu\beta}G^{\alpha\gamma}(\partial_\mu)_1\big(\astposwedgecos{2}{3}
f_{BCA}+\astposwedgesin{2}{3}d_{BCA}\big)A^A_\alpha A^B_\beta A^C_\gamma\\
&\negphantom{\qquad}\phantom{{}=}-\frac{1}{4g^2}G^{\alpha\gamma}G^{\beta\delta}\big(\astposwedgecos{1}{2}
f_{ABE}+\astposwedgesin{1}{2}d_{ABE}\big)\\
&\negphantom{\qquad}\phantom{={}-\frac{1}{4g^2}G^{\alpha\gamma}G^{\beta\delta}{}}\negphantom{{}\times{}}\times\big(\astposwedgecos{3}{4}
\fddu{C}{D}{E}+\astposwedgesin{3}{4}\dddu{C}{D}{E}\big)A^A_\alpha A^B_\beta A^C_\gamma A^D_\delta\pnt
\end{aligned}
\end{equation}
We have used the notation $(\partial)_i$ to indicate that the derivative acts
on the gauge field at the $i$th position in the product on the right hand 
side of the operator. 
To extract Feynman rules for this field theory means to find all connected 
tree-level Green functions. 
Comparing the above expression with \eqref{genaction} and 
\eqref{Kmatrixdelta}, one obtains
\begin{equation}
K_{AB}^{\alpha\beta}=-\frac{1}{g^2}\delta_{AB}\big(G^{\alpha\beta}\Box
-(1-\kappa^{-1})\partial^\alpha\partial^\beta\big)\pnt
\end{equation}
Going through the procedure of \eqref{Ftwopointfuncgeneral},
the $2$-point function in momentum space is found to be 
\begin{equation}\label{YMtwopoint}
\tilde G^{AB}_{\alpha\beta}(p,q)=-i\tilde\Delta^{AB}_{\alpha\beta}(p)(2\pi)^4\delta(p+q)\col\qquad
\tilde\Delta^{AB}_{\alpha\beta}(p)=\frac{g^2}{p^2}\delta^{AB}\Big(G_{\alpha\beta}
-(1-\kappa)\frac{p_\alpha p_\beta}{p^2}\Big)\pnt
\end{equation}
The proper $3$- and $4$-point vertices are found by comparing
the interaction parts of \eqref{LYMLGFexpand} with \eqref{genpotential}
and then using \eqref{FNpointtreegeneral} to get the
momentum space expressions where the momenta leave the interaction point. 
One obtains for the
$3$-point vertex 
\begin{equation}
\begin{aligned}
\tilde G_{ABC}^{\alpha\beta\gamma}(p,q,r)_\text{c}&=
\frac{1}{g^2}G^{\alpha\gamma}p^\beta\big(\astmomwedgecos{q}{r}
f_{BCA}-\astmomwedgesin{q}{r}d_{BCA}\big)(2\pi)^d\delta(p+q+r)\\
&\phantom{{}={}}
+\text{5 perm}\col
\end{aligned}
\end{equation}
where `perm' denotes the remaining permutations of the three momenta, Lorentz
and group indices.
The $4$-point vertex reads
\begin{equation}
\begin{aligned}
\tilde
G_{ABCD}^{\alpha\beta\gamma\delta}(p,q,r,s)_\text{c}&=-\frac{i}{g^2}G^{\alpha\gamma}G^{\beta\delta}\big(\astmomwedgecos{p}{q}
f_{ABE}-\astmomwedgesin{p}{q}d_{ABE}\big)\\
&\phantom{{}=-\frac{i}{g^2}G^{\alpha\gamma}G^{\beta\delta}}\negphantom{{}\times{}}\times\big(\astmomwedgecos{r}{s}
\fddu{C}{D}{E}-\astmomwedgesin{r}{s}\dddu{C}{D}{E}\big)\\
&\phantom{{}=-\frac{i}{g^2}G^{\alpha\gamma}G^{\beta\delta}}\negphantom{{}\times{}}\times(2\pi)^4\delta(p+q+r+s)+\text{5 perm}\pnt
\end{aligned}
\end{equation}
Here `perm' denotes the remaining permutations after dividing out 
the $4$-dimensional symmetry group $S$ of the vertex which includes the
permutations  
\begin{equation}
S=\big\{(\pi(1),\pi(2),\pi(3),\pi(4))\big\}=\big\{(1,2,3,4),(2,1,4,3),(3,4,1,2),(4,3,2,1)\big\}
\end{equation}
of the indices at the four legs $1$, $2$, $3$, $4$ (this removes the prefactor
$\frac{1}{4}$, see \eqref{FNpointtreegeneral}). 
\newlength{\eqofffiftytwo}
\setlength{\fboxsep}{0pt}
\settoheight{\eqoff}{\fbox{$=$}}
\setlength{\eqoff}{0.5\eqoff}
\setlength{\eqofftwenty}{\eqoff}
\setlength{\eqoffsixty}{\eqoff}
\setlength{\eqofffiftytwo}{\eqoff}
\addtolength{\eqofftwenty}{-20pt}
\addtolength{\eqoffsixty}{-60pt}
\addtolength{\eqofffiftytwo}{-52.055pt}
\begin{figure}
\begin{center}
\begin{equation*}
\begin{aligned}
\raisebox{\eqofftwenty}{%
\begin{picture}(140,40)(0,0)\scriptsize
\SetOffset(130,20)
\Line(-80,0)(-40,0)\Text(-90,0)[r]{$p,\alpha,A$}\Text(-30,0)[l]{$p,\beta,B$}
\end{picture}}%
&=-g^2\frac{i}{p^2}\delta^{AB}\Big(G_{\alpha\beta}
-(1-\kappa)\frac{p_\alpha p_\beta}{p^2}\Big)
\\%
\raisebox{\eqoffsixty}{%
\begin{picture}(155,120)(0,0)\scriptsize
\SetOffset(130,60)
\Line(-80,0)(-40,0)\Text(-90,0)[r]{$r,\gamma,C$}
\Line(-20,34.64)(-40,0)\Text(-15,43.3)[l]{$p,\alpha,A$}
\Line(-20,-34.64)(-40,0)\Text(-15,-43.3)[l]{$q,\beta,B$}
\Vertex(-40,0){1}
\end{picture}}%
&=%
\parbox[c][120pt][c]{175pt}{%
\begin{equation*}
\begin{aligned}
&-\frac{1}{g^2}\big(\astmomwedgecos{p}{q}
f_{ABC}-\astmomwedgesin{p}{q}d_{ABC}\big)\\
&\big(G^{\alpha\beta}(q-p)^\gamma+G^{\beta\gamma}(r-q)^\alpha+G^{\gamma\alpha}(p-r)^\beta\big)
\end{aligned}
\end{equation*}
}%
\\%
\raisebox{\eqofffiftytwo}{%
\begin{picture}(150.71,104.11)(0,0)\scriptsize
\SetOffset(115.355,52.055)
\Line(-11.716,28.284)(-40,0)\Text(-4.645,35.355)[l]{$p,\alpha,A$}
\Line(-68.284,28.284)(-40,0)\Text(-75.355,35.355)[r]{$q,\beta,B$}
\Line(-68.284,-28.284)(-40,0)\Text(-75.355,-35.355)[r]{$r,\gamma,C$}
\Line(-11.716,-28.284)(-40,0)\Text(-4.645,-35.355)[l]{$s,\delta,D$}
\Vertex(-40,0){1}
\end{picture}}%
&=%
\parbox[c][120pt][c]{179.29pt}{%
\begin{equation*}
\begin{aligned}
&-\frac{i}{g^2}\big(\astmomwedgecos{p}{q}
f_{ABE}-\astmomwedgesin{p}{q}d_{ABE}\big)\\
&\phantom{{}-\frac{i}{g^2}{}}\big(\astmomwedgecos{r}{s}
\fddu{C}{D}{E}-\astmomwedgesin{r}{s}\dddu{C}{D}{E}\big)\\
&\phantom{{}-\frac{i}{g^2}{}}\big(G^{\alpha\gamma}G^{\beta\delta}-G^{\alpha\delta}G^{\beta\gamma}\big)+
\text{2 perm} 
\end{aligned}
\end{equation*}
}%
\\%
\raisebox{\eqofftwenty}{%
\begin{picture}(140,40)(0,0)\scriptsize
\SetOffset(130,20)
\DashArrowLine(-40,0)(-80,0){1}\Text(-90,0)[r]{$p,A$}\Text(-30,0)[l]{$p,B$}
\end{picture}}%
&=\raisebox{\eqofftwenty}{%
\begin{picture}(140,40)(0,0)\scriptsize
\SetOffset(130,20)
\DashArrowLine(-80,0)(-40,0){1}\Text(-90,0)[r]{$p,A$}\Text(-30,0)[l]{$p_2,B$}
\end{picture}}%
=\frac{i}{p^2}\delta^{AB}
\\%
\raisebox{\eqoffsixty}{%
\begin{picture}(155,120)(0,0)\scriptsize
\SetOffset(130,60)
\DashArrowLine(-40,0)(-20,34.64){1}\Text(-15,43.3)[l]{$p,A$}
\Line(-80,0)(-40,0)\Text(-90,0)[r]{$q,\beta,B$}
\DashArrowLine(-20,-34.64)(-40,0){1}\Text(-15,-43.3)[l]{$r,C$}
\Vertex(-40,0){1}
\end{picture}}%
&=%
\parbox[c][120pt][c]{175pt}{%
\begin{equation*}
\begin{aligned}
p^\beta\big(\astmomwedgecos{p}{q}f_{ABC}
+\astmomwedgesin{p}{q}d_{ABC}\big)
\end{aligned}
\end{equation*}
}
\end{aligned}
\end{equation*}
\caption{Feynman rules \cite{Bonora:2000ga,Armoni:2000xr,Szabo:2001kg} 
for noncommutative $U(N)$ YM theory,
  $p_\rho\frac{1}{2}\theta^{\rho\sigma}q_\sigma=p\wedge q$. The straight lines
  denote gauge bosons and the dashed arrow lines pointing to a vertex or away
  from it denote ghosts and antighosts respectively. 
 Momentum conservation is understood and all momenta point to the vertices.}\label{fig:NCUNYMFrules}
\end{center}
\end{figure}

The ghost Lagrangian $\mathcal{L}_\FP$ reads, after adding a total derivative 
that modifies the interaction term 
\begin{equation}\label{LFPGexpand}
\begin{aligned}
\mathcal{L}_\FP&\cong-\bar C^A\delta_{AB}\partial^\mu\partial_\mu C^B\\
&\phantom{\cong{}}-(\partial^\beta)_1\Big(\astposwedgecos{3}{2}f_{CBA}
+\astposwedgesin{3}{2}d_{CBA}\Big)\bar C^AA^B_\beta C^C\pnt
\end{aligned}
\end{equation}
From this one finds for the $2$-point function of the ghost $C$ and antighost
$\bar C$ respectively
\begin{equation}\label{ghosttwopoint}
\tilde G^{AB}_{(C)}(p,q)=\tilde G^{AB}_{(\bar C)}(p,q)=\frac{i}{p^2}\delta^{AB}(2\pi)^d\delta(p+q)\pnt
\end{equation}
Again using \eqref{FNpointtreegeneral},
the $3$-point vertex that describes the antighost-gauge-ghost interaction 
can easily be read-off from \eqref{LFPGexpand} and is found to be
\begin{equation}
\tilde G_{(\bar
  CAC),ABC}^\beta(p,q,r)=-p^\beta\Big(\astmomwedgecos{q}{r}f_{CBA}-\astmomwedgesin{q}{r}d_{CBA}\Big)(2\pi)^d\delta(p+q+r)\col
\end{equation}
where $(\bar CAC)$ indicates that the momenta (leaving the
interaction point) and indices $(p,A)$,
$(q,\beta,B)$ and $(r,C)$ refer to the antighost, gauge field and ghost
respectively. Clearly the spacetime index $\beta$ is assigned to the gauge
field. Due to the fact that the three fields are different, 
no summation over permutations appears here.  

The rules look very similar to that of ordinary YM
theory. Modifications are due to the presence of the $\ast$-commutator in
\eqref{NCYMfieldstrength} instead of the ordinary commutator. This introduces
the symmetric $d_{ABC}$ and generates additional momentum
dependent trigonometric factors in the vertices which are responsible
for the UV/IR effect \cite{Minwalla:1999px}. We already described this effect
in the Introduction. It received a lot of attention
in particular with respect to its stringy origin and its implications for the
renormalization program. However, the UV/IR effect is not manifest in
$\theta^{\mu\nu}$-expanded  perturbation theory for $U(N)$.
For $G\neq U(N)$, besides a conjecture for $SO(N)$ in \cite{Bonora:2001ny},
Feynman rules in terms of the full noncommutative $A_\mu$ are not known. 
Therefore, our goal in Section \ref{FrulesggG} will be to get information on
these rules by studying some issues of partial summing the known
$\theta$-expanded rules. Such rules would allow one to study UV/IR mixing
similar to the $U(N)$ case. 

\section[Construction of NCYM theories with gauge groups $G\neq U(N)$]{Construction of noncommutative Yang-Mills theories with gauge
  groups $G\neq U(N)$}
\label{sec:NCYMggG}

In Section \ref{sec:NCYMgaugegroups} we have seen that noncommutativity 
requires a careful choice of the underlying gauge group and that one has to 
give up the gauge field and the gauge transformation parameters being Lie
algebra valued. Instead they take values in the enveloping algebra. 
For $U(N)$ gauge groups, however, the choice of a 
matrix representation enables one to introduce matrix multiplication and 
the anticommutator in addition to the commutator, and therefore one can avoid
to work with the enveloping algebra. In contrast to the $U(N)$ case the
formulation
of gauge theories with other gauge groups is less unique. One has to work 
with the full infinite dimensional enveloping algebra and thus to introduce
infinitely many degrees of freedom.
In the following we will discuss two different kinds of constructions which
avoid these problems.  
The one we describe first is perturbative in the noncommutativity parameter 
$\theta^{\mu\nu}$ and it is based on the enveloping algebra description. 
In the second approach, which is exact $\theta^{\mu\nu}$, one starts with 
a $U(N)$ theory and imposes additional constraints which are respected by 
the $\ast$-commutator.  The constraints define a subgroup of $U(N)$ (endowed
with the $\ast$-commutator). 

\subsection{The enveloping algebra approach}
\label{subsec:envalg}
In a series of papers noncommutative gauge theories were 
discussed performing a perturbation expansion in $\theta^{\mu\nu}$. 
In \cite{Madore:2000en} the authors formulate a general setup for defining
noncommutative gauge theories. Besides the canonical structure
\eqref{astproddef}
they analyzed the Lie algebra structure and the quantum space structure.
The guiding principle of this discussion is to reduce the infinite number of 
fields that arises as coefficients in the enveloping algebra to a finite 
number \cite{Jurco:2000ja}. The authors show that the 
enveloping algebra valued components (which have basis vectors
\eqref{envalggen} with $n>1$) can be expressed in terms of the 
finite number of Lie algebra valued ones ($n=1$ in \eqref{envalggen}) and
their derivatives. 
It turns out that there exists a solution in which the coefficient in front of 
the basis element \eqref{envalggen} formed from $n$ Lie algebra generators is of order $\mathcal{O}(\theta^{n-1})$.\footnote{In \cite{Jurco:2001rq} the
  authors remark that one can change this behaviour by using the freedom in
  the Seiberg-Witten map.} The coefficient of \eqref{envalggen} is then 
given by the corresponding $\mathcal{O}(\theta^{n-1})$ term in the
expansion of the 
Seiberg-Witten map. That means the ordinary gauge field $a_\mu$ is the leading
$n=1$ coefficient.
Some more details of this derivation are presented in Appendix
\ref{SWmapfromenvalg}.

In \cite{Jurco:2001my,Jurco:2001rq} the authors construct non-Abelian 
gauge theories in the enveloping algebra approach of \cite{Jurco:2000ja} and
determine the coefficients of the fields and the gauge transformation up to
$\mathcal{O}(\theta^2)$. As already mentioned, this corresponds to an expansion
of the Seiberg-Witten map up to the same order. See \cite{Barnich:2002pb} for
a related analysis 
which includes a discussion of ambiguities in the Seiberg-Witten map. 

From the above summarized references one can now extract the 
recipe to construct non-Abelian gauge theories with arbitrary gauge groups 
$G$ perturbatively to a certain order $\mathcal{O}(\theta^n)$.
One should take the fields and gauge parameter of an 
ordinary gauge theory with gauge group $G$ and insert them into the
Seiberg-Witten map expanded up to $\mathcal{O}(\theta^n)$.
This gives the noncommutative
fields and gauge parameter which then enter the action of the 
noncommutative gauge theory. 

\subsection{Subgroups of $U(N)$ via additional constraints}
\label{subsec:antiautoconstraint}
 The authors of \cite{Bonora:2000td} define gauge transformations 
in a subgroup of the noncommutative $U(N)$ gauge group (which is 
endowed with the $\ast$-commutator).
The corresponding infinitesimal gauge transformations and the gauge field then
do not belong to a subalgebra of the $\mathfrak{u}(N)$ Lie algebra
(both endowed with the standard commutator). This is not inconsistent 
because in Section \ref{sec:NCYMgaugegroups} we have seen 
that the noncommutative gauge field and gauge transformations
are not Lie algebra valued anyway. 
The subgroup is defined by setting up constraints on the gauge field and 
gauge transformation parameter.  Their construction works for $SO(N)$ and 
$Sp(N)$ 
subgroups, because in these cases an anti-automorphism of the noncommutative 
algebra of functions can be used to formulate the required constraints.
This approach is exact in the noncommutativity parameter
$\theta^{\mu\nu}$. For vanishing $\theta^{\mu\nu}$ one recovers the 
ordinary $SO(N)$ and $Sp(N)$ gauge theories.
It is a disadvantage of this approach that the formulation of the 
anti-automorphism
requires one to interpret the elements of the algebra not only as spacetime
dependent but also as functions of the noncommutativity parameter
$\theta^{\mu\nu}$, which then is treated as a variable and not as a (constant)
parameter. But if one allows for an expansion in $\theta^{\mu\nu}$ then the
dependence on $\theta^{\mu\nu}$ has an 
interpretation in the context of the Seiberg-Witten map \eqref{SWmapansatz}
where the noncommutative gauge fields are indeed given by a power series in
$\theta^{\mu\nu}$. 
The authors find that their constraint translates to the condition that the 
ordinary gauge field $a_\mu$, which is mapped to the noncommutative gauge
field $A_\mu$, takes values in the corresponding ordinary Lie subalgebra.
The $\theta^{\mu\nu}$-expanded version of this constraint thus is in perfect 
agreement with the enveloping algebra approach. 
Some more details about setting up the constraint and about its relation to
the enveloping algebra approach can be found in Appendix 
\ref{antiautoconstraints}.
For the $SU(N)$ case an alternative constraint has been
proposed in \cite{Chu:2001if}.

For completeness let us remark that the authors of \cite{Bonora:2000td}
furthermore construct a string theory 
setup from which their theories follow in the Seiberg-Witten limit \eqref{SWlimit}.
The conditions on the gauge fields and gauge transformations which are
formulated with the anti-automorphism correspond to an orientifold projection. 
The string theory background is given by $D_p$-branes on top of an
orientifold plane with a constant $B$-field which is parallel to and possesses
opposite signs on both sides of the orientifold plane. 

The above analysis was refined in \cite{Bars:2001iq} where the authors 
use a modified version of the anti-automorphism 
of \cite{Bonora:2000td} that allows for a relaxation of the condition that
the elements of the algebra depend on $\theta^{\mu\nu}$. On the string side
the authors discuss stable (supersymmetric) background configurations of
$\text{D}p$-branes  and orientifold planes that correspond to the constructed
gauge theories.

In \cite{Bonora:2001ny} the authors addressed the 
problem of how to compute 1-loop amplitudes in noncommutative $SO(N)$ gauge
theories from string theory. The setup they use corresponds to the setup in
\cite{Bonora:2000td} that we explained above.
After a review of the techniques to calculate the annulus and M\"obius
contribution in the ordinary case the authors extend their analysis to the 
noncommutative case. They find that the $\theta$-dependence is given 
by a simple factor multiplying the 1-loop amplitude (at $\theta^{\mu\nu}=0$) 
which is
the same behavior already found at tree level 
\cite{Bonora:2000td,Seiberg:1999vs}, see \eqref{VOcorrthetadependence}. 
The authors conjecture Feynman rules for the 
noncommutative $SO(N)$ theory from the tree level amplitudes. These rules 
are \emph{exact} in $\theta$ and therefore should be suitable for the study
of effects that require the full $\theta$-dependence (like UV/IR mixing
\cite{Minwalla:1999px}, see the Introduction and the 
end of Section \ref{sec:FrulesNCYM}). 
They observe, however, that there is a mismatch between the field theory
limit of the string 1-loop amplitudes and the field theory computation 
using these Feynman rules. 

\section[Feynman rules for NCYM theories with gauge groups $G\neq
U(N)$]{Feynman rules for noncommutative Yang-Mills theories with gauge groups $G\neq U(N)$}
\label{FrulesggG}
In the previous Section we have described in brief two proposals for a 
formulation of noncommutative YM theories with gauge groups $G\neq U(N)$.
In the enveloping algebra approach described in Subsection \ref{subsec:envalg},
it was found that the coefficients of the enveloping algebra valued basis elements are related to the Lie algebra valued terms via the Seiberg-Witten map
\eqref{SWmapansatz}. The noncommutative gauge field and gauge transformation
parameter of a gauge theory with group $G$ are the Seiberg-Witten map of the
ordinary gauge field and parameter that take values in $\mathfrak{g}$, the Lie
algebra of $G$. On the other hand the approach described in Subsection
\ref{subsec:antiautoconstraint} imposes  
constraints on the noncommutative $U(N)$ gauge field and parameter
to define gauge theories with groups $G\subset U(N)$. Using the Seiberg-Witten map to translate these constraints to the ordinary fields and parameter 
leads exactly to the same result as found in the enveloping algebra approach:
the ordinary fields and parameter take values in $\mathfrak{g}$ which then is 
a subalgebra of $\mathfrak{u}(N)$. 

Since the first approach is \emph{perturbative} 
in $\theta^{\mu\nu}$ and defines a
theory in terms of the ordinary fields, whereas the second is \emph{exact} in
$\theta^{\mu\nu}$ and uses the noncommutative fields, it is an interesting 
question to ask how both quantum field theoretical formulations are related.
We want to see what happens if one tries to resum the 
$\theta^{\mu\nu}$-expansion in the enveloping algebra approach. 
In particular it is important to clarify if and
how the Feynman rules for such a theory can be constructed and if one can
confirm the rules of \cite{Bonora:2001ny}. If they are consistent one should
be able to reproduce them because they were taken from the framework of 
\cite{Bonora:2000td} which is compatible with the enveloping algebra approach.
 
The aim of this Section, which is based on our work \cite{Dorn:2002ah}, 
is to get information on Feynman rules for 
noncommutative gauge theories with gauge groups $G\neq U(N)$ that include the
exact dependence on $\theta^{\mu\nu}$ and especially to decide if the rules for
$SO(N)$ given in \cite{Bonora:2001ny} are consistent. 
In principle one should work with 
the approach of Subsection \ref{subsec:antiautoconstraint} that uses a
constraint.
We have seen that this constraint has the interpretation that
under  the Seiberg-Witten map the ordinary fields and parameter are restricted
to a subalgebra of $\mathfrak{g}$. One can therefore resolve the constraint by
using the formulation of the theory in the ordinary fields which is obtained
by use of the full Seiberg-Witten map (exact in $\theta^{\mu\nu}$). 
Since the latter is not known in the 
non-Abelian case one is forced to work with its $\theta$-expansion, i.\ e.\ the
constraint is perturbative in $\theta$. 
We will show that one can nevertheless extract statements from such a setup
that are universal, i.\ e.\ that do not depend on the order to which the
constraint has been expanded. For this one has to analyze a resummation of
the $\theta$-expansion.
To be more precise the formulation is as follows:
first express the noncommutative $U(N)$ gauge field $A_\mu$ and
gauge transformation $\Lambda$ via the Seiberg-Witten map \eqref{SWmaplin}
in terms of the ordinary $U(N)$ gauge field $a_\mu$ and gauge
transformation $\lambda$, respectively. After that both $a_\mu$ and $\lambda$
are constrained to take values in the Lie subalgebra
$\mathfrak{g}\subset\mathfrak{u}(N)$ of $G\subset U(N)$.
The gauge theory with gauge group $G$ is now defined by the noncommutative 
Yang-Mills action \eqref{NCYMaction} together with the corresponding gauge
transformations \eqref{NCgaugetrafoWop}, \eqref{NCgaugetrafoinfinit} 
\emph{and} the constraint that
\eqref{SWmapansatz} is valid with $a_\mu,\lambda\in\mathfrak{g}$. 
After choosing the Lorenz gauge condition and the Feynman gauge
($\kappa=1$) we have from \eqref{NCLYMLGFLFPG} the complete
setup
\begin{equation}\label{NCSYMSGFSFPG}
S[A,C,\bar C]=-\frac{1}{2g^2}\int\de^dx\,\tr\big(F_{\mu\nu}F^{\mu\nu}+(\partial^\mu A_\mu)^2\big)-2\int\de^dx\,\tr\big(\bar C\partial^\mu(\partial_\mu C+i\astcomm{C}{A_\mu})\big)\col
\end{equation}
where according to \eqref{SWmaplin} and
\eqref{SWmapghostlin} one has
\begin{equation}\label{ALambdaCbarCSWmaplin}
\begin{aligned}
A_\mu[a]&=a_\mu-\frac{1}{4}\theta^{\alpha\beta}\acomm{a_\alpha}{\partial_\beta
  a_\mu+f_{\beta\mu}}+\mathcal{O}(\theta^2)\col\\
\Lambda[\lambda,a]&=\lambda+\frac{1}{4}\theta^{\alpha\beta}\acomm{\partial_\beta\lambda}{a_\alpha}+\mathcal{O}(\theta^2)\col\\
C[c,a]&=c+\frac{1}{4}\theta^{\alpha\beta}\acomm{\partial_\beta
  c}{a_\alpha}+\mathcal{O}(\theta^2)\col\\
\bar C&=c\pnt
\end{aligned}
\end{equation}
Inserting these transformation formulas into \eqref{NCSYMSGFSFPG} then leads
to an action for the ordinary gauge fields $a_\mu$ and ghosts $c$, $\bar c$
which is a power series in $\theta^{\mu\nu}$ and one can determine the Feynman
rules. Besides the standard
propagators and vertices like in the $U(N)$ theory in
Fig.\ \ref{fig:NCUNYMFrules} one has in general an
infinite set of additional vertices with an increasing number of legs,
derivatives, and powers of $\theta^{\mu\nu}$. For our further discussion it is
useful to stress that all these 
vertices are generated by the $\theta^{\mu\nu}$-expansion of the whole action, 
i.\ e.\ both the noncommutative kinetic and interaction term.
In the following we call this kind of perturbation theory the
$\theta$-{\it expanded perturbation theory} for the noncommutative
$G$ gauge theory. It is extensively studied in the references of Subsection
\ref{subsec:envalg}.  
On the other side for the $U(N)$ case it is straightforward to get directly
from \eqref{NCSYMSGFSFPG} Feynman rules in terms of
$A_\mu$, $C$ and $\bar C$ as already discussed in Section \ref{sec:FrulesNCYM}.
They are \emph{exact} in $\theta^{\mu\nu}$.
We will now follow the strategy:
\begin{enumerate}
\item
In Subsection \ref{subsec:pintconstrainedtheo} we will introduce the 
path integral formulation for the constrained theory and show how the 
constraint is resolved by a variable transformation inside the path
integral.
\item In Subsection \ref{subsec:soneperttheo} we will then extract the
  $\theta$-expanded Feynman
  rules for this formulation and discuss the behaviour of the theory under 
resummation of the $\theta$-expansion for general gauge groups $G$ and 
compare with the case of $U(N)$. 
\item In Subsection \ref{subsec:nptproof} we will prove that for $G\neq
  U(N)$ one generates infinitely many connected $n$-point Green functions and 
one therefore cannot define a consistent set (a finite number) of Feynman 
rules. 
\item In Subsection \ref{subsec:restrictedsource} we will discuss a
  counterexample that the Feynman rules for $SO(N)$ given in
  \cite{Bonora:2001ny} cannot be derived from the theory.

\end{enumerate}

\subsection{Path integral quantization of the constrained theory}
\label{subsec:pintconstrainedtheo}

We will now formulate the $\theta$-expanded perturbation theory and define 
our task in precise technical terms. The original noncommutative interactions
are kept as some of the vertices of our wanted Feynman
rules. The theory is summed with
respect to the vertices generated by the expansion of the kinetic term
only. We start with \eqref{NCSYMSGFSFPG}, the noncommutative $U(N)$ YM-theory
in Feynman gauge described with the noncommutative
gauge field $A_\mu$ and Faddeev-Popov ghosts $C$ and $\bar C$
 and separate 
\begin{equation}\label{7}
S[A,C,\bar C]=S_\text{kin}[A,C,\bar C]+S_I[A,C,\bar C]\col
\end{equation}
with
\begin{equation}\label{8}
S_\text{kin}[A,C,\bar C]=-\frac{1}{g^2}\int\de^dx\,\tr \partial_\mu
A_\nu
\partial^\mu A^\nu-\int\de^dx\,\partial_\mu\bar C\partial^\mu C\pnt
\end{equation}
According to \eqref{genfuncdef} the generating functional for noncommutative $G$ Green functions is given by
\begin{equation}\label{9}
Z_G[J,\bar{\eta},\eta]=\int_{a,c,\bar c\in\mathfrak{g}}\mathcal{D}A\,\mathcal{D}\bar C\,\mathcal{D}C\,\e^{i(S[A,C,\bar C ]+
A\cdot J+\bar{\eta}\cdot C+\bar C\cdot\eta )}\col
\end{equation}
where we have introduced sources $J$, $\eta$ and $\bar\eta$ for the gauge
field, ghost and anti-ghost respectively.
By the notation $\int_{a,c,\bar c\in\mathfrak{g}}$ we indicate the
integration over $A$, $C$, $\bar C$ with the constraint that their image
under the inverse Seiberg-Witten map is in $\mathfrak{g}$, i.\ e.\ $a,c,\bar c
\in\mathfrak{g}$. From now on we will work with the $N\times N$ matrix
representation of $\mathfrak{u}(N)$ that fulfills \eqref{Liebase}.
For $U(N)$ the constraint is trivially solved by $A_\mu=A^A_\mu T_A$ and free
integration over $A^A_\mu$, $C^A$, $\bar C^A$.

To explore the possibility
of noncommutative $G$ Feynman rules, which after some possible
projection work with the $U(N)$ vertices, we write \eqref{9} using \eqref{7}
as (see \eqref{genfuncpotextract})
\begin{equation}\label{10}
Z_G[J,\bar{\eta},\eta ]=\e^{iS_I[\prefuncderivi{J},\prefuncderivi{\bar{\eta}},\prefuncderivi{\eta}]}Z^\text{kin}_G
[J,\bar{\eta},\eta ]\col
\end{equation}
with
\begin{equation}\label{11}
Z^\text{kin}_G[J,\bar{\eta},\eta ]=\int_
{a,c,\bar c\in\mathfrak{g}}\mathcal{D}A\,\mathcal{D}\bar
C\,\mathcal{D}C\,\e^{i(S_\text{kin}[A,C,\bar C]+ A\cdot J+\bar{\eta}\cdot C+\bar
  C\cdot\eta )}\pnt
\end{equation} 
Denoting by $\mathcal{J}$ the functional determinant for changing the
integration variables from $A$, $C$, $\bar C$ to $a$, $c$, $\bar c$ we get 
\begin{equation}\label{12}
Z^\text{kin}_G[J,\bar{\eta},\eta ]=\int_{a,c,\bar
  c\in\mathfrak{g}}\mathcal{D}a\,\mathcal{D}\bar
c\,\mathcal{D}c\,\mathcal{J}[a,c,\bar c ]\e^{i(S_\text{kin}[a,c,\bar
  c]+s_1[a,c,\bar c ]+A[a]\cdot J+\bar{\eta}\cdot C[c,a]+\bar c\cdot\eta )}\pnt
\end{equation}
The new quantity $s_1[a,c,\bar c]$ appearing above is defined via \eqref{ALambdaCbarCSWmaplin} and \eqref{8} by
\begin{equation}\label{13}
S_\text{kin}[A[a],C[c,a],\bar c]=S_\text{kin}[a,
c,\bar c]+s_1[a,c,\bar c]\pnt
\end{equation}
Applying \eqref{cgenfuncdef} and \eqref{cGreenfuncdef} here, 
the logarithm of \eqref{12} divided by $Z_G^\text{kin}[0,0,0]$ 
is the generating functional for the connected
Green functions of the composites $A$, $C$, $\bar C$ in the field theory
with elementary fields $a,c,\bar c$ interacting via
$s_1-i\ln\mathcal{J}$. Therefore, it can be represented by 
(see \eqref{cGreengenfuncexpansion}) 
\begin{equation}\label{13a}
\ln\frac{Z_G^\text{kin}[J,\bar{\eta},\eta ]}{Z_G^\text{kin}[0]}=
\sum_n\frac{i^n}{n!}\int\de^dx_1\dots\de^dx_n\,
\langle A(x_1)\dots A(x_n)\rangle^\text{kin}_\text{c}J(x_1)\dots J(x_n)+\dots\col
\end{equation}
where $\langle A(x_1)\dots A(x_n)\rangle^\text{kin}_\text{c}$
stands for the $n$-point connected Green function of $A_\mu$ in this field
theory. The dots at the end represent the corresponding ghost and
mixed ghost and gauge field terms. 

Neglecting $\mathcal{J}$ (a justification will follow in Subsection \ref{subsec:soneperttheo}) these are just the
Green functions for the composites $A$, $C$, $\bar C$ obtained within
$\theta $-expanded perturbation theory by partial summation of all diagrams
built with vertices generated by the $\theta$-expansion of the noncommutative
kinetic term only.

In the $U(N)$-case $Z^\text{kin}_{U(N)}[J,\bar{\eta},\eta]$
as given by \eqref{11} is a trivial Gaussian integral and one obtains 
an expression similar to \eqref{genfunckin}. It is the generating
functional of Green functions for $A$, $C$, $\bar C$ treated as free fields.
Then in \eqref{13a} only the two point functions $\langle
AA\rangle^\text{kin}_\text{c}$ and 
$\langle C\bar C\rangle^\text{kin}_\text{c}$ are different from zero. In
addition they are equal to the free propagators as discussed in Section
\ref{sec:FrulesNCYM} (see \eqref{YMtwopoint}, \eqref{ghosttwopoint}). 

Starting with free fields and imposing a constraint in the generic case
generates an interacting theory. We want to decide what happens in our
case \eqref{11} for $G\subset U(N)$.  
By some special circumstance it could be that only the connected two-point
functions are modified. Another less restrictive possibility would be
that connected $n$-point functions beyond some finite $n_0>2$
vanish. In both 
cases from \eqref{10} we would get Feynman rules with a finite number
of building blocks.

For $U(N)$ the equivalent representation \eqref{12} is due to
a simple field redefinition of a free theory. Therefore, looking at
the $n$-point functions of the (in terms of $a,c,\bar c$, see
\eqref{ALambdaCbarCSWmaplin}) composite operators $A$, $C$, $\bar C$ the summation
of the perturbation theory with respect to $s_1[a,c,\bar c]-i\ln\mathcal{J}$
must yield the free field result guaranteed by
\eqref{11}.\footnote{This is a manifestation of the equivalence
  theorem \cite{Kamefuchi:1961sb}.}    

On the other side for $G\subset U(N)$ we cannot directly evaluate
\eqref{11} and are forced to work with \eqref{12}. It will turn out
to be useful to study both $U(N)$ and $G\subset U(N)$ in parallel.
Since the result for $U(N)$ is a priori known, one has some checks
for the calculations within the $s_1$-perturbation theory. 

\subsection{$s_1$-perturbation theory for $U(N)$ and $G\subset U(N)$}
\label{subsec:soneperttheo}
In both cases our gauge fields take values in the $N\times N$ matrix 
representation of the Lie 
algebra $\mathfrak{u}(N)$. 
We write
\begin{equation}\label{14}
A_\mu=A_\mu^AT_A\col
\end{equation}
where we have used \eqref{Liebase} for the generators $T^A$ of
$\mathfrak{u}(N)$. 
Then \eqref{ALambdaCbarCSWmaplin} implies
\begin{equation}\label{ACbarCSWmaplincomponents}
\begin{aligned}
A_\mu^M&=a_\mu^M-\frac{1}{2}\theta^{\alpha\beta}a_\alpha^P\partial_{\beta
}a_\mu^Q\dddu{P}{Q}{M}+\frac{1}{4}\theta^{\alpha\beta}a_\alpha^P\partial_\mu
a_{\beta}^Q\dddu{P}{Q}{M}
-\frac{1}{4}\theta^{\alpha\beta}a_\alpha^Pa_{\beta}^Qa_\mu^R\duud{M}{S}{P}f_{SQR}+O(\theta^2)\col\\
C^M&=c^M+\frac{1}{4}\theta^{\alpha\beta}\partial_\alpha c^Pa_{\beta }^Q
\dddu{P}{Q}{M}+O(\theta^2)\\
\bar C^M&=\bar c^M\pnt
\end{aligned}
\end{equation}
In the case $G\subset U(N)$, $G\neq U(M)$, $M<N$ \footnote{In the
  following we sometimes implicitly understand that $G\subset U(N)$
  excludes $U(M)$ subgroups.} we indicate the {\it
  generators spanning the Lie algebra $\mathfrak{g}$ of $G$} with a 
{\it lower case}
Latin index and the {\it remaining} ones with a {\it primed lower
  case} Latin index. Upper case Latin indices run over all $U(N)$ generators,
  see \eqref{Liebase} and \eqref{Liesubalgebrarel}.
Since $\mathfrak{g}$ is the Lie subalgebra of $\mathfrak{u}(N)$ that
  corresponds to the subgroup $G\subset U(N)$ we have according to
  \eqref{Liesubalgebrarel}
\begin{equation}\label{subalstruc}
f_{abc'}=0\col\quad\forall a,b,c'\qquad\text{and}\qquad d_{abc'}\neq
0\quad\text{for some }a,b,c'
\pnt
\end{equation}
As discussed in the previous Section, the noncommutative $G$ gauge
field theory is then defined by unconstrained functional integration
over $a_\mu^b,c^b,\bar c^b$ and by the condition
\begin{equation}\label{19}
a_\mu^{b'}=c^{b'}=\bar c^{b'}=0\pnt
\end{equation}
In spite of \eqref{19} via \eqref{ACbarCSWmaplincomponents} with \eqref{subalstruc} one has non-vanishing
$A_\mu^{b'}$ and $C^{b'}$.\\ 

We are interested in \eqref{12}, i.\ e.\ the Green functions of $A$, $C$, $\bar C$,
which are composites in terms of $a,c,\bar c$. For the diagrammatic
evaluation one gets from \eqref{ACbarCSWmaplincomponents} the $external$ vertices where all
momenta are directed to the interaction point, and a slash denotes a 
derivative of the field at the corresponding leg. (We write down the
$\propto\theta^0$ and $\propto\theta^1$ contributions only. Momentum
conservation at all vertices is understood.) 
\settoheight{\eqoff}{\fbox{$=$}}
\setlength{\eqoff}{0.5\eqoff}
\addtolength{\eqoff}{-40pt}
\begin{equation}
\raisebox{\eqoff}{%
\begin{picture}(120,80)(0,0)\scriptsize
\Vertex(40,40){1}\Text(30,40)[r]{$p,\mu,M$}
\Line(40,40)(80,40)\Text(90,40)[l]{$k,\alpha,A$}
\end{picture}}
=
\begin{cases}
\delta^\alpha_\mu\delta^M_A & \text{for $M=m$}\\
0 & \text{for $M=m'$}
\end{cases}\col
\label{20}
\end{equation}
\begin{equation}
\raisebox{\eqoff}{%
\begin{picture}(120,80)(0,0)\scriptsize
\Vertex(40,40){1}\Text(30,40)[r]{$p,\mu,M$}
\Line(40,40)(74.641,60)\Text(84.641,60)[l]{$k_2,\beta,B$}
\Line(40,40)(74.641,20)\Text(84.641,20)[l]{$k_1,\alpha,A$}
\Line(43.08,35.335)(45.58,39.665)
\end{picture}}
=i\left(\frac{1}{4}\theta^{\beta\alpha}\delta_\mu^\nu
-\frac{1}{2}\theta^{\beta\nu}\delta_\mu^\alpha\right)\dddu{A}{B}{M}(k
_1)_\nu\col
\label{21}
\end{equation}
\begin{equation}
\raisebox{\eqoff}{%
\begin{picture}(120,80)(0,0)\scriptsize
\Vertex(40,40){1}\Text(30,40)[r]{$p,\mu,M$}
\Line(40,40)(74.641,60)\Text(84.641,60)[l]{$k_3,\gamma,C$}
\Line(40,40)(80,40)\Text(90,40)[l]{$k_2,\beta,B$}
\Line(40,40)(74.641,20)\Text(84.641,20)[l]{$k_1,\alpha,A$}
\end{picture}}
=-\frac{1}{4}\theta^{\alpha\beta}\duud{M}{E}{A}f_{EBC}\delta^\gamma_\mu\col
\label{22}
\end{equation}
and
\begin{equation}
\settoheight{\eqoff}{\fbox{$=$}}
\setlength{\eqoff}{0.5\eqoff}
\addtolength{\eqoff}{-40pt}
\raisebox{\eqoff}{%
\begin{picture}(140,80)(0,0)\scriptsize
\Vertex(40,40){1}\Text(30,40)[r]{$p,M$}
\DashArrowLine(80,40)(40,40){1}\Text(90,40)[l]{$k,\alpha,A$}
\end{picture}}
=
\raisebox{\eqoff}{%
\begin{picture}(120,80)(0,0)\scriptsize
\Vertex(40,40){1}\Text(30,40)[r]{$p,M$}
\DashArrowLine(40,40)(80,40){1}\Text(90,40)[l]{$k,A$}
\end{picture}}
=
\begin{cases}
\delta^M_A & \text{for $M=m$}\\
0 & \text{for $M=m'$}
\end{cases}
\col
\label{23}
\end{equation}
\begin{equation}
\raisebox{\eqoff}{%
\begin{picture}(120,80)(0,0)\scriptsize
\Vertex(40,40){1}\Text(30,40)[r]{$p,M$}
\Line(40,40)(74.641,60)\Text(84.641,60)[l]{$k_2,\beta,B$}
\DashArrowLine(74.641,20)(40,40){1}\Text(84.641,20)[l]{$k_1,A$}
\Line(43.08,35.335)(45.58,39.665)
\end{picture}}
=\frac{i}{4}\theta^{\nu\beta}\dddu{A}{B}{M}(k_1)_\nu\pnt
\label{24}
\end{equation}
The insertion of \eqref{ACbarCSWmaplincomponents} into \eqref{13} yields
$s_1[a,c,\bar c]$ generating the {\it internal} vertices\\
\settoheight{\eqoff}{\fbox{$=$}}
\setlength{\eqoff}{0.5\eqoff}
\addtolength{\eqoff}{-40pt}
\begin{equation}\label{kinthreevertex}
\raisebox{\eqoff}{%
\begin{picture}(150,80)(0,0)\scriptsize
\SetOffset(35,0)
\Line(0,40)(40,40)\Text(-5,40)[r]{$k_1,\alpha,A$}
\Line(32.5,37.5)(32.5,42.5)
\Line(35,37.5)(35,42.5)
\Vertex(40,40){1}
\Line(40,40)(74.641,60)\Text(84.641,60)[lb]{$k_3,\gamma,C$}
\Line(40,40)(74.641,20)\Text(84.641,20)[lt]{$k_2,\beta,B$}
\Line(43.08,35.335)(45.58,39.665)
\end{picture}}
=\frac{1}{g^2}\left(\frac{1}{4}\theta^{\gamma\beta}G^{\nu\alpha}-\frac{1}{2}\theta^{\gamma\nu}G^{\alpha\beta}\right)d_{ABC}k_1^2(k_2)_\nu
\col
\end{equation}
\begin{equation}\label{kinfourvertex}
\raisebox{\eqoff}{%
\begin{picture}(155,80)(0,0)\scriptsize
\SetOffset(35,0)
\Line(0,40)(40,40)\Text(-5,40)[r]{$k_1,\alpha,A$}
\Line(32.5,37.5)(32.5,42.5)
\Line(35,37.5)(35,42.5)
\Vertex(40,40){1}
\Line(40,40)(74.641,60)\Text(84.641,60)[lb]{$k_4,\delta,D$}
\Line(40,40)(80,40)\Text(90,40)[l]{$k_3,\gamma,C$}
\Line(40,40)(74.641,20)\Text(84.641,20)[lt]{$k_2,\beta,B$}
\end{picture}}
=\frac{i}{4g^2}\theta^{\delta\gamma}\dddu{A}{D}{E}f_{ECB}G^{\alpha\beta} k_1^2\col
\end{equation}
\begin{equation}\label{kingaugeghostvertex}
\raisebox{\eqoff}{%
\begin{picture}(150,80)(0,0)\scriptsize
\SetOffset(35,0)
\DashArrowLine(40,40)(0,40){1}\Text(-5,40)[r]{$k_1,A$}
\Line(32.5,37.5)(32.5,42.5)
\Line(35,37.5)(35,42.5)
\Vertex(40,40){1}
\DashArrowLine(74.641,60)(40,40){1}\Text(84.641,60)[lb]{$k_3,C$}
\Line(40,40)(74.641,20)\Text(84.641,20)[lt]{$k_2,\beta,B$}
\Line(43.08,44.6655)(45.58,40.335)
\end{picture}}
=-\frac{1}{4}\theta^{\beta\nu}d_{ABC}k_1^2(k_3)_\nu\pnt
\end{equation}
The double slash stems from the derivatives in \eqref{8} after
partial integration and denotes the action of
$\Box=\partial_\mu\partial^\mu$ at the corresponding leg.

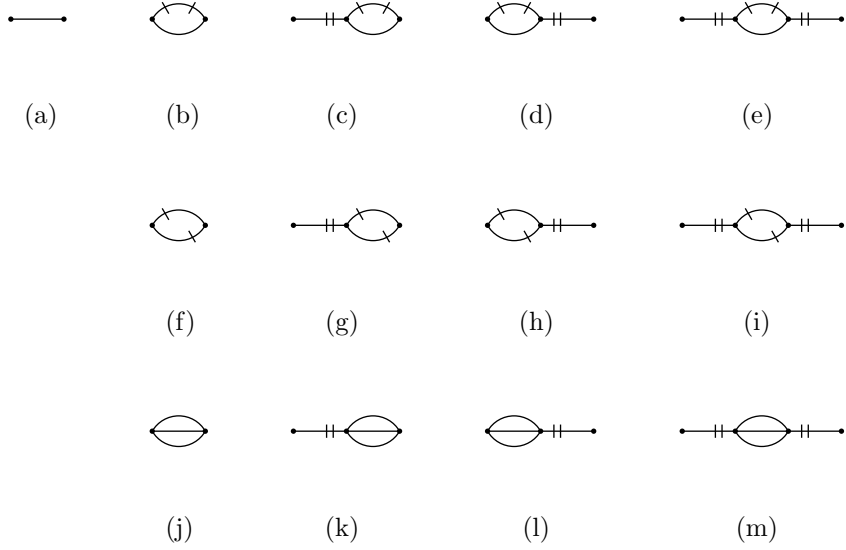
\begin{figure}
\begin{center}
\subfigure[]{%
\begin{picture}(30,30)(0,0)\scriptsize
\Vertex(5,15){1}
\Vertex(25,15){1}
\Line(5,15)(25,15)
\end{picture}\label{one}}\qquad
\subfigure[]{%
\begin{picture}(30,30)(0,0)\scriptsize
\Vertex(5,15){1}
\Vertex(25,15){1}
\CArc(15,9.2265)(11.547,30,150)
\Line(11.25,17.0615)(8.75,21.3915)
\Line(18.75,17.0615)(21.25,21.3915)
\CArc(15,20.7735)(11.547,-150,-30)
\end{picture}\label{two}}\qquad
\subfigure[]{%
\begin{picture}(50,30)(0,0)\scriptsize
\SetOffset(20,0)
\Vertex(-15,15){1}
\Line(-15,15)(5,15)
\Line(0,12.5)(0,17.5)
\Line(-2.5,12.5)(-2.5,17.5)
\Vertex(5,15){1}
\Vertex(25,15){1}
\CArc(15,9.2265)(11.547,30,150)
\Line(11.25,17.0615)(8.75,21.3915)
\Line(18.75,17.0615)(21.25,21.3915)
\CArc(15,20.7735)(11.547,-150,-30)
\end{picture}\label{three}}\qquad
\subfigure[]{%
\begin{picture}(50,30)(0,0)\scriptsize
\SetOffset(30,0)
\Vertex(15,15){1}
\Line(15,15)(-5,15)
\Line(0,12.5)(0,17.5)
\Line(2.5,12.5)(2.5,17.5)
\Vertex(-5,15){1}
\Vertex(-25,15){1}
\CArc(-15,9.2265)(11.547,30,150)
\Line(-11.25,17.0615)(-8.75,21.3915)
\Line(-18.75,17.0615)(-21.25,21.3915)
\CArc(-15,20.7735)(11.547,-150,-30)
\end{picture}\label{four}}\qquad
\subfigure[]{%
\begin{picture}(70,30)(0,0)\scriptsize
\SetOffset(20,0)
\Vertex(-15,15){1}
\Line(-15,15)(5,15)
\Line(0,12.5)(0,17.5)
\Line(-2.5,12.5)(-2.5,17.5)
\Vertex(5,15){1}
\Vertex(25,15){1}
\CArc(15,9.2265)(11.547,30,150)
\Line(11.25,17.0615)(8.75,21.3915)
\Line(18.75,17.0615)(21.25,21.3915)
\CArc(15,20.7735)(11.547,-150,-30)
\Vertex(45,15){1}
\Line(25,15)(45,15)
\Line(30,12.5)(30,17.5)
\Line(32.5,12.5)(32.5,17.5)
\end{picture}\label{five}}\\
%
%
\begin{picture}(30,30)(0,0)\scriptsize
\end{picture}\qquad
\subfigure[]{%
\begin{picture}(30,30)(0,0)\scriptsize
\Vertex(5,15){1}
\Vertex(25,15){1}
\CArc(15,9.2265)(11.547,30,150)
\Line(11.25,17.0615)(8.75,21.3915)
\CArc(15,20.7735)(11.547,-150,-30)
\Line(18.75,12.9385)(21.25,8.6085)
\end{picture}\label{six}}\qquad
\subfigure[]{%
\begin{picture}(50,30)(0,0)\scriptsize
\SetOffset(20,0)
\Vertex(-15,15){1}
\Line(-15,15)(5,15)
\Line(0,12.5)(0,17.5)
\Line(-2.5,12.5)(-2.5,17.5)
\Vertex(5,15){1}
\Vertex(25,15){1}
\CArc(15,9.2265)(11.547,30,150)
\Line(11.25,17.0615)(8.75,21.3915)
\CArc(15,20.7735)(11.547,-150,-30)
\Line(18.75,12.9385)(21.25,8.6085)
\end{picture}\label{seven}}\qquad
\subfigure[]{%
\begin{picture}(50,30)(0,0)\scriptsize
\SetOffset(30,0)
\Vertex(15,15){1}
\Line(15,15)(-5,15)
\Line(0,12.5)(0,17.5)
\Line(2.5,12.5)(2.5,17.5)
\Vertex(-5,15){1}
\Vertex(-25,15){1}
\CArc(-15,9.2265)(11.547,30,150)
\Line(-18.75,17.0615)(-21.25,21.3915)
\CArc(-15,20.7735)(11.547,-150,-30)
\Line(-11.25,12.9385)(-8.75,8.6085)
\end{picture}\label{eight}}\qquad
\subfigure[]{%
\begin{picture}(70,30)(0,0)\scriptsize
\SetOffset(20,0)
\Vertex(-15,15){1}
\Line(-15,15)(5,15)
\Line(0,12.5)(0,17.5)
\Line(-2.5,12.5)(-2.5,17.5)
\Vertex(5,15){1}
\Vertex(25,15){1}
\CArc(15,9.2265)(11.547,30,150)
\Line(11.25,17.0615)(8.75,21.3915)
\CArc(15,20.7735)(11.547,-150,-30)
\Line(18.75,12.9385)(21.25,8.6085)
\Vertex(45,15){1}
\Line(25,15)(45,15)
\Line(30,12.5)(30,17.5)
\Line(32.5,12.5)(32.5,17.5)
\end{picture}\label{nine}}\\
%
%
\begin{picture}(30,30)(0,0)\scriptsize
\end{picture}\qquad
\subfigure[]{%
\begin{picture}(30,30)(0,0)\scriptsize
\Vertex(5,15){1}
\Vertex(25,15){1}
\CArc(15,9.2265)(11.547,30,150)
\CArc(15,20.7735)(11.547,-150,-30)
\Line(5,15)(25,15)
\end{picture}\label{ten}}\qquad
\subfigure[]{%
\begin{picture}(50,30)(0,0)\scriptsize
\SetOffset(20,0)
\Vertex(-15,15){1}
\Line(-15,15)(5,15)
\Line(0,12.5)(0,17.5)
\Line(-2.5,12.5)(-2.5,17.5)
\Vertex(5,15){1}
\Vertex(25,15){1}
\CArc(15,9.2265)(11.547,30,150)
\CArc(15,20.7735)(11.547,-150,-30)
\Line(5,15)(25,15)
\end{picture}\label{eleven}}\qquad
\subfigure[]{%
\begin{picture}(50,30)(0,0)\scriptsize
\SetOffset(30,0)
\Vertex(15,15){1}
\Line(15,15)(-5,15)
\Line(0,12.5)(0,17.5)
\Line(2.5,12.5)(2.5,17.5)
\Vertex(-5,15){1}
\Vertex(-25,15){1}
\CArc(-15,9.2265)(11.547,30,150)
\CArc(-15,20.7735)(11.547,-150,-30)
\Line(-5,15)(-25,15)
\end{picture}\label{twelve}}\qquad
\subfigure[]{%
\begin{picture}(70,30)(0,0)\scriptsize
\SetOffset(20,0)
\Vertex(-15,15){1}
\Line(-15,15)(5,15)
\Line(0,12.5)(0,17.5)
\Line(-2.5,12.5)(-2.5,17.5)
\Vertex(5,15){1}
\Vertex(25,15){1}
\CArc(15,9.2265)(11.547,30,150)
\CArc(15,20.7735)(11.547,-150,-30)
\Line(5,15)(25,15)
\Vertex(45,15){1}
\Line(25,15)(45,15)
\Line(30,12.5)(30,17.5)
\Line(32.5,12.5)(32.5,17.5)
\end{picture}\label{thirteen}}
\end{center}
\caption{Contributions to $\langle A_\mu^MA_\nu^N\rangle^\text{\rm
    kin}_\text{\rm c}$ up to order $\theta^2$} 
\label{propcorr}
\end{figure}

\settoheight{\eqoff}{\fbox{$=$}}
\setlength{\eqoff}{0.5\eqoff}
\addtolength{\eqoff}{-15pt}
\begin{figure}
\begin{center}
\raisebox{\eqoff}{%
\begin{picture}(35,30)(0,0)\scriptsize
\SetOffset(5,0)
\Vertex(5,15){1}
\Vertex(25,15){1}
\Line(5,15)(0.67,22.5)
\Line(5,15)(9.67,22.5)
\CArc(5,25)(5,-30,210)
\Line(5,15)(25,15)
\end{picture}}
$\qquad,\qquad$
\raisebox{\eqoff}{%
\begin{picture}(50,30)(0,0)\scriptsize
\SetOffset(20,0)
\Vertex(-15,15){1}
\Line(-15,15)(5,15)
\Vertex(5,15){1}
\Vertex(25,15){1}
\CArc(15,9.2265)(11.547,30,150)
\Line(11.25,17.0615)(8.75,21.3915)
\Line(8.4631,15.587)(5.9631,19.917)
\CArc(15,20.7735)(11.547,-150,-30)
\Line(5,15)(25,15)
\end{picture}}
$\qquad,\qquad$
\raisebox{\eqoff}{%
\begin{picture}(70,30)(0,0)\scriptsize
\SetOffset(20,0)
\Vertex(-15,15){1}
\Line(-15,15)(5,15)
\Vertex(5,15){1}
\Vertex(25,15){1}
\DashCArc(15,9.2265)(11.547,30,150){1}
\Line(11.25,17.0615)(8.75,21.3915)
\Line(8.4631,15.587)(5.9631,19.917)
\Line(18.75,17.0615)(21.25,21.3915)
\DashCArc(15,20.7735)(11.547,-150,-30){1}
\Line(11.25,12.9385)(8.75,8.6085)
\Line(18.75,12.9385)(21.25,8.6085)
\Line(21.5369,14.413)(24.0369,10.083)
\Vertex(45,15){1}
\Line(25,15)(45,15)
\end{picture}}
$\qquad,\qquad\dots$
\end{center}
\caption{Some additional vanishing contributions to $\langle
  A_\mu^MA_\nu^N\rangle^\text{\rm kin}_\text{\rm c}$}
\label{detcontrib}
\end{figure}
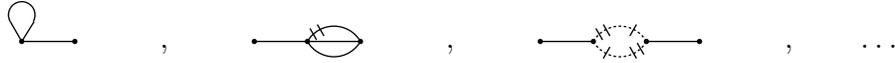

The propagators were given in Fig.\ \ref{fig:NCUNYMFrules} and simplify in Feynman 
gauge ($\kappa=1$) to 
\begin{equation}
-ig^2 G_{\alpha\beta}\delta^{AB}\frac{1}{k^2}\col\qquad
i\delta^{AB}\frac{1}{k^2} 
\label{28} 
\end{equation}
for the commuting gauge field and ghosts, respectively.

Up to now we have not taken into account the functional determinant $\mathcal{J}$
in \eqref{12}. To simplify the analysis we use dimensional regularization.
Then this determinant is equal to one, and all diagrams containing
momentum integrals not depending on any external momentum or mass
parameter (tadpole type)
are zero, see e.\ g.\ \cite{Zinn-Justin:1989mi}. 
For other regularizations these
tadpole type diagrams just cancel the determinant contributions, at least
in the $U(N)$ case.\\

After these preparations we consider the 2-point function  $\langle
A_\mu^MA_\nu^N\rangle^\text{kin}_\text{c}$
within the perturbation theory with respect to $s_1$, see \eqref{12}. 
Fig.\ \ref{propcorr} shows all diagrams
up to order $\theta^2$ which do not vanish in dimensional regularization.
To give an impression which diagrams are absent, Fig.\ \ref{detcontrib}
presents some of them. 

Let us first continue with the $U(N)$ case. Then a straightforward analysis
shows that the diagrams in Fig.\ \ref{two} - \ref{five} cancel among
each other. The same is true for Fig.\ \ref{six} - \ref{nine} and for
Fig.\ \ref{ten} - \ref{thirteen}.
 
\settoheight{\eqoff}{\fbox{$=$}}
\setlength{\eqoff}{0.5\eqoff}
\addtolength{\eqoff}{-40pt}
\begin{figure}
\begin{center}
\raisebox{\eqoff}{%
\begin{picture}(190,80)(0,0)\scriptsize
\SetOffset(35,0)
\Line(0,40)(40,40)\Text(-5,40)[r]{$p,\mu,M$}
\Vertex(0,40){1}
\Line(32.5,37.5)(32.5,42.5)
\Line(35,37.5)(35,42.5)
\Vertex(40,40){1}
\Line(40,40)(74.641,60)\Text(74.641,65)[rb]{$k_2,\beta,B$}
\Line(40,40)(74.641,20)\Text(74.641,15)[rt]{$k_1,\alpha,A$}
\Line(43.08,35.335)(45.58,39.665)
\GCirc(109.282,40){40}{0.7}
\end{picture}}
$\qquad+\qquad$
\raisebox{\eqoff}{%
\begin{picture}(150,80)(0,0)\scriptsize
\SetOffset(-5,0)
\Vertex(40,40){1}\Text(35,40)[r]{$p,\mu,M$}
\Line(40,40)(74.641,60)\Text(74.641,65)[rb]{$k_2,\beta,B$}
\Line(40,40)(74.641,20)\Text(74.641,15)[rt]{$k_1,\alpha,A$}
\Line(43.08,35.335)(45.58,39.665)
\GCirc(109.282,40){40}{0.7}
\end{picture}}
\end{center}
\caption{Canceling graphs from the $\propto a^2$ terms of the
  $\propto\theta$ terms of the SW map} 
\label{cancelmech1}
\end{figure}
\settoheight{\eqoff}{\fbox{$=$}}
\setlength{\eqoff}{0.5\eqoff}
\addtolength{\eqoff}{-40pt}
\begin{figure}
\begin{center}
\raisebox{\eqoff}{%
\begin{picture}(190,80)(0,0)\scriptsize
\SetOffset(35,0)
\Line(0,40)(40,40)\Text(-5,40)[r]{$p,\mu,M$}
\Vertex(0,40){1}
\Line(32.5,37.5)(32.5,42.5)
\Line(35,37.5)(35,42.5)
\Vertex(40,40){1}
\Line(40,40)(74.641,60)
\Line(40,40)(74.641,20)
\Line(40,40)(68.282,40)
\GCirc(109.282,40){40}{0.7}
\end{picture}}
$\qquad+\qquad$
\raisebox{\eqoff}{%
\begin{picture}(150,80)(0,0)\scriptsize
\SetOffset(-5,0)
\Vertex(40,40){1}\Text(35,40)[r]{$p,\mu,M$}
\Line(40,40)(74.641,60)
\Line(40,40)(74.641,20)
\Line(40,40)(68.282,40)
\GCirc(109.282,40){40}{0.7}
\end{picture}}
\end{center}
\caption{Canceling graphs from the $\propto a^3$ terms of the $\propto\theta$ terms of the SW map}
\label{cancelmech2}
\end{figure}

The cancellation mechanism is quite general. Let us denote by
$M(k_1,\alpha ,A\vert k_2,\beta ,B)$ an arbitrary subdiagram with two
marked legs, denoted by a shaded bubble in
Fig.\ \ref{cancelmech1}. Then the sum of the two diagrams in
Fig.\ \ref{cancelmech1} is equal to 
\begin{equation}
\begin{aligned}
&-\frac{ig^2}{p^2}G_{\mu\nu}\delta^{MN}\frac{1}{g^2}\left(\frac{1}{4}
\theta^{\beta\alpha}G^{\lambda\nu}-\frac{1}{2}\theta^{\beta\lambda}G^{\nu\alpha}\right)d_{NAB}p^2(k_1)_{\lambda}M(k_1,\alpha ,A\vert k_2,\beta ,B)\\
&+i\left(\frac{1}{4}\theta^{\beta\alpha}\delta_\mu^{\lambda}-\frac{1}{2}
\theta^{\beta\lambda}\delta^\alpha_\mu\right)\dddu{A}{B}{M}(k_1)_{\lambda}
M(k_1,\alpha ,A\vert k_2,\beta ,B)=0\pnt
\label{29}
\end{aligned}
\end{equation}
A similar general cancellation mechanism holds for diagrams of the
type shown in Fig.\ \ref{cancelmech2}.

Considering the $s_1$-perturbation theory we have convinced ourselves that 
for $U(N)$ the propagator of the composite fields assumes the form
\begin{equation}
\langle A^M_\mu A^N_\nu\rangle^\text{kin}_\text{c}=
-ig^2 G_{\mu\nu}\delta^{MN}\frac{1}{p^2}+O(\theta^3)\pnt
\label{30}
\end{equation}
Of course, from the representation \eqref{11} we know a priori that
there are in all orders of $\theta$ no corrections to the free propagator.
Nevertheless the above exercise was useful, since it unmasked the cancellation
mechanism for Fig.\ \ref{cancelmech1} and Fig.\ \ref{cancelmech2} as
being essential for establishing the 
already known result purely within $s_1$-perturbation theory. It is
straightforward to check also the vanishing of connected $n$-point
functions for $n>2$.\\

What changes if we switch from $U(N)$ to $G\subset U(N)$? First of all, then
we do not know the answer in advance and we have to rely only on
$s_1$-perturbation theory. Secondly, in this perturbation theory the
above cancellation mechanism is no longer present for external points
that carry a primed index and are thus related to elements of  
$\mathfrak{u}(N)$ that are not in $\mathfrak{g}$. 
Then according to \eqref{20} the external vertex
to start with in the first diagrams of Fig.\ \ref{cancelmech1} and
Fig.\ \ref{cancelmech2}  is zero, i.\ e.\ the partners to cancel the second
diagrams disappear. This observation is a strong hint that for
$G\subset U(N)$ there remain non-vanishing connected Green functions
$\langle A(x_1)\dots A(x_n)\rangle^\text{kin}_\text{c}$ for all
integer $n$. An explicit proof will be given in the next Section. 

\subsection{Non-vanishing $n$-point Green functions generated by 
$\ln Z_G^\text{kin}$} 
\label{subsec:nptproof}

The connected Green functions 
\begin{equation*}
G_{\text{c}\negphantom{\text{c}{}}\phantom{\mathrm{kin},}\mu_1\dots\mu_n}^{\mathrm{kin},m'_1\dots
  m'_n}(p_1,\dots,p_n)=\Big\langle A^{M_1}_{\mu_1}[a(x_1)]\dots A^{M_n}_{\mu_n}[a(x_n)]\Big\rangle^\text{kin}_\text{c}
\end{equation*}
are power series in $\theta$ and $g$. To prove their non-vanishing for
generic $\theta$ and $g$ it is sufficient to extract at least one
non-zero contribution to $G_\text{c}^\text{kin}$ of some fixed order in
$\theta$ and $g$.  

To find for our purpose the simplest tractable component of the Green
function it turns out to be   
advantageous to restrict all of the group indices $M_i$ to
primed indices that do not correspond to generators of the Lie algebra of
$G$. Then the Green function simplifies in first 
nontrivial order of the Seiberg-Witten map to:
\begin{equation}
\Big\langle\prod_{i=1}^n\Big(A^{(2){m'_i}}_{\phantom{(2)}\mu_i}[a(x_i)]+A^{(3){m'_i}}_{\phantom{(3)}\mu_i}[a(x_i)]\Big)\Big\rangle^\text{kin}_\text{c}\pnt
\label{fothetagreen}
\end{equation}
Here $A^{(2)}$, [$A^{(3)}$] denote the $\propto\theta$ part of the
Seiberg-Witten map  
\eqref{ACbarCSWmaplincomponents} with quadratic, [cubic] dependence on the ordinary field
$a_\mu$. Thus, the above function is $\mathcal{O}(\theta^n)$.   
Focusing now on the special contribution which is exactly $\propto\theta^n$,
it is clear that in addition to the external vertices further
$\theta$-dependence (e.\ g.\ higher order corrections to the Seiberg-Witten
map) is not allowed. That means this special part of the connected Green
function is universal with
respect to the $\theta$-expansion of the constraint
\eqref{ALambdaCbarCSWmaplin} where $a_\mu\in\mathfrak{g}$.

The special contribution to the Green function $\propto\theta^n$
then consists of $n$ to $\frac{3}{2}n$, [$\frac{3}{2}(n-1)+1$]
internal lines for $n$ 
even, [odd]. Two or three of these originate from each of the $n$
points (external vertices). There are no
further internal vertices present stemming from the interaction term
$s_1[a,c,\bar c ]$
in \eqref{12} since this would increase the power in $\theta$.  

In our
normalization where the coupling constant $g$ is absorbed into the fields,
each propagator enlarges the power of the
diagram in $g$ by $g^2$. Thus, for general coupling $g$ it is
sufficient to check the non-vanishing of all connected diagrams with
the same number of propagators. Here we choose the minimum case of $n$
propagators where we can neglect all contributions from $A^{(3)}$ in
\eqref{fothetagreen}. Then it follows that the connected
$\propto\theta^ng^{2n}$ contributions to the Green function are
given by the type of diagrams shown in Fig.\ \ref{diagtype}.
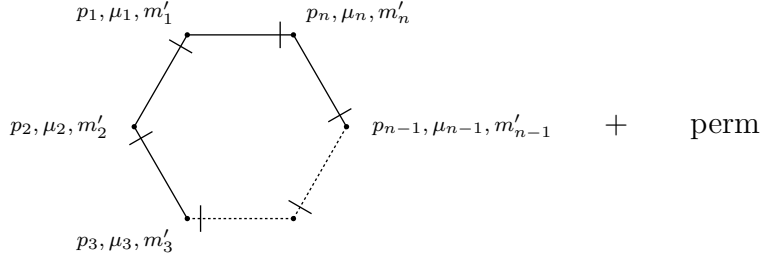
\begin{figure}
\begin{center}
\begin{picture}(330,120)(0,0)\scriptsize
\SetOffset(110,60)
\Line(-20,34.64)(20,34.64)\Text(-25,43.3)[r]{$p_1,\mu_1,m'_1$}
\Line(-40,0)(-20,34.64)\Text(-50,0)[r]{$p_2,\mu_2,m'_2$}
\Line(-20,-34.64)(-40,0)\Text(-25,-43.3)[r]{$p_3,\mu_3,m'_3$}
\DashLine(20,-34.64)(-20,-34.64){1}
\DashLine(40,0)(20,-34.64){1}\Text(50,0)[l]{$p_{n-1},\mu_{n-1},m'_{n-1}$}
\Line(20,34.64)(40,0)\Text(25,43.3)[l]{$p_n,\mu_n,m'_n$}
\Line(-26.83,32.81)(-18.17,27.81)
\Line(-41.83,-6.83)(-33.12,-1.83)
\Line(-15,-39.64)(-15,-29.64)
\Line(26.83,-32.81)(18.17,-27.81)
\Line(41.83,6.83)(33.12,1.83)
\Line(15,39.64)(15,29.64)
\Vertex(-40,0){1}
\Vertex(-20,34.64){1}
\Vertex(20,34.64){1}
\Vertex(40,0){1}
\Vertex(20,-34.64){1}
\Vertex(-20,-34.64){1}
\Text(137.5,0)[l]{\normalsize $+\qquad\text{perm}$}
\end{picture}
\end{center}
\caption{graphs $\propto\theta^ng^{2n}$ of the connected $n$-point Green function} 
\label{diagtype}
\end{figure}

These diagrams are 1PI in terms of the ordinary fields $a_\mu$, but 
we will further on denote them only as connected. One could imagine that 
with another field redefinition it might be possible to 
transform all these diagrams to connected, but not 1PI
diagrams. It could then happen that one can construct them from a finite set 
of building blocks. A redefined field in which the type of diagrams in
Fig.\ \ref{diagtype} could appear as connected 1PI diagrams would be given by
the remnant of the Seiberg-Witten map without the leading contribution.
Then, each external vertex in Fig.\ \ref{diagtype} would be the origin of a
single line and the diagram could possibly be 1PI.
Therefore, we are careful with our notation, but we have some arguments 
why such a field transformation is not relevant or possible here. 
Firstly, such a redefinition would not affect our analysis, because
such a finite set would describe a perturbation expansion in orders of 
$\theta^{\mu\nu}$, and this is what we do not want.
Secondly, the equivalence theorem \cite{Kamefuchi:1961sb,Zinn-Justin:1989mi}
is formulated for the field redefinitions that
have to start with a term linearly in the fields. In the remnant of the
Seiberg-Witten map which is the candidate for the required field redefinition,
the linear term is absent.   

We will now determine the total number of the diagrams in 
Fig.\ \ref{diagtype}. The
two lines starting at each point are distinguishable due to the
derivative at one leg. To construct all connected contributions we
connect the first leg of the first external vertex to one of the
$2n-2$ other legs that do not start at the same external point. The next
one is connected to one of the remaining $2n-4$ allowed legs, such
that no disconnected subdiagram is produced and so on. We thus have to
add-up $(2n-2)!!=(n-1)!2^{n-1}$ diagrams. All of them can be drawn
like the one shown in Fig.\ \ref{diagtype} by permuting the external
momenta, Lorentz and group indices and the internal legs.    

To sum-up all diagrams it is convenient to define two classes of
permutations: 
the first includes all permutations that interchange
the two distinct 
legs at one or more external vertices
with the distribution of the 
external momenta, Lorentz and group indices held fixed. The second contains all
permutations which interchange the external quantities 
such that this cannot be traced back to a permutation of the distinct
lines at the external vertices. We call its
elements proper permutations in the following. 

In total $2^n$ combinations exist, generated by interchanging the distinct
legs when the external points are fixed. The proper permutations are
the ones which cannot be mapped to each other by acting with the Dihedral
group $D_n$. $D_n$ is the symmetry group of an $n$-sided regular polygon. 
It is $2n$ dimensional and it is represented by
\begin{equation}
D_n=\big\{x,y\,|\,x^2=y^2=\mathds{1},\,(xy)^n=\mathds{1}\big\}=\big\{S^iR^j\,|\,S^n=R^2=\mathds{1},\,RSR^{-1}=S^{-1}\big\}\col
\end{equation}
where the $2n$ elements are given by reflections along $n$ of the symmetry
axes through the origin, $n-1$ rotations with angle $2\pi\frac{k}{n}$,
$k=1,\dots,n-1$ and the identity. 
There are $n!$ configurations of the external points and
with each one $2n-1$ others are identified by acting with $D_n$,
i.\ e.\  there are $\frac{n!}{2n}=\frac{(n-1)!}{2}$ proper permutations. 
This is consistent with the total number of diagrams.

The connected $\propto\theta^ng^{2n}$ contributions to the
momentum space Green function can thus be cast into the following
form:
\newcommand{\Ri}{(}
{\setlength{\fboxsep}{0pt}
\setlength{\fboxrule}{0pt}
\settowidth{\arlength}{\fbox{$\Ri$}}
\settoheight{\arheight}{\fbox{$\Ri$}}
\settodepth{\ardepth}{\fbox{$\Ri$}}
\addtolength{\arheight}{\ardepth}
\setlength{\shift}{0.5\arheight}
\addtolength{\shift}{0.2ex}
\addtolength{\shift}{-\ardepth}
\settoheight{\eqoff}{\fbox{$=$}}
\setlength{\eqoff}{0.5\eqoff}
\addtolength{\eqoff}{-60pt}
\begin{equation}
G_{\text{c}\negphantom{\text{c}{}}\phantom{\text{kin},}\mu_1\dots\mu_n}^{\text{kin},m'_1\dots
  m'_n}(p_1,\dots,p_n)\big|_{\propto\theta^ng^{2n}}=\sum_{\{i_1,\dots,i_n\}\in\frac{S_n}{D_n}}
\raisebox{\eqoff}{%
\begin{picture}(240,120)(0,0)\scriptsize
\SetOffset(110,60)
\SetOffset(108.75,62.165)
\Vertex(-20,34.64){1}
\Line(-26.25,23.815)(-20,34.64)
\Line(-20,34.64)(-7.5,34.64)\Text(-25,43.3)[r]{$p_{i_1},\mu_{i_1},m'_{i_1}$}
\Line(-15,37.14)(-15,32.14)
\SetOffset(111.25,57.835)
\Vertex(-20,34.64){1}
\Line(-25,25.98)(-20,34.64)
\Line(-20,34.64)(-10,34.64)
\Line(-24.665,31.56)(-20.335,29.06)
\SetOffset(107.5,60)
\Vertex(-40,0){1}
\Line(-33.75,-10.825)(-40,0)
\Line(-40,0)(-33.75,10.825)\Text(-50,0)[r]{$p_{i_2},\mu_{i_2},m'_{i_2}$}
\Line(-39.665,5.58)(-35.335,3.08)
\SetOffset(112.5,60)
\Vertex(-40,0){1}
\Line(-35,-8.66)(-40,0)
\Line(-40,0)(-35,8.66)
\Line(-39.665,-5.58)(-35.335,-3.08)
\SetOffset(108.75,57.835)
\Vertex(-20,-34.64){1}
\Line(-20,-34.64)(-26.25,-23.815)\Text(-25,-43.3)[r]{$p_{i_3},\mu_{i_3},m'_{i_3}$}
\Line(-7.5,-34.64)(-20,-34.64)
\Line(-24.665,-31.56)(-20.335,-29.06)
\SetOffset(111.25,62.165)
\Vertex(-20,-34.64){1}
\Line(-20,-34.64)(-25,-25.98)
\Line(-10,-34.64)(-20,-34.64)
\Line(-15,-37.14)(-15,-32.14)
\SetOffset(111.25,57.835)
\Vertex(20,-34.64){1}
\Line(26.25,-23.815)(20,-34.64)
\Line(20,-34.64)(7.5,-34.64)
\Line(15,-37.14)(15,-32.14)
\SetOffset(108.75,62.165)
\Vertex(20,-34.64){1}
\Line(25,-25.98)(20,-34.64)
\Line(20,-34.64)(10,-34.64)
\Line(24.665,-31.56)(20.335,-29.06)
\SetOffset(112.5,60)
\Vertex(40,0){1}
\Line(33.75,10.825)(40,0)
\Line(40,0)(33.75,-10.825)\Text(50,0)[l]{$p_{i_{n-1}},\mu_{i_{n-1}},m'_{i_{n-1}}$}
\Line(39.665,-5.58)(35.335,-3.08)
\SetOffset(107.5,60)
\Vertex(40,0){1}
\Line(35,8.66)(40,0)
\Line(40,0)(35,-8.66)
\Line(39.665,5.58)(35.335,3.08)
\SetOffset(111.25,62.165)
\Vertex(20,34.64){1}
\Line(20,34.64)(26.25,23.815)
\Line(7.5,34.64)(20,34.64)\Text(25,43.3)[l]{$p_{i_n},\mu_{i_n},m'_{i_n}$}
\Line(24.665,31.56)(20.335,29.06)
\SetOffset(108.75,57.835)
\Vertex(20,34.64){1}
\Line(20,34.64)(25,25.98)
\Line(10,34.64)(20,34.64)
\Line(15,37.14)(15,32.14)
\SetOffset(110,60)
\Line(-27.5,21.65)(-32.5,12.99)
\Text(-27.5,21.65)[]{\begin{rotate}{60}\raisebox{-0.5\shift}[\arheight][\arlength]{%
\hspace{-0.3\arlength}\fbox{$\Ri$}}\end{rotate}}
\Text(-32.5,12.99)[]{\begin{rotate}{-120}\raisebox{-0.5\shift}[\arheight][\arlength]{\hspace{-0.3\arlength}\fbox{$\Ri$}}\end{rotate}}
\Line(-27.5,-21.65)(-32.5,-12.99)
\Text(-27.5,-21.65)[]{\begin{rotate}{-60}\raisebox{-0.5\shift}[\arheight][\arlength]{\hspace{-0.3\arlength}\fbox{$\Ri$}}\end{rotate}}
\Text(-32.5,-12.99)[]{\begin{rotate}{120}\raisebox{-0.5\shift}[\arheight][\arlength]{\hspace{-0.3\arlength}\fbox{$\Ri$}}\end{rotate}}
\DashLine(-5,-34.64)(5,-34.64){1}
\Text(-5,-34.64)[]{\begin{rotate}{180}\raisebox{-0.5\shift}[\arheight][\arlength]{\hspace{-0.3\arlength}\fbox{$\Ri$}}\end{rotate}}
\Text(5,-34.64)[]{\begin{rotate}{0}\raisebox{-0.5\shift}[\arheight][\arlength]{\hspace{-0.3\arlength}\fbox{$\Ri$}}\end{rotate}}
\DashLine(27.5,-21.65)(32.5,-12.99){1}
\Text(27.5,-21.65)[]{\begin{rotate}{-120}\raisebox{-0.5\shift}[\arheight][\arlength]{\hspace{-0.3\arlength}\fbox{$\Ri$}}\end{rotate}}
\Text(32.5,-12.99)[]{\begin{rotate}{60}\raisebox{-0.5\shift}[\arheight][\arlength]{\hspace{-0.3\arlength}\fbox{$\Ri$}}\end{rotate}}
\Line(27.5,21.65)(32.5,12.99)
\Text(27.5,21.65)[]{\begin{rotate}{120}\raisebox{-0.5\shift}[\arheight][\arlength]{\hspace{-0.3\arlength}\fbox{$\Ri$}}\end{rotate}}
\Text(32.5,12.99)[]{\begin{rotate}{-60}\raisebox{-0.5\shift}[\arheight][\arlength]{\hspace{-0.3\arlength}\fbox{$\Ri$}}\end{rotate}}
\Line(-5,34.64)(5,34.64)
\Text(-5,34.64)[]{\begin{rotate}{180}\raisebox{-0.5\shift}[\arheight][\arlength]{\hspace{-0.3\arlength}\fbox{$\Ri$}}\end{rotate}}
\Text(5,34.64)[]{\begin{rotate}{0}\raisebox{-0.5\shift}[\arheight][\arlength]{\hspace{-0.3\arlength}\fbox{$\Ri$}}\end{rotate}}
\end{picture}}
\pnt
\label{fogreen}
\end{equation}}%
Here the brackets around the external vertices denote a sum over both
configurations where the two legs are interchanged. These $n$ sums are 
then multiplied, and describe exactly the $2^n$ permutations of the
distinct two legs at each vertex. 

The sum of the two permutations at one external vertex occurring $n$
times in \eqref{fogreen} reads 
\settoheight{\eqoff}{\fbox{$=$}}
\setlength{\eqoff}{0.5\eqoff}
\addtolength{\eqoff}{-40pt}
\begin{equation*}
\raisebox{\eqoff}{%
\begin{picture}(200,80)(0,0)\scriptsize
\Text(30,40)[r]{$p,\mu,M$}
\Vertex(40,40){1}
\Line(40,40)(74.641,60)
\Line(43.08,44.665)(45.58,40.335)
\Line(40,40)(74.641,20)
\Text(94.641,40)[]{\normalsize $+$}
\SetOffset(74.641,0)
\Vertex(40,40){1}
\Line(40,40)(74.641,60)
\Line(40,40)(74.641,20)
\Line(43.08,35.335)(45.58,39.665)
\Text(84.641,60)[l]{$q,\alpha,A$}
\Text(84.641,20)[l]{$r,\beta,B$}
\end{picture}}
=-\frac{i}{4}\dddu{A}{B}{M}\Big[2(\theta^{\beta\gamma}q_\gamma\delta_\mu^\alpha+\theta^{\alpha\gamma}r_\gamma\delta_\mu^\beta)+\theta^{\alpha\beta}(q_\mu-r_\mu)\Big]\pnt
\end{equation*}
Using this, the analytic expression for the
$\propto\theta^ng^{2n}$ part of the connected Green function is
given by
\begin{equation}
\begin{aligned}
& G_{\text{c}\negphantom{\text{c}{}}\phantom{\mathrm{kin},}\mu_1\dots\mu_n}^{\mathrm{kin},m'_1\dots m'_n}(p_1,\dots,p_n)\big|_{\propto\theta^ng^{2n}}\\
& \qquad
=\sum_{\{i_1,\dots,i_n\}\in\frac{S_n}{D_n}}\frac{g^{2n}}{4^n}\int\frac{\de^dk}{(2\pi)^d}\prod_{r=1}^n\duud{a_r}{m'_{i_r}}{a_{r+1}}\Big[-2\theta^{\alpha_r\gamma_r}(q_{r-1})_{\gamma_r}
G_{\mu_{i_r}\alpha_{r+1}}+2\theta_{\alpha_{r+1}}^{\phantom{\alpha_{r+1}}\gamma_r}(q_r)_{\gamma_r}
\delta_{\mu_{i_r}}^{\alpha_r}\\
&\phantom{\qquad
{}=\sum_{\{i_1,\dots,i_n\}\in\frac{S_n}{D_n}}\frac{g^{2n}}{4^n}\int\frac{\de^dk}{(2\pi)^d}\prod_{r=1}^nd^{a_ra_{r+1}m'_{i_r}}\Big[}-\theta^{\alpha_r}_{\phantom{\alpha_r}\alpha_{r+1}}(q_{r-1}+q_r)_{\mu_{i_r}}\Big]\frac{1}{q_{r-1}^2}\col
\end{aligned}
\label{greenkinexpl}
\end{equation}
where summation over $\alpha_r$ appearing twice in the sequence of
multiplied square brackets is understood. Thereby one has to
identify $a_{n+1}=a_1$, $\alpha_{n+1}=\alpha_1$, 
$p_{i_n}=-\sum_{r=1}^{n-1}p_{i_r}$. The $q_r$ are defined by
\begin{equation}
q_r=q_r(k,p_{i_1},\dots,p_{i_r})=k+\sum_{s=1}^rp_{i_s}\pnt
\label{qdef}
\end{equation} 
In Appendix \ref{appnptproof} we prove that this expression is indeed
non-zero at least for even $n$ and the most symmetric
non-trivial configuration of the external momenta, Lorentz and group indices. 
This means that non-vanishing connected $n$-point functions for arbitrary high
$n$ exist in
the kinetic perturbation theory, leading to infinitely many building
blocks in the $\theta$-summed case. In other words one needs infinitely many
elements to formulate Feynman rules for the noncommutative $G$ gauge
theory if one insists on keeping the noncommutative $U(N)$ vertices as 
components.       

Due to the fact that the expressions discussed above cannot be
affected by higher order corrections of \eqref{ACbarCSWmaplincomponents} 
this statement is universal, i.\ e.\  independent of the power in $\theta$ 
up to which the constraint $a_\mu\in\mathfrak{g}$ is implemented.

\subsection{The case with sources restricted to the Lie algebra of $G$}
\label{subsec:restrictedsource}
Up to now we have looked for Feynman rules working with the original
$U(N)$ vertices and sources $J^M$ taking values in the full $\mathfrak{u}(N)$
Lie algebra. This seemed to be natural since in the enveloping
algebra approach for $G\subset U(N)$ the noncommutative gauge $A^M$-field, 
although constrained, carries indices $M$ running over all generators
of $\mathfrak{u}(N)$.

There is still another option to explore. First one can restrict the sources
$J,\eta ,\bar{\eta}$ in \eqref{9} by hand to take values in $\mathfrak{g}$ 
only. Then instead of pulling out in \eqref{10} the complete
interaction $S_I$ one separates only those parts of $S_I$, which yield
vertices whose external legs carry lower case Latin indices referring to
$\mathfrak{g}$ exclusively. The remaining parts of $S_I$, generating
vertices with at least one leg owning a primed index, are kept under
the functional integral. The functional integration and the constraint
remain unchanged. We denote this splitting of $S_I$ by
\begin{equation}
S_I[A,C,\bar C]=S_i[A,C,\bar C]+S'_{i}[A,C,\bar C]
\label{bb1}
\end{equation}
and the sources by hatted quantities
\begin{equation}\label{restsources}
\hat J^{a'}=\hat{\bar{\eta}}^{a'}=\hat{\eta}^{a'}=0\pnt
\end{equation}
Then
\begin{equation}
Z_G[\hat J,\hat{\bar{\eta}},\hat\eta]=\e^{iS_{i}[\prefuncderivi{\hat J},\prefuncderivi{\hat{\bar{\eta}}},\prefuncderivi{\hat\eta}]}\hat Z_G[\hat J,\hat{\bar{\eta}},\hat\eta]
\label{bb3}
\end{equation}
and
\begin{equation}
\begin{aligned}
\hat Z_G[\hat J,\hat{\bar{\eta}},\hat\eta]&=\int_
{a,c,\bar c\in\mathfrak{g}}\mathcal{D}A\,\mathcal{D}\bar C\,\mathcal{D}C\,\e^{i(S_\text{kin}[A,C,\bar C]
+S'_{i}[A,C,\bar C] + A\hat J+\hat{\bar{\eta}}C+\bar C\hat\eta )}\\
&=\int_{a,c,\bar c\in\mathfrak{g}} \mathcal{D}a\,\mathcal{D}\bar c\,\mathcal{D}c\,\mathcal{J}\e^{i(S_\text{kin}[a,c,\bar c]+\hat s_1[a,c,\bar c ]+A[a]\hat J+\hat{\bar{\eta}}C[c,a]+\bar c\hat\eta )}\col
\label{bb4}
\end{aligned}
\end{equation}
where $\hat s_1[a,c,\bar c ]$ is defined by
\begin{equation}
S_\text{kin}[A[a],C[c,a],\bar c]+S'_{i}[A[a],C[c,a],\bar c]=S_\text{kin}[a,
c,\bar c]+\hat s_1[a,c,\bar c]\pnt
\label{bb5}
\end{equation}
If now the generating functional of connected Green functions
\begin{equation}
\ln\frac{\hat Z_G[\hat J,\hat{\bar{\eta}},\hat\eta ]}{\hat Z_G[0]}=\sum_n\frac{i^n}{n!}\int\de^dx_1\dots\de^dx_n\,
\langle A(x_1)\dots A(x_n)\rangle^{\text{kin}+S'_i}_\text{c}\hat J(x_1)\dots\hat J(x_n)+\dots\col
\label{bb6}
\end{equation} 
e.\ g.\ for $G=SO(N)$, would generate only the free
propagators (like \eqref{13a} does for $G=U(N)$), the $SO(N)$ Feynman rules conjectured in ref. \cite{Bonora:2001ny}
would have been derived via partial summation of the $\theta$-expanded 
perturbation theory in the enveloping algebra approach.
As was argued in \cite{Bonora:2000td} and explained before,
the constraint they use is equivalent to requiring
that the image under the inverse SW map is in $\mathfrak{SO}(N)$.
Thus, a generation of additional vertices would disprove the 
rules of \cite{Bonora:2001ny}.

In the remaining part of this Section we show that there are additional 
vertices.
For this purpose we consider $\langle A(x_1)\dots
A(x_n)\rangle^{\text{kin}+S'_i}_\text{c}$ and look at it
as a power series in $g^2$ and $\theta$. To prove that a contribution 
with $n>2$ to \eqref{bb6} it is not
identically zero, it is sufficient to find a particular non-vanishing
order in $g^2$ and $\theta$. 

New vertices arise from expressing the noncommutative
fields $A_\mu$ either in the original noncommutative kinetic term or 
$3$-point or $4$-point interactions in $S'_i$ (see \eqref{bb1} via 
\eqref{ACbarCSWmaplincomponents})
in terms of the ordinary field $a_\mu$.
To take into consideration only the vertices \eqref{kinthreevertex},
\eqref{kinfourvertex} and \eqref{kingaugeghostvertex} 
at $\mathcal{O}(\theta)$ generated by an expansion of the
kinetic term is not sufficient if one wants to avoid to work with higher
orders of the Seiberg-Witten map.
The reason is as follows. 
Because of \eqref{restsources} it is clear that only the components 
$\langle A^{m_1}(x_1)\dots A^{m_n}(x_n)\rangle^{\text{kin}+S'_i}_\text{c}$ 
(with unprimed lower case Latin indices) contribute to \eqref{bb6}.  
For $SO(N)$, where $d_{abc}=0$ \cite{deAzcarraga:1998ya,Fuchs:1997jv}, it
follows immediately that all  
$\mathcal{O}(\theta)$ contributions from the kinetic term vanish in this case 
and contributions can only start at $\mathcal{O}(\theta^2)$.
At this order, however, a mixing with the $\mathcal{O}(\theta^2)$ terms in the
Seiberg-Witten map occurs.

If one does not want to work with the higher order terms in the Seiberg-Witten
map, one has to look at diagrams that include the remnants of the $3$-
and $4$-point vertices in $S'_i$ of \eqref{bb1}. 
They are given by the diagrams of Fig.\ \ref{fig:NCUNYMFrules}, but carry at
least 
one primed index. It is clear that these primed indices have to be converted 
to unprimed lower case indices via $\theta$-dependent terms in the expansion 
of the noncommutative fields. 
Each primed index increases the order in
$\theta$ by at least one. 
We will now search for interactions of only gauge fields 
of lowest possible order in
$\theta$.\footnote{Note that the original 3-point or 4-point interactions Fig.\ \ref{fig:NCUNYMFrules} by themselves
are $\theta $-dependent via the $\ast$-product. But since we are searching
for the lowest order in $\theta$ this further $\theta$-dependence can be
disregarded.}
The number of primed indices then has to be minimized.    
As can be seen from Fig.\ \ref{fig:NCUNYMFrules} both pure gauge
vertices include 
factors with the structure constants. From \eqref{subalstruc} one can
immediately see that the contributions with only one primed index vanish for
both vertices so that two primed indices is the minimal number. 
The next step is to
ensure that these primed indices are only internal, i.\ e.\ one has
to look at the $\theta$-expansion of the noncommutative gauge fields and  
choose the $\theta$-dependent terms for the fields that are attached to the
two legs that carry primed indices. It is easy 
to see that the linear order terms of \eqref{ALambdaCbarCSWmaplin} are
sufficient to transform each primed index into two or three unprimed indices.
Although the lowest order in $\theta$ is now $\mathcal{O}(\theta^2)$,
no mixing with higher order contributions to the Seiberg-Witten map can 
occur in this case because one is forced to put $\theta$-dependent terms
on two legs.
With the help of a single $3$-vertex $S'_i$ one can now
construct a $5$-, $6$- of $7$-point function that carries only unprimed
lower case indices. A single $4$-vertex of $S'_i$ leads to an
$6$-, $7$- or $8$-point function of this kind. 
A tree level $n$-point function with $k$ vertices possesses $k-1$ internal
lines that connect the vertices. If $l$ of these vertices come from the
$\propto\theta$ terms of the  
Seiberg-Witten map and $k-l$ vertices are taken from $S'_i$ and thus of order
$g^{-2}$ the total order of the tree level $n$-point function is
$g^{2(n+l-1)}\theta^l$.
Even for generic $g$ and $\theta$ the $6$- and $7$-point functions constructed 
from the $3$- and $4$-point vertices mix because the Seiberg-Witten map at 
linear order generates two and three unprimed legs from a primed one.  
This could in principle lead to a cancellation between the two contributions 
even if both are separately non zero.
The situation is different for the $5$- and the $8$-point functions. There 
is no other tree level diagram of the same order that leads to a $5$-point
function and thus to show that it does not vanish is a proof that there are 
new interactions. For the $8$-point function there is a mixing but it cannot 
lead to serious cancellations because its the standard mixing in YM theory. 
The underlying 
$4$-point vertex mixes with
the $4$-point amplitude made of two $3$-point
vertices. This mechanism ensures gauge invariance of the full amplitude but
does not annihilate it. 
Hence, it is sufficient to check the non-vanishing of 
the $8$-point amplitude that is built with the 
$4$-point vertex alone.
In Appendix \ref{appB} for $SO(3)$ $\langle A^{m_1}(x_1)\dots
A^{m_8}(x_8)\rangle_\text{c}^{\text{kin}+S'_i}$ is explicitly computed in
lowest order in $g$ and $\theta$ and shown not
to vanish. This serves as a counterexample that disproves the $SO(N)$ 
Feynman rules of \cite{Bonora:2001ny}.

The more ambitious program to exclude rules based on the vertices
in $S_i$ and an arbitrary but finite number of additional building blocks
would require to show, similar to the previous Section, that there 
is no $n_0$ assuring vanishing connected $n$-point functions for $n>n_0$. 
Although we have practically no doubt concerning this conjecture, a
rigorous proof is beyond our capabilities since 
for increasing $n$ higher and higher orders of the SW-map contribute.
This happens because in contrast to the proof in Subsection
\ref{subsec:nptproof} 
one is forced to look at Green functions with all external group
indices referring to generators of the Lie algebra of $G$ since no
primed indices of the remaining generators spanning $U(N)$ are probed.  


\clearemptydoublepage
\part{The BMN limit of the AdS/CFT correspondence}
\label{BMN}
\clearemptydoublepage
\chapter{Relations between string backgrounds}
\label{chap:BMNAdSpw}
In this chapter we will briefly discuss some $p$-brane solutions of
supergravity. Then we will introduce anti-de Sitter (AdS) spacetimes, 
the product spaces $\AdS\times\text{S}$ and the pp wave spacetimes. 
We will show how $\AdS\times\text{S}$ arises as a near horizon limit
of certain $p$-brane solutions and we will discuss the Penrose-G\"uven limit
in more detail since they play an essential role in
the $\AdS/\text{CFT}$ correspondence and in its BMN limit.
We will mostly work in general dimensions. This will enable us to encompass
certain types of solutions without repeating similar calculations and it will
shed some light on the general structure. 
Special emphasis is put on the $(D=10)$-dimensional
solutions as concrete setups in the formulation of the $\AdS/\text{CFT}$
correspondence and its BMN limit. We will also present the corresponding
solutions in $D=11$.

\section{The backgrounds}
\label{backgrounds}
\subsection{Some $p$-brane solutions of supergravity}
\label{SUGRApbranesol}
That part of a generic supergravity action which is sufficient for our purpose
to present some $p$-brane solutions in various dimensions $D$ can be written 
down in closed form. 
First it is advantageous to define
\begin{equation}
d=p+1\col\qquad d'=D-p-3\col
\end{equation}
such that $D=d+d'+2$. One should keep in mind that a supergravity theory 
in $D$ dimensions restricts the allowed $d$ and hence the possible
$(d-1)$-brane solutions. 
The relevant (bosonic) part of a supergravity action in 
the Einstein-frame reads \cite{Duff:1995an}
\begin{equation}
S_D=\frac{1}{2\kappa^2}\int\de^Dx\,\sqrt{-g}\Big[\mathcal{R}-\frac{1}{2}(\partial \phi)^2-\frac{1}{2}e^{-a(d)\phi}|F_{d+1}|^2\Big]
\end{equation}
for the metric $g_{MN}$ (from now on we choose the signature `mostly
  plus'), the dilaton\footnote{There is no dilaton in
  $11$-dimensional supergravity. One simply has to ignore it if one wants to
  deal with this case. Only the solutions with $a(d)=0$ are then relevant.} 
$\phi$ and a $(d+1)$-form field strength
$F_{d+1}$ which is defined as $F_{d+1}=\de A_d$. We have used the
abbreviations  
\begin{equation}
|F_{d+1}|^2=\frac{1}{(d+1)!}g^{M_1N_1}\dots g^{M_{d+1}N_{d+1}}F_{N_1\dots
 N_{d+1}}F_{M_1\dots M_{d+1}}\pnt 
\end{equation}
and
\begin{equation}\label{aofp}
a(d)^2=4-2\frac{dd'}{d+d'}\pnt
\end{equation}
The equations of motions which derive from the above given action read
\begin{equation}\label{SUGRAeom}
\begin{aligned}
\mathcal{R}^{MN}-\frac{1}{2}g^{MN}\mathcal{R}&=\frac{1}{2}\Big(\partial^M\phi\partial^N\phi-\frac{1}{2}g^{MN}(\partial\phi)^2\Big)\\
&\phantom{{}=}
+\frac{1}{2}\Big(\frac{1}{d!}F^M_{\phantom{M}P_1\dots
P_d}F^{NP_1\dots P_d}
-\frac{1}{2}g^{MN}|F_{d+1}|^2\Big)e^{-a(d)\phi}\col\\
\partial_M\big(\sqrt{-g}e^{-a(d)\phi}F^{MP_1\dots P_d}\big)&=0\col\\
\partial_M\big(\sqrt{-g}g^{MN}\partial_N\phi\big)&=-\frac{a(d)}{2}\sqrt{-g}e^{-a(d)\phi}|F_{d+1}|^2\col
\end{aligned}
\end{equation}
and they admit the so called extremal solutions of the form 
\cite{Duff:1995an,Petersen:1999zh} 
\begin{equation}\label{pbranesol}
\begin{aligned}
\de s_{d-1}^2&=H(y)^{-\frac{d'}{d+d'}}\de x^\mu\de
x_\mu+H(y)^\frac{d}{d+d'}\de y^i\de y_i\col\qquad\\
e^{2\phi}&=H(y)^{a(d)}\col\qquad H(y)=1+\Big(\frac{R_2}{y}\Big)^{d'}\col\\
F_{d+1}&=\vol(\mathds{R}^{1,d-1})\wedge\de H^{-1}(y)\col
\end{aligned}
\end{equation}
where $\mu=0,\dots,d-1$ and $i=d,\dots,d+d'+1$. The coordinate
$y=\sqrt{y^iy_i}$ measures the transverse separation from the brane that
resides at $y=0$, and $\vol(\mathds{R}^{1,d-1})$ 
is the volume form of $R^{1,p}$. The parameter $R_2$ is related to $\kappa$ and
the $\text{D}(d-1)$-brane tension $T_{d-1}$ as \cite{Duff:1995an}
\begin{equation}\label{L}
R_2^{d'}=\frac{2\kappa^2T_{d-1}}{d'\Omega_{d'+1}}=g_s(4\pi)^{\frac{d'}{2}-1}\alpha'^{\frac{d'}{2}}\Gamma\big(\tfrac{d'}{2}\big)\col
\end{equation}
where $\Omega_{d'+1}$ is the volume of the unit $\text{S}^{d'+1}$
\eqref{Svolume}. Furthermore, we have used that $\kappa$ and $T_p$ can be
expressed in terms of the squared string length $\alpha'$ and the string coupling $g_s$ as\footnote{In
  $D=11$ where the dilaton is absent one has to set $g_s=1$.} \cite{Petersen:1999zh}
\begin{equation}
\begin{aligned}
2\kappa^2&=(2\pi)^{d+d'-1}\alpha'^{\frac{d+d'}{2}}g_s^2\col\\
T_{d-1}&=\frac{1}{g_s(2\pi)^{d-1}\alpha'^{\frac{d}{2}}}\pnt
\end{aligned}
\end{equation}
In $D=d+d'+2$ dimensions with $g$ denoting the determinant of the metric, 
we now use the conventions 
\begin{equation}
\varepsilon_{0\dots D-1}=1\col\qquad\varepsilon^{0\dots D-1}=\frac{1}{g}
\end{equation}
for the total antisymmetric tensor density and define the Hodge-duality
operator as 
\begin{equation}
\star(\de x^{M_1}\wedge\dots\wedge\de
x^{M_d})=-\frac{\sqrt{-g}}{(D-d)!}\varepsilon^{M_1\dots
  M_d}_{\phantom{M_1\dots M_d}M_{d+1}\dots M_D}\de x^{M_{d+1}}\wedge\dots\wedge\de
x^{M_D}\pnt
\end{equation}
After rewriting the second term in the metric \eqref{pbranesol} in polar 
coordinates 
\begin{equation}
\de y^i\de y_i=\de y^2+y^2\de\Omega_{d'+1}^2\col
\end{equation}
one then finds for the differential form that occurs in $F_{d+1}$ of
\eqref{pbranesol}
\begin{equation}
\star\big(\vol(\mathds{R}^{1,d-1})\wedge\de y\big)=H(y)^{2-\frac{a^2(d)}{2}}y^{d'+1}\vol(\Omega_{d'+1})
\end{equation}
with $\vol(\Omega_{d'+1})$ denoting the volume form of the 
unit $\text{S}^{d'+1}$, see \eqref{Svolumeform}.
The Hodge-dual field strength then reads
\begin{equation}\label{dualfieldstrength}
\star F_{d+1}=-H(y)^{-\frac{a^2(d)}{2}}y^{d'+1}\partial_y H(y)\vol(\Omega_{d'+1})\pnt
\end{equation}

The above given solutions are $(d-1)$-branes embedded in $D=d+d'+2$ dimensions.
They have an isometry group $SO(1,d-1)\times SO(d'+2)$.
For $D=10$ and $d=4$ one obtains the $\text{D}3$-brane solution of type
$\twob$ supergravity with the self-dual $5$-form flux, i.\ e.\ the flux has to
fulfill the relation
\begin{equation}
F_5=\star F_5\pnt
\end{equation}
This can be achieved by replacing
\begin{equation}\label{selfdualitysolution}
F_5\to\frac{1}{2}(F_5+\star F_5)
\end{equation}
if one remembers that in case of Minkowski signature the action of the 
Hodge-star on a $p$-form $\omega_p$ fulfills
\begin{equation}
\star\star\omega_p=-(-1)^{p(D-p)}\omega_p\pnt
\end{equation} 
For $D=11$ and $d=3$ or $d=4$ one obtains the $\text{M}2$- and 
$\text{M}5$-branes where 
the $4$-form field strength of the theory is given by the last line in 
\eqref{pbranesol} for the $\text{M}2$- brane and by 
\eqref{dualfieldstrength} for the
$\text{M}5$-brane respectively. 

At the end let us discuss the measurement of energy in presence of a
gravitational source like the $(d-1)$-brane \eqref{pbranesol}.
The warp factor that multiplies the time differential $-\de x_0^2$ in
the metric \eqref{pbranesol} determines the gravitational redshift that 
is caused by the $(d-1)$-brane. An observer infinitely far away from the
$p$-branes measures energy with the operator $E_\infty=-i\partial_0$. 
At finite distance $y$ he uses
$E_y=-iH(y)^{\frac{d'}{2(d+d')}}\partial_0$. 
The energy of a particle at $y$  measured by an observer at infinity is thus redshifted by 
\begin{equation}\label{redshift}
E_\infty=H(y)^{-\frac{d'}{2(d+d')}}E_y\pnt
\end{equation}

\subsection{Anti-de Sitter spacetime}
\label{AdSspacetime}
$(d+1)$-dimensional Anti de-Sitter spacetime $\AdS_{d+1}$ 
with radius $R_1$ is a solution to the Einstein equations \eqref{Einsteineq}
with cosmological constant
\begin{equation}
\Lambda=-\frac{1}{2R_1^2}d(d-1)\pnt
\end{equation}
The Riemann and Ricci tensors and the scalar curvature read
\begin{equation}
\mathcal{R}_{\mu\nu\rho\sigma}=-\frac{1}{R_1^2}(g_{\mu\rho}g_{\nu\sigma}-g_{\mu\sigma}g_{\nu\rho})\col\qquad
\mathcal{R}_{\mu\nu}=-\frac{d}{R_1^2}g_{\mu\nu}\col\qquad
\mathcal{R}=-\frac{d(d+1)}{R_1^2}\pnt
\end{equation}
It is a maximally symmetric spacetime. That means
its metric possesses as many symmetries (Killing vectors) as 
$(d+1)$-dimensional flat space, namely $\frac{(d+1)(d+2)}{2}$. The space can be
realized as a coset
\begin{equation}
\AdS_{d+1}=\frac{SO(2,d)}{SO(1,d)}\col
\end{equation}
where $SO(2,d)$ and $SO(1,d)$ are the isometry group and the stabilizer
respectively.  
Furthermore, it can be constructed as an embedding of the hyperboloid 
\begin{equation}\label{defeq}
-X_0^2-X_{d+1}^2+\sum_{i=1}^d X_i^2=-R_1^2
\end{equation}
into the flat $(d+2)$-dimensional $\mathds{R}^{2,d}$ with metric
\begin{equation}\label{R2dmetric}
\de s_{\mathds{R}^{2,d}}^2=-\de X_0^2-\de X_{d+1}^2+\sum_{i=1}^2\de X_i^2\pnt
\end{equation}
With the coordinates
\begin{equation}
0\le t\le 2\pi\col\qquad 0\le\rho\col\qquad 0\le\bar\rho<\frac{\pi}{2}
\end{equation}
one can realize the embedding as follows
\begin{equation}
\begin{aligned}\label{Xinglobalcoord}
X_0&=R_1\cosh\rho\cos t=R_1\sec\bar\rho\cos t\col\\
X_{d+1}&=R_1\cosh\rho\sin t=R_1\sec\bar\rho\sin t\col\\
X_i&=R_1\sinh\rho\,\omega_i=R_1\tan\bar\rho\,\omega_i\col\qquad
i=1,\dots,d\col\qquad\sum_i\omega_i^2=1\col
\end{aligned}
\end{equation}
where we have presented two alternative choices $\rho$ and $\bar\rho$. 
These coordinates with the above given ranges cover the hyperboloid exactly
once and are therefore denoted as global coordinates. 
But what we mean with $\AdS_{d+1}$ (if not otherwise stated) is the universal 
covering of the hyperboloid with the coordinate ranges 
\begin{equation}
0\le t\col\qquad 0\le\rho\col\qquad 0\le\bar\rho<\frac{\pi}{2}\col
\end{equation}
i.\ e.\ one allows a multiple 
wrapping of the hyperboloid in the global time direction. The induced
metric on the hyperboloid becomes in the two sets of global coordinates
\begin{equation}\label{AdSmetricglobalcoord}
\begin{aligned}
\de s_\AdS^2&=R_1^2\big(-\cosh^2\rho\de t^2+\de\rho^2+\sinh^2\rho\de\Omega_{d-1}^2\big)\\
&=R_1^2\sec^2\bar\rho\big(-\de t^2+\de\bar\rho^2+\sin^2\bar\rho\de\Omega_{d-1}^2\big)\col
\end{aligned}
\end{equation}
where $\de\Omega_{d-1}^2$ is the metric of the unit $\text{S}^{d-1}$.
In the second line the term in parenthesis describes a space with topology 
$\mathds{R}\times\text{S}^d$. It is the $(d+1)$-dimensional Einstein static
universe (ESU). From the second line it is then easy to read off the conformal 
boundary of $\AdS_{d+1}$. One simply has to check where the conformal factor
diverges. This happens at $\bar\rho=\frac{\pi}{2}$ or equivalently 
$\rho\to\infty$. 
The structure of the boundary is found from the terms in parenthesis by
computing it at these values of the coordinates.
In the second line of \eqref{AdSmetricglobalcoord}, 
the factor in front of the unit $\text{S}^{d-1}$
is one and hence the boundary itself is given by 
$\mathds{R}\times\text{S}^{d-1}$.
In addition, two points $i^-$ and $i^+$ of timelike past and future infinity
belong to the boundary structure of $\AdS_{d+1}$ \cite{Hawking:1973}. 
The reason why one has to include 
these points is that one cannot make the time coordinate finite without  
pinching off the spatial directions.

The metric of $\AdS_{d+1}$ can be cast into a very simple form if one 
introduces a patchwise coordinate system that covers only one half of 
$\AdS_{d+1}$. The embedding reads in these coordinates
$(s,x_0,\vec x)$, where $s>0$
\begin{equation}\label{prePcoord}
\begin{aligned}
X_0&=\frac{1}{2s}\big(1+s^2(R_1^2+\vec x^2-x_0^2)\big)\col\\
X_{d+1}&=R_1\,s\,x_0\col\\
X_i&=R_1\,s\,x_i\col\qquad i=1,\dots,d-1\col\\
X_d&=\frac{1}{2s}\big(1-s^2(R_1^2-\vec x^2+x_0^2)\big)\col
\end{aligned}
\end{equation}
and the metric is given by
\begin{equation}
\de s_\AdS^2=R_1^2\Big(\frac{\de s^2}{s^2}+s^2(-\de x_0^2+\de\vec x^2)\Big)\pnt
\end{equation}
One can then perform a coordinate change $x_\perp=s^{-1}$, $0< x_\perp$ 
to obtain the metric in the so called Poincar\'e coordinates
\begin{equation}\label{AdSmetricPcoord}
\de s_\AdS^2=\frac{R_1^2}{x_\perp^2}(\de x_\perp^2-\de x_0^2+\de\vec x^2)\pnt
\end{equation}
It is easy to identify the region of the hyperboloid that is covered by 
these coordinates. Subtract the embedding coordinates $X_0$ and 
$X_d$ in global as well as in the Poincar\'e coordinates
\begin{equation}
X_0-X_d=R_1(\sec\bar\rho\cos t-\tan\bar\rho\,\omega_d)=R_1^2\frac{1}{x_\perp}
\end{equation}
and use the condition $0< x_\perp$ which then leads to the restriction
\begin{equation}\label{Ppatchineq}
\cos t>\sin\bar\rho\,\omega_d\pnt
\end{equation}
With the coordinate ranges $0\le\rho<\frac{\pi}{2}$ and $-1\le\omega_d\le 1$
one finds that the Poincar\'e patch looks different for the two hemispheres
with $\omega_d<0$ and $\omega_d\ge 0$. For the case of $\AdS_2$, where
$\omega_1\in S^0=\{-1,1\}$, the patch is shown in Fig.\ \ref{fig:hyperboloid}.
\begin {figure}
\begin{center}
\hspace*{-2.25cm}
\text{\epsfig{file=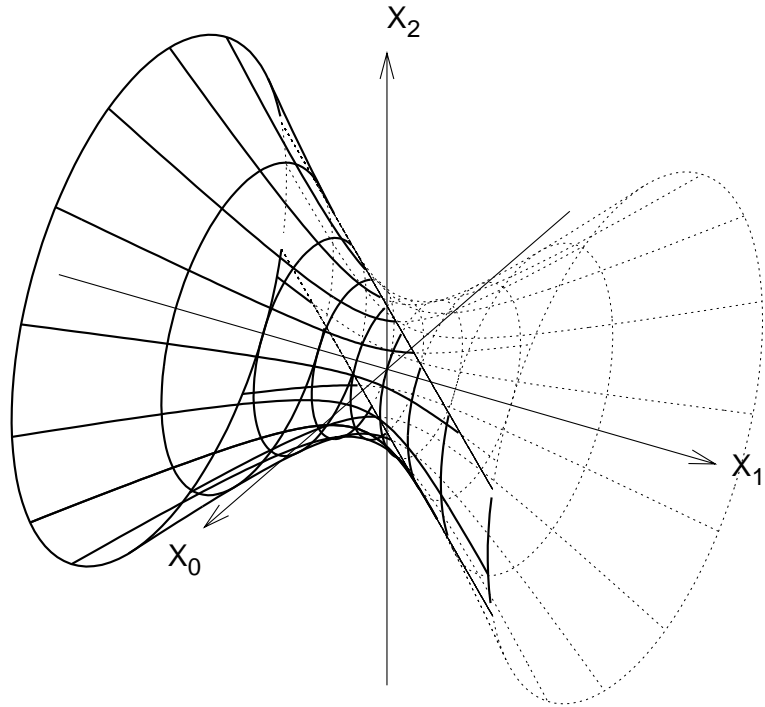, width=0.9\textwidth, angle=270}}
\caption{$\AdS_2$ embedded as a hyperboloid into $\mathds{R}^{2,1}$. In this 
picture constant global coordinates $\rho$, $\bar\rho$ are represented as 
circles and constant $t$ are given by the remaining curves. For generic
$\AdS_{d+1}$ with $d>1$ only one half of the hyperboloid can be drawn because
$X_1$ has to be replaced by $|\vec X|$ and every point then represents a
$(d-1)$-dimensional sphere.
The part drawn with full lines denotes the Poincar\'e patch that fulfills
\eqref{Ppatchineq}.}\label{fig:hyperboloid}
\end{center}
\end{figure}

With the help of the embedding one can define the so called chordal distance,
defined as the square length of the straight line that connects two points on 
the submanifold. Using the metric \eqref{R2dmetric} and the embedding
\eqref{defeq} one finds in global as well as in Poincar\'e
coordinates
\begin{equation}\label{AdSchordaldist}
\begin{aligned}
u(x,x')=(X(x)-X(x'))^2&=2R_1^2\Big[-1+\cosh\rho\cosh\rho'\cos(t-t')-\sinh\rho\sinh\rho'\,\omega^i\omega'_i\Big]\\
&=\frac{R_1^2}{x_\perp x'_\perp}\Big[(x_\perp-x'_\perp)^2-(x_0-x'_0)^2+(\vec
x-\vec x')^2\Big]\pnt
\end{aligned}
\end{equation}

An important point is the conformal mapping of $\AdS_{d+1}$ to other
spacetimes.
The metric without the conformal factor in the second line of 
\eqref{AdSmetricglobalcoord} 
describes one half of the Einstein static universe (ESU) which has topology
$\mathds{R}\times\text{S}^{d}$. $\AdS_{d+1}$ can therefore be conformally 
mapped to one half of the $(d+1)$-dimensional ESU. The fact that only one 
half is covered is caused by the coordinate range $0\le\bar\rho<\frac{\pi}{2}$
instead of $0\le\bar\rho\le\pi$ that covers the full ESU. The spatial part of
the boundary at $\bar\rho=\frac{\pi}{2}$ is mapped to a $(d-1)$-dimensional 
subsphere within the $\text{S}^d$ of the ESU.
From \eqref{AdSmetricPcoord} it follows immediately that $\AdS_{d+1}$ is 
conformally flat.

\subsection{The product spaces $\AdS\times\text{S}$}
\label{AdSSspacetime}
The direct product space of of $(d+1)$-dimensional anti-de Sitter space with
radius $R_1$ and of
a $(d'+1)$-dimensional sphere with radius $R_2$ is \emph{not} a solution to
the Einstein equations \eqref{Einsteineq} with cosmological constant. This can
be easily seen from the components of the Ricci tensor and 
the scalar curvature written down in \eqref{AdSScurvature} below.
But geometries of this type arise as near horizon limits of $p$-brane
solutions in supergravity theories. This type of spacetime is of
interest for us because it is an essential ingredient in the formulation of 
the $\AdS/\text{CFT}$ correspondence.
The non-vanishing components of the Riemann and Ricci tensors and of the scalar
curvature read 
\begin{equation}\label{AdSScurvature}
\begin{aligned}
\mathcal{R}_{\mu\nu\rho\sigma}&=-\frac{1}{R_1^2}(g_{\mu\rho}g_{\nu\sigma}-g_{\mu\sigma}g_{\nu\rho})\col\qquad
&\mathcal{R}_{\mu\nu}&=-\frac{d}{R_1^2}g_{\mu\nu}\col\\ 
\mathcal{R}_{mnrs}&=\frac{1}{R_2^2}(g_{mr}g_{ns}-g_{ms}g_{nr})\col
&\mathcal{R}_{mn}&=\frac{d'}{R_2^2}g_{mn}\col\\
\end{aligned}\qquad
\mathcal{R}=-\frac{d(d+1)}{R_1^2}+\frac{d'(d'+1)}{R_2^2}\pnt
\end{equation}
It is not a maximally symmetric spacetime
because the number of Killing vectors is given by
$\frac{(d+1)(d+2)}{2}+\frac{(d'+1)(d'+2)}{2}$ instead of
$\frac{(d+d'+2)(d+d'+3)}{2}$ for the corresponding flat
$(d+d'+2)$-dimensional space. As a group manifold it reads
\begin{equation}\label{AdSScoset}
\AdSS=\frac{SO(2,d)}{SO(1,d)}\times\frac{SO(d'+2)}{SO(d'+1)}\pnt
\end{equation} 
The sphere itself can be described by the embedding 
\begin{equation}
\sum_{i=1}^{d'+2}Y_i^2=R_2^2\col
\end{equation}
in the flat $(d'+2)$-dimensional $\mathds{R}^{d'+2}$ with the standard
Euclidean metric.
One choice of coordinates that is very useful is given by
\begin{equation}\label{Yinglobalcoord}
\begin{aligned}
Y_1&=R_2\cos\vartheta\cos\psi\col\\
Y_2&=R_2\cos\vartheta\sin\psi\col\\
Y_i&=R_2\sin\vartheta\,\hat\omega_i\col\qquad
i=3,\dots,d+2\col\qquad\sum_i\hat\omega_i^2=1\pnt
\end{aligned}
\end{equation}
Some more details about these coordinates can be found in Appendix
\ref{app:SandSH} .
The induced metric then assumes the form
\begin{equation}
\de s_\text{S}^2=R_2^2\de\Omega_{d'+1}^2=R_2^2\big(
\cos^2\vartheta\de\psi^2+\de\vartheta^2+\sin^2\vartheta\de\hat\Omega_{d'-1}^2\big)\col
\end{equation}
and the chordal distance reads
\begin{equation}\label{Schordaldist}
v(y,y')=(Y(y)-Y(y'))^2=2R_2^2\Big[1-\cos\vartheta\cos\vartheta'\cos(\psi-\psi')-\sin\vartheta\sin\vartheta'\,\hat\omega^i\hat\omega'_i\Big]\pnt
\end{equation}
The chordal distance of a direct product space like $\AdSS$ is then given with
\eqref{AdSchordaldist} as
the direct sum $u+v$ of the distances in both spaces (remember that the chordal
distances are squared distances). 

The metric of the complete product space
$\AdSS$ reads in the global or Poincar\'e coordinates for $\AdS_{d+1}$ respectively
\begin{equation}\label{AdSSmetric}
\begin{aligned}
\de s_{\AdS\times\text{S}}^2&=R_1^2\sec^2\bar\rho\Big(-\de
t^2+\de\bar\rho^2+\sin^2\bar\rho\de\Omega_{d-1}^2+\frac{R_2^2}{R_1^2}\cos^2\bar\rho\de\Omega_{d'+1}^2\Big)\\
&=R_1^2\frac{1}{x_\perp^2}\Big(-\de x_0^2+\de x_\perp^2+\de\vec x^2+\frac{R_2^2}{R_1^2}x_\perp^2\de\Omega_{d'+1}^2\Big)\pnt
\end{aligned}
\end{equation}
Both expressions indicate that $R_1=R_2$ is a special value.
At $R_1=R_2$ the metric in the first line is up to the conformal factor the
 $(d+d'+2)$-dimensional Einstein static universe (ESU) with 
the topology $\mathds{R}\times\text{S}^{d+d'+1}$. 
In contrast to pure $\AdS$ that is conformal to only one half of an ESU, 
$\AdSS$ is conformal to a complete ESU.  
In the Poincar\'e coordinates 
the $(d'+1)$-dimensional sphere and the coordinate $x_\perp$ 
together form a flat $(d'+2)$-dimensional space with $x_\perp$ as radial 
coordinate. The expression in parenthesis then is the metric of
$\mathds{R}^{1,d+d'+1}$ and in this case $\AdSS$ is conformally flat as a
whole, and this is not only valid for 
its factors $\AdS_{d+1}$ and $\text{S}^{d+1}$ separately.
A computation of the Weyl tensors in Appendix \ref{app:Weyltens} shows that 
$R_1=R_2$ is the necessary and sufficient condition for the above
statements for $d,d'>0$. 

The conformal boundary of $\AdSS$ is the same as for pure $\AdS_{d+1}$, see
the discussion around \eqref{AdSmetricglobalcoord}.
This can be easily seen by the fact that at the boundary, where
$\bar\rho=\frac{\pi}{2}$, the factor in front
of the part that refers to $\text{S}^{d'+1}$ in the metric becomes zero. 
Hence, the sphere shrinks to a point at the boundary and has no influence on
the boundary structure. Like in the case of pure $\AdS_{d+1}$ one has to 
add two points $i^-$ and $i^+$ representing timelike past and future infinity
to obtain the complete boundary structure.  
\subsection{pp wave and plane wave spacetimes}
\label{ppwavespacetime}
The `plane fronted waves with parallel rays' (pp waves in
  short) contain the plane waves as a subclass. The latter
were originally discussed as solutions of the vacuum
Einstein equations that describe gravitational waves far away from their
sources, see for instance \cite{Einstein:1937qu}.
Brinkmann \cite{Brinkmann:1923,Brinkmann:1925}
introduced the `plane fronted wave' and Rosen assumed that spacetime filling
  plane waves do not exist due to singularities. He rejected them as 
unphysical. Robinson \cite{Robinson:1956}, however,  
found that the singularities are
caused by the choice of the coordinate system and are not physical
  singularities. In \cite{Bondi:1959aj} the authors discussed
plane waves in general and defined them as solutions of the vacuum Einstein
equations with as many symmetries as found for an electromagnetic plane wave. 
Penrose \cite{Penrose:1965} showed that a plane wave admits no 
spacelike hypersurface which would be adequate for the global specification of
  Cauchy data and that it is therefore impossible to
  embed the plane wave globally into a pseudo-Euclidean space.  
Later Cahen and Wallach \cite{Cahen:1970} classified all Lorentzian symmetric 
spaces. The plane wave metric is of that type \cite{Figueroa-O'Farrill:2001nz}.

The pp wave / plane wave spacetimes attract new attention as solutions
 of the
  $11$-dimensional \cite{Kowalski-Glikman:1984wv,Hull:1984vh} and of type
  $\twob$ supergravity \cite{Blau:2001ne}. 
The most general $D$-dimensional pp wave metric is given by \cite{Blau:2003rt}
\begin{equation}
\de s^2=-4\de z^+\de z^-+H(z^+,z)(\de z^+)^2+2A_i(z^+,z)\de z^i\de z^++\de \vec
z^2\pnt
\end{equation}
It has flat transverse ($i=1,\dots,D-2$) $(D-2)$-dimensional space (`plane
fronted') and a covariantly constant null killing vector $\partial_{z^-}$ 
(`parallel rays').
In the following we will consider the case without $A_i(z^+,z)$. In
\cite{Blau:2003rt} it was shown that one can remove $A_i(z^+,z)$ by a field
redefinition if it is of the form  $A_i(z^+,z)=A_{ij}(z^+)z^j$.
The metric then has the form
\begin{equation}\label{ppwavemetric}
\de s^2=-4\de z^+\de z^-+H(z^+,z)(\de z^+)^2+\de \vec
z^2\pnt
\end{equation}
and the non-vanishing components of the Christoffel connection
\eqref{connection} read
\begin{equation}\label{pwconnection}
\Gudd{-}{+}{+}=-\frac{1}{4}\partial_+H\col\qquad\Gudd{-}{+}{i}=\Gudd{-}{i}{+}=-\frac{1}{4}\partial_iH\col\qquad\Gudd{i}{+}{+}=-\frac{1}{2}\partial_iH\pnt
\end{equation}
Using \eqref{curvaturetensor} and \eqref{Riccitensor} one finds
\begin{equation}\label{ppwavecurvature}
\Ruddd{-}{i}{+}{j}=\frac{1}{4}\partial_i\partial_jH\col\qquad\mathcal{R}_{++}=-\frac{1}{2}\Delta
H\col\qquad\mathcal{R}=0 
\end{equation}
for the only independent components of the Riemannian curvature tensor and for
the complete Ricci tensor and the curvature scalar. The
Laplacian in the $(D-2)$-dimensional transverse space is denoted with
$\Delta$.  The pp wave thus obeys the Einstein equations \eqref{Einsteineq} in
the vacuum ($\Lambda=0$) if $H$ is a harmonic function in the
$(D-2)$-dimensional transverse space.

For the subclass of the plane waves the function $H$ in 
\eqref{ppwavemetric} is quadratic in the transverse coordinates $z^i$
\cite{Stephani:2003tm,Hubeny:2002zr}
\begin{equation}
H(z^+,z)=H_{ij}(z^+)z^iz^j\pnt
\end{equation}
Hence, the plane wave is Ricci flat if 
the matrix $H_{ij}$ is traceless
\begin{equation}
\delta^{ij}H_{ij}=0\pnt
\end{equation}
Another special case is given by 
\begin{equation}\label{Hdelta}
H_{ij}(z^+,z)=H(z^+)\delta_{ij}z^iz^j
\end{equation}
which is relevant in what follows. It is easy to see that in this case the 
pp wave is conformally flat. The Ricci-flat solution is denoted as purely
gravitational and the conformally flat one as purely electromagnetic
respectively \cite{Penrose:1965}.  

With the metric \eqref{ppwavemetric} one can construct $(D=10)$-dimensional
string and $(D=11)$-dimensional supergravity 
backgrounds by switching on some flux
such that \eqref{SUGRAeom} is obeyed after inserting \eqref{ppwavecurvature}.
For type $\twob$ one finds solutions of the form
\begin{equation}
\begin{aligned}
\de s^2&=-4\de z^+\de z^-+H(z^+,z)(\de z^+)^2+\de\vec z^2\col\\
F_5&=\de z^+\wedge\varphi(z^+,z)\col
\end{aligned}
\end{equation}
where $H$ has to fulfill
\begin{equation}
\Delta H=-|\varphi|^2\col\qquad|\varphi|^2=\frac{1}{4!}\varphi_{ijkl}\varphi^{ijkl}
\end{equation}
and $\Delta$ is the Laplacian in the transverse $(D-2=8)$-dimensional space.
For $11$-dimensional supergravity these solutions look rather similar with
\begin{equation}
\begin{aligned}
\de s^2&=-4\de z^+\de z^-+H(z^+,z)(\de z^+)^2+\de\vec z^2\col\\
F_4&=\de z^+\wedge\varphi(z^+,z)\col
\end{aligned}
\end{equation}
where $H$ has to fulfill
\begin{equation}
\Delta H=-\varphi^2\col\qquad\varphi^2=\frac{1}{3!}\varphi_{ijk}\varphi^{ijk}
\end{equation}
and $\Delta$ is the Laplacian in the transverse $(D-2=9)$-dimensional space.

An important subclass of the above given solutions in both cases is again 
of the plane wave type with an $H$ of the form
\begin{equation}
H(z^+,z)=H(z)=H_{ij}z^iz^j\col
\end{equation}
where $H_{ij}$ 
is a constant symmetric matrix and $i,j=1,\dots,D-2$ are indices in
the $8$- and $9$-dimensional transverse subspaces for type $\twob$ and
$11$-dimensional supergravity respectively. The corresponding spacetimes are
called homogeneous plane waves. They are Lorentzian symmetric spaces
 generically discussed in \cite{Cahen:1970}
and are denoted as Cahen-Wallach (CW) spaces in the literature
\cite{Figueroa-O'Farrill:2001nz,Blau:2001ne,Blau:2002mw}. 
The $5$- and $4$-form fluxes are homogeneous in these cases and null, i.\ e.\ 
$|F_5|^2=0$ and $|F_4|^2=0$ respectively and
the complete solutions is denoted as homogeneous pp (Hpp) wave
\cite{Figueroa-O'Farrill:2001nz}. 

The above solutions have at least $16$ Killing spinors and therefore preserve
at least one half of supersymmetry. For a special chosen
matrix $H_{ij}$ one finds that in both cases, type $\twob$ and $11$-dimensional
supergravity, there exists one solution that admits $32$ Killing spinors and
thus preserves maximal supersymmetry.
For type $\twob$ the solution becomes \cite{Blau:2001ne}\footnote{Note that in 
\cite{Blau:2001ne} there is an additional
  factor of $2$ in $\varphi$. The result presented here 
coincides with \cite{Russo:2004kr}.}
\begin{equation}\label{typeIIBpw}
H_{ij}=-\mu^2\delta_{ij}\col\qquad\varphi=2\mu(\de z^1\wedge\de z^2\wedge\de z^3\wedge\de z^4+\de z^5\wedge\de z^6\wedge\de z^7\wedge\de z^8)\pnt
\end{equation}
Since $H$ is of the form
\eqref{Hdelta} the spacetime is conformally flat.  
In $D=11$ dimensions one finds \cite{Kowalski-Glikman:1984wv}
\begin{equation}\label{11dpw}
H_{ij}=\begin{cases}
-\frac{1}{9}\mu\delta_{ij} & i,j=1,\dots,3 \\
-\frac{1}{36}\mu\delta_{ij} & i,j=4,\dots,9
\end{cases}
\col\qquad\varphi=\mu\de z^1\wedge\de z^2\wedge\de z^3\col
\end{equation}
which is denoted as Kowalski-Glikman (KG) solution in the literature 
\cite{Figueroa-O'Farrill:2001nz,Blau:2001ne,Blau:2002mw}. It is important to
remark that a common factor in $H_{ij}$  can be absorbed in a simultaneous
rescaling of $z^+$ and $z^-$.
In particular this means that the parameter
$\mu$ can be set to an arbitrary non-zero value.

It has been shown by Penrose \cite{Penrose:1965} that it is impossible to
globally embed the plane wave spacetimes 
into a pseudo-Euclidean spacetime. 
However, an isometric
embedding of the $D$-dimensional CW spaces with metric
\begin{equation}\label{CWmetric}
\de s^2=-4\de z^+\de z^-+H_{ij}z^iz^j(\de z^+)^2+\de \vec
z^2
\end{equation}
in $\mathds{R}^{2,D}$ is possible \cite{Blau:2002mw}.
The metric of $\mathds{R}^{2,D}$ \eqref{R2dmetric} via the 
coordinate transformations
\begin{equation}
Z_+^1=\frac{1}{2}(Z_0+Z_d)\col\quad Z_-^1=\frac{1}{2}(Z_0-Z_d)\col\quad Z_+^2=\frac{1}{2}(Z_{d+1}+Z_{d-1})\col\quad Z_-^2=\frac{1}{2}(Z_{d+1}-Z_{d-1})
\end{equation}
can be transformed to    
 \begin{equation}
\de s^2=-4\sum_{k=1}^2\de Z_+^k\de Z_-^k+\sum_{i=1}^D\de Z_i^2\pnt
\end{equation}
If the hypersurface is defined as
\begin{equation}
\sum_{k=1}^2Z_+^kZ_+^k=1\col\qquad H_{ij}Z^iZ^j+4\sum_{k=1}^2Z_+^kZ_-^k=0
\end{equation}
and parameterized as follows
\begin{equation}
\begin{aligned}
Z_+^1&=\cos z^+\col\\
Z_-^1&=-z^-\sin z^+-\frac{1}{4}H_{ij}z^iz^j\cos z^+\col\\
Z_+^2&=\sin z^+\col\\
Z_-^2&=z^-\cos z^+-\frac{1}{4}H_{ij}z^iz^j\sin z^+\col\\
Z_i&=z_i\col
\end{aligned}
\end{equation}
one finds that the induced metric is given by \eqref{CWmetric}.
The chordal distance in the plane wave reads
\begin{equation}\label{pwchordaldist}
\begin{aligned}
\Phi(z,z')&=-4\sum_{k=1}^2(Z_+^k(z)-Z_+^k(z'))(Z_-^k(z)-Z_-^k(z'))+\sum_{i=1}^D(Z_i(z)-Z_i(z'))^2\\
&=-4(z^--z'^-)\sin(z^+-z'^+)+2H_{ij}(z^iz^j+z'^iz'^j)\sin^2\tfrac{z^+-z'^+}{2}+(\vec
z-\vec z')^2\pnt
\end{aligned}
\end{equation}

\section{Limits of spacetimes}
\label{limitsofspacetimes}
In this Section we discuss two limiting processes that relate the previously
discussed spacetimes with each other and that are of relevance in the
$\AdS/\text{CFT}$ correspondence and in its BMN limit. 
The first so called near horizon limit applies to the 
$p$-brane solutions of supergravity and, for special choices of the spacetime
dimensions, produces a background of the form $\AdSS$.  
The second so called Penrose-G\"uven limit applies to any supergravity
solution, transforming the geometry into a plane wave and additionally acting
on the other fields of the background. 

\subsection{The near horizon limit}
\label{nhlimit}
In the $\AdS/\text{CFT}$ correspondence the near horizon limit
\cite{Gibbons:1993sv} 
of the $p$-brane backgrounds plays an important role.
In the setup of type $\twob$ string theory in the presence of a stack of
$\text{D}3$-branes, one has closed strings in the bulk and open
strings that describe the excitation of the branes ending on the branes. 
In the near
horizon limit these two different string sectors decouple, and two alternative 
descriptions of the near horizon regions are related 
via the $\AdS/\text{CFT}$ correspondence. 
The near horizon limit can also be applied to other geometries e.\ g.\ given
by a configuration of different branes or 
to black holes 
\cite{Maldacena:1998re,Kallosh:1992ta,Chamseddine:1996pi,Ferrara:1997tw}. 
In the former case it extracts information about the
shape of the background close to the branes. In particular, the 
metric in the limit describes how the spacetime looks like in this region. 
For some special brane configurations one obtains geometries of the form
$\AdSS\times M^{D-(d+d'+2)}$, where $M$ is a $(D-(d+d'+2))$-dimensional
Euclidean manifold.
Here we will restrict ourselves to the case a single (stack of) $(d-1)$-branes
where the limiting spacetime is a direct product of only two factors
and show under which conditions the limiting geometry becomes
$\AdS_{d+1}\times\text{S}^{d'+1}$. 

The starting point is the $(d-1)$-brane solution of supergravity given 
in \eqref{pbranesol}. We remember that the coordinate $y=\sqrt{y^iy_i}$
describes the distance 
from the $(d-1)$-brane with `extension' $R_2$. To focus into the region 
close to the brane therefore means to discuss the solution for $R_2\gg y$.
For general $d$ and $d'$, however, one does not find an
$\AdS_{d+1}\times\text{S}^{d'+1}$ geometry close to the brane. The condition
on the dimensions turns out to be 
\begin{equation}\label{dimcond}
dd'=2(d+d')\Leftrightarrow a(d)^2=0\col
\end{equation}
where $a(d)$ is defined in \eqref{aofp} and from \eqref{pbranesol} one
immediately finds that all these solutions have a constant dilaton $\phi$.
The metric in this case reads for $R_2\gg y$
\begin{equation}
\begin{aligned}
\de s^2&=\Big(\frac{R_2}{y}\Big)^{-\frac{d'^2}{d+d'}}\de x^\mu\de
x_\mu+\Big(\frac{R_2}{y}\Big)^\frac{dd'}{d+d'}(\de y+y^2\de\Omega_{d'+1}^2)\\
&=\Big(\frac{R_2}{y}\Big)^{-\frac{2d'}{d}}\de
x^\mu\de x_\mu+\Big(\frac{R_2}{y}\Big)^2\de y^2+L^2\de\Omega_{d'+1}^2\pnt
\end{aligned}
\end{equation} 
One now has to replace $y$ by a new variable such that the first two terms
combine to the metric of $\AdS_{d+1}$ (in the Poincar\'e coordinates).
The redefinition reads
\begin{equation}\label{ytoxperp}
\frac{y}{R_2}=
\Big(\frac{\gamma R_2}{x_\perp}\Big)^\frac{d}{d'}\col\qquad \frac{\de
  y}{y}=-\frac{d}{d'}\frac{\de x_\perp}{x_\perp}\col
\end{equation} 
where $\gamma$ is a real parameter that has to be determined. 
The metric now becomes
\begin{equation}
\de s^2=\Big(\frac{\gamma R_2}{x_\perp}\Big)^2\Big[\Big(\frac{d}{d'\gamma}\Big)^2\de x_\perp^2+\de x^\mu\de x_\mu\Big]+L^2\de\Omega_{d'+1}^2\pnt
\end{equation}
A comparison with \eqref{AdSmetricPcoord} shows that $\gamma$ has to fulfill
\begin{equation}\label{radiiratio}
\gamma=\frac{R_1}{R_2}=\frac{d}{d'}\col
\end{equation}
in order for the first part of the metric to describe $\AdS_{d+1}$ ((in
the Poincar\'e coordinates).
Then it is clear that $R_1$ and $R_2$ are the 
radii of $\AdS_{d+1}$ and of $\text{S}^{d'+1}$ respectively.

What remains to be done is to determine the near horizon limit of the fluxes.
In polar coordinates one has $\de H^{-1}=\partial_y H^{-1}\de y$, and one finds
in the limit $R_2\gg y$ using the variable substitution \eqref{ytoxperp}
and $\gamma R_2=R_1$
\begin{equation}
\de H^{-1}=-d\Big(\frac{R_1}{x_\perp}\Big)^d\frac{\de
  x_\perp}{x_\perp}\pnt 
\end{equation}
With the volume form of $\AdS_{d+1}$ which in the Poincar\'e coordinates reads
\begin{equation}
\vol(\AdS_{d+1})=\Big(\frac{R_1}{x_\perp}\Big)^{d+1}\de x_\perp\wedge \de
x^0\wedge\dots\wedge\de x^{d-1}
\end{equation}
one then finds for the flux
\begin{equation}\label{nhlimitfieldstrength}
F_{d+1}=\frac{d}{R_1}\vol(\AdS_{d+1})
\end{equation}
and for its Hodge-dual \eqref{dualfieldstrength} with $a(d)=0$
\begin{equation}\label{nhlimitdualfieldstrength}
\star F_{d+1}=\frac{d'}{R_2}\vol(\text{S}^{d'+1})\col
\end{equation}
where $\text{S}^{d'+1}$ denotes the sphere with radius $R_2$.

At the end we discuss the measurements of energies in the near horizon
geometry.  
The exponential factor in the variable redefinition 
\eqref{ytoxperp} is strictly positive and therefore $R_2\gg y$ corresponds
to $R_2\ll x_\perp$. 
According to 
the discussion around \eqref{redshift} 
the deeper a particle is located in the
interior of AdS the more redshifted its wavelength appears to an observer at
some fixed position further outside (with smaller $x_\perp$).
The energy of a particle at $x_\perp$ in AdS 
measured by an observer at infinity reads 
\begin{equation}\label{AdSredshiftPcoord}
E_\infty=\frac{R_1}{x_\perp}E_{x_\perp}\pnt
\end{equation}

\subsection{The Penrose-G\"uven limit}
\label{PGlimit}
One of the main ingredients that lead to the formulation of the BMN limit
of the $\AdS/\text{CFT}$ correspondence is the observation that the
$10$-dimensional 
plane wave background of type $\twob$ string theory arises as a
Penrose-G\"uven limit of the $\AdS_5\times\text{S}^5$ background. 
In \cite{Penrose:1976} Penrose describes that the expansion of the metric
around a null geodesic in a congruence of null geodesics without conjugate 
(focal) points leads to a plane wave metric. The absence of conjugate (focal)
points (that are defined as points where infinitesimally neighboured null
geodesics intersect) is important because otherwise the coordinate system
breaks down. Intuitively this expansion is the one-dimensional analog
of the expansion of a spacetime around a single point which results in the
flat tangent space at that point. Before the work of Penrose appeared it was
already observed by Pirani \cite{Pirani:1959} that the 
gravitational field of a fast moving particle the more approaches a plane wave
the higher its velocity. 

The work of Penrose was later extended by G\"uven
\cite{Gueven:2000ru}, who included all massless 
fields into the limiting process. This enables one to apply the limiting 
procedure to a complete string background and then to arrive at a new 
consistent background.\footnote{The homogeneity of the actions of the theories
under the required rescaling guarantees that the limit of a consistent
background again is itself a consistent background.}  

Let us in brief review the Penrose-G\"uven limit. 
A general string background in $D$ dimensions consists of all bosonic fields 
including the
metric $g_{MN}$, the NS-NS fields $B_{MN}$ and $\phi$, a background gauge
field $A_M$ and 
the $(p+1)$-form R-R potentials $A_{p+1}$.
One first chooses a coordinate system so that the metric takes the form
\begin{equation}
\de s^2=2\de z^+(-2\de z^-+\alpha\de z^++\beta_i\de z^i)-C_{ij}\de z^i\de z^j\col
\end{equation}
where $C_{ij}$ is a symmetric positive definite $(D-2)\times(D-2)$ matrix.
This coordinate system breaks down at the nearest conjugate point where 
$\det C=0$. The above choice of coordinates singles out a congruence of null
geodesics with $z^+,z^i=\text{const.}$ and $z^-$ the affine parameter along the
geodesics.
The next step is to fix a particular gauge. We will collectively deal with 
all fields $B_{MN}$, $A_M$ and $A_{p+1}$ and denote them as $A_{p+1}$. 
The formulas for $B$ and the gauge field $A$ follow from the expressions for 
$A_{p+1}$ by simply setting $p=1$ and $p=0$ respectively.
The $(p+1)$-form possesses a gauge freedom of the form
\begin{equation}
A_{p+1}\to A'_{p+1}=A_{p+1}+\de\Lambda_p\col\qquad\Lambda_p\in\Omega^p\col
\end{equation}
where $\Lambda_p$ is a $p$-form. It is used to set
\begin{equation}
(A_{p+1})_{-i_1\dots i_{p-1}}=0\pnt
\end{equation}
The next step is to rescale the coordinates with a positive real number
$\Omega$ in the following way
\begin{equation}
z^-=\bar z^-\col\qquad z^+=\Omega^2\bar z^+\col\qquad z^i=\Omega\bar z^i
\end{equation}
and to define new fields according to
\begin{equation}\label{fieldrescaling}
\begin{gathered}
\bar
g_{\mu\nu}(\Omega)=\Omega^2g_{\mu\nu}(\Omega)\col\qquad\bar\phi(\Omega)=\phi(\Omega)\col\\
\bar A_{p+1}(\Omega)=\Omega^{-p-1}A_{p+1}(\Omega)\col\qquad\bar 
F_{p+2}(\Omega)=\de\bar A_{p+1}(\Omega)=\Omega^{-p-1}F_{p+2}(\Omega)\pnt 
\end{gathered}
\end{equation}
One then finds in the $\Omega\to0$ limit
\begin{equation}
\begin{aligned}
\de\bar s^2&=-4\de\bar z^+\de\bar z^--C_{ij}\de\bar z^i\de\bar z^j\col\\
\bar\phi&=\bar\phi(\bar z^-)\\
\bar A_{p+1}&=\frac{1}{(p+1)!}\bar A(\bar z^-)_{i_1\dots i_{p+1}}\de\bar z^{i_1}\wedge\dots\wedge\de\bar z^{i_{p+1}}+\text{gauge}\\
\bar F_{p+2}&=\frac{1}{(p+1)!}\bar F(\bar z^-)_{-i_1\dots i_{p+1}}\de\bar z^-\wedge\de\bar z^{i_1}\wedge\dots\wedge\de\bar z^{i_{p+1}}\col
\end{aligned}
\end{equation}
which is a representation of the plane wave metric and of the fields in Rosen
coordinates. One can transform to Brinkmann (or harmonic) coordinates
$(u,v,y^i)$ with the relations
\begin{equation}
\bar z^-=u\col\qquad\bar z^+=v-\frac{1}{4}C'_{ij}(\bar z^-)Q^i_{\phantom{i}k}(\bar z^-)Q^j_{\phantom{j}l}(\bar z^-)y^ky^l\col\qquad\bar z^i=Q^i_{\phantom{i}j}(\bar z^-)y^j\col
\end{equation}
where a prime denotes a derivative w.\ r.\ t.\ the argument and the matrix $Q^i_{\phantom{i}j}$ fulfills
\begin{equation}
C_{ij}Q^i_{\phantom{i}k}Q^j_{\phantom{j}l}=\delta_{jl}\col\qquad 
C_{ij}\big((Q')^i_{\phantom{i}k}Q^j_{\phantom{j}l}-Q^i_{\phantom{i}k}(Q')^j_{\phantom{j}l}\big)=0\pnt
\end{equation}
If one then defines 
\begin{equation}
h_{kl}=-\big(C'_{ij}Q^j_{\phantom{j}l}+C_{ij}(Q'')^j_{\phantom{j}l}\big)Q^i_{\phantom{i}k}
\end{equation}
one obtains the final expression for the plane wave background
\begin{equation}
\begin{aligned}
\de\bar s^2&=2\de u\de v-h_{kl}(u)y^ky^l\de u^2-\delta_{kl}\de y^k\de y^l\col\\
\bar\phi&=\bar\phi(u) \\
\bar F_{p+2}&=\frac{1}{(p+1)!}\bar F(u)_{ui_1\dots i_{p+1}}\de u\wedge\de
y^{i_1}\wedge\dots\wedge\de y^{i_{p+1}}\pnt
\end{aligned}
\end{equation}
Note that the field strengths but not the potentials retain their form under 
the coordinate transformation.

Before we discuss special cases let us collect what are called the hereditary
issues in the literature \cite{Geroch:1969,Blau:2002mw}. Due to Geroch 
\cite{Geroch:1969}   
a property is called hereditary if, whenever a family of spacetimes has this 
property, all limits of this family have this property, too. Here we will
follow the definition of \cite{Blau:2002mw} where a property is called
hereditary if it is valid for an
initial consistent background (not necessarily for a complete family of
spacetimes) and remains valid in the Penrose-G\"uven limit (not necessarily in
all limits) and not only for the metric but for all fields of the background. 
The hereditary properties are \cite{Blau:2002mw}
\begin{enumerate}
\item The Penrose limit of a Ricci flat spacetime is Ricci flat.
\item The Penrose limit of a conformally flat spacetime is conformally flat.
\item The Penrose limit of an Einstein space is always Ricci flat.
\item The Penrose limit of a locally symmetric space (e.\ g.\ the CW spaces)
is locally symmetric.
\item The Penrose-G\"uven limit of a solution of the supergravity 
equations of motion remains a solution.
\item The dimension of the symmetry algebra does not decrease in the
  Penrose-G\"uven limit.
\item The number of conserved supersymmetries does not decrease in the
  Penrose-G\"uven limit. 
\end{enumerate}
We will now review the Penrose limits of $\AdS_{d+1}$ and $\AdSS$.

Since $\AdS_{d+1}$ is a conformally flat locally symmetric Einstein space it 
follows from the above given list of hereditary properties that its Penrose 
limit must be a conformally flat locally symmetric Ricci flat space which is 
of course isometric to $\mathds{R}^{1,d}$. 
 
For $\AdSS$ the situation is more subtile but it can nevertheless be discussed
for general dimensions \cite{Blau:2002mw}. Remember that $\AdSS$ 
can be represented as the group manifold \eqref{AdSScoset}. 
Since the space is homogeneous we can 
go to an arbitrary point $p_0$ of this spacetime and obtain the different 
Penrose limits by taking all possible different null geodesics that cross
$p_0$. It is clear that two null geodesics that are now completely specified
by their directions can only lead to different Penrose
limits if one cannot map the tangent vectors (which are null) 
of these geodesics at $p_0$ into each
other by using the isometries that keep $p_0$ as a fixed point. These
isometries are precisely given by the elements of the stabilizer which
is $H=SO(1,d)\times SO(d'+1)$. The number of different Penrose limits is
therfore given by the number of different orbits of null directions under the
action of $H$.
Since $\text{S}^{d'+1}$ has only spacelike directions it is clear that the 
component of the tangential vector in the $\AdS_{d+1}$ part can either be 
null or timelike. If it is null it has to lie on an $\text{S}^{d-1}$ and the 
geodesic is stationary in the sphere. If, however, the AdS part of the 
tangential vector is timelike then it can point in any timelike direction 
and is not confined to the lightcone. Its tip lies on the ball $\text{B}^d$ of
future pointing timelike directions. If one has chosen the component in AdS 
one can only choose the direction in $\text{S}^{d'+1}$, i.\ e.\ the component
of the tangent vector in $\text{S}^{d'+1}$ is then confined to end on an
$\text{S}^{d'}$. The AdS part of the stabilizer is $SO(1,d)$. One can
therefore map all vectors starting at $p_0$ with fixed length to each other 
by a spatial rotation and then identify timelike vectors with different
lengths via a Lorentz boost. In $\text{S}^{d'+1}$ the second factor of the 
stabilizer $SO(d'+1)$ allows one to identify with each other 
all vectors with equal length that start at $p_0$ . 
That means there are only two 
distinct orbits and therefore two different Penrose limits in $\AdSS$. 
One is obtained by expanding around a null geodesic which has no movement in
the sphere $\text{S}^{d'+1}$. The other corresponds to a geodesic that has a
non-vanishing movement in $\text{S}^{d'+1}$. 
The first choice is not very interesting. The Penrose limit is the same as in
case of pure $\AdS_{d+1}$, that means flat Minkowski space. 
The limit for the second choice of geodesic will be worked out in the
following. 

The simplest way to obtain the plane wave spacetimes from $\AdSS$ is to cast
\eqref{AdSmetricglobalcoord} into the form 
\begin{equation}\label{AdSSmetricglobalcoord}
\de s^2=R_1^2\Big(-\cosh^2\rho\de t^2+\de\rho^2+\sinh^2\rho\de\Omega_{d-1}^2+\frac{R_2^2}{R_1^2}\big(\cos^2\vartheta\de\psi^2+\de\vartheta^2+\sin^2\vartheta\de\hat\Omega_{d'-1}^2\big)\Big)\pnt
\end{equation}
Going to lightcone coordinates
\begin{equation}
z^+=\frac{1}{2}\Big(t+\frac{R_2}{R_1}\psi\Big)\col\qquad z^-=\frac{R_1^2}{2}\Big(t-\frac{R_2}{R_1}\psi\Big)\col
\end{equation}
one replaces all coordinates as follows
\begin{equation}\label{BMNcoordinates}
t=z^++\frac{z^-}{R_1^2}\col\qquad\psi=\frac{R_1}{R_2}\Big(z^+-\frac{z^-}{R_1^2}\Big)\col\qquad\rho=\frac{r}{R_1}\col\qquad\vartheta =\frac{y}{R_2}\col
\end{equation}
and expands in powers of $R_1^{-1}$ and  $R_2^{-1}$.
Neglecting terms $\mathcal{O}(R_1^{-1})$ and $\mathcal{O}(R_2^{-1})$
the metric reads
\begin{equation}\label{AdSSmetricexpansion}
\begin{aligned}
\de s^2&=R_1^2\Big(-\Big[(\de z^+)^2+\frac{r^2}{R_1^2}(\de
z^+)^2+\frac{2}{R_1^2}\de z^+\de z^-\Big]+\frac{\de
  r^2}{R_1^2}+\frac{r^2}{R_1^2}\de\Omega_{d-1}^2\\
&\phantom{{}=R_1^2\Big(}+\Big[(\de z^+)^2-\frac{y^2}{R_2^2}(\de
z^+)^2-\frac{2}{R_1^2}\de z^+\de z^-\Big]+\frac{\de
  y^2}{R_1^2}+\frac{y^2}{R_1^2}\de\hat\Omega_{d'-1}^2\Big)\\
&=-4\de z^+\de z^--\Big(r^2+\frac{R_1^2}{R_2^2}y^2\Big)(\de
z^+)^2+\de r^2+r^2\de\Omega_{d-1}^2+\de y^2+y^2\de\hat\Omega_{d'-1}^2\pnt
\end{aligned}
\end{equation}
This result is exact if we take the limit $R_1,R_2\to\infty$,
$\frac{R_1}{R_2}=\text{fixed}$, which clearly corresponds to the plane wave 
limit. Indeed, the above result has the form of a plane wave spacetime if we 
take the embedding space coordinates $\omega$ and $\hat\omega$ respectively
for the subspheres of $\AdS_{d+1}$ and $\text{S}^{d'+1}$ and define Cartesian
coordinates  
\begin{equation}\label{transvspacepolarcoord}
\begin{aligned}
x_i&=r\,\omega_i\col&\qquad i&=1,\dots,d\col&\qquad\sum_i\omega_i^2&=1\col \\
y_{i'}&=y\,\hat\omega_{i'}\col&\qquad i'&=1,\dots,d'\col&\qquad\sum_{i'}\hat\omega_{i'}^2&=1\pnt 
\end{aligned}
\end{equation}
The metric then becomes with $\vec z=(\vec x,\vec y)$
\begin{equation}
\begin{aligned}
\de s^2&=-4\de z^+\de z^-+H_{ij}z^iz^j(\de z^+)^2+\delta_{ij}\de z^i\de
z^j\col\\
H_{ij}&=\begin{cases}-\delta_{ij} & i,j=1,\dots,d \\
                    -\frac{R_1^2}{R_2^2}\delta_{ij} & i,j=d+1,\dots,d+d'
\end{cases}
\end{aligned}\pnt
\end{equation} 
This metric includes precisely the ones of the $10$-dimensional plane wave 
solutions of type $\twob$ \eqref{typeIIBpw}, where $R_1=R_2$, $d=d'=4$ and of
the $11$-dimensional KG solution \eqref{11dpw}
with $R_2=2R_1$ and $d=3$, $d'=6$ or $R_2=\frac{1}{2}R_1$ and $d=3$, $d'=6$ 
after changing a common factor of $H_{ij}$ 
by a redefinition of $z^+$ and $z^-$.

At the end we have to compute the flux in the the Penrose-G\"uven limit.
From \eqref{nhlimitfieldstrength} and \eqref{nhlimitdualfieldstrength} one
can see that the rescaling \eqref{fieldrescaling} is necessary 
because the $(d+1)$-form flux is of order $\mathcal{O}(R_2^d)$. With our
coordinate choice \eqref{BMNcoordinates} where the scale factor is included,
 the radii dependence should cancel out up to finite ratios $\frac{R_1}{R_2}$.
The volume forms in global coordinates read
\begin{equation}
\begin{aligned}
\vol(\AdS_{d+1})&=R_1^{d+1}\cosh\rho(\sinh\rho)^{d-1}\de t\wedge\de
\rho\wedge\vol(\Omega_{d-1})\col\\
\vol(\text{S}^{d'+1})&=R_2^{d'+1}\cos\theta(\sin\theta)^{d'-1}\de\psi\wedge\de
\theta\wedge\vol(\hat\Omega_{d'-1})\col
\end{aligned}
\end{equation}
where $\vol(\Omega_{d-1})$, $\vol(\hat\Omega_{d'-1})$ are volume forms on the 
corresponding unit spheres.
Changing the coordinates according to \eqref{BMNcoordinates} and 
only keeping the leading contribution then results in 
\begin{equation}
\vol(\AdS_{d+1})=R_1\,r^{d-1}\de z^+\wedge\de
r\wedge\vol(\Omega_{d-1})\col\quad\vol(\text{S}^{d'+1})=R_1\,y^{d'-1}\de
z^+\wedge\de y\wedge\vol(\hat\Omega_{d'-1})\pnt 
\end{equation}
Expressed in the variables \eqref{transvspacepolarcoord},
one then finds for the $(d+1)$-form flux 
\begin{equation}
F_{d+1}=d\de z^+\wedge\vol(\mathds{R}^d)
\end{equation}
and the dual flux becomes with $D=d+d'+2$
\begin{equation}
\star F_{d+1}=d'\frac{R_1}{R_2}\de z^+\wedge\vol(\mathds{R}^{d'})=d\de z^+\wedge\vol(\mathds{R}^{d'})\col
\end{equation}
where the last equality follows with \eqref{radiiratio}.
According to \eqref{selfdualitysolution} the selfdual $5$-form field strength 
now reads
\begin{equation}
F_5=2\de z^+\wedge(\vol_x(\mathds{R}^4)+\vol_y(\mathds{R}^4))\col
\end{equation}
where the subscript denotes which set of coordinates one has to 
take from \eqref{transvspacepolarcoord}.
We have found the flux of the $10$-dimensional plane wave background
\eqref{typeIIBpw}.


\clearemptydoublepage
\chapter{Holography}
\label{chap:BMNholo}
In this Chapter we will first briefly describe the
$\AdS/\text{CFT}$ correspondence. Special emphasis will be put on the
concrete realization of holography in which the bulk-to-boundary propagator 
plays an important role. As a crucial point for the motivation of the analysis
in Chapter \ref{chap:propaga} we will show how this propagator is determined
from  
the bulk-to-boundary propagator.
Secondly, we will give a short description of the BMN limit of the
$\AdS/\text{CFT}$ correspondence. Then we will summarize the proposals 
about the realization of holography in this limit. On that basis we will then 
fix the direction of the analysis in the Chapters \ref{chap:boundgeo} and 
\ref{chap:propaga}.

\section{The $\AdS/\text{CFT}$ correspondence and holography}
\label{AdSCFTholo}
In the $\AdS/\text{CFT}$ correspondence \cite{Maldacena:1998re} 
string theory on an AdS space is related to a CFT on the boundary of this
space. Here we will mainly describe
the correspondence where the string theory is type $\twob$ on  
$\AdS_5\times\text{S}^5$. The boundary CFT then is the superconformal
$\mathcal{N}=4$ super Yang-Mills (SYM) theory in $4$ dimensions.
The following review will be rather short. For more details we refer the 
reader to the reviews 
\cite{Aharony:1999ti,D'Hoker:2002aw,Klebanov:2000me,DiVecchia:1999du,Petersen:1999zh}. 

Let us first describe the string theory setup in brief. We have seen in 
Subsection \ref{SUGRApbranesol}
that $10$-dimensional type $\twob$ supergravity admits a $3$-brane 
solution with non-vanishing self dual $5$-form flux $F_5$. 
This $3$-brane corresponds to a $\text{D}3$-brane
in the full string theory and thus type $\twob$ supergravity in its $3$-brane
background should be a low energy description of type $\twob$ string 
theory with a
$\text{D}3$-brane. On the other hand
\cite{Leigh:1989jq} the effective action can be
formulated including type $\twob$ supergravity in the bulk and the DBI action
on the branes. 
Instead of a single $\text{D}3$-brane one then takes a stack
of $N$ $\text{D}3$-branes where in general the parameter \eqref{L} of the 
$(d-1)$-brane solution \eqref{pbranesol} is replaced by
\begin{equation}
R_2^{d'}\to NR_2^{d'}\col
\end{equation} 
where $D=d+d'+2$. The gauge group on the branes is extended to the non-Abelian
$U(N)$. In the near horizon limit described in Subsection \ref{nhlimit} the
following now happens \cite{Aharony:1999ti,Akhmedov:1999rc,DiVecchia:1999yr}. 
The modes in the throat region, which is the regime close to the brane,
become trapped and decouple from the dynamics outside this 
region.\footnote{This
  is not completely true. The modes which describe the center of mass position
  of the stack of branes correspond to the $U(1)$ degrees of freedom and do not
 decouple \cite{Aharony:1999ti}.}  
The action of type $\twob$ supergravity in its $3$-brane background 
thus splits into 
two decoupled parts, the near horizon part which is type $\twob$ supergravity
in $\AdS_5\times\text{S}^5$ and a part which lives outside the horizon.
In the second description the interaction of the
$\text{D}3$-brane with the bulk vanishes in the limit 
and one finds two decoupled systems similar to the situation above. 
One part is $\mathcal{N}=4$ SYM with gauge group $SU(N)$ on the worldvolume of 
the brane and the 
other part is exactly the same already found in the near horizon limit of the 
first description. Since both descriptions should be equivalent Maldacena
\cite{Maldacena:1998re} 
conjectured that supergravity in $\AdS_5\times\text{S}^5$ should be a dual
description of the $4$-dimensional $\mathcal{N}=4$ $SU(N)$ SYM gauge theory 
and he extended this conjecture to full string theory. The parameters of both
theories are related in the following way, see \eqref{L} and \eqref{radiiratio}
\begin{equation}\label{AdS5S5para}
R_1=R_2=R\col\qquad R^4=\lambda\alpha'^2\col\qquad\lambda=g^2N\col\qquad g^2=4\pi g_s\col
\end{equation}
where $g$ is the YM coupling and $\lambda$ the 'tHooft coupling constant. 
For the time being, however, it seems hopeless to work with the full string
theory on $\AdS_5\times\text{S}^5$. The reason is that the background 
includes the non-vanishing Ramond-Ramond flux $F_5$ and one is thus 
forced to work with the Green-Schwarz superstring action which up to now could 
not be quantized successfully in this case. The approximation one now uses is
type $\twob$ supergravity instead of the full string theory. 
It is faithful as long as the curvature radius of the background is large
compared to the string length. To avoid string loop corrections one should
furthermore take $g_s\to 0$. This can be realized by the following
sequence of limits 
\begin{equation}
N\to\infty\col\quad\text{with}\quad\lambda=\text{fixed}\col\quad\text{then}\quad\lambda\to\infty\pnt
\end{equation}
In this limit the correspondence then
relates classical supergravity in $\AdS_5\times\text{S}^5$ to the large $N$ limit of strongly coupled $\mathcal{N}=4$ SYM.  

If we regard the sphere as simply
generating the complete Kaluza-Klein (KK) tower of massive modes, the
conjecture then states that a $5$-dimensional supergravity theory should be 
equivalent to a $4$-dimensional CFT. Since equivalent means that both include
the same information, this correspondence is an example of the holographic
principle: higher dimensional information can be stored in a lower dimensional
description. Thereby the lower dimensional theory should not contain more than
one degree of freedom per Planck area \cite{'tHooft:1999bw,Susskind:1998dq}. 
In the following we will explain more precisely how holography is understood
in the $\AdS/\text{CFT}$ correspondence.

The $\AdS/\text{CFT}$ 
correspondence relates the correlation functions of
operators $\mathcal{O}_\Delta$ with conformal dimension $\Delta$ in the 
$d$-dimensional CFT to 
the classical value of the action $S_\twob$ 
of (dimensionally reduced) type $\twob$ 
supergravity in $\AdS_{d+1}$
evaluated with boundary condition $\bar\phi_\Delta(\bar x)$, $\bar x=(x_0,\vec x)$ as
follows \cite{Gubser:1998bc,Witten:1998qj} 
\begin{equation}\label{AdSCFTcorrcos}
\bigg\langle\exp{\int\de^d\bar x\,\bar\phi_\Delta(\bar x)\mathcal{O}_\Delta(\bar x)}\bigg\rangle_\text{CFT}=\e^{-S_\twob[\phi_\Delta[\bar\phi_\Delta]]}\pnt
\end{equation}
We remark that the above formulation is given for Euclidean $\AdS_{d+1}$ and
that we will keep dealing with the Euclidean case to the end of this Section.
The Lorentzian formulation can be found in 
\cite{Balasubramanian:1998sn,Balasubramanian:1998de}.
The metric with Euclidean signature in Poincar\'e coordinates is given by
\eqref{AdSmetricPcoord}  
but with the minus sign in front of $\de x_0^2$ converted to plus.
The boundary of $\AdS_{d+1}$ is situated at $x_\perp=0$. 
The supergravity modes $\phi_\Delta(x)$, $x=(x_\perp,\bar x)$ 
that correspond to the operators in the CFT  
are given by the non-normalizable 
modes\footnote{The normalizable modes determine the vacuum structure (VEVs) of
  the CFT. See \cite{Witten:1998qj} and  \cite{Balasubramanian:1998sn} for
differences between the Euclidean and Lorentzian case.} 
which scale 
as \cite{Freedman:1998tz,D'Hoker:2002aw}
\begin{equation}\label{phiboundbehaviour}
\bar\phi_\Delta(\bar x)=\lim_{x_\perp\to 0}\phi_\Delta(x)x_\perp^{\Delta-d}\pnt
\end{equation}
One can regard the bulk supergravity fields $\phi_\Delta$ 
as functionals of the
corresponding boundary values by computing the convolutions 
\begin{equation}\label{AdSboundvalsol}
\phi_\Delta[\bar\phi_\Delta](x)=\int\de^d\bar x'\,K_\Delta\big(x,\bar
x'\big)\bar\phi_\Delta(\bar x')\col
\end{equation}
where $K_\Delta(x,\bar x')$ denotes the corresponding bulk-to-boundary
propagator which for a scalar field is defined as 
\cite{Freedman:1998tz,D'Hoker:2002aw}
\begin{equation}\label{bulkboundAdSprop}
K_\Delta\big(x,\bar
x'\big)=\frac{\Gamma(\Delta)}{\pi^\frac{d}{2}\Gamma(\Delta-\frac{d}{2})}\Big(\frac{x_\perp}{x_\perp^2+(\bar
  x-\bar x')^2}\Big)^\Delta\pnt
\end{equation}
It behaves near the boundary like
\begin{equation}
\lim_{x_\perp\to 0}K_\Delta\big(x,\bar x'\big)x_\perp^{\Delta-d}=\delta^d(\bar
x-\bar x')\pnt
\end{equation}
The given prescription can now be interpreted as follows
\cite{Witten:1998qj}. The $d$-dimensional CFT lives on the $d$-dimensional
boundary of $\AdS_{d+1}$. Sources that couple to 
the operators of the CFT are identified with the boundary
values for the corresponding non-normalizable supergravity modes 
which solve the supergravity equations of motion. The bulk supergravity 
modes are constructed via a convolution
of the boundary value with the corresponding bulk-to-boundary propagator.
According to \eqref{AdSCFTcorrcos} an $n$-point correlation function of CFT
operators can now be evaluated by taking the corresponding $n$ functional
derivatives w.\ r.\ t.\ the boundary values $\bar\phi_\Delta$ 
(and after that setting $\bar\phi_\Delta=0$).

One can represent 
the different contributions to the $n$-point function by so called Witten 
diagrams \cite{Witten:1998qj}, which are Feynman diagrams for the
$\AdS/\text{CFT}$ correspondence.
$\AdS_{d+1}$ is represented by a disk and its $d$-dimensional boundary becomes
the boundary circle of the disk. The $n$ points are now distributed along the
circle and from each point originates a line into the interior of the disk. 
These lines represent the bulk-to-boundary propagators. The $n$ lines in the
 interior are now combined into vertices found in the type
 $\twob$ supergravity action. If a diagram contains several vertices in the
 interior they have
 to be connected with lines that correspond to the
 bulk-to-bulk propagator in $\AdS_{d+1}$.\footnote{Such a diagram is generated
   if the exponential function on the R.\ H.\ S.\ of 
   \eqref{AdSCFTcorrcos} is expanded to at least quadratic order in the
   action.}  

The bulk-to-bulk propagator $G_\Delta(x,x')$ is defined as a solution of
the differential equation
\begin{equation}\label{AdSbulktobulkpropdiffeq}
(\Box_x-m^2)G_\Delta(x,x')=-\frac{1}{\sqrt{g_\AdS}}\delta(x,x')\col
\end{equation} 
where $\Box_x$ is the Laplace operator on Euclidean $\AdS_{d+1}$ 
acting on the first
argument of $G_\Delta(x,x')$, and $g_\AdS$ is the
determinant of the metric. The solution reads
\cite{Burgess:1985ti,D'Hoker:2002aw} 
 \begin{equation}\label{AdSprop}
G_\Delta(x,x')=\frac{\Gamma(\Delta)}{R_1^{d-1}2\pi^\frac{d}{2}\Gamma(\Delta-\frac{d}{2}+1)}\Big(\frac{\xi}{2}\Big)^\Delta\hypergeometric{\tfrac{\Delta}{2}}{\tfrac{\Delta}{2}+\tfrac{1}{2}}{\Delta-\tfrac{d}{2}+1}{\xi^2}\col\quad\xi=\frac{2R_1^2}{u+2R_1^2}\col
\end{equation}
where $u=u(x,x')$ is the chordal distance given in \eqref{AdSchordaldist} and
hence in Poincar\'e coordinates $\xi$ has the form  
\begin{equation}\label{xiinPcoord}
\xi=\frac{2x_\perp x'_\perp}{x_\perp^2+x'^2_\perp+(\bar x-\bar x')^2}\pnt
\end{equation}
It becomes zero for $x'_\perp\to 0$.

As is shown in Appendix \ref{app:bulkboundproprel}
the bulk-to-boundary propagator $K_\Delta(x,\bar x')$
can be obtained from the bulk-to-bulk propagator $G_\Delta(x,x')$ 
as follows (see \eqref{bulktoboundarypropagatorrel})
\begin{equation}\label{bboundpropinbbprop}
K_\Delta(x,\bar x')=-R_1^{d-1}
\big[(d-\Delta)x'^{-\Delta}_\perp-x'^{1-\Delta}_\perp\partial'_\perp
\big] G_\Delta(x,x')\big|_{x'_\perp=0}\pnt
\end{equation}
Thus, using $G_\Delta(x,x')$, the boundary value problem can be solved.

We have seen that the $\AdS/\text{CFT}$ correspondence is holographic in the
sense that a theory of gravity in $\AdS_{d+1}$ can be described by 
a conformal field theory on the $d$-dimensional boundary of that space. 
It was shown \cite{Susskind:1998dq} that the boundary theory respects the 
restriction that only one bit of information per Planck area is stored and
hence it provides a true holographic description. 
This result is related to the fact that 
infrared effects in the bulk theory correspond to ultraviolet 
effects in the boundary theory, known as UV/IR connection.
 
The role of the coordinate perpendicular to the boundary, which is called
holographic direction, is of central importance.
In the Poincar\'e coordinates, $x_\perp$ is the holographic coordinate and
$x_\perp=0$ is the position of the boundary where the metric
\eqref{AdSmetricPcoord} diverges.
Bulk computations thus require a cutoff $\delta>0$
\cite{Witten:1998qj,Susskind:1998dq}, where now $x_\perp\ge\delta$,
 to regularize the infinite size of
the volume. From the perspective of the bulk theory, $\delta$ is an infrared
cutoff. For its interpretation in the boundary theory we remember that 
due to \eqref{AdSredshiftPcoord} 
an observer on the boundary measures the lower energy the more the 
source of this energy lies in the interior of AdS. 
The infrared region of 
the boundary theory therefore corresponds to the deep interior of the AdS,
whereas the ultraviolet regime is dominated by the region close to the 
boundary. 
Hence, the infrared cutoff $\delta$ in the bulk theory becomes an
ultraviolet cutoff in the boundary theory \cite{Susskind:1998dq}.

\section{The BMN correspondence and holography}
\label{BMNholo}
In Section \ref{PGlimit} we have seen that the $10$-dimensional maximally
supersymmetric plane wave background arises as 
a Penrose-G\"uven limit of the $\AdS_5\times\text{S}^5$ background 
of type $\twob$ string theory. Berenstein, Maldacena and
 Nastase \cite{Berenstein:2002jq} then translated the limit to the gauge
 theory side and formulated a new limit of the full $\AdS/\text{CFT}$
 correspondence, which is technically tractable even beyond the supergravity
 approximation. In the following we
will denote the correspondence in this limit as BMN correspondence and the 
limit that has to be taken on both sides of the correspondence collectively 
as the BMN limit.

As we have seen in Section \ref{AdSCFTholo}, 
the $\AdS/\text{CFT}$ correspondence obeys
the holographic principle and hence it is natural to investigate how 
holography can be established in its BMN limit. 
Before we start to present a summary of some work that appeared in this
context, we will in brief describe the BMN correspondence.
More detailed reviews are \cite{Plefka:2003nb,Sadri:2003pr,Russo:2004kr}.

The proposal of Berenstein, Maldacena and Nastase \cite{Berenstein:2002jq} 
starts with the discussion of how the Penrose-G\"uven limit translates to 
the gauge theory side. We have seen in Subsection \ref{PGlimit} that the
Penrose-G\"uven limit of $\AdS_5\times{S}^5$ can be realized by sending the 
common embedding radius $R$ to infinity. The result is given in
\eqref{typeIIBpw} and the parameter $\mu$ will be kept in the following. 
The requirement to have finite lightcone momenta 
\begin{equation}\label{lcmom}
p^-=\frac{\mu}{2}(E-J)\col\qquad p^+=\frac{E+J}{2\mu R^2}
\end{equation} 
in the limit leads to the condition  that the energy $E$ and the angular momentum $J$ of a
supergravity mode both scale with $R^2$ but that $E-J=\text{finite}$.    
$E$ and $J$ are identified with the conformal dimension $\Delta$ and 
with the charge of a $U(1)$ subgroup of the $SO(6)$ $R$-symmetry group 
respectively. The $U(1)$ subgroup is the one singled out in the Penrose-G\"uven
limit in the $\text{S}^5$.
 According to \eqref{AdS5S5para} the limit on the gauge 
theory side becomes
\begin{equation}
N\to\infty\col\qquad\frac{J}{\sqrt{N}}=\text{fixed}\col\qquad g=\text{fixed}\pnt
\end{equation}
The operators that survive the limit have to obey
\begin{equation}
\Delta-J=\text{finite}\ge0\pnt
\end{equation}
A chiral primary operator (CPO) $\mathcal{O}_k$ in the scalar sector of
$\mathcal{N}=4$ SYM that has conformal dimension $\Delta=k$ reads
\begin{equation}
\mathcal{O}_k=C^{a_1\dots a_k}\tr[\phi_{a_1}\dots\phi_{a_k}]\col
\end{equation}
where $C^{a_1\dots a_k}$ are traceless symmetric tensors (that correspond to
the spherical harmonics on $\text{S}^5$) and $\phi_a$, $a=1,\dots,6$ denote
the scalars of $\mathcal{N}=4$ SYM.
The field combination $Z=\frac{1}{\sqrt{2}}(\phi_5+i\phi_6)$ 
carries definite scaling dimension $\Delta=1$ and charge $J=1$. 
The remaining four scalars
$\phi_a$ and covariant derivatives $D_\mu Z$, $\mu=1,\dots,4$
carry $\Delta-J=1$.
They are denoted as `impurities' that are inserted into the composite
operators of the string state operator mapping. A sample of the dictionary
reads \cite{Berenstein:2002jq,Plefka:2003nb}
\begin{equation}\label{BMNstringstateopmap}
\begin{aligned}
|0,p^+\rangle&\leftrightarrow\frac{1}{\sqrt{JN^J}}\tr[Z^J]\col\\
\alpha_0^{\dagger
  i}|0,p^+\rangle&\leftrightarrow\frac{1}{\sqrt{N^J}}\tr[\Phi_i Z^J]\col\\
\alpha_n^{\dagger i}\alpha_{-n}^{\dagger
  j}|0,p^+\rangle&\leftrightarrow\frac{1}{\sqrt{JN^J}}\sum_{l=0}^J\tr[\Phi_i
Z^l\Phi_jZ^{J-l}]\e^\frac{2\pi inl}{J}\pnt
\end{aligned}
\end{equation} 
Here $|0,p^+\rangle$ denotes the string groundstate and we have listed only
the bosonic operators with up to two impurities  
\begin{equation}
\Phi_i=(D_1Z,\dots,D_4Z,\phi_1,\dots\phi_4)\col\qquad i=1,\dots,8\pnt
\end{equation} 
The phase factor in \eqref{BMNstringstateopmap} is chosen such that, according
to \eqref{lcmom}, the 
anomalous `twist' $\Delta-J$ of the composite operators match with the
energy eigenvalues of the plane wave string Hamiltonian in light cone gauge
\cite{Metsaev:2002re,Berenstein:2002jq}
\begin{equation}\label{Hlc}
H_\text{lc}=2p^-=\sum_{n=-\infty}^\infty
N_n\mu\sqrt{1+\frac{n^2}{(\mu p^+\alpha')^2}}\pnt 
\end{equation}
Here $N_n$ is the occupation number at level $n$.
From \eqref{lcmom} one finds with \eqref{AdS5S5para} that the parameters in the
limit are related as follows
\begin{equation}
\frac{1}{(\mu p^+\alpha')^2}=\frac{g^2N}{J^2}=\lambda'
\end{equation}
and we have defined the expansion parameter $\lambda'$ that is the effective 
coupling constant in the BMN limit.
One now immediately sees that string states only built with zero-mode
oscillators $\alpha_0^{\dagger i}$ (see \eqref{BMNstringstateopmap}) 
are protected against corrections in powers of 
$\lambda'$. The non-protected states contain a minimum number of two
oscillators with $n\neq0$. They correspond to the two-impurity operators with
a non-trivial phase factor that depends on their separation within the trace.  
The phase factor leads to the matching
between the anomalous twist $\Delta-J$ and the light cone energy in the planar
limit
\begin{equation}
\Delta-J=2\sqrt{1+\lambda'n^2}=2+\lambda'n^2+\mathcal{O}(\lambda'^2)\pnt
\end{equation}
At one loop this is demonstrated in \cite{Berenstein:2002jq}. 
The two loop result is presented
in \cite{Gross:2002su} and a complete reproduction to all orders was obtained 
in \cite{Santambrogio:2002sb}.  

This is of course not the end of the story. Up to now we have only discussed
that the planar part of the BMN gauge theory matches with the free string
spectrum. The correspondence should also work if we include string
interactions. In the gauge theory they are argued to correspond to 
non-planar contributions. It was shown in
\cite{Kristjansen:2002bb,Constable:2002hw} that the 
naively expected suppression of non-planar diagrams does not hold.
The reason is that the number of elementary fields in the BMN operators 
grows with $J$ and hence the number of diagrams at each order compensate 
the $\frac{1}{N}$ suppression. In this way a certain class of 
non-planar contributions
survives the limit \cite{Kristjansen:2002bb}, and their non-planarity is
controlled by the effective genus counting parameter
\begin{equation}
g_2=\frac{J^2}{N}\pnt
\end{equation}
A genus $h$ contribution then is of order $(g_2)^{2h}$. 
To describe interactions on the string theory side, light cone string field 
theory is required \cite{Plefka:2003nb,Pankiewicz:2003pg}. 
There exist two different proposals for the cubic interaction vertex.
The first one is worked out in
\cite{Spradlin:2002ar,Spradlin:2002rv,Pankiewicz:2002tg,Pankiewicz:2003kj,Pankiewicz:2003ap}.
For this vertex a relation like $H_\text{lc}=\mu(\Delta-J)$ holds even for
non-planar diagrams if one interpretes it as an operator relation, where
both sides act in different Hilbert spaces.
The picture behind this is that multiple string interactions correspond to the 
two point functions of multi-trace operators on the gauge theory side 
\cite{Verlinde:2002ig,Beisert:2002bb}.
The situation is different for the second vertex which is constructed in
\cite{DiVecchia:2003yp}. It is proposed to correspond to the 
gauge theory $3$-point function of BMN operators in the original
basis\footnote{This is the basis before one has taken into account 
operator mixing at non-planar level.}, 
because a relation like in the case of the first vertex does not hold. 
Compared to the first proposal, this proposal becomes problematic beyond tree level, because the gauge theory $3$-point function in the original basis of
operators then is no longer of the form dictated by conformal invariance
Furthermore, it is unclear how this proposal should be extended to more than
$3$-point interactions \cite{Beisert:2002bb}.
A clear decision between the two proposals should be possible as soon as it has
become clear if or in which sense the $3$-point function enters the duality.

After this short summary of some aspects of the BMN correspondence we will now
in brief present some proposals of how holography could be realized in 
this limit. 

In \cite{Das:2002cw} the authors come to the conclusion that the gauge theory
dual is Euclidean and lives in a $4$-dimensional subspace that is formed by 
four of the eight transversal coordinates $z_i$ in \eqref{typeIIBpw}.
One of their arguments for the Euclidean signature in this case is
that the duality relates the operators of the lower dimensional theory to the
transverse oscillators (in light cone gauge). In their setup the 
holographic coordinate is given by $z^+$. 

In \cite{Kiritsis:2002kz} the authors argue that the holographic direction in
the plane wave background \eqref{typeIIBpw} is $z^+$ 
by analyzing the behaviour of 
particles in the background. In particular they identify possible non-compact 
directions and analyze which of these directions effectively 
do not confine particles. They find that $z^+$ is effectively non-compact in
this sense. Furthermore,
via a coordinate transformation the conformal
flatness of the plane wave \eqref{typeIIBpw} becomes manifest. In this new
coordinates the metric is given by flat Minkowski space with a conformal
factor that depends only on $x^+=-\cot\frac{z^+}{2}$.
This coordinate system is seen as analogously to the Poincar\'e
patch for AdS. The gauge theory dual then should live on a slice of constant
$x^+\to-\infty$. Since the wave equation in these coordinates is only of 
first order in the light cone coordinates $x^+$ and in $x^-$,
only the non-normalizable modes are found. 
The absence of the normalizable modes lead the authors to the proposal
that one has to impose boundary conditions on both slices (constant
$x^+\to-\infty$ and $x^-\to-\infty$) and that the  
data on the slice of constant $x^-$ should then determine the vacuum
structure.  
This situation is different compared to the  full Lorentzian
$\AdS/\text{CFT}$ correspondence, where normalizable and non-normalizable modes
exist \cite{Balasubramanian:1998sn,Balasubramanian:1998de}. 
The authors give a concrete proposal of how to compute 
correlation functions with the bulk-to-boundary propagator in their setup.
They determine the bulk-to-bulk and bulk-to-boundary propagator directly
by using the eigenmodes of the wave equation in their specific coordinate
system.  
The concrete case of bosonic strings in $\AdS_3\times\text{S}^3\times M^{20}$
supported by NS-NS $3$-form flux is dealt with in \cite{Bianchi:2004vf} and is
in agreement with the proposal of \cite{Kiritsis:2002kz}.

A different proposal is worked out in \cite{Leigh:2002pt}. 
The authors argue 
that the holographic direction is given by the first radial coordinate that 
corresponds to $r$ in \eqref{AdSSmetricexpansion} and originates from the 
AdS part. The holographic screen is identified as the corresponding sphere
plus the lightcone direction $z^+$. 
The boundary theory should be the original $\mathcal{N}=4$ SYM
one, living at $r\to\infty$. The setup is very similar to the one 
in the $\AdS/\text{CFT}$ correspondence.
In their approach both normalizable and non-normalizable modes are
present and the allowed modes carry positive lightcone momentum
$p^+$.

The approach of \cite{Berenstein:2002sa} starts with working out the structure
of the 
conformal boundary of the plane wave which turns out to be a $1$-dimensional 
null line parameterized by the coordinate $z^+$ in \eqref{typeIIBpw}.  
Since this coordinate contains the time direction of the AdS part, the authors
propose that the boundary theory should be a one-dimensional quantum
mechanical system that can be realized as a matrix model. It
should describe the lowest KK modes of the $4$-dimensional $\mathcal{N}=4$
SYM theory after compactification on $\text{S}^3$. 
The authors furthermore state that the
observables of the boundary theory should be finite time transition matrix
elements between states that describe multiple strings. 
However a comparison with the string
calculation then requires care because of the following reason.
A naive interpretation of $z^+$ as a time variable is not justified if one
defines the lightcone coordinates as in \eqref{BMNcoordinates}. The
periodicity of $\psi$ then implies a periodicity of $z^+$. Instead
\cite{Russo:2002rq} one can define $z^+=t$ and keep $z^-$ as in
\eqref{BMNcoordinates}. Then only $z^-$ becomes periodic with a large period of
$2\pi R_1R_2$. The authors argue that the effect of this compactification 
should be considered in the normalization of the wavefunctions as follows:
bulk modes should be normalized considering $z^-$ 
as non-compact because one uses 
the plane wave metric \eqref{typeIIBpw} which is the exact limit 
$R_1=R_2\to\infty$ of $\AdS_5\times\text{S}^5$. But in the boundary theory one
should normalize the states considering $z^-$ as compact. 
The authors of \cite{Berenstein:2002sa} furthermore comment on the proposals
of \cite{Das:2002cw,Kiritsis:2002kz,Leigh:2002pt}. 
They remark that \cite{Das:2002cw} does not 
consider that the BMN correspondence is a limit of the $\AdS/\text{CFT}$
correspondence.
Their argument is that to put the gauge theory on a Euclidean space
does not follow from the original $\AdS/\text{CFT}$ correspondence where the
Euclidean version only occurs after Wick rotating to Euclidean AdS. 
They criticize \cite{Kiritsis:2002kz} with the 
argument that it is not allowed to work in a patch along the lines of the 
Poincar\'e
patch in AdS. They state that the translation to an Euclidean version is
not possible in the plane wave case, however that it is essential in the full
$\AdS/\text{CFT}$ correspondence to allow working with the Poincar\'e patch.
Their comment on the proposal \cite{Leigh:2002pt} is that one should not forget
that the null geodesic around which one expands is in the center of AdS and that
,blowing up its neighbourhood, the old boundary of AdS, where the original
$\mathcal{N}=4$ SYM lives, lies outside the plane wave.  

This result of \cite{Berenstein:2002sa} that the conformal boundary is a
one-dimensional null line was confirmed by \cite{Marolf:2002ye} where the 
authors used the construction of Geroch, Kronheimer and Penrose
\cite{Geroch:1972,Hawking:1973} 
to attach a causal boundary to a spacetime, see also \cite{Hubeny:2002zr}. 
The construction uses the
indecomposable past (IP) and future (IF) sets. Proper indecomposable past
(PIP) and future (PIF) sets are the pasts and futures of points 
in the spacetime, whereas 
terminal indecomposable past (TIP) and future (TIF) sets are not the pasts 
and futures of points in the spacetime. 
The idea is to regard TIPs and TIFs as representing 
points of the causal boundary of the spacetime. We do not want to give
all exact definitions here but physically motivate what is happening. 
TIPs can be seen as the pasts of future inextendible timelike curves. A future 
inextendible timelike curve is one without a future endpoint. 
Precisely, a future endpoint $p_0$ to a
 future directed (parameter $\gamma$ increases in the
future direction) non spacelike curve is defined by taking a neighbourhood 
around $p_0$ and 
demanding that by increasing the parameter $\gamma$ along that curve
one always
enters the neighbourhood at $\gamma=\gamma_0$ but never leaves it again for
arbitrary $\gamma>\gamma_0$. It is then clear that a future inextendible
timelike curve is one on which for every point $p$ on the curve one can
find a point $q$ that lies in the future of $p$. The PIP of $p$ is thus 
included in the PIP of $q$ but there is no PIP that includes all other PIPs of
points on the curve. Hence, it is a TIP that includes all PIPs of points on the
curve. This TIP now represents the future endpoint of the curve that is an
element of the causal boundary of the spacetime. 
This motivates why one regards TIPs and TIFs to represent points of the
causal boundary of the spacetime. 
It is interesting to note that for AdS and the plane wave
 spacetimes, points of the causal boundary are represented by 
TIPs \emph{and} TIFs 
(see \cite{Hawking:1973} for AdS and \cite{Marolf:2002ye} for the
plane wave).\footnote{We disregard some special points here and in the
 following.} 
In less technical words, the boundaries of these spacetimes allow
for an \emph{exchange} of information with the interior. 
This is in contrast i.\ e.\
 to Minkowski-space which boundary points are \emph{either} 
represented by TIPs \emph{or}
 TIFs and therefore can \emph{either} be influenced by \emph{or} 
influence the bulk but not both simultaneously.

The problem, where the lower dimensional holographic partner should reside, 
was addressed in \cite{Siopsis:2002vw}
from a different perspective.  The idea is to introduce a
holographic screen that is not necessarily connected with the boundary 
structure of the space. The author discusses the Penrose limit of pure
AdS which is flat Minkowski space. He shows that a projection on scale
invariant states fixes a de Sitter (dS) hypersurface of codimension one in AdS 
which in the flat space limit becomes Minkowski space and the isometry group
$SO(2,d)$ then becomes the conformal group on it. The analysis has been
extended in \cite{Siopsis:2002nx} to $\AdSS$ where the Penrose limit is taken
along a null geodesic with movement in the sphere.

In \cite{Akhmedov:2002ms} the author argues that the dual gauge theory in the 
BMN limit is not an effective one-dimensional theory but the full
$4$-dimensional SYM theory.  
The underlying picture is that already in the full 
$\AdS/\text{CFT}$ correspondence the SYM theory should be seen regarded as
living on any hypersurface with constant holographic coordinate 
($\rho$ in global and $x_\perp$ in Poincar\'e coordinates), where its constant
value is related to the energy scale of the SYM theory. According to
\eqref{AdSSmetricexpansion}, 
in the plane wave background each hypersurface has constant $r$ and is given
by the $3$-dimensional sphere from the AdS part with $z^+$ as time direction. 

In \cite{Yoneya:2003mu} the author proposes that one should interpret the
bulk-boundary connection in the BMN correspondence as a tunneling phenomenon. 
The proposal starts from the observation that particles that move in the
$\text{S}^5$ 
of the original $\AdS_5\times\text{S}^5$ background never reach the conformal 
boundary. They are confined in a region in the interior of AdS. Hence, a theory
on the boundary of $\AdS_5\times\text{S}^5$ seems to have no influence on or
be influenced by this region.   
As we have seen in Section \ref{PGlimit}, the Penrose limit that leads 
to the plane wave spacetime zooms into the neighbourhood of 
a null geodesic with movement in $\text{S}^5$ and therefore, applying the above
considerations, its bulk theory seems to be disconnected from the AdS boundary.
The author 
observes that with a purely imaginary action one finds trajectories that
connect two boundary points by entering the bulk of AdS. He argues that
the tunneling picture follows with a double Wick rotation that leads to 
Euclidean time in AdS and transforms the angle coordinate $\psi$ of 
\eqref{BMNcoordinates}, which parameterizes the null geodesic in the
$\text{S}^5$, the role of a time coordinate. 
The author proposes to relate the Euclidean S-matrix to the operator product
expansion of BMN operators on the boundary. 

The authors of \cite{Mann:2003qp} analyze the $3$-point correlation function of
two BMN and one non-BMN chiral primary operator (CPO) in the $\AdS/\text{CFT}$
correspondence. They compare the CFT and supergravity calculations before and 
after the BMN limit. They find the expected agreement \cite{Lee:1998bx} in the
full $\AdS/\text{CFT}$ context, even using the coordinates of 
\eqref{BMNcoordinates}
and taking the $R\to\infty$ limit. They argue, however, that in a holographic
 setup where one wants to relate the amplitude to a correlation function of
 local operators in the dual theory, the amplitude should be truncated. In
 their proposal one
 should remove the part that describes the propagation from the AdS boundary 
to the geodesic along which the Penrose limit is taken and replace it by a
$\delta$-function. To be more precise the time (or in the variables of
\eqref{BMNcoordinates} $z^+$-dependence 
of the amplitude which has 
poles with a period of $\pi$ should be replaced by a $\delta$-function with a
single pole to guarantee locality in time. This leads to a mismatch between 
the string and the gauge theory result. The authors give possible explanations
 for the mismatch such that this does not necessarily lead to an exclusion of a holographic principle in the BMN limit. 

A completely different point of view is taken in \cite{Gubser:2002tv}.
There, the authors find that the 
BMN limit is a concrete realization of general considerations 
in \cite{Polyakov:2001af} about operators with large spin.
It can be embedded into the more general
framework of finding classical solutions for single strings on
$\AdS_5\times\text{S}^5$ and computing the quantum fluctuations around them.
The BMN limit then has the following interpretation
\cite{Gubser:2002tv,Frolov:2002av}: the null geodesic around which the Penrose 
limit is taken on the string side is a classical pointlike string 
solution.\footnote{In \cite{Frolov:2002av} it is shown that there exist two pointlike string solutions in $\AdS_5\times{S}^5$. Both move along an equator of
$\text{S}^5$. One does, whilst the other does not move in spatial
directions of $\AdS_5$. Both are equivalent by a coordinate transformation in 
$\AdS_5\times\text{S}^5$. This fits perfectly with the considerations in
Subsection \ref{PGlimit} that all configurations with different velocity
components in $\AdS_5$ and in $\text{S}^5$ are equivalent as long as the
velocity in $\text{S}^5$ is different from zero.}
The quantum fluctuations around this solution that enter the action in 
quadratic order can then be interpreted as the embedding coordinates of 
a string in the $10$-dimensional plane wave background. 
In \cite{Tseytlin:2002ny} the author now argues that to try to find a 
holographic setup in the BMN limit might be misleading because the 
limit simply represents the $1$-loop approximation of the $\sigma$-model.   

The above given brief descriptions show that holography in the
BMN limit of the $\AdS/\text{CFT}$ correspondence is by far less understood 
than in the $\AdS/\text{CFT}$ correspondence itself. Although the above 
proposals are different, they may not necessarily exclude each other,
and they might be equivalent descriptions \cite{Kiritsis:2002kz}.
In principle one can follow two ways to find a holographic setup 
in the BMN correspondence. One can either work directly in the BMN 
limit and disregard the fact that it follows from the $\AdS/\text{CFT}$
correspondence, or one can try to get information on the holographic principle 
by following the limiting process from the full $\AdS/\text{CFT}$
to the BMN correspondence, and thereby observe what happens to 
the ingredients in a holographic setup. 
The advantage of the second approach should be that it excludes 
holographic formulations which are not related to holography in the full
$\AdS/\text{CFT}$ correspondence. 

In the following we will work in the spirit of this second proposal and first 
discuss geometrical quantities in Chapter \ref{chap:boundgeo}. 
In Section \ref{AdSpwboundaries} we will investigate
the behaviour of the boundary structure of $\AdS_5\times\text{S}^5$ in the
Penrose limit to the plane wave. This extends the
analysis of \cite{Berenstein:2002sa} where it is discussed
\emph{after} the limit has been taken.
To analyze the causal structure in connection with a holographic picture,
geodesics, and in particular null geodesics, reaching the holographic screen 
out of the bulk, play a central role. In Section \ref{AdSpwgeodesics}
we will determine all possible geodesics in $\AdS_5\times{S}^5$ and in the
$10$-dimensional plane wave. 
Section \ref{AdSpwgeorel} then deals with the question of how the boundary
reaching null geodesics in $\AdS_5\times{S}^5$ are translated to the plane
wave geodesics in the Penrose limit.  
 
After having discussed the geometrical quantities, we refer to the 
observations in Section \ref{AdSCFTholo} and argue that propagators 
should play an essential role in the realization of holography. 
Hence, in Chapter \ref{chap:propaga} we will focus on the scalar propagator in 
a generic $\AdSS$ background and analyze for $d=d'=4$ the behaviour when
taking the Penrose limit.
In Section \ref{waveeqapp} we will use the differential equation for the
propagator to
work out under which circumstances simple powerlike solutions can be found. 
The generalization to arbitrary dimensions and curvature radii is useful to
identify the general mechanism that leads to these kind of solutions.
In Section \ref{weylana} they will then be rederived from the solutions in
flat  space, using a conformal map.
An interpretation of the result including global properties will be given in 
Section \ref{relESU}. 
In Section \ref{modesummation} the propagator will be explicitly constructed 
by summing up the Kaluza-Klein modes. 
For the particular case of $\AdS_5\times\text{S}^5$ the Penrose limit will
then be discussed in Section \ref{pwlimit}.


\clearemptydoublepage
\chapter{Boundaries and geodesics in $\AdS\times\text{S}$ and in the plane
  wave}
\label{chap:boundgeo}
We have mentioned in Section \ref{BMNholo} that the boundaries of 
$\AdS_5\times\text{S}^5$ and of the $10$-dimensional plane wave 
exchange information with the
interior. This is equivalent to saying that each point on these boundaries
is represented by a TIP as well as 
TIF.\footnote{From now on we will neglect that the
  two separate points $i^-$ and $i^+$ of timelike past and
  future infinity are part of the conformal boundary.} 
A light ray travels from the interior to the boundary, 
is reflected and travels back into the interior in finite time.
It is interesting to investigate how this picture is translated in the Penrose
limiting process from $\AdS_5\times\text{S}^5$ to the plane wave background.
Therefore, in Section \ref{AdSpwboundaries} 
we will first analyze how the boundary of
$\AdS_5\times\text{S}^5$ approaches the plane wave boundary in the limiting
process. The analysis will be carried out in the two sets of coordinates
introduced in \cite{Berenstein:2002jq} and \cite{Berenstein:2002sa} which from
now on we denote as BMN and BN coordinates respectively.
In Section \ref{AdSpwgeodesics} 
we will then determine all geodesics in $\AdS_5\times\text{S}^5$
and in the plane wave. The boundary reaching null geodesics in both 
spacetimes will then be of particular interest for us because they are the
light rays traveling between bulk and boundary.  
Locally, $\AdS_5\times\text{S}^5$ geodesics converge
to plane wave geodesics because the Penrose limit which connects both 
spacetimes is realized by sending a parameter (that the metric depends on) 
to infinity. 
But our results will be useful for analyzing global
aspects. In particular, in Section \ref{AdSpwgeorel} we will work out how 
differently some null geodesics approach in both spaces their corresponding 
conformal boundary and in which sense
these $\AdS_5\times\text{S}^5$ null geodesics approach in the Penrose limit 
the null geodesics running to the plane wave conformal boundary. 
This Chapter is based on our work \cite{Dorn:2003ct}.

\section{Common description of the conformal boundaries}
\label{AdSpwboundaries}
In this Section we will relate the conformal boundaries of
 $\AdS_5\times\text{S}^5$ and of the plane wave in a suitable coordinate
 system. 
Since the angular coordinate $\psi$ in the coordinates
 \eqref{AdSSmetricglobalcoord} for $\AdS_5\times\text{S}^5$  is constrained 
by $-\pi\leq\psi\leq\pi$, one finds from \eqref{BMNcoordinates} that the
 coordinates $z^+$ and $z^-$ are restricted to 
\begin{equation}\label{bound0}
R^2z^+-\pi R^2\leq z^-\leq R^2z^++\pi R^2\col
\end{equation}
where we have set $R=R_1=R_2$ for the radii of $\AdS_5$ and of $\text{S}^5$
respectively.  
This is a strip in the $(z^+,z^-)$-plane bounded by the two parallel 
straight lines with slope $R^2$ and crossing the $z^+$-axis at $-\pi$ and
$\pi$, respectively. For $R\to\infty$ this strip becomes
the coordinate range $-\infty <z^-<\infty$, $-\pi\leq z^+\leq\pi$.
Taking the limit for the metric, the identification of the two
boundaries of the strip is given up, and it makes sense to extend
to the whole $(z^+,z^-)$-plane. If one wants to avoid the restriction
to the strip already for finite $R$, one has to puncture $\text{S}^5$ at its poles
and to go then to the universal covering obtained by allowing $\psi$
to take any real value.

The sequence of coordinate transformations, done in \cite{Berenstein:2002sa} to analyze
the conformal boundary of the plane wave geometry \eqref{intro4}, can be
summarized as follows. Writing $\de\vec z^2=z^2\de\Omega_7^2$, the $\Omega_7$ coordinates remain untouched. 
Then in a first step one transforms\footnote{Note that our definitions for
  $z^{\pm}$ follow \cite{Berenstein:2002jq} and thus slightly differ from \cite{Berenstein:2002sa}.}
in the patch $z^+\in(-\frac{\pi}{2},\frac{\pi}{2})$ the coordinates $z^+$,
$z^-$, $z$ to $\theta$, $\varphi$, $\zeta$
\begin{equation}\label{bound1}
\begin{aligned}
\cot\theta&=\frac{\big((1-z^2)\tan z^+-4z^-\big)\cos z^+}{2z}\col\\
\tan\tfrac{\varphi\pm\zeta}{2}&=\frac{1}{2}(1+z^2)\tan z^++2z^-\pm \frac{z}{\sin\theta\cos z^+}\pnt
\end{aligned}
\end{equation}
The  new coordinates are constrained by 
\begin{equation}
0\leq\theta\leq\pi\col\qquad 0\leq\zeta\leq\pi\col\qquad|\varphi\pm
\zeta|\leq\pi\pnt  
\end{equation}
The second step uses
the periodicity properties 
of the trigonometric functions to glue the other $z^+$-strips, resulting
in the final coordinate range
\begin{equation}\label{bound2}
0\leq\theta\leq\pi\col\qquad 0\leq\zeta\leq\pi\col\qquad-\infty <\varphi <\infty\pnt
\end{equation}
Then the plane wave metric, up to a conformal factor, turns out to be
that of the Einstein static universe $\mathds{R}\times\text{S}^9$. The 
analysis of singularities
of the conformal factor, determining the conformal boundary of the plane wave,
becomes most transparent after a change of parameterization of $\text{S}^9$. Let
denote $(z_1,z_2,\vec z)$ Cartesian coordinates in an embedding
$\mathds{R}^{10}$, then the parameterization by $\theta$, $\zeta$ is related to
that by $\alpha$, $\beta$ via\footnote{We shift $\alpha $ to
  $\alpha-\frac{\pi}{2}$ and $\beta$ to $\beta-\pi$ relative to
  \cite{Berenstein:2002sa}.} 
\begin{equation}\label{bound3}
\begin{aligned}
z_1&=\cos\zeta=\sin\alpha \cos\beta\col\nonumber\\
z_2&=\cos\theta \sin\zeta=\sin\alpha \sin\beta\col\nonumber\\
|\vec z|&=\sin\theta\sin\zeta=\cos\alpha\pnt
\end{aligned}
\end{equation} 
The range for $\alpha$, $\beta$ is
\begin{equation}\label{bound4}
0\leq\alpha\leq\frac{\pi}{2}\col\qquad 0\leq\beta\leq 2\pi\pnt
\end{equation} 
Now the plane wave metric in these BN coordinates takes the form \cite{Berenstein:2002sa}
\begin{equation}\label{bound5}
\de s^2_\text{pw}=\frac{1}{|\e^{i\varphi}+\sin\alpha\e^{i\beta}|^2}\left (-\de\varphi^2+\de\alpha^2+\sin^2\alpha \de\beta^2+\cos^2\alpha \de\Omega_7^2\right )\pnt
\end{equation}
The conformal factor is singular if and only if 
$\alpha=\frac{\pi}{2}$ and $\varphi
=\beta+(2k+1)\pi$, $k\in\mathds{Z}$. Since at $\alpha=\frac{\pi}{2}$ the $\text{S}^7$
part due the $\cos^2\alpha $ factor in front of $\de\Omega_7^2$ shrinks
to a point, the conformal boundary\footnote{Here and for the
  $\AdS_5\times\text{S}^5$ case below, while speaking about the conformal
  boundary, we omit the two isolated points $i^\pm$ for timelike infinity.}
 of the plane wave is one-dimensional, see also Fig.\ \ref{cboundpwBNcoord}
\begin {figure}
\begin{center}
\setlength{\unitlength}{0.240900pt}
\begin{picture}(1500,900)(0,0)
\footnotesize
\put(762,649){\makebox(0,0)[l]{$\scriptstyle\alpha=0$}}
\put(958,649){\makebox(0,0)[l]{$\scriptstyle\alpha=\frac{\pi}{2}$}}
\put(803,457){\makebox(0,0)[l]{$\scriptstyle\delta$}}
\put(749,763){\makebox(0,0)[l]{$\varphi$}}
\thinlines \path(749,472)(939,458)
\put(939,458){\line(0,1){45}}
\thinlines \path(749,472)(848,427)
\put(945,294){\line(0,1){355}}
\put(554,294){\line(0,1){355}}
\thinlines \path(609,535)(890,408)\put(890,408){\vector(2,-1){0}}

\thinlines \path(506,435)(993,508)\put(993,508){\vector(4,1){0}}

\put(749,280){\vector(0,1){458}}
\thinlines \path(651,338)(631,342)
\thinlines \path(631,342)(612,345)
\thinlines \path(612,345)(595,347)
\thinlines \path(595,347)(581,349)
\thinlines \path(581,349)(570,350)
\thinlines \path(570,350)(561,351)
\thinlines \path(561,351)(556,352)
\thinlines \path(556,352)(554,353)
\thinlines \path(554,353)(554,354)
\thinlines \path(554,354)(558,354)
\thinlines \path(558,354)(565,355)
\thinlines \path(565,355)(575,357)
\thinlines \path(575,357)(588,358)
\thinlines \path(588,358)(603,360)
\thinlines \path(603,360)(621,363)
\thinlines \path(621,363)(641,366)
\thinlines \path(641,366)(662,370)
\thinlines \path(662,370)(685,375)
\thinlines \path(685,375)(709,380)
\thinlines \path(709,380)(734,386)
\thinlines \path(734,386)(759,393)
\thinlines \path(759,393)(783,401)
\thinlines \path(783,401)(807,410)
\thinlines \path(807,410)(831,419)
\thinlines \path(831,419)(853,430)
\thinlines \path(853,430)(873,441)
\thinlines \path(873,441)(891,452)
\thinlines \path(891,452)(907,464)
\thinlines \path(907,464)(921,477)
\thinlines \path(921,477)(931,490)
\thinlines \path(931,490)(939,503)
\thinlines \path(939,503)(944,517)
\thinlines \path(944,517)(946,530)
\thinlines \path(946,530)(944,544)
\thinlines \path(944,544)(939,558)
\thinlines \path(939,558)(932,571)
\thinlines \path(932,571)(921,584)
\thinlines \path(921,584)(908,597)
\thinlines \path(908,597)(892,609)
\thinlines \path(892,609)(874,620)
\thinlines \path(874,620)(853,631)
\thinlines \path(853,631)(831,642)
\thinlines \path(831,642)(808,651)
\thinlines \path(808,651)(784,660)
\thinlines \path(784,660)(759,668)
\thinlines \path(759,668)(735,675)
\thinlines \path(735,675)(710,681)
\thinlines \path(710,681)(686,687)
\thinlines \path(686,687)(663,691)
\thinlines \path(663,691)(652,693)
\thinlines \path(651,693)(631,690)
\thinlines \path(631,690)(612,686)
\thinlines \path(612,686)(595,681)
\thinlines \path(595,681)(581,675)
\thinlines \path(581,675)(570,670)
\thinlines \path(570,670)(561,663)
\thinlines \path(561,663)(556,657)
\thinlines \path(556,657)(554,651)
\thinlines \path(554,651)(554,644)
\thinlines \path(554,644)(558,638)
\thinlines \path(558,638)(565,632)
\thinlines \path(565,632)(575,626)
\thinlines \path(575,626)(588,620)
\thinlines \path(588,620)(603,615)
\thinlines \path(603,615)(621,610)
\thinlines \path(621,610)(641,606)
\thinlines \path(641,606)(662,603)
\thinlines \path(662,603)(685,601)
\thinlines \path(685,601)(709,599)
\thinlines \path(709,599)(734,598)
\thinlines \path(734,598)(759,598)
\thinlines \path(759,598)(783,599)
\thinlines \path(783,599)(807,600)
\thinlines \path(807,600)(831,602)
\thinlines \path(831,602)(853,605)
\thinlines \path(853,605)(873,609)
\thinlines \path(873,609)(891,614)
\thinlines \path(891,614)(907,619)
\thinlines \path(907,619)(921,624)
\thinlines \path(921,624)(931,630)
\thinlines \path(931,630)(939,636)
\thinlines \path(939,636)(944,642)
\thinlines \path(944,642)(946,649)
\thinlines \path(946,649)(944,655)
\thinlines \path(944,655)(939,662)
\thinlines \path(939,662)(932,668)
\thinlines \path(932,668)(921,674)
\thinlines \path(921,674)(908,679)
\thinlines \path(908,679)(892,684)
\thinlines \path(892,684)(874,689)
\thinlines \path(874,689)(853,693)
\thinlines \path(853,693)(831,696)
\thinlines \path(831,696)(808,698)
\thinlines \path(808,698)(784,699)
\thinlines \path(784,699)(759,700)
\thinlines \path(759,700)(735,700)
\thinlines \path(735,700)(710,699)
\thinlines \path(710,699)(686,698)
\thinlines \path(686,698)(663,695)
\thinlines \path(663,695)(641,692)
\thinlines \path(641,692)(622,688)
\thinlines \path(622,688)(604,683)
\thinlines \path(604,683)(588,678)
\thinlines \path(588,678)(576,673)
\thinlines \path(576,673)(566,667)
\thinlines \path(566,667)(558,661)
\thinlines \path(558,661)(554,654)
\thinlines \path(554,654)(554,648)
\thinlines \path(554,648)(556,641)
\thinlines \path(556,641)(561,635)
\thinlines \path(561,635)(570,629)
\thinlines \path(570,629)(581,623)
\thinlines \path(581,623)(595,618)
\thinlines \path(595,618)(611,613)
\thinlines \path(611,613)(630,608)
\thinlines \path(630,608)(651,605)
\thinlines \path(651,605)(673,602)
\thinlines \path(673,602)(696,600)
\thinlines \path(696,600)(721,598)
\thinlines \path(721,598)(746,598)
\thinlines \path(746,598)(770,598)
\thinlines \path(770,598)(795,599)
\thinlines \path(795,599)(819,601)
\thinlines \path(819,601)(841,604)
\thinlines \path(841,604)(862,607)
\thinlines \path(862,607)(882,611)
\thinlines \path(882,611)(899,616)
\thinlines \path(899,616)(914,621)
\thinlines \path(914,621)(926,627)
\thinlines \path(926,627)(935,633)
\thinlines \path(935,633)(942,639)
\thinlines \path(942,639)(945,646)
\thinlines \path(945,646)(945,652)
\thinlines \path(945,652)(942,658)
\thinlines \path(942,658)(936,665)
\thinlines \path(936,665)(927,671)
\thinlines \path(927,671)(915,676)
\thinlines \path(915,676)(900,682)
\thinlines \path(900,682)(883,686)
\thinlines \path(883,686)(864,691)
\thinlines \path(864,691)(843,694)
\thinlines \path(843,694)(821,697)
\thinlines \path(821,697)(797,699)
\thinlines \path(797,699)(772,700)
\thinlines \path(772,700)(748,700)
\thinlines \path(748,700)(723,700)
\thinlines \path(723,700)(698,699)
\thinlines \path(698,699)(675,696)
\thinlines \path(675,696)(653,694)
\thinlines \path(651,516)(631,512)
\thinlines \path(631,512)(612,508)
\thinlines \path(612,508)(595,503)
\thinlines \path(595,503)(581,498)
\thinlines \path(581,498)(570,492)
\thinlines \path(570,492)(561,486)
\thinlines \path(561,486)(556,480)
\thinlines \path(556,480)(554,473)
\thinlines \path(554,473)(554,467)
\thinlines \path(554,467)(558,460)
\thinlines \path(558,460)(565,454)
\thinlines \path(565,454)(575,448)
\thinlines \path(575,448)(588,442)
\thinlines \path(588,442)(603,437)
\thinlines \path(603,437)(621,433)
\thinlines \path(621,433)(641,429)
\thinlines \path(641,429)(662,426)
\thinlines \path(662,426)(685,423)
\thinlines \path(685,423)(709,421)
\thinlines \path(709,421)(734,420)
\thinlines \path(734,420)(759,420)
\thinlines \path(759,420)(783,421)
\thinlines \path(783,421)(807,423)
\thinlines \path(807,423)(831,425)
\thinlines \path(831,425)(853,428)
\thinlines \path(853,428)(873,432)
\thinlines \path(873,432)(891,436)
\thinlines \path(891,436)(907,441)
\thinlines \path(907,441)(921,447)
\thinlines \path(921,447)(931,452)
\thinlines \path(931,452)(939,459)
\thinlines \path(939,459)(944,465)
\thinlines \path(944,465)(946,471)
\thinlines \path(946,471)(944,478)
\thinlines \path(944,478)(939,484)
\thinlines \path(939,484)(932,490)
\thinlines \path(932,490)(921,496)
\thinlines \path(921,496)(908,502)
\thinlines \path(908,502)(892,507)
\thinlines \path(892,507)(874,511)
\thinlines \path(874,511)(853,515)
\thinlines \path(853,515)(831,518)
\thinlines \path(831,518)(808,520)
\thinlines \path(808,520)(784,522)
\thinlines \path(784,522)(759,523)
\thinlines \path(759,523)(735,523)
\thinlines \path(735,523)(710,522)
\thinlines \path(710,522)(686,520)
\thinlines \path(686,520)(663,517)
\thinlines \path(663,517)(641,514)
\thinlines \path(641,514)(622,510)
\thinlines \path(622,510)(604,506)
\thinlines \path(604,506)(588,501)
\thinlines \path(588,501)(576,495)
\thinlines \path(576,495)(566,489)
\thinlines \path(566,489)(558,483)
\thinlines \path(558,483)(554,477)
\thinlines \path(554,477)(554,470)
\thinlines \path(554,470)(556,464)
\thinlines \path(556,464)(561,457)
\thinlines \path(561,457)(570,451)
\thinlines \path(570,451)(581,445)
\thinlines \path(581,445)(595,440)
\thinlines \path(595,440)(611,435)
\thinlines \path(611,435)(630,431)
\thinlines \path(630,431)(651,427)
\thinlines \path(651,427)(673,424)
\thinlines \path(673,424)(696,422)
\thinlines \path(696,422)(721,421)
\thinlines \path(721,421)(746,420)
\thinlines \path(746,420)(770,421)
\thinlines \path(770,421)(795,422)
\thinlines \path(795,422)(819,424)
\thinlines \path(819,424)(841,426)
\thinlines \path(841,426)(862,430)
\thinlines \path(862,430)(882,434)
\thinlines \path(882,434)(899,438)
\thinlines \path(899,438)(914,444)
\thinlines \path(914,444)(926,449)
\thinlines \path(926,449)(935,455)
\thinlines \path(935,455)(942,462)
\thinlines \path(942,462)(945,468)
\thinlines \path(945,468)(945,474)
\thinlines \path(945,474)(942,481)
\thinlines \path(942,481)(936,487)
\thinlines \path(936,487)(927,493)
\thinlines \path(927,493)(915,499)
\thinlines \path(915,499)(900,504)
\thinlines \path(900,504)(883,509)
\thinlines \path(883,509)(864,513)
\thinlines \path(864,513)(843,517)
\thinlines \path(843,517)(821,519)
\thinlines \path(821,519)(797,521)
\thinlines \path(797,521)(772,522)
\thinlines \path(772,522)(748,523)
\thinlines \path(748,523)(723,522)
\thinlines \path(723,522)(698,521)
\thinlines \path(698,521)(675,519)
\thinlines \path(675,519)(653,516)
\thinlines \path(651,338)(631,335)
\thinlines \path(631,335)(612,331)
\thinlines \path(612,331)(595,326)
\thinlines \path(595,326)(581,320)
\thinlines \path(581,320)(570,315)
\thinlines \path(570,315)(561,308)
\thinlines \path(561,308)(556,302)
\thinlines \path(556,302)(554,296)
\thinlines \path(554,296)(554,289)
\thinlines \path(554,289)(558,283)
\thinlines \path(558,283)(565,276)
\thinlines \path(565,276)(575,271)
\thinlines \path(575,271)(588,265)
\thinlines \path(588,265)(603,260)
\thinlines \path(603,260)(621,255)
\thinlines \path(621,255)(641,251)
\thinlines \path(641,251)(662,248)
\thinlines \path(662,248)(685,246)
\thinlines \path(685,246)(709,244)
\thinlines \path(709,244)(734,243)
\thinlines \path(734,243)(759,243)
\thinlines \path(759,243)(783,243)
\thinlines \path(783,243)(807,245)
\thinlines \path(807,245)(831,247)
\thinlines \path(831,247)(853,250)
\thinlines \path(853,250)(873,254)
\thinlines \path(873,254)(891,259)
\thinlines \path(891,259)(907,264)
\thinlines \path(907,264)(921,269)
\thinlines \path(921,269)(931,275)
\thinlines \path(931,275)(939,281)
\thinlines \path(939,281)(944,287)
\thinlines \path(944,287)(946,294)
\thinlines \path(946,294)(944,300)
\thinlines \path(944,300)(939,307)
\thinlines \path(939,307)(932,313)
\thinlines \path(932,313)(921,319)
\thinlines \path(921,319)(908,324)
\thinlines \path(908,324)(892,329)
\thinlines \path(892,329)(874,334)
\thinlines \path(874,334)(853,337)
\thinlines \path(853,337)(831,341)
\thinlines \path(831,341)(808,343)
\thinlines \path(808,343)(784,344)
\thinlines \path(784,344)(759,345)
\thinlines \path(759,345)(735,345)
\thinlines \path(735,345)(710,344)
\thinlines \path(710,344)(686,342)
\thinlines \path(686,342)(663,340)
\thinlines \path(663,340)(641,337)
\thinlines \path(641,337)(622,333)
\thinlines \path(622,333)(604,328)
\thinlines \path(604,328)(588,323)
\thinlines \path(588,323)(576,318)
\thinlines \path(576,318)(566,312)
\thinlines \path(566,312)(558,305)
\thinlines \path(558,305)(554,299)
\thinlines \path(554,299)(554,293)
\thinlines \path(554,293)(556,286)
\thinlines \path(556,286)(561,280)
\thinlines \path(561,280)(570,274)
\thinlines \path(570,274)(581,268)
\thinlines \path(581,268)(595,262)
\thinlines \path(595,262)(611,258)
\thinlines \path(611,258)(630,253)
\thinlines \path(630,253)(651,250)
\thinlines \path(651,250)(673,247)
\thinlines \path(673,247)(696,245)
\thinlines \path(696,245)(721,243)
\thinlines \path(721,243)(746,243)
\thinlines \path(746,243)(770,243)
\thinlines \path(770,243)(795,244)
\thinlines \path(795,244)(819,246)
\thinlines \path(819,246)(841,249)
\thinlines \path(841,249)(862,252)
\thinlines \path(862,252)(882,256)
\thinlines \path(882,256)(899,261)
\thinlines \path(899,261)(914,266)
\thinlines \path(914,266)(926,272)
\thinlines \path(926,272)(935,278)
\thinlines \path(935,278)(942,284)
\thinlines \path(942,284)(945,290)
\thinlines \path(945,290)(945,297)
\thinlines \path(945,297)(942,303)
\thinlines \path(942,303)(936,310)
\thinlines \path(936,310)(927,316)
\thinlines \path(927,316)(915,321)
\thinlines \path(915,321)(900,327)
\thinlines \path(900,327)(883,331)
\thinlines \path(883,331)(864,336)
\thinlines \path(864,336)(843,339)
\thinlines \path(843,339)(821,342)
\thinlines \path(821,342)(797,344)
\thinlines \path(797,344)(772,345)
\thinlines \path(772,345)(748,345)
\thinlines \path(748,345)(723,345)
\thinlines \path(723,345)(698,343)
\thinlines \path(698,343)(675,341)
\thinlines \path(675,341)(653,339)
\thinlines \path(812,443)(825,446)
\thinlines \path(825,446)(837,448)
\thinlines \path(837,448)(848,451)
\thinlines \path(848,451)(857,455)
\thinlines \path(857,455)(864,458)
\thinlines \path(864,458)(869,462)
\end{picture}

\vspace{-1.5cm}
\end{center}
\caption{Part of the boundary of $\AdS_5\times S^5$ and of  the plane wave in
  BN coordinates ($\delta=\beta+\pi$)}\label{cboundpwBNcoord} 
\end {figure}
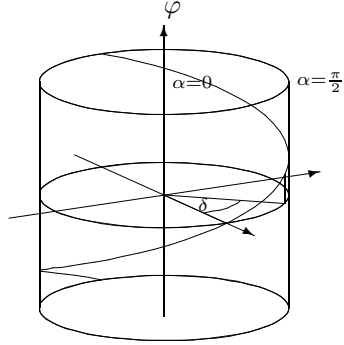

To avoid confusion in comparing Fig.\ \ref{cboundpwBNcoord} with similar
looking pictures for 
$\AdS_5\times\text{S}^5$, where the half of some Einstein static universe is
depicted, it is appropriate to stress that Fig.\ \ref{cboundpwBNcoord} 
represents the
whole Einstein static universe $R\times S^9$ although the radius variable
of the cylinder runs from zero to $\frac{\pi}{2}$ only. This range for $\alpha$
is due to its special role in the parameterization of $\text{S}^9$ in \eqref{bound3}.

The coordinate transformations just discussed for the identification
of the conformal boundary of the plane wave of course can also be applied 
to the $\AdS_5\times\text{S}^5$ metric. A priori these new coordinates are not 
a favourite choice to give any special insight into the $\AdS_5\times\text{S}^5$ geometry. In particular they are not well suited to find the conformal boundary. 

But we can turn the argument around. Since we know already the conformal
boundary of $\AdS_5\times\text{S}^5$, we can look where this boundary is situated
in the new coordinates and hope to find some illuminating picture for its
degeneration in the $R\to\infty$ limit which produces the plane wave metric. 

As we have discussed in connection with the metrics
\eqref{AdSmetricglobalcoord} and \eqref{AdSSmetricglobalcoord}, the
conformal boundary of
respectively $\AdS_{d+1}$ and $\AdSS$ is at $\rho\to\infty$ with all other
coordinates kept fixed at arbitrary finite values. 

Translating this into the
coordinates \eqref{BMNcoordinates}, \eqref{transvspacepolarcoord} it is at
$r\to\infty$ and $z^+$, $z^-$, $y$, $\omega_i$, $\tilde\omega_{i'}$, 
$i,i'=1,\dots 3$  fixed at arbitrary finite values. Before applying    
\eqref{bound1} we define $z$ for the $\AdS_5\times\text{S}^5$ case
by
\begin{equation}
r=z\cos\chi\col\qquad y=z\sin\chi\col
\label{bound6}
\end{equation}
i.\ e.\  $z^2=r^2+y^2$.
In the following coordinate transformation according to \eqref{bound1} $\chi$, 
$\omega_i$, $\tilde\omega_{i'}$ remain untouched. The conformal
$\AdS_5\times\text{S}^5$ boundary is now at 
$z\to\infty$, 
$\chi\to 0$, $z^+$, $z^-$, $\omega_i$, $\tilde\omega_{i'}$ fixed
at arbitrary finite values.
The expansion of the first line of \eqref{bound1} yields
\begin{equation}
\cot\theta=-\frac{z}{2}\sin z^++\mathcal{O}(z^{-1})\pnt
\end{equation}
From this one finds (as above we again start with $|z^+|\leq\frac{\pi}{2}$ and
glue the other $z^+$ patches afterwards) 
\begin{equation}
\lim_{\overset{z\to\infty}{\underset{z^+>0}{}}}\theta=\pi\col\qquad\lim_{\overset{z\to\infty}{\underset{z^+<0}{}}}\theta=0\col\qquad\lim_{\overset{z\to\infty}{\underset{z^+=0}{}}}\theta=\frac{\pi}{2}\pnt
\label{bound7}
\end{equation}
Furthermore, by coupling $z^+\to 0$ in a suitable way with $z\to\infty$ one can reach any $\theta\in(0,\pi)$
\begin{equation}
\lim_{\overset{z\to\infty}{\underset{z^+=c/z}{}}}\theta=\arctan\big(-\tfrac{2}{c}\big)\pnt
\label{bound8}
\end{equation}
In the second equation of \eqref{bound1} one has to insert
\begin{equation}\label{sinincot}
\frac{1}{\sin\theta}=
\frac{z}{2}|\sin z^+|\sqrt{1+\frac{4}{z^2\sin^2z^+}}\pnt
\end{equation}
At fixed $z^+$ the expansion of the second equation of \eqref{bound1} then
reads
\begin{equation}
\tan\tfrac{\varphi\pm\zeta}{2}=\big(1\pm\epsilon(\sin
z^+)\big)\frac{z^2}{2}\tan z^+\pm\frac{1}{|\sin z^+|\cos
  z^+}+\mathcal{O}(z^{-1})\col 
\end{equation}
where $\epsilon$ is the sign function defined in \eqref{signfuncdef}. From the
above result one finds (again for $|z^+|\le\frac{\pi}{2}$)
\begin{equation}
\bigg\{\begin{array}{cc}z^+<0&(\text{i.\ e.\
    }\theta\to 0)\\ z^+>0&(\text{i.\ e.\
    }\theta\to\pi)\end{array}\bigg\} 
\Longrightarrow\lim_{z\to\infty}\tan\tfrac{\varphi+\zeta}{2}
=\bigg\{\begin{array}{c}\text{finite}\\\infty\end{array}\bigg\}
\col\quad\lim_{z\to\infty}\tan\tfrac{\varphi-\zeta}{2}
=\bigg\{\begin{array}{c}\infty\\\text{finite}\end{array}\bigg\}\pnt
\label{bound9}
\end{equation} 
In addition one gets for $z\to\infty$ coupled as in \eqref{bound8}
with $z^+\to 0$ that \eqref{sinincot} becomes
\begin{equation}
\frac{1}{\sin\theta}=\sqrt{\frac{c^2}{4}+1}+\mathcal{O}(z^{-1})\pnt
\end{equation}
The second equation of \eqref{bound1} then reads
\begin{equation}
\tan\tfrac{\varphi\pm\zeta}{2}=z\bigg(\pm\sqrt{\frac{c^2}{4}+1}+\frac{c}{2}\bigg)+\mathcal{O}(z^0)\col
\end{equation}
such that one gets in this case
\begin{equation}
0<\theta<\pi \Longrightarrow
\lim_{\overset{z\to\infty}{\underset{z^+=c/z}{}}}\tan\tfrac{\varphi\pm\zeta}{2}=\pm\infty\pnt 
\label{bound10}
\end{equation}
Putting together \eqref{bound7}-\eqref{bound10}, we see that in the projection
onto the three coordinates $\varphi$, $\theta$, $\zeta$ the conformal boundary
of the ($|z^+|<\frac{\pi}{2}$)-patch of $\AdS_5\times\text{S}^5$
is mapped to the one-dimensional line starting at 
$(\varphi,\zeta,\theta)=(-\pi,0,0)$, 
running first with $\theta=0$ and $\zeta-\varphi=\pi$
to $(\varphi,\zeta,\theta)=(0,\pi,0)$, then with $\varphi=0$ and $\zeta=\pi$
to $(\varphi,\zeta,\theta)=(0,\pi,\pi)$ and finally with $\theta=\pi$ and
$\varphi+\zeta=\pi$ to $(\varphi,\zeta,\theta)=(\pi,0,\pi)$, see also
Fig.\ \ref{cboundpwanglecoord}
\begin{figure}
\begin{center}
\begin{picture}(150,230)(0,0)
\SetOffset(60,120)
\LongArrow(0,0)(-35,-35)\Text(-40,-40)[]{$\zeta$}
\LongArrow(0,0)(60,0)\Text(70,0)[]{$\theta$}
\LongArrow(0,0)(0,90)\Text(0,100)[]{$\varphi$}
\DashLine(40,-90)(40,90){4}\Text(45,-5)[]{$\scriptsize\pi$}
\DashLine(0,-90)(0,90){4}\Text(-7.5,40)[]{$\scriptsize\pi$}
\DashLine(-20,-110)(-20,70){4}\Text(-25,-15)[]{$\scriptsize\pi$}
\DashLine(20,-110)(20,70){4}
\SetWidth{1}
\DashLine(-20,60)(0,40){2}
\DashLine(0,40)(40,40){2}
\Line(40,40)(20,-20)
\Line(20,-20)(-20,-20)
\Line(-20,-20)(0,-40)
\DashLine(0,-40)(40,-40){2}
\DashLine(40,-40)(20,-100){2}
\end{picture}
\end{center}
\caption{Part of the boundary of $\AdS_5\times\text{S}^5$ and of the plane
  wave in $(\varphi,\zeta,\theta)$ coordinates}\label{cboundpwanglecoord}
\end{figure}
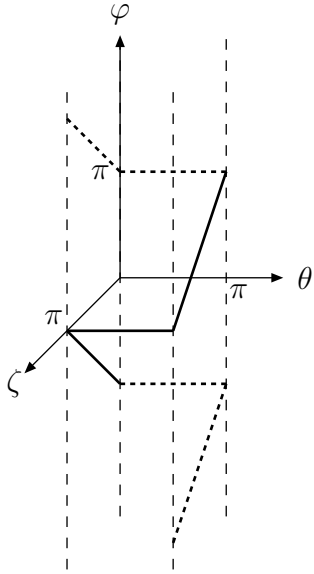

Translating this via \eqref{bound3} into the coordinates
  $(\varphi,\alpha,\beta)$ we find the line\footnote{Note that the piece  from
  $(\varphi,\zeta,\theta)=(0,\pi,0)$ to $(\varphi,\zeta,\theta)=(0,\pi,\pi)$
  with $\varphi=0$ and $\zeta=\pi$ is mapped to one point 
$(\varphi,\alpha,\beta )=(0,\frac{\pi}{2},\pi)$.}  
$\alpha=\frac{\pi}{2}$, $\beta=\pi+\varphi$, $-\pi<\varphi<+\pi$.
After gluing the other $z^+$-patches we can conclude:

The projection onto the coordinates $(\varphi,\alpha,\beta)$ of the conformal
boundary of $\AdS_5\times\text{S}^5$ coincides with that of a part of the
conformal boundary of the $10$-dimensional plane wave (see
\eqref{ppwavemetric} with $H(z^+,z)=-\delta_{ij}z^iz^j$) 
\begin{equation}
\de s^2_\text{pw}=-4\de z^+\de z^--\vec z^{2}(\de z^+)^2+\de\vec z^2\pnt
\label{intro4}
\end{equation}
That in this projection only a part of the plane wave boundary line appears as the $\AdS_5\times\text{S}^5$ boundary is due to the restriction to the $\AdS_5\times\text{S}^5$-strip \eqref{bound0}. Note that
this restriction can be circumvented as discussed at the beginning of this Section.

Taking into account the other seven coordinates, the $\AdS_5\times\text{S}^5$ boundary
is of course not one-dimensional. But by using the same coordinates
both for $\AdS_5\times\text{S}^5$ \emph{and} the plane wave, we now have visualized
the degeneration of the conformal boundary in the process of approaching the
plane wave limit. In the projection to three of the BN coordinates 
($\varphi,\alpha,\beta $) the boundary stays throughout this process at the same location. The extension in the remaining coordinates degenerates to a point in the limit.

The picture is more involved if one compares the two boundaries in the
BMN coordinates $(z^+,z^-,z)$. As noted in \cite{Berenstein:2002sa}, due to the singularity
of the coordinate transformation on the boundary line in 
$(\varphi,\alpha,\beta)$, the limits in $(z^+,z^-,z)$ which map to this
boundary line are not unique. Besides
\begin{equation}
\text{Limit (i):}\qquad z\to \infty\col\qquad z^+,z^-=\text{finite}\col
\label{bound11}
\end{equation}
just discussed above, the second limit is
\begin{equation}
\text{Limit (ii):}\qquad z^-\to \pm \infty\col\qquad z^+,z=\text{finite}\col
\label{bound12}
\end{equation}
as trivially seen from \eqref{bound1}.

Now the situation looks a little bit cumbersome. The conformal boundary  of $\AdS_5\times\text{S}^5$  is realized via limit (i), that of the plane wave via both
limits (i) \emph{and} (ii), although the the plane wave itself is a limit of
 $\AdS_5\times\text{S}^5$.

To get a better understanding of this situation, we are now asking what
happens in the $R\to\infty$ limit with geodesics, in particular null
geodesics, which reach the conformal $\AdS_5\times\text{S}^5$ boundary. After
an explicit construction of all the  geodesics both for
$\AdS_5\times\text{S}^5$ and the plane wave \eqref{intro4} in the next
Section, we come back to this question.

\section{Geodesics in $\AdS_5\times\text{S}^5$ and in the plane wave}
\label{AdSpwgeodesics}
\subsection{Geodesics in $\AdS_5\times\text{S}^5$}
\label{AdSgeodesics}
We start with the $\AdS_5\times\text{S}^5$ metric, where the $\AdS_5$ part is
given in the global by coordinates in the first line of 
\eqref{AdSmetricglobalcoord}
\begin{equation}\label{geo1}
\de s^2=R^2(-\de t^2\cosh^2\rho+\de\rho^2+\sinh^2\rho \de\Omega^2_3+\de\Omega^2_5)\pnt
\end{equation}
The geodesic equations for the $\AdS_5$ and $\text{S}^5$ coordinates decouple. Geodesics on $\text{S}^5$ are great circles. Whether the geodesic in the total manifold 
$\AdS_5\times\text{S}^5$ moves
in $\text{S}^5$ or stays at a fixed $\text{S}^5$ position has consequences for the overall causal property (spacelike, timelike, null) only. There is no effect on the $\AdS_5$
coordinates. Therefore, we can concentrate on the $\AdS_5$ part. 
It can be regarded as a warped product of a two-dimensional space with
coordinates $t$ and $\rho$ and a $3$-dimensional sphere. In Appendix
\ref{app:geoinwarpedspace} we present the details how the geodesic equations 
can be simplified in this case. The warp factor in front of the spherical part
has only influence on the parameterization of the geodesics, but not on their
shape. They are still given by great circles on the subsphere.
The simplified geodesic equations for $t$, $\rho$ and $f$, where $f$
captures the effects of the movement in the $3$-dimensional unit sphere, 
are given by \eqref{AdSgeoeq} and read
\begin{align}
\ddot t +2\dot\rho\dot t\tanh\rho&=0\col\label{geo2}\\
\ddot\rho+(\dot t^2-\dot f^2)\sinh\rho\cosh\rho&=0\col\label{geo8}\\
\ddot f+2\dot\rho\dot f\coth\rho&=0\pnt\label{geo5}
\end{align}
To be more precise, $\dot f$ is given in terms of the metric
$g(\Omega_3)_{mn}$ and the angle velocities $\dot y^m$ of the unit
$\text{S}^3$ as 
\begin{equation}\label{geo7}
\dot f^2=g(\Omega_3)_{mn}\dot y^m\dot y^n\pnt
\end{equation}
Hence, $f(\tau)$ itself is the angle between the position vectors along the 
geodesic at $\tau=0$ and at $\tau$ and $\dot f$ has the interpretation as
velocity in the $3$-dimensional unit sphere. Remark that there is some freedom
in the definition of $f$ such that the above result corresponds to a particular
choice, see Appendix \ref{app:geoinwarpedspace} for more details.   
Summarizing the discussion so far, the coordinates in the $\text{S}^3$ either
remain constant ($\dot f=0$) or describe a movement on a great circle ($
\dot f\neq 0$).

We will now solve the geodesic equations for $t$, $\rho$ and $f$.
Straightforward integration of \eqref{geo2} and \eqref{geo5} yields  
\begin{equation}\label{geo9}
\dot t=\frac{b}{\cosh^2\rho}\col\qquad\dot f=\frac{\tilde b}{\sinh^2\rho}\col
\qquad b,\tilde b\quad\text{constant}\pnt
\end{equation}
With \eqref{geo8} one derives an equation for $\rho$ alone
\begin{equation}\label{geo10}
\ddot{\rho}+\frac{b^2}{\cosh^3\rho}\sinh\rho-\frac{\tilde b^2}{\sinh^3\rho}\cosh\rho=0\pnt
\end{equation}
This equation can be integrated after multiplication with $\dot\rho$. 
If one introduces the integration constant $c$ one finds
\begin{equation}\label{geo11}
\dot{\rho}^2=\frac{c}{R^2}-\frac{\tilde b^2}{\sinh^2\rho}+\frac{b^2}{\cosh^2\rho}\pnt
\end{equation}
This is precisely the condition for the parameter to be an affine one for the
full metric \eqref{geo1}. In fact, if we denote the coordinates on $\AdS_5$
with $x$, the condition that $\tau$ is an affine parameter reads
\begin{equation}
G_{\mu\nu}\dot x^\mu\dot x^\nu=R^2(-\dot t^2\cosh^2\rho+\dot\rho^2+\dot f^2\sinh^2\rho)=c\col
\end{equation}
and one finds \eqref{geo11} with the help of \eqref{geo9}. 

Are all solutions of \eqref{geo11} also solutions of \eqref{geo10}?
At first the constancy of the scalar product of the tangential vector with itself is of course a much weaker condition than the geodesic equations. But in
writing down \eqref{geo11} we already have implied the geodesic equations
for all coordinates, except for $\rho$. Under these circumstances, at least
as long as $\dot\rho\neq 0$, the constant scalar product condition
is equivalent to the geodesic equation for the last coordinate $\rho$.
If $\rho=\text{const.}\neq 0$ it follows immediately from \eqref{geo8} that 
$\dot t=\dot f=\text{const.}$ If $\rho=0$ the only possibilities are 
geodesics that move only in the time direction of $\AdS_5$ or stay at a point.

Since $\dot{\rho}^2$ is a non-negative quantity, from \eqref{geo11} and
$$\frac{c}{R^2}-\frac{\tilde b^2}{\sinh^2\rho}+\frac{b^2}{\cosh^2\rho}\leq \frac{c}{R^2}+b^2\col\qquad\forall\rho\col$$
as a byproduct, we find a constraint on $c$ and the integration constant $b$
\begin{equation}
\frac{c}{R^2}+b^2\geq 0\pnt\label{geo12}
\end{equation}

For further analyzing the consequences of the positiveness of both sides of 
\eqref{geo11} we introduce the abbreviations
\begin{equation}
A=\frac{c}{R^2}\col\qquad B=b^2+\frac{c}{R^2}-\tilde b^2\col\qquad C=-\tilde b^2\pnt
\label{geo13}
\end{equation}
Then first of all, by these definitions and the inequality \eqref{geo12} the constants 
$A$, $B$, $C$ are universally constrained by
\begin{equation}
C\leq 0\col\qquad B\geq A+C\col\qquad B\geq C\pnt
\label{geo13a}
\end{equation}
In addition, checking whether there are real $\rho$-values for which
the R.\ H.\ S.\  of \eqref{geo11} is non-negative, it turns out that only
four classes of ranges\footnote{The special case $A=B=C=0$ corresponds to a
  point, not to a curve.} 
of the constants $A,B,C$  are allowed. Integrating case by case first \eqref{geo11} and then \eqref{geo9} for the four classes one finds:
\\[2mm]
\underline{\it type I}
\begin{equation}\label{geo15}
\begin{gathered}
A>0\col\\
0\leq \frac{\sqrt{B^2-4AC}-B}{2A}\leq\sinh^2\rho\col
\end{gathered}
\end{equation}
\\
\begin{equation}\label{geo19}
\begin{aligned}
\rho&=\arsh\sqrt{\frac{1}{4A}\Big(\e^{\pm 2\sqrt A(\tau+\tau_0)}+
(B^2-4AC)\e^{\mp 2\sqrt A(\tau+\tau_0)}\Big)-\frac{B}{2A}}\col
\\
t&=\pm\arctan\bigg(\frac{\e^{\pm 2\sqrt A(\tau+\tau_0)}+2A-B}{2b\sqrt A}\bigg)+t_0\col\\
f&=\pm\arctan\bigg(\frac{\e^{\pm 2\sqrt A(\tau+\tau_0)}-B}{2\tilde b\sqrt{A}}\bigg)+f_0\col
\end{aligned}
\end{equation}
\\[2mm]
\underline{\it type~II}
\begin{equation}\label{geo16}
\begin{gathered}
A<0\col\qquad B^2-4AC>0\col\qquad B>0\col\\
0\leq\frac{B-\sqrt{B^2-4AC}}{-2A}\leq\sinh^2\rho\leq\frac{B+\sqrt{B^2-4AC}}{-2A}\col
\end{gathered}
\end{equation}
\\
\begin{equation}\label{geo21}
\begin{aligned}
\rho&=\arsh\sqrt{\frac{1}{-2A}\big(B\pm\sqrt{B^2-4AC}\sin(2\sqrt{-A}(\tau+\tau_0))
\big)}\col\\
t&=\pm\frac{1}{2}\arccot\bigg(\frac{2\sqrt{-A}b\cos(2\sqrt{-A}(\tau+\tau_0))}{
\sqrt{B^2-4AC}\pm(B-2A)\sin(2\sqrt{-A}(\tau+\tau_0))}\bigg)+t_0\col\\
f&=\pm\frac{1}{2}\arccot\bigg(\frac{2\sqrt{-A}\tilde b\cos(2\sqrt{-A}(\tau+\tau_0))}{\sqrt{B^2-4AC}\pm B\sin(2\sqrt{-A}(\tau+\tau_0))}\bigg)+f_0\col
\end{aligned}
\end{equation}
\\[2mm]
\underline{\it type III}
\begin{equation}\label{geo17}
\begin{gathered}
A=0\col\qquad B>0\col\\
0\leq\frac{-C}{B}\leq\sinh^2\rho\col
\end{gathered}
\end{equation}
\\
\begin{equation}
\begin{aligned}
\rho&=\arsh\sqrt{B(\tau+\tau_0)^2-\frac{C}{B}}\col\\
t&=\arctan\Big(\frac{B(\tau+\tau_0)}{b}\Big)+t_0\col\\
f&=\arctan\Big(\frac{B(\tau+\tau_0)}{\tilde b}\Big)+f_0\col
\end{aligned}
\label{geo23}
\end{equation}
\\[2mm]
\underline{\it type~IV}
\begin{equation}\label{geo18}
\begin{gathered}
A<0\col\qquad B^2-4AC=0\col\qquad B\geq 0\col\\
\sinh^2\rho=\frac{B}{-2A}\col
\end{gathered}
\end{equation}
\\
\begin{equation}\label{geo25}
\begin{aligned}
\rho&=\arsh\sqrt{\frac{B}{-2A}}\col\\
t&=\sqrt{-A}\tau+t_0\col\\
f&=\pm\sqrt{-A}\tau+f_0\pnt   
\end{aligned}
\end{equation}
\\[2mm]

Perhaps it is useful to stress, that in the absence of any movement in
the $\text{S}^3$, i.\ e.\  for $C=0$, the formulas \eqref{geo19},
\eqref{geo21} and \eqref{geo23} for $\rho$ simplify to\\[2mm]
\underline{\it type I with C=0}
\begin{equation}
\rho=\arsh\bigg(\sqrt{\frac{|B|}{A}}\big|\sinh\big(\sqrt{A}(\tau+\tau'_0)\big)\big|\bigg)\col
\label{geo27}
\end{equation}
\underline{\it type II with C=0}
\begin{equation}
\rho=\arsh\bigg(\sqrt{\frac{B}{-A}}\big|\sin\big(\sqrt{-A}(\tau+\tau'_0)\big)\big|\bigg)\col
\label{geo28}
\end{equation}
\underline{\it type III with C=0}
\begin{equation}
\rho=\arsh\big(\sqrt B|\tau+\tau_0|\big)\pnt
\label{geo29}
\end{equation}
The $\pm$ alternative in \eqref{geo19} and \eqref{geo21} has been
absorbed into the shift of the integration constant $\tau_0$ to $\tau'_0$.

The causal properties of the geodesics and their relation to the conformal
boundary (note footnote 5) can be summarized in the following table.
\begin{center}
\begin{tabular}{|p{10mm}|p{40mm}|p{40mm}|p{40mm}|}
\hline
type&causal properties \vfill w.\ r.\ t.\ $\AdS_5$&causal properties \vfill w.\ r.\ t.\ $\AdS_5\times\text{S}^5$&reaches conf.\ bound.\ of $\AdS_5\times\text{S}^5$\\
\hline
\hline
I&space-like&space-like&yes\\
\hline
II&time-like&all&no\\
\hline
III&null&null or space-like&yes\\
\hline
IV&time-like&all&no\\
\hline
\end{tabular}
\end{center}
For later use it is important to stress, that null geodesics in the sense of full $\AdS_5\times\text{S}^5$ reaching the boundary have to be of {\it type III}. For them no movement
in $\text{S}^5$ is allowed while a movement in $\text{S}^3$ is possible  
as long as $b^2>\tilde b^2$.
\subsection{Geodesics in the plane wave}
\label{pwgeodesics}
Here the metric is given by the $10$-dimensional version of
\eqref{ppwavemetric} with $H(z^+,z)=-\delta_{ij}z^iz^j$
\begin{equation}\label{10dimppwavemetric}
\de s^2=-4\de z^+\de z^--\vec z^{2}(\de z^+)^2+\de\vec z^2\col
\end{equation}
such that one finds for the geodesic equations \eqref{geodesicequations}
with the Christoffel connection \eqref{pwconnection} 
\begin{align}
\ddot z^+&=0\col\label{geo31}\\
\ddot z^-+\frac{1}{2}\dot z^+\frac{\de}{\de\tau}\vec
z^{2}&=0\col\label{geo32}\\ 
\ddot z^i+(\dot z^+)^2z^i&=0\pnt\label{geo33}
\end{align}
Then \eqref{geo31} implies linear dependence of $z^+$ on the the affine
parameter $\tau$
\begin{equation}\label{geo34}
z^+=\alpha\tau+z_0^+\pnt
\end{equation}
Obviously now the geodesics fall into two classes, {\it type A} with $\alpha=0$ and {\it type B} with $\alpha\neq 0$.\\[2mm]
\underline{\it type A}\\[2mm]
\begin{equation}\label{geo35}
z^+=\text{const.}\col\qquad z^-=\beta\tau+z^-_0\col\qquad z^i=\gamma^i\tau+z^i_0\pnt
\end{equation}
The scalar product of their tangential vector with itself is given by
$(\gamma^i)^2$ (see Appendix \eqref{app:chordalvsgeodist} for a relation
between the geodesic and the chordal distances).
This implies:

All {\it type A} geodesics are null or space-like. 
Space-like {\it type A} geodesics
reach infinity in the transversal coordinates $\vec z$. {\it Type A} null 
geodesics are given by constant $z^+$ and $z^i$ as well as $z^-$ running 
between
$\pm\infty$.\\[2mm]
If we choose $\alpha\neq 0$ in \eqref{geo34} we find after the integration
 of \eqref{geo32} and of \eqref{geo33} the second type of geodesics given
 by\\[2mm] 
\underline{\it type B}\\[2mm]  
\begin{equation}\label{typeBgeo}
\begin{aligned}
z^+&=\alpha\tau+z_0^+\col\\
z^-&=\frac{1}{8}\sum_i(\beta^i)^2\sin\big(2\alpha(\tau+\tau^i_0)\big)+\gamma\tau+z^-_0\col\\
z^i&=\beta^i\sin\big(\alpha(\tau+\tau_0^i)\big)\pnt
\end{aligned}
\end{equation}
The scalar product of the tangential vector with itself is now equal to
$-4\alpha\gamma$ (see Appendix \eqref{app:chordalvsgeodist} for a relation
between the geodesic and the chordal distances),
and we conclude:

All {\it type B} geodesics either stay at $z^i=0$ (for $\beta^i=0$) or 
oscillate in the transversal coordinates $z^i$ (for $\beta^i\neq 0$).
All space or time-like {\it type B} geodesics ($\gamma\neq 0$) reach 
$\pm\infty$
both in $z^+$ and $z^-$. {\it Type B} null geodesics ($\gamma=0$) reach $\pm\infty$ \emph{only} with respect to $z^+$. Furthermore, they stay at fixed $\vec z$ and $z^-$ ($\vec\beta=0$)  or oscillate both in $z^i$ and
$z^-$ ($\beta^i\neq 0$).

In conclusion null geodesics reaching the conformal boundary of the plane
wave, see \eqref{bound11}, \eqref{bound12}, are necessarily of {\it type A}.
There are no null geodesics reaching the conformal boundary within the
asymptotic regime of limit (i).

Closing this Section we comment on a simple discussion of the plane wave 
null geodesics in using the BN coordinates of \eqref{bound5}. In general
null geodesics are invariant under a Weyl transformation. Such a 
transformation only effects the choice of affine parameters along the null 
geodesics. Null geodesics with respect to \eqref{bound5} without the Weyl
factor are given by great circles in $\text{S}^9$ accompanied by a compensating
movement along the time-like direction $\varphi $. If we discuss $\text{S}^9$ as
an embedding in $\mathds{R}^{10}$, reaching $\alpha=\frac{\pi}{2}$ 
is equivalent to reaching the $(z_1,z_2)$-plane. There are of course great
circles within this plane. They correspond to null geodesics either winding at
 $\alpha=\frac{\pi}{2}$ in constant distance to the conformal plane wave
boundary around the cylinder in Fig.\ \ref{cboundpwBNcoord} 
up to $\varphi\to\pm\infty$ 
or they wind in the orthogonal direction crossing the conformal plane wave
boundary. In the sense of $\mathds{R}\times\text{S}^9$ there is nothing
special  
with such a crossing. But going back to the metric including the Weyl factor, starting from an inside point, the boundary is reached at infinite affine
parameter. Furthermore, there are of course great circles staying completely
away from the $(z_1,z_2)$-plane (i.\ e.\  $\alpha=\frac{\pi}{2}$). They 
correspond to null geodesics generically oscillating in
$0<\alpha<\frac{\pi}{2}$ and running up to $\varphi\to\pm\infty$. 
Finally, great 
circles can also intersect the $(z_1,z_2)$-plane. Then they correspond to
null geodesics oscillating in $\alpha$ and touching $\alpha=\frac{\pi}{2}$.
Obviously some of them reach the conformal boundary line of the plane
wave. According to the above analysis in BMN coordinates they are of
{\it type A}, too. 
\section{Conformal boundaries and geodesics}
\label{AdSpwgeorel}
As discussed in Section \ref{AdSgeodesics}, only null-geodesics of {\it type
  III} reach the 
conformal boundary of $\AdS_5\times\text{S}^5$. They necessarily stay at fixed
$\text{S}^5$-position. Translating \eqref{geo23} into the coordinates of 
\eqref{BMNcoordinates}
we get with $R=R_1=R_2$
\begin{equation}
\begin{aligned}
z^+&=\frac{1}{2}\Big(\arctan\Big(\frac{B(\tau+\tau_0)}{b}\Big)
+t_0+\psi\Big)\col\\
z^-&=\frac{R^2}{2}\Big(\arctan\Big(\frac{B(\tau+\tau_0)}{b}\Big)
+t_0-\psi\Big)\col\\
r&=R\arsh\sqrt{B(\tau+\tau_0)^2-\frac{C}{B}}\col\\
f&=\arctan\Big(\frac{B(\tau+\tau_0)}{\tilde b}\Big)
+f_0\col\\
y&=R\vartheta\pnt
\end{aligned}
\label{bg1}
\end{equation}
Our goal is to find in the $R\to\infty$ limit a correspondence
to null geodesics of the plane wave. Therefore, our $\AdS_5\times\text{S}^5$
geodesics have to stay at least partially within the range of finite
$z^+,z^-,r,y$. Taking $R\to\infty$ at fixed $\tau$ would send all
$z^-$ to infinity. But of course the affine parameter itself is determined
only up to a constant rescaling. Therefore, the best procedure 
is to eliminate the affine parameter completely.

First from \eqref{bg1} we conclude, that along the full range of a {\it type III} null geodesic, i.\ e.\  for
($-\infty<\tau<\infty$), the coordinate $z^+$ runs within an interval of length $\frac{\pi}{2}$:
 $z^+\in\big(\frac{1}{2}(t_0+\psi)-\frac{\pi}{4},\frac{1}{2}(t_0+\psi)+\frac{\pi}{4}\big)$ and $z^-$ runs  within an interval of length $\frac{\pi}{2}R^2$:
$z^-\in\big(R^2(\frac{t_0-\psi}{2}-\frac{\pi}{4}),R^2(\frac{t_0-\psi}{2}+\frac{\pi}{4})\big)$. To ensure that the $z^-$ interval for $R\to\infty$ stays at least partially within the range of finite values both endpoints of the interval have to have the opposite sign. Thus, we have
to restrict $t_0$ and $\psi$ by
\begin{equation}
-\frac{\pi}{2}<t_0-\psi<\frac{\pi}{2}\pnt
\label{bg2}
\end{equation}
In addition one has universally
\begin{equation}
|f(\tau=+\infty)-f(\tau=-\infty)|=\pi\pnt
\label{bg3}
\end{equation}
As already mentioned around
\eqref{geo7}, $f$ can be understood as the angle along the great
circle in $\text{S}^3$ on which our null geodesics is running. Therefore, 
for {\it type III} geodesics the positions for $\tau=-\infty$
and $\tau=+\infty$ within the $\text{S}^3$ are always antipodal to each
another.\footnote{In the limiting case, where the null geodesics goes through $r=0$, $f$ becomes a step function.}

After these preparations we now eliminate the affine parameter and
express $z^+$, $r$ and $f$ in terms of $z^-$
(note that for {\it type III} we have $A=0$ and $B=b^2-\tilde b^2$)
\begin{equation}
\begin{aligned}
z^+&=\frac{z^-}{R^2}+\psi\col\\
r&=R\arsh\sqrt{\frac{\tan^2(\frac{2z^-}{R^2}-t_0+\psi)+\frac{\tilde b^2}{b^2}}{1-\frac{\tilde b^2}{b^2}}}\col\\
f&=f_0+\arctan\Big(\frac{b}{\tilde b}\tan\Big(\frac{2z^-}{R^2}-t_0+\psi\Big)\Big)\col\\
y&=R\vartheta\pnt
\end{aligned} 
\label{bg4}
\end{equation} 
The minimal value for $r$ is
\begin{equation} 
r_\text{min}=R\arsh\Bigg(\frac{|\frac{\tilde
      b}{b}|}{\sqrt{1-\frac{\tilde b^2}{b^2}}}\Bigg)\pnt
\end{equation}
Since we insist on finite $r_\text{min}$ for $R\to\infty$ we have to rescale ($r_0=\lim_{R\to\infty}r_\text{min}$)
\begin{equation}
\frac{|\tilde b|}{|b|}=\frac{r_0}{R}\pnt
\label{bg5}
\end{equation}
Although we have now realized finite $r_\text{min}$, the
$z^-$ value where $r_\text{min}$ is reached stays finite for 
$R\to\infty$ only
if \eqref{bg2} is replaced by the stronger rescaling condition
\begin{equation}
t_0-\psi=\frac{a}{R^2}\pnt
\label{bg6}
\end{equation}

Altogether, to stay at least with part of the {\it type III} null geodesics
within the range of finite BMN coordinates, it is mandatory to perform
the rescalings \eqref{bg5}, \eqref{bg6} and to keep $y$ fixed. The
remaining parameters replacing $t_0,\psi ,b,\tilde b,f_0,\vartheta$
are $\psi,a,b,r_0,f_0,y$. 

Considering now at fixed $z^-$ the $R\to\infty$ limit of \eqref{bg4}
one arrives at
\begin{equation}
\begin{aligned}
z^+&=\psi+\mathcal{O}(R^{-2})\col\\
r&=r_0+\mathcal{O}(R^{-2})\col\\
f&=f_0+\mathcal{O}(R^{-2})\col\\
y&=\text{const.}
\end{aligned}
\label{bg7}
\end{equation}
Constant $r,y$ via \eqref{bound6} give constant $z$. In addition, constant
$f$, i.\ e.\  no movement in the $\text{S}^3$, and the a priori absence of any
movement in $\text{S}^5$ leads to constant $\vec z$. This together with the
constancy of $z^+$ implies:

An $\AdS_5\times\text{S}^5$ null geodesics, reaching the conformal boundary,
for any finite $z^-$-interval at $R\to\infty$ converges uniformly to a
{\it type A} null geodesics of the plane wave. 

However, the approach of the $\AdS_5\times\text{S}^5$ null geodesics to the
conformal boundary of $\AdS_5\times\text{S}^5$ is realized within the
asymptotic regime of limit (i), see \eqref{bound11}, \eqref{bg1},  but that of
the plane wave null geodesics within the regime of limit (ii), see
\eqref{bound12} and text after \eqref{geo35}. That means even for large $R$,
after a region of convergence, on their way to the boundary they diverge from
one another at the very end (in the $z^+,z^-,z$ coordinates under
discussion). 

In a global setting the situation is most simply illustrated for {\it type III}
null geodesics crossing the origin of the transverse BMN coordinates $\vec z$,
i.\ e.\  $r_0=y=0$. We also put $a=0$, the case $a\neq 0$ can be simply recovered
by the replacement $z^-\to z^--\frac{a}{2}$. Then first
of all $z^-$ runs between $\pm\frac{\pi}{4}R^2$. Furthermore, \eqref{bg4} implies 
\begin{equation}
\begin{aligned}
\frac{r}{R}&=F\big(\tfrac{z^-}{R^2}\big)\col\quad\text{with}\quad F(z)=\arsh|\tan(2z)|\col\\
f&=f_0\pm\frac{\pi}{2}\epsilon(z^-)\col
\end{aligned}
\label{bg8}
\end{equation}
where $\epsilon$ is the sign function \eqref{signfuncdef}.
The plane wave geodesic is at $z=y=r=0,-\infty<z^-<\infty$. It is the uniform
limit for $R\to\infty$ in the region 
$|z^-|<R^{1-\varepsilon}$, $\varepsilon>0$. 
This convergence is due to the different powers of
$R$ on the l.h.s. and in the argument of the function $F$ on the R.\ H.\ S.\
of \eqref{bg8}, see also Fig.\ \ref{fig:convergence}. 

The picture in Fig.\ \ref{fig:convergence} has to be completed by the freedom
to choose a point 
on $\text{S}^3$ to fix the direction in the space of the $\vec z$ coordinates. 
This completely specifies the
{\it type III} null geodesics under discussion. Then the conformal boundary
reaching null geodesics of $\AdS\times\text{S}^5$ crossing the origin of the
transversal BMN coordinates $\vec z=0$ form a cone with base $\text{S}^3$. 
The three parameters
to specify the $\text{S}^3$ position together with $\psi$ nicely correspond to
the four-dimensionality of the $\AdS\times\text{S}^5$ boundary. In the
$R\to\infty$ limit this cone degenerates.  
\begin{figure}
\begin{center}
\text{\epsfig{file=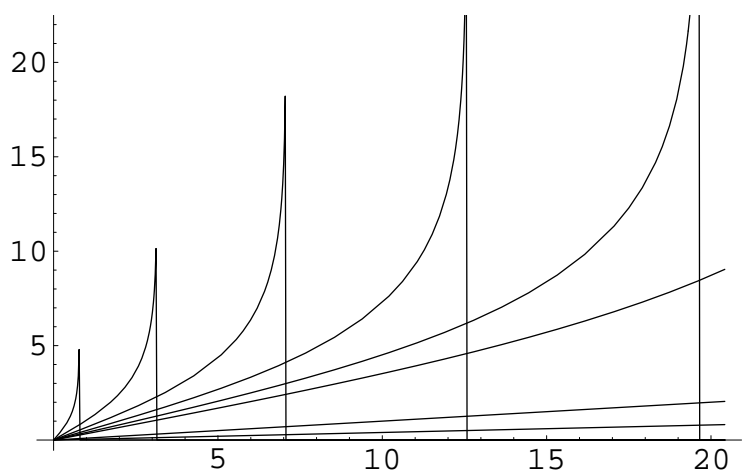, width=100mm}}
\caption{Approach of boundary reaching $\AdS_5\times\text{S}^5$ null geodesics
  to a boundary 
reaching null geodesics of the plane wave. The plane wave null geodesics
runs along the horizontal axis up to infinity. The plot shows $r$ versus $z^-$ 
for $\AdS_5\times\text{S}^5$ null geodesics \eqref{bg8} in the cases
$R=1,2,3,4,5,6,20,50$.}  
\label{fig:convergence}
\end{center}
\end{figure}

\clearemptydoublepage
\chapter{The scalar bulk-to-bulk propagator in $\AdS\times\text{S}$ and in the
plane wave}
\label{chap:propaga}
In this Chapter we will focus on the scalar
bulk-to-bulk propagator $G(z,z')$ in $\AdS_{d+1}\times\text{S}^{d'+1}$ 
defined as solution of a differential
equation similar to \eqref{AdSbulktobulkpropdiffeq} in pure $\AdS_{d+1}$. 
From the first sight it might be
less obvious why we deal especially with this propagator. It appears to be  
more natural to analyze the bulk-to-boundary propagator instead, because it 
determines the bulk supergravity fields from their boundary values via
\eqref{AdSboundvalsol} and thus enter directly the holographic
description. Furthermore, it was shown \cite{D'Hoker:1999ni} 
that the bulk-to-bulk propagator is no longer needed in
computing any correlation functions in the $\AdS/\text{CFT}$
setup.

However, the bulk-to-boundary propagator cannot be directly used to 
get information about Penrose limiting process from $\AdS_5\times\text{S}^5$
to the $10$-dimensional plane wave.
As we have learned in Section \ref{AdSpwgeorel}, 
only a limited region around the particular null geodesics, at which the
Penrose limit is taken, converges to the plane wave spacetime. 
The $\AdS_5\times\text{S}^5$ boundary is not part of the plane wave spacetime. 
Hence, a reasonable limit (where both points are part of
the plane wave spacetime)
of the bulk-to-boundary propagator in $\AdS_5$ does not exist.

The situation is different for the bulk-to-bulk propagator in the full
$\AdS_5\times\text{S}^5$ spacetime. Here one 
can choose both points within the region of convergence such that the  
limit of the propagator is well defined. The discussion in Section
\ref{AdSCFTholo} has shown that its study can be useful 
to get information about a holographic setup. There we have seen 
that the bulk-to-boundary propagator 
in $\AdS_{d+1}$ (see \eqref{bulkboundAdSprop}) 
is related to the bulk-to-bulk propagator (see \eqref{AdSprop}) 
if the boundary has codimension one compared to the bulk dimension.  
It was sufficient to work with the $\AdS_{d+1}$ propagators and deal with the 
additional dimensions from the sphere in a Kaluza-Klein
decomposition. In this case the relation \eqref{bboundpropinbbprop} 
holds between the bulk-to-bulk and the bulk-to-boundary propagator. 

The situation is somewhat different in case of the BMN correspondence, where 
the direct product structure of the underlying spacetime breaks down and
one has to deal with all dimensions on equal footing.   
The hope is that, depending on how holography is effectively realized in
the plane wave, one can nevertheless 
find some relation that allows one to compute the
appropriate bulk-to boundary from the bulk-to-bulk propagator in the 
plane wave similarly to \eqref{bboundpropinbbprop} in the case of $\AdS_{d+1}$.
Furthermore, the scalar bulk-to-bulk propagator in the BMN plane wave has been
constructed in \cite{Mathur:2002ry} by a direct approach 
leaving the issue of its
derivation via a limiting process 
as an open problem. 

Motivated by this perspective, and because it is an interesting problem in its 
own right, we will present in this Chapter the construction of the scalar
propagator on $\AdSS$ spaces with the respective embedding radii $R_1$ and
$R_2$, and then discuss for $\AdS_5\times\text{S}^5$ and $R_1=R_2$ 
its behaviour in the Penrose limiting process to
the $10$-dimensional plane wave. 
Allowing for generic dimensions $d$ and $d'$ as well as generic curvature
radii  $R_1$ and $R_2$ is very helpful to understand the general mechanism for
the construction of the propagator. Of course only some of these spacetimes
are parts of consistent supergravity backgrounds, see Sections
\ref{SUGRApbranesol} and \ref{nhlimit}. This Chapter is mainly based on our
work \cite{Dorn:2003au}.
    
In Section \ref{waveeqapp} we will focus on the differential equation defining
the scalar propagator in generic $\AdSS$ backgrounds. 
Within this Section we will be able to find the propagator
in conformally flat situations, i.\ e.\ for equal embedding radii of 
$\AdS_{d+1}$ and $\text{S}^{d'+1}$  and for masses corresponding
to Weyl invariant actions. For a comparison we will also discuss the case 
of pure $\AdS_{d+1}$. In the Weyl invariant case, by
using the conformal transformation that relates the Poincar\'e patch to flat
space, the solutions of the differential equations can 
alternatively be obtained from the well-known propagators in flat space.
We will demonstrate this in Section \ref{weylana}.
However, since only patches of $\AdS_{d+1}$ and $\AdSS$ are conformal to flat
space, it is not appropriate for a discussion of global aspects of 
the solutions. 
But as we have seen in 
Sections \ref{AdSspacetime} and \ref{AdSSspacetime}, $\AdS_{d+1}$ and 
$\AdSS$ can be globally conformally mapped respectively to one half and to the 
complete ESU of corresponding dimension. 
We will use this to discuss some global issues of the propagators in Section 
\ref{relESU}.   
With the hope to get the propagator for generic masses,
in Section \ref{modesummation} we
study its KK mode sum. We will be able to perform the sum, whenever
a linear relation holds between the conformal dimension of the KK mode
and the quantum number parameterizing the eigenvalue of the Laplacian on
the sphere. Beyond the cases treated in the previous Sections
this applies to certain additional mass values, but fails to solve
the full generic problem. As a byproduct, the comparison with
the result of Section \ref{waveeqapp} yields a theorem on the summation of
certain products of Gegenbauer and Legendre functions.  

In Section \ref{pwlimit} we will apply the plane wave limit to  
$\AdS_5\times\text{S}^5$. We will explicitly show that the massless
propagator indeed reduces to the expression of 
\cite{Mathur:2002ry}. Furthermore, we will present the limit of the full
differential equation which is fulfilled by the propagator of massive scalar
fields given in \cite{Mathur:2002ry}. 

\section{The differential equation for the propagator and its solution}
\label{waveeqapp}
\subsection{The scalar propagator on $\AdSS$}
\label{AdSSderiv}
The scalar propagator is defined as the solution 
of
\begin{equation}\label{waveeq}
(\Box_z-M^2)G(z,z')=\frac{i}{\sqrt{-g}}\delta(z,z')\col
\end{equation}
with suitable boundary conditions at infinity. $\Box_z$ denotes the d'Alembert
operator on $\AdSS$, acting on the first argument of the propagator $G(z,z')$.
Again, we denote the coordinates referring to the $\AdS_{d+1}$ factor by
$x$ and those referring to the $\text{S}^{d'+1}$  factor by $y$,
i.\ e.\ $z=(x,y)$. In a
  continuation to Euclidean space the R.\ H.\ S.\ of \eqref{waveeq} is changed
  to $-\frac{1}{\sqrt{g}}\delta (z,z')$ such that it is consistent with
  \eqref{AdSbulktobulkpropdiffeq}. The change is due to the procedure to fix
  the normalization by integrating the L.\ H.\ S.\ over space(time). 
In the Lorentzian case the additional factor $-i$ is generated by the required
Wick-rotation.

We first look for solutions at $z\neq z'$ and discuss the behaviour at $z=z'$
afterwards. 
The constructions of $\AdS_{d+1}$ and $\text{S}^{d'+1}$ via the embeddings in
$\mathds{R}^{2,d}$ and in $\mathds{R}^{d'+2}$ respectively were presented
in Sections \ref{AdSspacetime} and \ref{AdSSspacetime}.\footnote{Remember that
  with $\AdS_{d+1}$ we always mean the universal covering, if not otherwise
indicated.} The embeddings endow us with the chordal
distances that can be used to measure distances between two points on the 
manifolds. For $\AdS_{d+1}$ and $\text{S}^{d'+1}$ they are defined in 
\ref{AdSchordaldist} and \ref{Schordaldist} and denoted as $u(x,x')$ and
$v(y,y')$, respectively. 
The chordal distance $u$ is a unique function of $x$ and $x'$ if one restricts
oneself to the hyperboloid. On the
universal covering it is continued as a periodic function.
For later use we note that on the hyperboloid and on the sphere the 
antipodal points $\tilde x$ and $\tilde y$ to given points $x$ and $y$ are
defined by changing the sign of the embedding coordinates $X$ and $Y$
respectively. 
From \eqref{AdSchordaldist} and \eqref{Schordaldist} one then finds with
$\tilde u=u(x,\tilde x')$, 
$\tilde v=v(y,\tilde y')$
\begin{equation}\label{uvaprel}
u+\tilde u=-4R_1^2\col\qquad v+\tilde v=4R_2^2\pnt
\end{equation}
Using the homogeneity and isotropy of both $\AdS_{d+1}$ and $\text{S}^{d'+1}$
it is clear that the propagator can depend on $z,z'$ only via the chordal
distances $u(x,x')$ and $v(y,y')$. Strictly speaking this at first applies
only if $\AdS_{d+1}$ is restricted to the hyperboloid. Up to subtleties due 
to time ordering (see the end of Section \ref{relESU}) 
this remains true also on the universal covering.
The d'Alembert operator then simplifies to
\begin{equation}\label{dalemchordal}
\begin{aligned}
\Box_z&=\Box_x+\Box_y\col\\
\Box_x&=2(d+1)\Big(1+\frac{u}{2R_1^2}\Big)\preparderiv{u}+\Big(\frac{u^2}{R_1^2}+4u\Big)\doublepreparderiv{u}\col\\
\Box_y&=2(d'+1)\Big(1-\frac{v}{2R_2^2}\Big)\preparderiv{v}-\Big(\frac{v^2}{R_2^2}-4v\Big)\doublepreparderiv{v}\pnt
\end{aligned}
\end{equation}

One can now ask for a solution of \eqref{waveeq} at $z\neq z'$ that only
depends on the total chordal distance $u+v$. Indeed, using
\eqref{dalemchordal}, it is easy to derive that such a solution exists if and
only if
\begin{equation}\label{conditions}
R_1=R_2=R\col\qquad M^2=\frac{d'^2-d^2}{4R^2}\pnt
\end{equation}
Furthermore, it is necessarily powerlike and given by
\begin{equation}
G(z,z')\propto (u+v)^{-\frac{d+d'}{2}}\pnt
\end{equation}
Extending this to $z=z'$ we find just the right power for the short
distance singularity to generate the $\delta$-function on the R.\ H.\ S.\ of 
\eqref{waveeq}. Hence, after fixing the normalization we end up with
\begin{equation}\label{AdSSprop}
G(z,z')=\frac{\Gamma(\frac{d+d'}{2})}{4\pi^{\frac{d+d'}{2}+1}}\frac{1}{(u+v+i\varepsilon(t,t'))^\frac{d+d'}{2}}\pnt
\end{equation}
Note that due to \eqref{uvaprel} besides the singularity at $z=z'$ there is 
another one at the total antipodal point where 
$z=\tilde z'=(\tilde x', \tilde y')$. 
We have introduced an $i\varepsilon$-prescription by replacing $u\to
u+i\varepsilon$, where $\varepsilon$ depends explicitly on time. 
We will comment on this in Section \ref{relESU}. In particular, we will see
that on the universal covering of the hyperboloid the singularity at the 
total antipodal point does not lead to 
an additional $\delta$-source on the R.\ H.\ S.\ of \eqref{waveeq}.

Scalar fields with mass $m^2$ in $\AdS_{d+1}$ via the $\AdS/\text{CFT}$ correspondence
are related to CFT fields with conformal dimension
\begin{equation}\label{conformaldim}
\Delta_\pm(d,m^2)=\frac{1}{2}\Big(d\pm\sqrt{d^2+4m^2R_1^2}\Big)\pnt
\end{equation}
Note that the exponent of $(u+v)$ in the denominator of the propagator
\eqref{AdSSprop} 
is just equal to $\Delta_+(d,M^2)$. From the $\AdS_{d+1}$ point of view
the $(d+d'+2)$-dimensional mass $M^2$ is the mass of the KK zero mode
of the sphere. We will come back to these issues in Section 
\ref{modesummation}.

For completeness let us add another observation. Disregarding for a moment the 
source structure, 
under the conditions \eqref{conditions} there is a solution of \eqref{waveeq},
that depends only on $(u-v)$. 
The explicit form is
\begin{equation}\label{AdSSmirprop}
\tilde G(z,z')\propto\frac{1}{(u-v+4R^2+i\varepsilon(t,t'))^\frac{d+d'}{2}}\pnt
\end{equation}
It has the same asymptotic falloff as \eqref{AdSSprop}. But due to
\eqref{uvaprel} it has singularities only at the semi-antipodal points
where $z=z'_\text{s}=(x',\tilde y')$ and 
$z=\tilde z'_\text{s}=(\tilde x',y')$. 
We will say more on $\tilde G(z,z')$ in Sections \ref{weylana} and
\ref{relESU}.

At the end of this Subsection we give a simple interpretation of the
conditions \eqref{conditions}. We have already seen from the discussion
referring to 
the $\AdSS$ metric \eqref{AdSSmetric} that $R_1=R_2$ is exactly the condition
for conformal flatness of the complete product space $\AdSS$ as a whole and
that it is then conformal to the complete Einstein static universe.
Furthermore, the mass condition just singles out the case of a scalar
field coupled in Weyl invariant manner to the gravitational background.
The corresponding $D$-dimensional action is 
\begin{equation}
S=-\frac{1}{2}\int\de^Dz\,\sqrt{-g}\Big[g^{\mu\nu}\partial_\mu\phi\partial_\nu\phi+\frac{D-2}{4(D-1)}\mathcal{R}\phi^2\Big]\pnt
\end{equation}
Inserting the constant curvature scalar $\mathcal{R}$ for $\AdSS$
\eqref{AdSScurvature} 
with equal radii one gets for the mass just the value in 
\eqref{conditions}.

Altogether in this Subsection we have constructed the scalar $\AdSS$
propagator for the case of Weyl invariant coupling to the metric 
in conformally flat situations. The Weyl invariant coupled field is the 
natural generalization of the massless field in flat space. 

\subsection{A remark on the propagator on pure $\AdS_{d+1}$}
\label{AdSderiv}
Having found for $\AdSS$ such a simple expression for
the scalar propagator, one is wondering whether the well known AdS propagators
can also be related to simple powers of the chordal distance.

In \eqref{AdSprop} we have given the general massive scalar propagator on pure
$\AdS_{d+1}$ space corresponding to the two distinct conformal dimensions
$\Delta_{\pm}$ defined in \eqref{conformaldim}.

Again, here a powerlike solution of 
the differential equation \eqref{AdSbulktobulkpropdiffeq} 
(continued to Minkowski signature) on pure
$\AdS_{d+1}$ with the d'Alembert operator given in \eqref{dalemchordal} 
exists for the Weyl invariant coupled mass 
value 
\begin{equation}\label{AdSWeylmass}
m^2=\frac{1-d^2}{4R_1^2}\pnt
\end{equation}
There is no condition on $R_1$ because in Subsection \ref{AdSspacetime} we have
seen that pure AdS spaces are always conformally flat.
The related value for the conformal dimension from \eqref{conformaldim}
is then $\Delta_{\pm}=\frac{d\pm 1}{2}$. The powerlike solution is given by
\begin{equation}\label{AdSprop2}
G(x,x')=\frac{\Gamma(\frac{d-1}{2})}{4\pi^{\frac{d+1}{2}}}\frac{1}{(u+i\varepsilon(t,t'))^\frac{d-1}{2}}\pnt
\end{equation}%
In contrast to the $\AdSS$ case here the exponent of $u$ is given by
$\Delta_-(d,m^2)$. We have again kept the option of a time dependent
$i\varepsilon(t,t')$ and will comment on it in Section \ref{relESU}.

The above solution can indeed be obtained from \eqref{AdSprop} by taking the
sum of the expressions for $\Delta_+$ and $\Delta_-$. 
In addition one finds another simple structure by taking the difference. 
They are given by 
\begin{equation}\label{AdSsuperpos}
\begin{aligned}
\frac{1}{2}(G_{\Delta_-}+G_{\Delta_+})&=\frac{\Gamma(\frac{d-1}{2})}{4\pi^{\frac{d+1}{2}}}\frac{1}{(u+i\varepsilon(t,t'))^\frac{d-1}{2}}\col\\
\frac{1}{2}(G_{\Delta_-}-G_{\Delta_+})&=\frac{\Gamma(\frac{d-1}{2})}{4\pi^{\frac{d+1}{2}}}\frac{1}{(u+4R^2+i\varepsilon(t,t'))^\frac{d-1}{2}}\pnt
\end{aligned}
\end{equation}
Both expressions are derived by using \eqref{AdSprop} and 
\eqref{hypergeom1} of Appendix 
\ref{app:userelHyper}. 
The first combination has the right short distance singularity to be a solution
of \eqref{waveeq}. The second combination resembles \eqref{AdSSmirprop}.
We will say more on these linear combinations in Sections \ref{weylana} and
\ref{relESU}. 

\subsection{Comment on masses and conformal dimensions on $\AdS_{d+1}$}
\label{BFbounds}
On AdS spaces one has to respect the Breitenlohner-Freedman 
bounds \cite{Breitenlohner:1982bm,Breitenlohner:1982jf}.
The first bound is derived from the condition that there is no energy flux
through the boundary of $\AdS_{d+1}$. This condition with the requirement to
have real energies effectively is a restriction on $\Delta_\pm$ to be real. 
One finds that the masses have to obey
\begin{equation}\label{BFbound}
m^2\geq -\frac{d^2}{4R_1^2}\pnt
\end{equation}
Furthermore, the so called unitarity bound requires
\begin{equation}\label{splitbound}
\Delta >\frac{d-2}{2}\pnt
\end{equation}
This implies that for $-\frac{d^2}{4R_1^2}\leq m^2 < \frac{4-d^2}{4R_1^2}$
both, $\Delta _+$ and $\Delta _-$ are allowed, since they are both
normalizable. On the other side for $\frac{4-d^2}{4R_1^2}\leq m^2$ only 
$\Delta _+$ is allowed. 

The masses for Weyl invariant coupling are $\frac{1-d^2}{4R_1^2}$ and
$\frac{d'^2-d^2}{4R_1^2}$ for $\AdS_{d+1}$ and $\AdSS$,
respectively. Hence, in our Weyl invariant cases for pure AdS 
$\Delta _+$ and $\Delta _-$ are allowed while for $\AdSS$ with $d'>1$
only $\Delta _+$ is allowed. 

\section{Derivation of the propagator from the flat space one}
\label{weylana}
In the previous Section we have shown that a simple powerlike solution of
\eqref{waveeq} can be found if the underlying spacetime is $\AdS_{d+1}$ or a
conformally flat product space $\AdSS$  and if the corresponding scalar field 
is Weyl invariant coupled to the curvature of the background. Both properties
allow for a mapping of the differential equation, the scalar field and the
propagator to flat space. The other way around, one can use Weyl invariance in
this special case to construct the propagator of Weyl invariant coupled fields
on conformally flat backgrounds from the flat space massless propagator.  

We will use this standard construction to rederive the $\AdSS$ expressions
 \eqref{AdSSprop} and \eqref{AdSSmirprop} from the flat space solutions.

The relevant Weyl transformation in a $D$-dimensional manifold is
\begin{equation}\label{weyltraf1}
g_{\mu\nu}\to \varrho\,g_{\mu\nu}\col\qquad
\phi\to\phi'=\varrho^\frac{2-D}{4}\phi\pnt 
\end{equation}
If then the metric is of the form $g_{\mu\nu}(z)=\varrho(z)\,\eta_{\mu\nu}$ one
finds the following relation between the propagator in curved and flat space
\begin{equation}
G(z,z')=\big(\varrho(z)\,\varrho(z')\big)^{\frac{2-D}{4}}\,G_\text{flat}(z,z')\col\qquad
G_\text{flat}(z,z')=\frac{\Gamma(\frac{D-2}{2})}{4\pi^{\frac{D}{2}}}\frac{1}{((z-z')^2+i\epsilon)^{\frac{D-2}{2}}}
\pnt
\end{equation}
It can be derived either by formal manipulations with the corresponding
functional integral or by using the covariance properties of the defining
differential equation.\footnote{Of course, the discussion has to be completed
  by considering also the boundary conditions.}

Applying the formula first to pure AdS one gets with the metric
\eqref{AdSmetricPcoord}, that the propagator in Poincar\'e coordinates reads 
\begin{equation}
G(x,x')=\frac{\Gamma(\frac{d-1}{2})}{R_1^{d-1}4\pi^{\frac{d+1}{2}}}
\Big(\frac{1}{x_\perp x'_\perp}\big[(x_\perp-x'_\perp)^2-(x^0-x'^0)^2+(\vec
x-\vec x')^2+i\epsilon\big]\Big)^{\frac{1-d}{2}}\pnt
\end{equation}
With the help of \eqref{AdSchordaldist} it is easy to see that this expression
is equal to \eqref{AdSprop2}. 

The Poincar\'e patch of pure $\AdS_{d+1}$, shown in Fig.\ \ref{fig:hyperboloid}
for $d=1$, 
is conformal to a flat half space with $x_\perp\ge0$.
$x_\perp=0$ corresponds to the conformal boundary of AdS.
Let us first disregard that the flat half space represents only one half
of $\AdS_{d+1}$ and discuss global issues later. 
We can then implement either Dirichlet or Neumann boundary
conditions by the standard mirror charge method. 
To $x=(x_\perp,x^0,x^1,\dots,x^{d-1})$ we relate the mirror 
point\footnote{Using $x_\perp<0$ for parameterizing the second Poincar\'e 
patch the mirror point is at the antipodal position on the hyperboloid (see
\eqref{prePcoord} and Fig.\ \ref{fig:hyperboloid}).}
\begin{equation}\label{mirror}
\tilde x=(-x_\perp,x^0,x^1,\dots,x^{d-1})
\end{equation}
and the mirror propagator by
\begin{equation}
\tilde G_\text{flat}(x,x')=G_\text{flat}(x,\tilde x')\pnt
\end{equation}
Then $\frac{1}{2}(G_{\Delta_-}-G_{\Delta_+})$ in the second line of
\eqref{AdSsuperpos} turns out to be just the Weyl transformed version of
$\tilde G_\text{flat}(x,x')$. Equivalently we can state, that
$G_{\Delta_+}$ and $G_{\Delta_-}$ are the Weyl transformed versions
respectively of the
Dirichlet and Neumann propagator in the flat halfspace.

The situation is different for $\AdSS$ spacetimes. According to
\eqref{AdSSmetric}, $x_{\perp}\ge0$ becomes a
radial coordinate of a full $(d'+2)$-dimensional flat subspace of a total 
space with coordinates 
\begin{equation}\label{zcoord}
z=\big(x_0,\vec x,x_\perp\tfrac{\vec Y}{R}\big)\col
\end{equation}
where $\vec Y^2=R^2$
are the embedding coordinates of $S^{d'+1}$. The boundary of
the  AdS part is mapped to the origin of the $(d'+2)$-dimensional subspace. 
Similarly to the pure AdS case, $G(z,z')$ from
\eqref{AdSSprop} is the Weyl transform of $G_\text{flat}(z,z')$. 
To see this one has to cast the length square on the $(d'+2)$-dimensional
subspace, which appears in the denominator of the propagator, into the form
\begin{equation}
\frac{1}{R^2}(x_\perp\vec Y-x'_\perp\vec
  Y')^2=x_\perp^2+x'^2_\perp-2\frac{x_\perp x'_\perp}{R^2}\vec Y\vec Y'
=(x_\perp-x'_\perp)^2+\frac{x_\perp x'_\perp}{R^2}v\col
\end{equation} 
where we have used \eqref{Schordaldist} and remember that $u(x,x')$ is
given by \eqref{AdSchordaldist}. 
In addition, with 
\begin{equation}\label{twistapzcoord}
z_\text{s}=
\big(x_0,\vec x,-x_\perp\tfrac{\vec Y}{R}\big)\col\qquad\tilde
G_\text{flat}(z,z')=G_\text{flat}(z,z'_\text{s})
\end{equation}
we find that the second simple solution \eqref{AdSSmirprop} is the Weyl
transformed version of $\tilde G_\text{flat}(z,z')$.

The coordinates \eqref{zcoord} and
\eqref{twistapzcoord} are related by replacing $\vec Y$ by $-\vec Y$, i.\ e.\
$z_\text{s}$ is related to $z$ by going to the antipodal point 
in the sphere, according to the definition of $z_\text{s}$ after 
\eqref{AdSSmirprop}. 
The two points $z$, $z_\text{s}$ are elements of $\mathds{R}^{d+d'+2}$ 
lying in the first Poincar\'e patch where $x_\perp\ge0$.

As we mentioned before, one has to be careful with global issues. We work in
the Poincar\'e patch that only covers points with $x_\perp\ge0$.
It is easy to see that the coordinates \eqref{zcoord} of $z$ and
\eqref{twistapzcoord} of $z_\text{s}$ 
remain unchanged if one simultaneously replaces $x_\perp$
by $-x_\perp$ and $\vec Y$ by $-\vec Y$. 
This operation switches from $z$ and $z_\text{s}$ respectively to the total
antipodal positions $\tilde z$ and $\tilde z_\text{s}$, that
are covered by a second Poincar\'e patch with $x_\perp<0$. 
Thus, the latter points, being elements of the complete manifold,
are not covered by the first Poincar\'e patch.
In the context of pure $\AdS_{d+1}$, the mirror point $\tilde x$ in
\eqref{mirror} related to $x$ is outside of the first Poincar\'e patch 
but it is still a point in $\AdS_{d+1}$ covered by the second Poincar\'e patch.
Hence, $\tilde x$ is not an element of the flat half space that is 
conformal to the first Poincar\'e patch.
We will now analyze the global issues more carefully by
working with the corresponding ESU.

\section{Relation to the ESU}
\label{relESU}
As discussed in Subsections \ref{AdSspacetime} and \ref{AdSSspacetime},
$\AdS_{d+1}$ and $\AdSS$ with $R_1=R_2$ 
are conformal to respectively one half and to the 
full ESU of the corresponding dimension.
This conformal relation 
has been used in \cite{Avis:1978yn} at $d=3$ to find consistent
quantization schemes on $\AdS_4$. In case of the Weyl invariant mass value
\eqref{AdSWeylmass} the quantization prescription
on the ESU leads to two different descriptions for pure AdS. One can either 
choose transparent boundary conditions or reflective boundary conditions 
at the image of the AdS boundary. The reflectivity
of the boundary is guaranteed for either Dirichlet or Neumann boundary
conditions. This is realized by choosing a subset of modes with definite 
symmetry properties, whereas in the transparent case all modes are used. 
Quantization in the reflective case leads one to the solutions 
$G_{\Delta_\pm}$.  
These results motivate why we will work on the ESU in the following. 
We will find the antipodal points and see how the mirror charge construction 
works. Then we will discuss what this implies for the well known propagators 
in $\AdS_{d+1}$ and our solutions for $\AdSS$ in the Weyl invariant cases.

A convenient global coordinate system with coordinate $\bar\rho$, where the
conformal equivalence between $\AdS_{d+1}$ or $\AdSS$ and the corresponding
ESU is obvious, was defined in \eqref{Xinglobalcoord}.
In these coordinates a point $\tilde x$ 
antipodal to the point $x=(t,\bar\rho,x_\Omega)$ in $\AdS_{d+1}$ is given by
\begin{equation}\label{AdSantipodglobalcoord}
\tilde x=(t+\pi,\bar\rho,\tilde x_\Omega)\col  
\end{equation} 
where $x_\Omega$ denotes the angles of the $(d-1)$-dimensional subsphere of 
$\AdS_{d+1}$ with embedding coordinates $\omega_i$ (see\eqref{Xinglobalcoord}),
such that one finds\footnote{See \eqref{Santipodal} for an explicit relation
  between the angles.}
\begin{equation}\label{AdSsubsphereantipodalrel}
\omega_i(\tilde x_\Omega)=-\omega_i(x_\Omega)\pnt
\end{equation}
The above relation \eqref{AdSantipodglobalcoord} must not be confused with the
relation between two points  
that are antipodal to each other on the sphere of the ESU at fixed time. 

We now want to visualize the above relation on the sphere of the ESU. 
For convenience we choose $\AdS_2$ such that the ESU has topology
$\mathds{R}\times\text{S}^1$. The subsphere of $\AdS_2$ is given by
$\text{S}^0=\{-1,1\}$ such that we have $\omega=\pm1$. 
Hence, the transformation of $x_\Omega$ as prescribed in
\eqref{AdSantipodglobalcoord} becomes a flip between the two points of the
$\text{S}^0$. The information contained in $\text{S}^0$ can be traded for an 
additional sign information of $\bar\rho$, and therefore 
the transformation from  $x_\Omega$ to $\tilde x_\Omega$ 
simply corresponds to an reflection at $\bar\rho=0$. 
We will now describe the time shift. 
After the transformation of the spatial coordinates
is performed, one has found the antipodal event at time $t+\pi$. 
To relate it to an event at the original time $t$ one simply travels back
in time along any null geodesics that crosses the spatial position of the
antipodal event. On the ESU these null geodesics are clearly great circles.
They meet at two points on the sphere. One is at the spatial position of the 
event and the other point is the antipodal point on the sphere of the ESU.
The time it takes for a massless particle to travel between these two points 
is given by $\pi$, see Fig.\ \ref{fig:antipodalpntinESU}. 
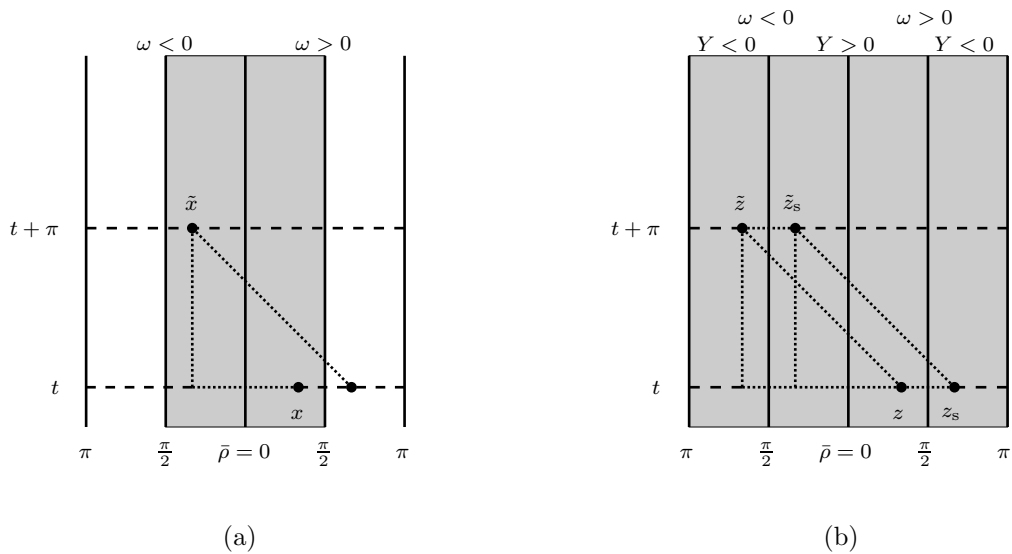
\begin{figure}
\begin{center}
\subfigure[]{\label{fig:AdSESU}%
\begin{picture}(200,180)(0,0)\scriptsize
\SetWidth{1}
\SetOffset(0,0)
\SetScale{1}
\SetWidth{0}
\GBox(70,20)(130,160){0.8}
\SetWidth{1}
\Text(40,10)[]{$\pi$}
\Text(70,10)[]{$\frac{\pi}{2}$}
\Text(100,10)[]{$\bar\rho=0$}
\Text(130,10)[]{$\frac{\pi}{2}$}
\Text(160,10)[]{$\pi$}
\Text(70,165)[]{$\omega<0$}
\Text(130,165)[]{$\omega>0$}
\Line(40,20)(40,160)
\Line(70,20)(70,160)
\Line(100,20)(100,160)
\Line(130,20)(130,160)
\Line(160,20)(160,160)
\DashLine(40,35)(80,35){4}\Text(30,35)[cr]{$t$}
\DashLine(120,35)(160,35){4}
\DashLine(40,95)(160,95){4}\Text(30,95)[cr]{$t+\pi$}
\DashLine(120,35)(80,35){1}
\DashLine(80,35)(80,95){1}
\DashLine(80,95)(140,35){1}
\Vertex(120,35){2}\Text(120,25)[]{$x$}
\Vertex(80,95){2}\Text(80,105)[]{$\tilde x$}
\Vertex(140,35){2}
\end{picture}}
\qquad
\subfigure[]{\label{fig:AdSSESU}%
\begin{picture}(200,180)(0,0)\scriptsize
\SetWidth{1}
\SetOffset(-200,0)
\SetScale{1}
\SetWidth{0}
\GBox(240,20)(360,160){0.8}
\SetWidth{1}
\Text(240,10)[]{$\pi$}
\Text(270,10)[]{$\frac{\pi}{2}$}
\Text(300,10)[]{$\bar\rho=0$}
\Text(330,10)[]{$\frac{\pi}{2}$}
\Text(360,10)[]{$\pi$}
\Text(270,175)[]{$\omega<0$}
\Text(255,165)[]{$Y<0$}
\Text(330,175)[]{$\omega>0$}
\Text(300,165)[]{$Y>0$}
\Text(345,165)[]{$Y<0$}
\Line(240,20)(240,160)
\Line(270,20)(270,160)
\Line(300,20)(300,160)
\Line(330,20)(330,160)
\Line(360,20)(360,160)
\DashLine(240,35)(260,35){4}\Text(230,35)[cr]{$t$}
\DashLine(340,35)(360,35){4}
\DashLine(240,95)(260,95){4}\Text(230,95)[cr]{$t+\pi$}
\DashLine(280,95)(360,95){4}
\DashLine(260,95)(280,95){1}
\DashLine(320,35)(260,35){1}
\DashLine(260,35)(260,95){1}
\DashLine(260,95)(320,35){1}
\DashLine(320,35)(340,35){1}
\DashLine(280,35)(280,95){1}
\Vertex(320,35){2}\Text(320,25)[]{$z$}
\Vertex(260,95){2}\Text(260,105)[]{$\tilde z$}
\DashLine(280,95)(340,35){1}
\Vertex(280,95){2}\Text(280,105)[]{$\tilde z_\text{s}$}
\Vertex(340,35){2}\Text(340,25)[]{$z_\text{s}$}
\end{picture}}
\caption{$\AdS_2$ (Fig.\ \ref{fig:AdSESU}) and
  $\AdS_2\times\text{S}^0$ (Fig.\ \ref{fig:AdSSESU}) conformally mapped to the
  corresponding ESU. The regions that are covered are displayed as gray-filled
  regions. The ESU is given by a cylinder such
that one has to identify the two boundaries of the strip where $\rho=\pi$.
The two points of the $S^0$ within $\AdS_2$ and of the extra factor $S^0$ in
  the product space are $\omega=\pm1$ and $\frac{Y}{R}=\pm1$, respectively.
$\tilde x$ and $\tilde z$, $z_\text{s}$, $\tilde z_\text{s}$ are 
the antipodal points to $x$ and $z$ in respectively $\AdS_2$ and
$\AdS_2\times\text{S}^0$. They are 
constructed by following the lines with small dashsize. The horizontal 
direction
corresponds to the transformation in the space coordinates and the vertical 
one is associated to the time shift. The diagonal lines then point to the 
source at the corresponding conjugate point where null geodesics intersect. 
The conjugate points can be regarded as effective time shifted sources with
  the same time coordinate as the original event $x$ or $z$.}
\label{fig:antipodalpntinESU}
\end{center}
\end{figure}
In this way one now arrives at an event that can have caused the event at later
time $t+\pi$, and that has the same time coordinate as $x$, and its 
coordinate value $\bar\rho$ is given by a reflection at 
$\bar\rho=\frac{\pi}{2}$ on $\text{S}^1$. 
As $\bar\rho=\frac{\pi}{2}$ is the position of the AdS boundary, the mirror 
image to $x$ is situated outside of the region that corresponds to AdS. 
The effect of the original source at $x$ in combination with 
the mirror source either at $\tilde x$ as given in
\eqref{AdSantipodglobalcoord} or 
at equal times mirrored at the boundary is that a light ray that travels to the
boundary of AdS is reflected back into the interior.

Let us now discuss what happens in the case of $\AdSS$. The point
$z=(t,\bar\rho,x_\Omega,y)$ possesses 
the total antipodal point $\tilde z$ and the two semi-antipodal points 
$z_\text{s}$ and $\tilde z_\text{s}$ given by
\begin{equation}\label{AdSSantipodglobalcoord}
\tilde z=(t+\pi,\bar\rho,\tilde x_\Omega,\tilde y)\col\qquad
z_\text{s}=(t,\bar\rho,x_\Omega,\tilde y)\col\qquad
\tilde z_\text{s}=(t+\pi,\bar\rho,\tilde x_\Omega,y)\col
\end{equation}
where $x_\Omega$ is as in the pure $\AdS_{d+1}$ case and fulfills 
\eqref{AdSsubsphereantipodalrel} and $y$ are all angle coordinates of 
$S^{d'+1}$.

In Fig.\ \ref{fig:antipodalpntinESU} the case of $\AdS_2\times\text{S}^0$, 
is shown. The effect of the factor $\text{S}^0$ can be alternatively described
by adding to the range $0\le\bar\rho\le\frac{\pi}{2}$ the interval  
$\frac{\pi}{2}\le\bar\rho\le\pi$. This is possible because in the ESU at
$\bar\rho=\frac{\pi}{2}$ the $\text{S}^0$ shrinks to a point.
The complete ESU is now covered by the image of 
$\AdS_2\times\text{S}^0$. 
The map to an antipodal position within the $\AdS_2$ factor is as before, 
one finds the spatial coordinates by reflecting at $\bar\rho=0$. 
Within the $\text{S}^0$ factor, the antipodal position is found by reflecting
at $\bar\rho=\frac{\pi}{2}$.
Using this, it can be seen that w.\ r.\ t.\ the point $z$,
the point $\tilde z$ is at the antipodal position on the $\text{S}^1$ of 
the ESU.
Traveling back in time from $t+\pi$ to $t$ along a null geodesic,
one arrives at $z$ from where one started. 
In the same way, the two semi-antipodal points 
$z_\text{s}$, $\tilde z_\text{s}$ are connected with each other by light
rays. On the sphere of the ESU the $z$ and $z_\text{s}$ are
related by a reflection at $\bar\rho=\frac{\pi}{2}$. Here, in contrast to 
the case of $\AdS_2$, even the mirror events at equal times are 
situated within the image of $\AdS_2\times\text{S}^0$. The above results 
are straightforwardly generalized to arbitrary dimensions.

Coming back to the discussion in Section \ref{weylana}, we can now make more 
precise statements about the mirror charge method to impose definite
boundary conditions at $\bar\rho=\frac{\pi}{2}$. A linear combination of 
the two solutions like in \eqref{AdSsuperpos} does not necessarily 
generate additional   
$\delta$-sources on the R.\ H.\ S.\ of the differential equation
\eqref{AdSbulktobulkpropdiffeq}, although both powerlike solutions in
\eqref{AdSsuperpos} have singularities within $\AdS_{d+1}$, the expression in
the first line has one at $x=x'$ and the expression in the second line has one
at $x=\tilde x'$. The singularity of the second expression only appears at
$t=t'+\pi$, and its contribution to the R.\ H.\ S.\ of the
differential equation \eqref{AdSbulktobulkpropdiffeq}
depends on the time ordering prescription.
In the cases where the $\theta$-function used for time ordering has
an additional step at $t=t'+\pi$, a second  $\delta$-function is generated
(see \cite{Avis:1978yn} for a discussion of $\AdS_4$). 
With the standard time ordering one finds that
$G_{\Delta_\pm}$ are solutions with a source at $x=x'$ only.
For $\AdS_4$ this was obtained in \cite{Dullemond:1985bc}.

The situation is different for $\AdSS$, where the propagator
\eqref{AdSSprop} has singularities at $z=z'$, $z=\tilde z'$ and 
the second solution \eqref{AdSSmirprop} has singularities at
$z=z'_\text{s}$, $z=\tilde z'_\text{s}$. 
Again, whether 
the singularities at $z=\tilde z'$ and $z=\tilde z'_\text{s}$ appear 
as $\delta$-sources 
on the R.\ H.\ S.\ of the differential equation \eqref{waveeq}, depends 
on the chosen time ordering. However in contrast to the pure $\AdS_{d+1}$
case, the singularity of the second solution \eqref{AdSSmirprop} at
$z=z'_\text{s}$ always leads to a $\delta$-source on the 
R.\ H.\ S.\ of \eqref{waveeq} but at the wrong position.
This result corresponds to the above observation on the ESU that the mirror
sources at equal times are not part of the image of $\AdS_{d+1}$ but of 
$\AdSS$. 

At the end let us give some comments on the
$i\varepsilon(t,t')$-prescription. First of all, one has to introduce it in all
expressions \eqref{AdSSprop}, \eqref{AdSSmirprop} and \eqref{AdSsuperpos},
since all of them have singularities at coincident or antipodal positions.
Secondly, as worked out for $\AdS_4$,
a time independent $\varepsilon(t,t')=\epsilon$ refers to taking the step
function $\theta(\sin(t-t'))$ for time ordering \cite{Avis:1978yn} which is 
appropriate if one restricts oneself to the hyperboloid. 
Standard time ordering with $\theta(t-t')$, being appropriate on the universal
covering, yields a time dependent
$\varepsilon(t,t')=\epsilon\sgn((t-t')\sin(t-t'))$ \cite{Dullemond:1985bc}.
As mentioned in Section \ref{waveeqapp}, due to the time dependence of
$\varepsilon(t,t')$, the coordinate dependence of the solutions is not entirely
included in $u$ and $v$.

\section{Mode summation on $\AdSS$}
\label{modesummation}
In this Section we will use the propagator on
pure $\text{AdS}_{d+1}$ given by \eqref{AdSprop}
and the spherical harmonics on $\text{S}^{d'+1}$ to construct the
propagator on $\AdSS$ via its mode expansion, summing up all the KK modes. 
We will be able to perform the sum only
for special mass values where the conformal dimensions $\Delta_\pm$ of the
scalar modes are linear functions of $l$, with $l$ denoting the $l$th  mode in
the KK tower. Even a mixing of several scalar modes of this kind is allowed. 
The mixing case is interesting because it occurs in
supergravity theories on $\AdS_{d+1}\times\text{S}^{d+1}$ backgrounds
\cite{Corley:1999uz,Michelson:1999kn,Deger:1998nm,Kim:1985ez,Gunaydin:1985fk}.
For example in type $\twob$ supergravity in $\text{AdS}_5\times\text{S}^5$ the
mass eigenstates of the mixing matrix for scalar modes
\cite{Kim:1985ez,Gunaydin:1985fk} correspond to the bosonic
chiral primary and descendant operators in the $\AdS/\text{CFT}$ dictionary
\cite{Lee:1998bx}.
For these modes  $\Delta_\pm$ depend linearly on $l$.

The main motivation for investigating the mode summation was the hope to find
the propagator for generic mass values. But forced to stay in a regime of a
linear $\Delta_\pm$ versus $l$ relation we can give up the condition of
conformal flatness, but remain restricted to special mass values. We
nevertheless present this study since several interesting aspects are found
along the way. Furthermore, in the literature it is believed that an explicit
computation of the KK mode summation is too cumbersome \cite{Mathur:2002ry}. 
We will show how to deal with the mode summation by discussing the
$\AdS_3\times\text{S}^3$ case first, allowing for unequal radii but
necessarily a special mass value. The result will then be compared to the
expressions in the previous Sections by specializing to equal embedding
radii. 

Having discussed this special case we will comment on the modifications which
are necessary to deal with generic $\AdSS$ spacetimes.  

The results of the previous Sections in connection with the expression for the
mode summation in the conformally flat and Weyl invariant coupled case lead to
the formulation of a summation rule for a product of Legendre functions and
Gegenbauer polynomials. An independent proof of this rule is given in Appendix
\ref{app:HyperGsummationproof}. With this it is possible to discuss the
results in generic dimensions without doing all the computations
explicitly. Furthermore, the sum rule might be useful for other applications,
too.  

For the solution of \eqref{waveeq} we make the following 
ansatz\footnote{This ansatz is designed to generate a solution that 
corresponds to \eqref{AdSSprop}. If one wants to generate a solution
corresponding to \eqref{AdSSmirprop} one has to replace either $y$ or $y'$ by
the corresponding antipodal coordinates $\tilde y$ or $\tilde y'$.} 
\begin{equation}\label{AdSSpropmodeexp}
G(z,z')=\frac{1}{R_2^{d'+1}}\sum_{I}G_I(x,x')Y^I(y)Y^{\ast I}(y')\col
\end{equation} 
where we sum over the multiindex $I=(l,m_1,\dots,m_{d'})$ such that $l\ge
m_1\ge\dots\ge m_{d'-1} \ge | m_{d'}|\ge 0$, $Y^I$ denote the spherical
harmonics on $\text{S}^{d'+1}$, and `$\ast$' means complex conjugation. 
Some useful relations for the spherical
harmonics can be found in Appendix \ref{app:SandSH}.

The mode dependent Green function on $\AdS_{d+1}$ then fulfills
\begin{equation}
\Big(\Box_x
-M^2-\frac{l(l+d')}{R_2^2}\Big)G_I(x,x')=\frac{i}{\sqrt{-g_\text{AdS}}}\delta
(x,x')\col 
\end{equation} 
which follows when decomposing the d'Alembert operator like in
\eqref{dalemchordal} and using \eqref{SHcasimir}. 
The solution of this equation\footnote{As explained around \eqref{waveeq}, the
  R.\ H.\ S. of the equation deviates from the one in
  \eqref{AdSbulktobulkpropdiffeq} due to the continuation procedure 
between the Lorentzian and the Euclidean case.} was already given in
\eqref{AdSprop}, into which the (now KK mode dependent) conformal dimensions
enter. They were already defined in \eqref{conformaldim}, and the AdS mass is a
function of the mode label $l$ 
\begin{equation}\label{mkksphere}
m^2=M^2+m_\text{KK}^2=M^2+\frac{l(l+d')}{R_2^2}\pnt
\end{equation}

In the following as a simple example we will present the derivation of the propagator on $\AdS_3\times\text{S}^3$ via the KK mode summation. Compared to the physically more interesting $\AdS_5\times\text{S}^5$ background the expressions are easier and the general formalism becomes clear.

Evaluating \eqref{AdSprop} for $d=d'=2$ the $\AdS_3$ propagator for the $l$th
KK mode is given by 
\begin{equation}\label{AdS3propexpl}
G_\Delta(x,x')=\frac{1}{R_12^{\Delta+1}\pi}\xi^\Delta\hypergeometric{\tfrac{\Delta}{2}}{\tfrac{\Delta}{2}+\tfrac{1}{2}}{\Delta}{\xi^2}=\frac{1}{R_14\pi}\frac{1+\sqrt{1-\xi^2}}{\sqrt{1-\xi^2}}\Big[\frac{\xi}{1+\sqrt{1-\xi^2}}\Big]^\Delta\pnt
\end{equation}
From \eqref{hypergeomstart}, \eqref{conformaldim} and \eqref{mkksphere} one
finds that the mode dependent positive branch of the conformal dimension reads 
\begin{equation}\label{AdS3S3conformaldim}
\Delta=\Delta_+=1+\frac{R_1}{R_2}\sqrt{\frac{R_2^2}{R_1^2}+l(l+2)+M^2R_2^2}\pnt
\end{equation}

The spherical part follows from \eqref{SHcomplete} of Appendix
\ref{app:SandSH} where we discuss it in more detail and is given by
\begin{equation}
\sum_{\scriptscriptstyle m_1\ge| m_2|\ge0 } ^l Y^I(y) Y^{\ast I}(y') = 
\frac{(l+1)}{2 \pi^2}
C_l^{(1)}(\cos\Theta)\col\qquad\cos\Theta=\frac{Y\cdot
  Y'}{R_2^2}=1-\frac{v}{2R_2^2}\pnt  
\end{equation}
Remember that the $C_l^{(\beta)}$ denote the Gegenbauer polynomials and $Y$,
$Y'$ in the formula for $\Theta$ are the embedding space coordinates of the
sphere, compare with \eqref{Yinglobalcoord} and \eqref{Schordaldist}. 
One thus obtains from \eqref{AdSSpropmodeexp}
\begin{equation}
G(z,z')=\frac{1}{8\pi^3R_1R_2^3}\frac{1+\sqrt{1-\xi^2}}{\sqrt{1-\xi^2}}\sum_{l=0}^\infty(l+1)\Big[\frac{\xi}{1+\sqrt{1-\xi^2}}\Big]^\Delta
C_l^{(1)}(\cos\Theta)\pnt 
\end{equation} 
In this formula $\Delta$ is a function of the mode parameter $l$ and we can
explicitly perform the sum only for special conformal dimensions which are
linear functions of $l$ 
\begin{equation}
\Delta=\Delta_+=\frac{R_1}{R_2}l+\frac{R_1+R_2}{R_2}\col
\end{equation}
following from \eqref{AdS3S3conformaldim} after choosing the special mass value
\begin{equation}\label{AdS3S3specmass}
M^2=\frac{1}{R_2^2}-\frac{1}{R_1^2}\pnt
\end{equation} 

The sum then simplifies and can explicitly be evaluated by a reformulation of
the $l$-dependent prefactor as a derivative and by using \eqref{Ggenfunc} 
\begin{equation}\label{sumcalc}
\sum_{l=0}^\infty(l+1)q^l
C_l^{(1)}(\eta)=\Big(q\preparderiv{q}+1\Big)\sum_{l=0}^\infty q^l
C_l^{(1)}(\eta)=\frac{1-q^2}{(1-2q\eta+q^2)^2}\pnt 
\end{equation}
With the replacements
\begin{equation}\label{qetadef}
q=\Big[\frac{\xi}{1+\sqrt{1-\xi^2}}\Big]^\frac{R_1}{R_2}\col\qquad
\eta=\cos\Theta 
\end{equation}
one now finds after some simplifications
\begin{equation}
G(z,z')=\frac{1}{8\pi^3R_1R_2^3}\frac{1}{\sqrt{1-\xi^2}}\xi^{1+\frac{R_1}{R_2}}\frac{(1+\sqrt{1-\xi^2})^\frac{R_1}{R_2}-(1-\sqrt{1-\xi^2})^\frac{R_1}{R_2}}{\Big[(1+\sqrt{1-\xi^2})^\frac{R_1}{R_2}-2\xi^\frac{R_1}{R_2}\cos\Theta+(1-\sqrt{1-\xi^2})^\frac{R_1}{R_2}\Big]^2}\pnt
\end{equation}
For the conformally flat case $R_1=R_2=R$, where \eqref{AdS3S3specmass}
becomes the mass generated by the Weyl invariant coupling to the background,
the above expression simplifies to 
\begin{equation}
G(z,z')=\frac{1}{4\pi^3R^4}\frac{\xi^2}{(2-2\xi\cos\Theta)^2}=\frac{1}{4\pi^3}\frac{1}{(u+v+i\varepsilon(t,t'))^2}\col
\end{equation}
where we have restored the $i\varepsilon(t,t')$-prescription. 
This result exactly matches \eqref{AdSSprop}.

The way to perform the KK mode summation on generic $\AdSS$ backgrounds is
very similar to the one presented above. 
One finds a linear relation between $l$ and $\Delta$ 
\begin{equation}
\Delta_\pm=\pm\frac{R_1}{R_2}l+\frac{dR_2\pm d'R_1}{2R_2}
\end{equation}
at the $(d+d'+2)$-dimensional mass value
\begin{equation}
M^2=\frac{d'^2R_1^2-d^2R_2^2}{4R_1^2R_2^2}\pnt
\end{equation}
This expression is a generalization of \eqref{AdS3S3specmass} and it 
reduces to \eqref{conditions} in the conformally flat case.
For generic dimension the way of computing the propagator is very similar to
the one presented for the $\AdS_3\times\text{S}^3$ background. However the
steps \eqref{AdS3propexpl} to express the hypergeometric function in the AdS
propagator and \eqref{sumcalc} to compute the sum become more tedious. For
dealing with the hypergeometric functions see the remarks in Appendix
\ref{app:userelHyper}. 
The sum generalizes in the way, that higher derivatives and more terms enter 
the expression \eqref{sumcalc}. 

Next we discuss the mode summation in the conformally flat case $R_1=R_2$ at
the Weyl invariant mass value but for generic $d$ and $d'$. 
In this case with the corresponding conformal dimensions
\begin{equation}\label{conformaldimconf}
\Delta=\Delta_+=l+\frac{d+d'}{2}\col
\end{equation}
using \eqref{AdSprop} and \eqref{SHcomplete}, the propagator is expressed as
\begin{equation}\label{modesumprop}
\begin{aligned}
G(z,z')&=\frac{\Gamma(\frac{d'}{2})}{4\pi}\Big(\frac{\xi}{2\pi
  R^2}\Big)^{\frac{d+d'}{2}}\\
&\phantom{={}}
\times\sum_{l=0}^\infty\frac{\Gamma(l+\frac{d+d'}{2})}{\Gamma(l+\frac{d'}{2})}\Big(\frac{\xi}{2}\Big)^l\hypergeometric{\tfrac{l}{2}+\tfrac{d+d'}{4}}{\tfrac{l}{2}+\tfrac{d+d'}{4}+\tfrac{1}{2}}{l+\tfrac{d'}{2}+1}{\xi^2}C_l^{(\frac{d'}{2})}(1-\tfrac{v}{2R^2})\pnt
\end{aligned}
\end{equation} 
This equality together with the solution \eqref{AdSSprop} has lead us to
formulate a sum rule for the above given functions at generic $d$ and
$d'$. The above series should exactly reproduce \eqref{AdSSprop}. 
In Appendix \ref{app:HyperGsummationproof} we give an independent direct proof
of the sum rule. 

Considering the mode summation one finds an interpretation of the asymptotic
behaviour of \eqref{AdSSprop} observed in Subsection \ref{AdSSderiv}. 
The asymptotic regime $u\to\infty$ corresponds to $\xi\to0$. As the
contribution of the $l$th mode is proportional to $\xi^{\Delta_+}\sim\xi^l$,
the conformal dimension of the zero mode determines the asymptotic behaviour.  

Note also that the additional singularity of \eqref{AdSSprop} at the total
antipodal position $z=\tilde z'$ can be seen already in
\eqref{AdSSpropmodeexp}.
Under antipodal reflection in $\AdS_{d+1}$ the pure AdS propagator fulfills
$G_{\Delta_\pm}(x,\tilde x')=(-1)^{\Delta_\pm}G_{\Delta_\pm}(x,x')$. On the 
sphere the spherical harmonics at antipodal points are related via
$Y^I(y)=(-1)^lY^I(\tilde y)$. Hence, in case that $\Delta_\pm$ is given by
\eqref{conformaldimconf}, replacing $z'$ by the total antipodal point 
$\tilde z'$ leads to the same expression for the mode sum up to an
$l$-independent phase factor. 

One final remark to the choice of $\Delta_+$. What happens if one performs the
mode expansion with AdS propagators based on $\Delta_-$? First in any case for
high enough KK modes $\Delta_-$ violates the unitarity bound
\eqref{splitbound}. But ignoring this condition from physics one can
nevertheless study the mathematical issue of summing with $\Delta_-$. The
corresponding series is given by \eqref{sumcalc} after replacing $q$ by
$q^{-1}$. It is divergent since for real $u$ the variable $q$ in
\eqref{qetadef} obeys $|q|\le 1$ (case $R_1=R_2$). One can give meaning to the
sum by the following procedure. $q$ as a function of $\xi$ has a cut between
$\xi=\pm 1$. If $|q|\le 1$ on the upper side of the cut, then $|q|\ge 1$ on
the lower side. Hence, it is natural to define the sum with $\Delta_-$ as the
analytic continuation from the lower side. By this procedure we found both for
$\AdS_3\times\text{S}^3$ and  $\AdS_5\times\text{S}^5$ up to an overall factor
$-1$ the same result as using $\Delta_+$. The sign factor can be understood as
a consequence of the continuation procedure.  

\section{The plane wave limit}
\label{pwlimit}
The propagator in the $10$-dimensional plane wave was constructed in
\cite{Mathur:2002ry}. We will now demonstrate how this propagator in the
massless case arises as a limit of our $\AdS_5\times\text{S}^5$ propagator
\eqref{AdSSprop} by following the limiting process.
As an additional consistency check we will take the $R\to\infty$ limit of the
differential equation \eqref{waveeq} using \eqref{dalemchordal} to obtain the
equation on the plane wave background and find that it is fulfilled by the
massive propagator given in \cite{Mathur:2002ry}. 

We first take the chordal distances $u$ and $v$ introduced in
\eqref{AdSchordaldist} and \eqref{Schordaldist}
\begin{equation}
\begin{aligned}
u&=2R^2\Big[-1+\cosh\rho\cosh\rho'\cos(t-t')-\sinh\rho\sinh\rho'\,\omega^i\omega'_i\Big]\col\\
v&=2R^2\Big[+1-\cos\vartheta\cos\vartheta'\cos(\psi-\psi')-\sin\vartheta\sin\vartheta'\,\hat\omega^i\hat\omega'_i\Big]\col
\end{aligned}
\end{equation} 
then we apply the variable transformations \eqref{BMNcoordinates} and 
\eqref{transvspacepolarcoord} such that 
one gets at large $R=R_1=R_2$ up to terms vanishing for $R\to\infty$
\begin{equation}\label{chordalpw}
\begin{aligned}
u&=2R^2\Big[-1+\cos\Delta z^++\frac{1}{R^2}\Big(-(\vec x^2+\vec
x{\hspace{0.5pt}}'^2)\sin^2\tfrac{\Delta z^+}{2}+\frac{(\Delta\vec
  x)^2}{2}-\Delta z^-\sin\Delta z^+\Big)\Big]\col\\
v&=2R^2\Big[+1-\cos\Delta z^++\frac{1}{R^2}\Big(-(\vec y^2+\vec
y{\hspace{1pt}}'^2)\sin^2\tfrac{\Delta z^+}{2}+\frac{(\Delta\vec
  y)^2}{2}-\Delta z^-\sin\Delta z^+\Big)\Big]\col
\end{aligned}
\end{equation}
where $\Delta z^\pm=z^\pm-z'^\pm$, $\Delta\vec x=\vec x-\vec
x{\hspace{0.5pt}}'$, $\Delta\vec y=\vec y-\vec
y{\hspace{0.5pt}}'$ and $\vec x=r\,\vec\omega$, $\vec y=y\,\vec{\hat\omega}$.
In the $R\to\infty$ limit the sum of both chordal distances is thus given by
\begin{equation}\label{totalchordalpw}
\Phi=\lim_{R\to\infty}(u+v)=-2(\vec z^2+\vec
z{\hspace{1pt}}'^2)\sin^2\tfrac{\Delta z^+}{2}+(\vec z-\vec
z{\hspace{1pt}}')^2-4\Delta z^-\sin\Delta z^+\col
\end{equation}
where $\vec z=(\vec x,\vec y)$, 
$\vec z{\hspace{1pt}}'=(\vec x{\hspace{0.5pt}}',\vec y{\hspace{1pt}}')$ and
$\Phi$ refers 
to the notation of \cite{Mathur:2002ry}. $\Phi$ is precisely the $R\to\infty$
limit of the total chordal distance on $\AdS_5\times\text{S}^5$, which is the
chordal distance in the plane wave, compare with \eqref{pwchordaldist} using
$H_{ij}=-\delta_{ij}$. It remains
finite as both $\sim R^2$ terms in \eqref{chordalpw} cancel. This happens due
to the expansion around a \emph{null} geodesic.

The massless propagator in the plane wave background in the $R\to\infty$ limit
of \eqref{AdSSprop} with $d=d'=4$ thus becomes 
\begin{equation}
G_\text{pw}(z,z')=\frac{3}{2\pi^5}\frac{1}{(\Phi+i\varepsilon(z^+,z{\hspace{1pt}}'^+))^4}\col
\end{equation}  
which agrees with \cite{Mathur:2002ry}. 

In addition we checked the massive propagator of \cite{Mathur:2002ry} which
fulfills the differential equation on the plane wave background. This equation
can be obtained from \eqref{waveeq} and \eqref{dalemchordal} by taking the
$R\to\infty$ limit. In the limit the sum of both chordal distances is given in
\eqref{totalchordalpw}. The difference is given by 
\begin{equation}\label{diffchordalpw}
\lim_{R\to\infty}\frac{u-v}{R^2}=4(\cos\Delta z^+-1)
\end{equation}
this has to be substituted into \eqref{dalemchordal}. 
Finally, one obtains the differential equation
\begin{equation}
\Big[4\cos\Delta
z^+\Big(5\preparderiv{\Phi}+\Phi\doublepreparderiv{\Phi}\Big)+4\sin\Delta
z^+\preparderiv{\Phi}\preparderiv{\Delta
  z^+}-M^2\Big]G_\text{pw}(z,z')=\frac{i}{\sqrt{-g_\text{pw}}}\delta
(z,z')\col
\end{equation}
which is fulfilled by the expression given in \cite{Mathur:2002ry}.
As already noticed in Section \ref{waveeqapp}, in contrast to the massless
propagator the massive one depends not only on the total chordal distance
$\Phi$ but in addition on \eqref{diffchordalpw}. 


\clearemptydoublepage 
\part{Summary and conclusions}
\clearemptydoublepage
%
This thesis addressed two topics. The first topic dealt with the formulation 
of Feynman rules for noncommutative YM theories with general gauge groups.
In the second part, ingredients for the formulation of the holographic
principle in the $\AdS/\text{CFT}$ correspondence and in its BMN limit were
analyzed. 

In part \ref{NC} we have discussed some issues of 
the noncommutative version of pure YM theory.
We have first reviewed how the noncommutative $U(1)$ theory 
arises from string theory in a background with constant $B$-field and 
how this led to the formulation of the Seiberg-Witten map.
The latter played an essential role in the formulation of
noncommutative gauge theories with arbitrary gauge groups $G$.

We have focused on the task of obtaining 
information about the Feynman rules for noncommutative YM theories with gauge
groups $G$.
We have shown how the Seiberg-Witten map between the sets of ghost fields 
can be extracted from the Faddeev-Popov gauge fixing procedure. 
Before we went to the crucial problem of analyzing the noncommutative YM
theories with gauge groups $G\neq U(N)$, we
rederived the well known Feynman rules for $G=U(N)$. 

To get information about the Feynman rules in the  
case of general $G\subset U(N)$,
we then started from the path integral formulation, imposing a constraint on
the integral over the noncommutative fields. 
In terms of the Seiberg-Witten map the latter was interpreted as
the restriction that the noncommutative fields are mapped to ordinary fields 
of an ordinary YM theory with gauge group $G$. 
This constraint was then resolved by using the power series of the Seiberg
Witten map to replace the constrained integration over the noncommutative
fields by an unconstrained one over the ordinary fields. 
In this way we arrived at the enveloping algebra approach, where additional 
interaction vertices were generated from the $\theta$-expansion of the
Seiberg-Witten map. 

To get information about the Feynman rules without $\theta$-expansion, we
studied the issue of partial summation of the above described 
$\theta$-expanded perturbation
theory. In this analysis we kept the noncommutative YM vertices already found
for $U(N)$ and focused on the remaining kinetic part of the perturbation
theory. 
For $G=U(N)$, we found agreement with the expected result for an unconstrained
integration, that  
only the connected $2$-point Green functions of the noncommutative fields
should be generated. We found that this was guaranteed by a cancellation
mechanism between two types of diagrams that are different w.\ r.\ t.\ one
of their interaction vertices. In one type of diagrams the latter is generated
by the expansion of the kinetic term in the action, in the second type of
diagrams 
the vertex has its origin in an expansion of the corresponding source term.  

For arbitrary $G\subset U(N)$, $G\neq U(M)$, $M<N$ this cancellation mechanism
breaks down because the leading contributions in the expansion of the source
terms vanishes and hence the first type of diagrams is absent. 
The number of legs of non-vanishing connected Green
functions generated by the remaining uncanceled parts is not
bounded from above. Hence, there are no Feynman rules based on the
noncommutative $U(N)$ YM vertices and, besides perhaps suitably modified
propagators, at most a finite number of additional building blocks
with gauge field or ghost legs.
 
As usual in the case of no go theorems one has to be very carefully
in stressing the input made. Our negative statement is bound to the 
a priori decision to work with the noncommutative $U(N)$ YM vertices.
Of course, at this stage we cannot exclude the existence of rules
that contain the exact $\theta$-dependence and that are based on some 
clever modification of these vertices.
We also cannot exclude that the infinite set of building blocks with
gauge field and ghost legs by means of some additional auxiliary field
could be resolved into rules with only a finite number of building blocks. 

To make contact with the conjectured rules for $SO(N)$,
we have then modified our setup extracting only the pure $SO(N)$
components of the noncommutative $U(N)$ YM vertices.
If the rules hold, the remaining parts must generate nothing 
beyond a connected two point
function. Taking the $SO(3)$ case as a counterexample, we were
able to show explicitly that there is a non-vanishing connected
8-point function. Hence, the conjectured Feynman rules are inconsistent 
with the framework in which they were defined. 

Our analysis has shown that a lot of
effort is required to obtain some information about the
Feynman rules with full $\theta$-dependence for arbitrary gauge groups $G$. 
It has furthermore shown that the cases $G=U(N)$ and $G\subset U(N)$ are 
essentially different: the first leads to an unconstrained path integral
formulation where it is straightforward to extract the Feynman rules, 
while the latter requires a constrained path integral leading to a more
involved 
analysis. Many questions remain unanswered or arise new from our analysis.
For instance, the task to find a general formalism to derive 
Feynman rules in case of
arbitrary $G$ remains unsolved. Of course, this is closely connected to the
question if one can formulate a more general no go theorem that excludes any 
clever attempt to formulate Feynman rules, including the possibility to modify
the original vertices and to introduce additional auxiliary fields.

In Part \ref{BMN} we have analyzed some ingredients that are important for 
a formulation of the holographic principle. 
We have reviewed the $\AdS/\text{CFT}$ correspondence and its 
BMN limit. Thereby, we discussed the connection between the underlying 
$\AdSS$ and plane wave backgrounds in detail.  
We especially focused on the realization of holography in the 
 $\AdS/\text{CFT}$ correspondence and summarized proposals for the less 
understood BMN case.
From this discussion it turned out that the boundary structure and the 
geodesics are important geometrical ingredients in a holographic setup. 
We have furthermore
shown that the bulk-to-boundary propagator plays an essential role 
in a holographic formulation, and we have derived its relation to the
bulk-to-bulk propagator. 
 We have motivated that, to get information about holography in the BMN limit, 
one should observe how the aforementioned quantities behave in the limiting
procedure from the $\AdS/\text{CFT}$ to the BMN
correspondence. 
In particular, this meant that we had to analyze the the boundaries, geodesics
and propagators in $\AdS_5\times\text{S}^5$ and observe their behaviour 
in the $10$-dimensional plane wave which arises in a Penrose limit.

For the discussion of the boundaries we have worked in two different
coordinate systems:
the first set that we denoted as the BMN coordinates 
were introduced to obtain the plane wave spacetime directly in the limit
of infinite embedding radii from the $\AdSS$ geometry. They are
the Brinkmann coordinates in the plane wave case. 
The second set which we called the BN coordinates are
convenient for finding the conformal boundary of the plane wave. 
 
We have shown that in the BN coordinate system the coordinates of the boundary
of 
$\AdS_5\times\text{S}^5$, in the projection
to 
three coordinates appears to be 
located at the same spiraling line as the plane wave boundary. 
Of course for $\AdS_5\times\text{S}^5$ on this line the extension 
with respect to the other $7$ coordinates is not degenerated to a point.
But we have generated a perhaps useful intuitive picture: The boundary
is always at the same line, taking the limit $R\to\infty$
the extension in the remaining $7$ coordinates shrinks to a point.
This then implies also the degeneration of the $3$ remaining
dimensions of the conformal boundary of $\AdS_5\times\text{S}^5$.
In the BMN coordinates it turned out
that, due to the singularity of the coordinate transformation
at the boundary line, the approach to this line is realized within
two different asymptotic regimes that we called 
(i) and (ii). Only limit (i) corresponds to the
conformal boundary of $\AdS_5\times\text{S}^5$.

We have then given a complete classification
for geodesics, both for the original full $\AdS_5\times\text{S}^5$ and
the plane wave and we identified the boundary reaching null geodesics.
In $\AdS_5\times\text{S}^5$  we found four different types of geodesics, but
only one type of null geodesics reaches the
conformal boundary. They stay at constant position in the $\text{S}^5$ and 
have to approach the boundary in the limit (i).
In contrast to the $\AdS_5\times\text{S}^5$ case, the boundary reaching null
geodesics of the plane wave approach it within limit (ii). 
This implied that for $R\to\infty $
the convergence of $\AdS_5\times\text{S}^5$ geodesics to plane wave geodesics
is not uniform outside the region $|z^-|<R^{1-\varepsilon}$, $\varepsilon>0$.
Hence, the naive picture is supported
that in BMN coordinates the $\AdS_5\times\text{S}^5$ space up to the order
of magnitude of $R$ looks like a plane wave.
Furthermore, we have found that at each point with finite BMN coordinates, the
null geodesics of $\AdS_5\times\text{S}^5$ reaching the conformal boundary
form a cone with base $\text{S}^3$. 
For $R\to\infty$, in the range where the BMN 
coordinates are fixed or grow slower than $R$, this cone degenerates to the
single plane wave null geodesic crossing the point under consideration and
reaching the plane wave conformal boundary. Therefore, all points in this range
effectively notice a degeneration of the boundary.   

We have then studied the the bulk-to-bulk propagator of a
scalar field in $\AdSS$ backgrounds. With the help of the previously presented
results we have explained that an analysis of the 
bulk-to-boundary propagator itself is of little use if one wants to observe the
behaviour in the Penrose limit. The reason is that the point on the boundary
lies outside of the region of convergence. 
However, we explained why the behaviour of the bulk-to-bulk
propagator can be studied in the plane wave limit. At least in the case where 
the holographic screen has codimension one compared to the bulk it is 
related to the corresponding propagator with one point in the bulk and the
other on the holographic screen. 

We have analyzed the propagator in $\AdSS$ backgrounds with generic
dimensions $d$, $d'$ and generic embedding radii $R_1$ and $R_2$ for both
factors, such that we could extract general statements about its construction.
   First, we have discussed the defining wave equation for the propator with
$\delta$-source in this background. On conformally flat backgrounds for Weyl
invariant coupled fields the propagator turned out to be 
simply powerlike in the sum of both chordal distances. 
In this case a further powerlike solution to the wave equation was found to
exist. It depends only on the difference of the chordal distances and
has singularities if the points are antipodal to each other either in the
$\AdS_{d+1}$ or in the $\text{S}^{d'+1}$ part.
To make contact with the Weyl invariant coupled case in pure AdS space,  
we have in brief presented a simple
powerlike solution and a solution with a singularity at the antipodal point
that are linear combinations of the well known AdS propagators with the
corresponding $\Delta_\pm$ values.

An alternative construction from the well known propagator in flat space 
was given by using the Weyl invariance that admits the required conformal 
mapping. 
Only the Poincar\'e patch that covers one half of AdS could be
dealt with in this way. This has prevented us from studying
global issues of the solutions in flat space. 

To analyze some global properties, we have then used the fact that
$\AdS_{d+1}$ and $\AdSS$ can be conformally mapped to respectively one half
and to the full corresponding ESU. 
We focussed on the source structure
of the previously found solutions of the corresponding differential equation.
It turned out that in $\AdS_{d+1}$ 
one of the two solutions has a singularity if the two points are 
antipodal to each other. 
We found that it depends on the time ordering prescription 
whether this leads to a contribution  
to the $\delta$-sources on the R.\ H.\ S.\ of the equation. 
In $\AdSS$, one finds that one of the solutions has singularities if both 
points coincide or if one point is at the total antipodal position. The other
solution has singularities if one of the points is at the antipodal position
either in $\AdS_{d+1}$ or in $\text{S}^{d'+1}$. In this case both solutions
contribute to the $\delta$-sources on the R.\ H.\ S.\ of the differential 
equation. The chosen time ordering can only influence if the other
singularities in both solutions lead to additional $\delta$-sources.
Hence, in contrast to the $\AdSS$ case, in $\AdS_{d+1}$ one has the option to
construct two independent solutions for the propagator without changing the
canonical source structure. 
 
In addition for $\AdSS$ we have investigated the KK decomposition of the
propagator using spherical harmonics. We have noted that the summation can be
performed even in non conformally flat backgrounds, but only for special mass
values. The relevant condition is that the conformal dimension of the field
mode is a linear function of the KK mode parameter.  
In the conformally flat case for a Weyl invariant coupled field the uniqueness
of the solution of the differential equation in combination with the KK
decomposition led to the formulation of a theorem that sums up a product of
Legendre functions and Gegenbauer polynomials. We presented an independent 
proof for this theorem.  

For $\AdS_5\times\text{S}^5$ we explicitly performed the Penrose
limit on our expression for the propagator to find the result on the plane
wave background. We found agreement with the result obtained by an explicit
construction in the plane wave
and got an interpretation for the spacetime
dependence of this result. It simply depends on the $R\to\infty$ limit of the
sum of both chordal distances on the original $\AdS_5\times\text{S}^5$, that is
the chordal distance on the $10$-dimensional plane wave. 
In the general massive case there is an additional dependence on the suitable
rescaled difference of both chordal distances. 
We formulated the differential equation in the limit and checked
that the well known massive propagator on the plane wave background 
is a solution.

From the above summarized observations 
one could draw the following rough picture of
what might happen in the limit from the $\AdS/\text{CFT}$ correspondence to 
its BMN limit. The fact that only a limited region in the interior of
$\AdS_5\times\text{S}^5$ converges to the $10$-dimensional plane wave, together
with the fact that only a subset of nearly protected operators survives on 
the boundary theory seems to indicate that the limit is accompanied by a
projection. One could speculate that the bulk region between the 
part that converges to the plane wave and the boundary of
$\AdS_5\times\text{S}^5$ is responsible for this projection. In this limit the
old boundary is blown apart. This could lead to a selection process that an
observer in the geometry converging to the plane wave can only measure and 
influence some of the degrees of freedom of the boundary theory.
On the level of the corresponding sources the selection process 
could be realized as follows: only those sources are present which 
correspond to modes decaying sufficiently slow for increasing $R$ 
when travelling through the region between the $\AdS_5\times\text{S}^5$
boundary and the region that converges to the plane wave.

Furthermore, one of the most important points is to find the right
holographic screen in the BMN correspondence.
We have seen that several proposals with holographic screens of various
dimensions exist. In particular if one regards the boundary as the holographic
screen the dual theory should be $1$-dimensional. The type of holographic 
screen then determines the further steps of how the propagator with one point
in the bulk and one point on the screen can be derived from our result for the
bulk-to-bulk propagator.


\clearemptydoublepage
\chapter*{Acknowledgements}
\addcontentsline{toc}{chapter}{Acknowledgements}
\label{ack}
First of all, let me express my deepest gratitude to Dr. Harald Dorn and 
Prof. Dieter L\"ust who offered me the opportunity to do my PhD at the
Humboldt-University of Berlin.

In addition, I am deeply indebted to Dr. Harald Dorn for 
the supervision of my thesis and
for a very pleasant collaboration. I thank him very much 
for his advice and the intensive discussions 
from which I  benefited from a great deal. 
I would also like to express my fullest thanks to 
Mario Salizzoni for his collaboration during the second 
year of my PhD, for many useful discussions and 
for proof reading parts of my PhD thesis.

Let me furthermore express my fullest gratitude to James Babington for 
proof reading my PhD thesis and  
delivering many useful comments and improving remarks. 
I would like to express my deepest thanks to Ingo Runkel and
Stephan Stieberger for very enlighting discussions that helped me to 
refine the presentation of some Sections in this thesis.  
I would also like to thank them for proof reading parts of my PhD thesis.
It gives me a great pleasure to thank Alessandro Torrielli (University of 
Padua) for valuable discussions about noncommutative 
geometry during his visit to Berlin and during our meeting in Barcelona.

I would like to thank my former and present office mates
Matthias Br\"andle, Volker Braun, Stefano Chiantese, Florian Gmeiner, 
Lars G\"orlich, Claus Jeschek, Dan Oprisa, and Tassilo Ott for 
many discussions about physics and other topics in life. 

I would like to thank the former and present members of Prof.\ L\"ust's 
group and of the associated Emmi-Noether group for many enlighting 
discussions and a productive working atmosphere. I would like to thank:
Oleg Andreev, Ralph Blumenhagen, Gottfried Curio, Gianguido
Dall'Agata, Danilo Diaz, Johanna Erdmenger, Johannes Gro\ss{}e, Zachary
Guralnik, Andrea Gregori, Robert Helling, Ingo Kirsch, Albrecht Klemm, 
Georgios Kraniotis, Karl Landsteiner, Calin Lazaroiu, 
Gabriel Lopes Cardoso, Andr\'e Miemiec,
Aalok Misra, Robert Schmidt, Zachary Guralnik, Nikolaos Prezas, Susanne
Reffert.

I am very grateful to Hans-J\"org Otto for maintaining my computer system 
and to our secretary Sylvia Richter for her administrative assistance.
I would also like to thank my flatmate Nina Zimmermann for useful comments 
about the introduction of my PhD thesis. 

Above all let me express my love to my parents for all their love,
encouragement and support throughout my life. 

This work was supported by the `Graduiertenkolleg 271'.

 
\clearemptydoublepage
\begin{appendix}
\chapter{Appendix to Part \ref{NC}}
\label{app:NC}
\section{Path integral quantization of quantum field theories}
\label{app:pathintQFT}
In this Appendix we will shortly review the path integral approach
\cite{Feynman:1948ur} that allows us to quantize a given field theory.
It is convenient for a discussion of symmetries and especially for 
understanding the covariant gauge fixing procedure in non-Abelian theories.
This Appendix should be regarded as a brief review of some of the tools that we
will need for our own analysis. A more detailed introduction to the
method can for instance be found in \cite{Muta:1998vi}.

\subsection{The path integral approach}

In quantum field theories one wants to compute correlation functions of the 
fields. The theory is usually defined by an action $S$.
In the path integral approach the extraction of an $n$-point function 
of the fields which we collectively denote with $\phi$ is given by
\begin{equation}\label{npointinpath}
\big\langle0\big|\T\big[\hat\phi(x_1)\dots\hat\phi(x_n)\big]\big|0\big\rangle=
N\int\mathcal{D}\phi\,\e^{iS[\phi]}\phi(x_1)\dots\phi(x_n)\col\qquad
N=\Big[\int\mathcal{D}\phi\,\e^{iS[\phi]}\Big]^{-1}\pnt
\end{equation}
The L.\ H.\ S.\ of this expression shows the $n$-point Green function in the
canonical  
operator formalism, where $\big|0\big\rangle$ is the vacuum, $\T$ denotes time
ordering and $\hat\phi(x)$ is a field operator at position $x$.
The above integral on the R.\ H.\ S.\
is given for a theory in Minkowski space and it is related to the version in
Euclidean space by performing a Wick rotation which replaces $i$ in front of
the action in the exponent by $-1$. The fields $\phi$ under the integral are 
classical functions. 
In the following we will omit the time 
ordering symbol, the operator-hat and the zeros in writing vacuum matrix
elements. 
It is worth remarking that the calculations below can be intuitively
understood if one interpretes the path integral as the continuum limit of a
number of integrals over $\phi(x_i)$ at discrete lattice points $x_i$. 

The functional derivative in $d$ spacetime dimensions is defined as 
\begin{equation}
\prefuncderiv{\phi(x')}\mathcal{F}[\phi(x)]=\lim_{\varepsilon\to0}\frac{1}{\varepsilon}\big(\mathcal{F}[\phi(x)+\varepsilon\delta^d(x-x')]-\mathcal{F}[\phi(x)]\big)\pnt
\end{equation} 
In particular one finds 
\begin{equation}
\prefuncderiv{\phi(x')}\phi(x)=\delta^d(x-x')
\end{equation}
and the chain rule  
\begin{equation}\label{funcderivchainrule}
\begin{aligned}
\prefuncderiv{\phi(x')}\mathcal{F}\big[\mathcal{G}[\phi(x)]\big]&=\int\de^dy\,\funcderiv{\mathcal{F}\big[\mathcal{G}[\phi(x)]\big]}{\mathcal{G}[\phi(y)]}\funcderiv{\mathcal{G}[\phi(y)]}{\phi(x')}
=\funcderiv{\mathcal{F}\big[\mathcal{G}[\phi(x)]\big]}{\mathcal{G}}\cdot\funcderiv{\mathcal{G}}{\phi(x')}
\col
\end{aligned}
\end{equation}
where we have defined the abbreviation $\cdot$ to indicate that possible 
indices are contracted and that the spacetime dependence is integrated over. 
In particular this means
\begin{equation}
f\cdot g=\int\de^dx\,f_I(x)g^I(x)\col
\end{equation}
and
\begin{equation}
f\cdot\mathcal{M}\cdot g=\int\de^dx\de^dy\,f^I(x)\mathcal{M}_{IJ}(x,y)g^J(y)\col
\end{equation}
for indices $I,J$ that collectively denote spacetime and group indices. 

The functional derivative allows one to define a generating functional $Z[J]$
from 
which all Green functions can be obtained. One introduces external sources
$J(x)$ for the fields and defines
\begin{equation}\label{genfuncdef}
Z[J]=\int\mathcal{D}\phi\,
\e^{iS[\phi]+i\phi\cdot J}\col
\end{equation}
such that $Z[0]=N^{-1}$ gives the normalization factor $N$ in
\eqref{npointinpath}. 
The $n$-point function can now be obtained by taking $n$ functional
derivatives of the generating functional and setting $J=0$ afterwards
\begin{equation}\label{nGfunctionfromgenfunc}
G(x_1,\dots,x_n)=\big\langle\phi(x_1)\dots\phi(x_n)\big\rangle=\frac{1}{Z[0]}\prefuncderivi{J(x_1)}\cdots\prefuncderivi{J(x_n)}Z[J]\bigg|_{J=0}\pnt
\end{equation}
The generating functional can therefore be written as a series expansion
in $J$
\begin{equation}
\frac{Z[J]}{Z[0]}=\sum_n\frac{i^n}{n!}\int\de^dx_1\dots\de^dx_n\,\big\langle\phi(x_1)\dots\phi(x_n)\big\rangle
J(x_1)\dots J(x_n)\pnt
\end{equation}
In principle a theory can now be specified by giving the action, the
generating functional, or all possible Green functions. 

There is, however, some redundancy because the above Green functions contain
pieces which factorize in independent parts. 
An $n$-point function in general includes terms 
which are already known from correlators with less field insertions. 
The removal of these contributions leads to the definition of the connected 
$n$-point Green function which consists only of the non-factorizable part.
The generating functional of the connected Green functions $W[J]$ is related 
to the functional $Z[J]$ via
\begin{equation}\label{cgenfuncdef}
Z[J]=\e^{iW[J]}\pnt
\end{equation}
We thus have   
\begin{equation}\label{cGreenfuncdef}
G(x_1,\dots,x_n)_\text{c}=\big\langle\phi(x_1)\dots\phi(x_n)\big\rangle_\text{c}=\prefuncderivi{J(x_1)}\cdots\prefuncderivi{J(x_n)}\ln\frac{Z[J]}{Z[0]}\bigg|_{J=0}\col
\end{equation} 
where the subscript $c$ indicates the connected part. 
The series expansion is given by
\begin{equation}\label{cGreengenfuncexpansion}
\ln\frac{Z[J]}{Z[0]}=\sum_n\frac{i^n}{n!}\int\de^dx_1\dots\de^dx_n\,\big\langle\phi(x_1)\dots\phi(x_n)\big\rangle_\text{c}
J(x_1)\dots J(x_n)\pnt
\end{equation}
That \eqref{cgenfuncdef} in terms of Feynman graphs describes the
connected part can be easily checked by computing some examples. An argument
based on the cluster property of $W[J]$ can be found in
\cite{Zinn-Justin:1989mi}.

But even the connected Green functions contain redundant information.
They are all given by tree diagrams of so called proper Green functions. 
The intuitive definition of a tree diagram in terms of Feynman graphs is that 
it is decomposed into two
parts by cutting one internal line. In contrast to this, 
the proper Green functions do not decay into two parts by cutting only one
internal line. They are one particle irreducible (1PI).
The generating functional of the proper Green functions $\Gamma[\phi]$
is defined as the Legendre transform of $W[J]$
\begin{equation}
\Gamma[\phi]=W[J]-\int\de^dx\,\phi(x)J(x)\col
\end{equation}
where $\phi(x)$ is the field in presence of the source $J(x)$, i.\ e.\ 
\begin{equation}
\phi(x)=\prefuncderiv{J(x)}W[J]\col\qquad
J(x)=-\prefuncderiv{\phi(x)}\Gamma[J]\pnt 
\end{equation}
One can obtain $\Gamma[\phi]$ from the above equations by first inserting 
a power series of $W[J]$ in the first equation, then inverting the first 
equation to obtain $J$ as a functional of $\phi$ and then equating this
with the second equation. This procedure requires the inverse of the connected 
$2$-point function which is the proper $2$-point function. One can convince
oneself in this way that $\Gamma[\phi]$ generates the proper Green functions. 
A proof along the lines that $\Gamma[\phi]$ is the effective action can be
found in \cite{Weinberg:1996kr}. An alternative proof that adds a disconnected 
piece with an infinitesimal parameter $\varepsilon$ to the propagator and 
then shows that the $\mathcal{O}(\varepsilon)$ contribution to $\Gamma[\phi]$
is connected is presented in \cite{Zinn-Justin:1989mi}. 

The knowledge of all proper Green functions up to a given order in the
perturbation expansion is sufficient for a computation of 
all Green functions up to the same order as tree diagrams. This is the
statement that $\Gamma[\phi]$ is the effective action of the underlying
quantum field theory.

To be more precise, the proper Green functions are amputated, i.\ e.\ their
external lines are removed. In a tree diagram the proper $n$-point functions
with $n>2$ are joined with the connected $2$-point function. External legs 
are restored by using the connected $2$-point function, too.  

We can now define two different sets of building blocks from which all 
Green functions can be computed. One set is extracted from the action $S[\phi]$
itself. The second set is extracted from the effective 
action $\Gamma[\phi]$ and it contains the effective building blocks that 
already include the quantum corrections. With this set only tree diagrams 
have to be computed.

At the end of this introduction let   
us discuss the physical meaning of the different kinds of Green functions.
The amputated $n$-point Green function with all external
momenta being on-shell is the $n$-point contribution to the S-matrix.
It is the amplitude that describes 
the scattering of $n$ particles with each other. One assumes that the 
interaction takes place at a finite region in space and time and the incoming
and outgoing particles are produced and respectively measured infinitely far
away from the interaction region. This is implemented by the amputation 
and the choice of on-shell external momenta of the $n$-point Green function.
The amputated Green function contains
contributions where not all particles really interact with each other. One
particle could enter the interaction region and leave it without interacting
with all the other particles. On the level of Green functions this means that 
the contribution to the $n$-point Green function describing such a scattering
process factorizes into a $2$- and an $n-2$-point function. As the $n$-point
Green function includes contributions from all possible scattering scenarios
of $n$ particles it encompasses contributions that factorize into two and more
Green functions with less legs. However, besides these 
contributions  there is one piece where
all $n$ particles interact and which therefore cannot be split into
independent factors. This part contains new information. The other
parts are already known as products of Green functions with less than $n$
legs. 

\subsection{Feynman graphs from path integrals}
\label{subsec:Frulespint}
In this Subsection we will describe how a path integral can be evaluated 
in perturbation theory. The discussion will be rather general and 
it will lead us to a prescription how to find the momentum space Feynman rules.

A theory of several fields (or field components) $\phi^I$ is given by an action
\begin{equation}\label{genaction}
\begin{aligned}
S[\phi]=-\frac{1}{2}\int\de^dx\de^dy\,\phi(x)^I\mathcal{K}_{IJ}(x,y)\phi^J(y)-\int\de^dx\,V[\phi(x)]\col
\end{aligned}
\end{equation}
where $\mathcal{K}$ describes an operator acting to the right, $I,J$ are
multi-indices and $V$ is a
potential term. We will now show how to deal with the path integral (see \eqref{genfuncdef}) perturbatively in powers of the potential, assuming that the
potential  describes a small perturbation to the quadratic part of the action.
First, remove the potential term inside the path integral by rewriting it 
in terms of functional derivatives
\begin{equation}\label{genfuncpotextract}
Z[J]=\e^{-i\int\de^dx\,
V[\prefuncderivi{J}]}\int\mathcal{D}\phi\,\e^{-\frac{i}{2}\phi\cdot
  \mathcal{K}\cdot \phi+i\phi\cdot J}\pnt
\end{equation}
The path integral with the remaining integrand can now be evaluated. 
The field redefinition
\begin{equation}
\phi'^I(x)=\phi^I(x)-\int\de^dy\,\Delta^{IJ}(x,y)J_J(y)\col
\end{equation}
where $\Delta$ is the classical two point function which is the inverse of the 
operator $\mathcal{K}_{IJ}$
\begin{equation}\label{KDeltarel}
\int\de^dy\,\mathcal{K}_{IJ}(x,y)\Delta^{JI'}(y,x')=\delta_I^{I'}\delta^d(x-x')
\end{equation}
transforms the integral into one of Gaussian type. 
The integral is independent of $J$ and
therefore contributes only to the normalization. 
The dependence on the sources $J$ is completely included in the term that
remaines from completing the square in the exponent.
We denote the free generating functional with $Z^\text{kin}$ and it is given by
\begin{equation}\label{genfunckin}
Z_\text{kin}[J]=\int\mathcal{D}\phi\,\e^{-\frac{i}{2}\phi\cdot
  \mathcal{K}\cdot \phi+i\phi\cdot J}=\e^{\frac{i}{2} J\cdot\Delta\cdot J}
\pnt
\end{equation}
The complete generating functional then reads 
\begin{equation}\label{genfunc}
Z[J]=\e^{-i\int\de^dx\,V[\prefuncderivi{J}]}\e^{\frac{i}{2} J\cdot\Delta\cdot J}
\pnt
\end{equation}

One can now determine the $n$-point Green functions to arbitrary order in the
perturbation expansion as follows.
First one expands the exponential function up to order $n$ in $V$. 
Then one acts with the derivatives inside the $V$ on the second exponential
factor. The last step is to project out from this result, all terms which are
of order $n$ in $J$.
The projection is performed by acting with $n$ functional derivatives, 
like in \eqref{nGfunctionfromgenfunc}, and setting $J=0$ afterwards.

\subsubsection{The free theory}
In the case of a free theory, where the first factor in \eqref{genfunc} is
absent, the generating functional is given by
\begin{equation}
Z[J]=Z_\text{kin}[J]=\e^{\frac{i}{2} J\cdot\Delta\cdot J}\pnt
\end{equation}
From the definition \eqref{cgenfuncdef} the generating functional for the
connected Green functions is as follows
\begin{equation}
W_\text{kin}[J]=\frac{1}{2}\int\de^dx\de^dy\,J_I(x)\Delta^{IJ}(x,y)J_J(y)\pnt
\end{equation}
The consequence of this result is that (using \eqref{cGreenfuncdef}) 
in the free case the only connected 
Green function is the $2$-point function
\begin{equation}\label{twopointfuncgeneral}
G^{I_1I_2}(x_1,x_2)=\big\langle\phi(x_1)\phi(x_2)\big\rangle_\text{c}=-i\Delta^{I_1I_2}(x_1,x_2)\pnt
\end{equation}
 A generic $n$-point Green
function is therefore either $0$ if $n$ is odd or it is a sum over all
possibilities to factorize the $n$-point function into a product of
$\frac{n}{2}$ $2$-point functions if $n$ is 
even\footnote{We exclude the possibility of 
non-vanishing vacuum expectation values of 
the fields and of a coupling to a background field}.

From now on assume that the operator $\mathcal{K}$ of \eqref{genaction}
has the following form
\begin{equation}\label{Kmatrixdelta}
\mathcal{K}_{IJ}(x,y)=K_{IJ}\delta^d(x-y)\col
\end{equation}
where $K_{IJ}$ is a matrix-valued differential operator that acts on $x$ to the right. The relation \eqref{KDeltarel} then becomes
\begin{equation}\label{KDeltarelspec}
K_{IJ}\Delta^{JI'}(x,x')=\delta_I^{I'}\delta^d(x-x')\pnt
\end{equation}
Introduce the Fourier transform of a general $n$-point function and its
inverse as follows
\begin{equation}\label{GFouriertransform}
\begin{aligned}
\tilde G^{I_1\dots I_n}(p_1,\dots,p_n)&=\int\de^dx_1\,\e^{-ip_1\cdot
  x_1}\cdots\int\de^dx_n\,\e^{-ip_n\cdot x_n}G^{I_1\dots
  I_n}(x_1,\dots,x_n)\col\\
G^{I_1\dots I_n}(x_1,\dots,x_n)&=\int\frac{\de^dp_1}{(2\pi)^d}\e^{ip_1\cdot
  x_1}\cdots\int\frac{\de^dp_n}{(2\pi)^d}\e^{ip_n\cdot x_n}\tilde G^{I_1\dots I_n}(p_1,\dots,p_n)\col
\end{aligned}
\end{equation}
where $p\cdot x$ denotes the ordinary scalar product. This refers to the
convention that all momenta are \emph{incoming} momenta
\cite{Itzykson:1980rh}, i.\ e.\ momentum $p_i$ flows to the point $x_i$. 
The $\delta$-functions 
are then represented as usual 
\begin{equation}\label{Fourierdelta} 
\delta^d(x)=\int\frac{\de^dp}{(2\pi)^d}\e^{ip\cdot x}\col\qquad\delta^d(p)=\int\frac{\de^dx}{(2\pi)^d}\e^{-ip\cdot x}\pnt
\end{equation}
One then finds for \eqref{KDeltarelspec}
\begin{equation}\label{Ftwopointfuncgeneral}
\tilde\Delta^{I_1I_2}(p_1,p_2)=\Big(\frac{1}{K_{\partial_{x_1}\to ip_1}}\Big)^{I_1I_2}(2\pi)^d\delta^d(p_1+p_2)=\tilde\Delta^{I_1I_2}(p_1)(2\pi)^d\delta^d(p_1+p_2)\col
\end{equation}
where $\big(K^{-1}_{\partial_{x_1}\to ip_1}\big)^{I_1I_2}$ indicates the expression
which one obtains from the differential operator $K_{I_1I_2}$ via replacing the
derivatives $\preparderiv{x_1}$ with $ip_1$ and then taking the inverse of the
matrix.

\subsubsection{The interacting theory}
Let us now discuss the interacting theory with the functional \eqref{genfunc},
and with a potential given by
\begin{equation}\label{genpotential}
V[\phi]=g\mathcal{G}\big[\phi^{I_1},\dots,\phi^{I_N}\big]\col
\end{equation}
where $g$ is
the coupling constant and $\mathcal{G}$ is a functional that is linear in all
its arguments.
Here we will not discuss the general perturbative expansion to all
orders in $g$ but instead focus on the extraction of the Feynman rules for the 
theory. The fundamental building blocks (or Feynman rules) from which all 
diagrams in a perturbative expansion can be built are the connected $2$-point
function that we have already found in \eqref{Ftwopointfuncgeneral}, and the
proper $n$-point tree level Green functions with $n>2$ in momentum space.
It is clear that from the potential \eqref{genpotential} one only finds an 
$N$-point vertex. Its exact expression in momentum space will now be
determined.

First expand \eqref{genfunc} in the lowest nontrivial order in the coupling
constant $g$. This here gives 
\begin{equation}
Z[J]=\Big\{1-ig\int\de^dz\,\mathcal{G}\big[\tprefuncderivi{J_{I_1}},\dots,\tprefuncderivi{J_{I_N}}\big]+\dots\Big\}\exp\Big\{\frac{i}{2}\int\de^dx\de^dy\,J_I(x)\Delta^{IJ}(x,y)J_J(y)\Big\}\pnt
\end{equation}
After acting with the $N$ derivatives in $\mathcal{G}$ the term that
contribute to the $N$-point function have to be of the order $N$.
The only relevant term in the expansion of the second exponential factor 
therefore is the one proportional to $(J\cdot\Delta\cdot J)^N$. 
One obtains
\begin{equation}\label{Npointconnectedtreegeneral}
\begin{aligned}
G^{J_1\dots J_N}(x_1,\dots,x_N)_\text{p'}&=\prefuncderivi{J_{J_1}(x_1)}\cdots\prefuncderivi{J_{J_N}(x_n)}(-i)g\int\de^dz\,\mathcal{G}
\big[\Delta\cdot J,\dots,\Delta\cdot J\big]\\
&=(-i)^{N+1}g\int\de^dz\sum_{\pi\in S_N}\mathcal{G}\big[\Delta^{I_1J_{\pi(1)}}(z,x_{\pi(1)}),\dots,\Delta^{I_NJ_{\pi(N)}}(z,x_{\pi(N)})\big]\col
\end{aligned}
\end{equation} 
where the subscript $\text{p'}$ denotes the proper (but not truncated) part
and where we have used the abbreviation 
\begin{equation}
(\Delta\cdot J)^I(z) =\int\de^dy\,\Delta^{IJ}(z,y)J_J(y)\pnt
\end{equation}
The sum in \eqref{Npointconnectedtreegeneral} runs over $N!$ permutations of
the permutation group $S_N$ . It is generated because there are
$N!$ possibilities for the $N$ functional
derivatives to act on the $N$ sources $J$. The Green function is therefore 
symmetric under permutations of the indices $(J_i,x_i)$, $i=1,\dots,N$ at each
leg with the indices at any other leg.
To find the Feynman rule for this vertex one now has to transform to momentum
space and to amputate the diagram.
First, insert the momentum space expressions for $\Delta$ given by
\begin{equation}
\Delta^{II'}(x,x')=\int\frac{\de^dp}{(2\pi)^d}\e^{-ip\cdot(x-x')}\tilde\Delta^{II'}(p)\col
\end{equation} 
which follows from \eqref{Ftwopointfuncgeneral} if $p$ points from $x$ to $x'$,
to obtain
\begin{equation}\label{FNpointconnectedtreegeneral}
\begin{aligned}
G^{J_1\dots J_N}(x_1,\dots,x_N)_\text{p'}&=(-i)^{N+1}g\int\frac{\de^dp_1}{(2\pi)^d}\e^{ip\cdot
  x_1}\cdots\int\frac{\de^dp_N}{(2\pi)^d}\e^{ip\cdot
  x_N}\int\de^dz\,\e^{-i(p_1+\dots+p_N)z}\\
&\qquad\sum_{\pi\in S_N}\mathcal{G}_{(\partial)_j\to ip_{\pi(j)}}\big[\tilde\Delta^{I_1J_{\pi(1)}}(p_{\pi(1)}),\dots,\tilde\Delta^{I_NJ_\pi(N)}(p_{\pi(N)})\big]\pnt
\end{aligned}
\end{equation}
Here $\mathcal{G}_{(\partial)_j\to ip_j}$ denotes the expression which is
obtained from the functional $\mathcal{G}$ by replacing all derivatives that
act on the $j$th argument by $ip_j$ ($j=1,\dots,N$). The functional then is 
no longer a functional. It becomes a function of the momenta $p_j$ and reads
\begin{equation}
\mathcal{G}_{(\partial)_j\to
  ip_j}\big[\tilde\Delta^{I_1J_1}(p_1),\dots,\tilde\Delta^{I_NJ_N}(p_N)\big]=\big(\mathcal{G}_{(\partial)_j\to
  ip_j}\big)_{I_1\dots
  I_N}\tilde\Delta^{I_1J_1}(p_1),\dots,\tilde\Delta^{I_NJ_N}(p_N)\pnt
\end{equation} 
The proper (truncated) Green function (denoted with subscript $\text{p}$) 
is now obtained by simply removing the $N$ factors
$-i\tilde\Delta$. Furthermore, we assume that $\mathcal{G}$ does not explicitly
depend on the spacetime coordinates $z$. Then
in \eqref{FNpointconnectedtreegeneral} the $z$-integration can be
carried out and one finds by comparing with \eqref{GFouriertransform}
and using \eqref{Fourierdelta} that
\begin{equation}\label{FNpointtreegeneral}
\begin{aligned}
\tilde G^{I_1\dots I_N}(p_1,\dots,p_N)_\text{p}&=-ig\sum_{\pi\in
  S_N}\big(\mathcal{G}_{(\partial)_j\to ip_{\pi(j)}}\big)^{I_{\pi(1)}\dots
  I_{\pi(N)}}(2\pi)^d\delta^d(p_1+\dots+p_N)\\
&=-ig\dim(S)\sum_{[\pi]\in\frac{S_N}{S}}\big(\mathcal{G}_{(\partial)_j\to
  ip_{\pi(j)}}\big)^{I_{\pi(1)}\dots
  I_{\pi(N)}}(2\pi)^d\delta^d(p_1+\dots+p_N)\pnt
\end{aligned}
\end{equation}
In the second line $S$ denotes the symmetry group of $\mathcal{G}$ and the sum
runs over one element of each orbit $[\pi]=\big\{\pi'\in
S_N\,|\,\pi'=S\pi\big\}$.  
It is clear that this simplifies a concrete evaluation, for instance if
$\mathcal{G}$ is symmetric in all arguments then $S$ is the full permutation
group with $\dim(S)=N!$ and the sum only consists of one element. 
 
The above expression directly produces the momentum space 
Feynman rule for the $N$-point interaction vertex \eqref{genpotential} 
with all momenta $p_j$ leaving the vertex. 
If one wants to have all momenta to point to the vertex then one has to 
replace $p_j\to-p_j$, see Fig.\ \ref{fig:NpointFrulegeneral}. 
Let us remark that if one wants to compute Feynman diagrams from the 
above rules one still has to deal with symmetry factors that depend 
on the concrete diagram. 

In Chapter \ref{chap:NCYM} the above given expressions are used  
to determine the Feynman rules for the noncommutative YM theories.
\setlength{\fboxsep}{0pt}
\settoheight{\eqoff}{\fbox{$=$}}
\setlength{\eqoff}{0.5\eqoff}
\setlength{\eqoffsixty}{\eqoff}
\setlength{\eqofftwenty}{\eqoff}
\addtolength{\eqofftwenty}{-20pt}
\addtolength{\eqoffsixty}{-60pt}
\begin{figure}
\begin{center}
\begin{equation*}
\begin{aligned}
\raisebox{\eqofftwenty}{%
\begin{picture}(140,40)(0,0)\scriptsize
\SetOffset(130,20)
\Line(-80,0)(-40,0)\Text(-90,0)[r]{$p,I$}\Text(-30,0)[l]{$p,J$}
\end{picture}}%
&=-i\Big(\frac{1}{K_{\partial_x\to ip}}\Big)^{IJ}
\\%
\raisebox{\eqoffsixty}{%
\begin{picture}(155,120)(0,0)\scriptsize
\SetOffset(130,60)
\Line(-60,34.64)(-40,0)\Text(-65,43.3)[r]{$p_1,I_1$}
\Line(-80,0)(-40,0)\Text(-90,0)[r]{$p_2,I_2$}
\Line(-60,-34.64)(-40,0)\Text(-65,-43.3)[l]{$p_3,I_3$}
\DashLine(-20,-34.64)(-40,0){1}
\DashLine(0,0)(-40,0){1}
\Line(-20,34.64)(-40,0)\Text(-15,43.3)[l]{$p_N,I_N$}
\Vertex(-40,0){1}
\end{picture}}%
&=%
-ig\dim(S)\sum_{[\pi]\in\frac{S_N}{S}}\big(\mathcal{G}_{(\partial)_j\to
  -ip_{\pi(j)}}\big)^{I_{\pi(1)}\dots
  I_{\pi(N)}}
\end{aligned}
\end{equation*}
\caption{General Feynman rules for the theory \eqref{genaction} with one
  $N$-point vertex \eqref{genpotential}. Momentum conservation is understood. The momentum $p$ of the propagator
 enters the point $x$ and all momenta of the vertex point to the vertex.}\label{fig:NpointFrulegeneral}
\end{center}
\end{figure}
\section{Invariance of the DBI action}
\label{DBIlagequivcheck}
Let $M$ be a quadratic invertible matrix and $\delta M$ a small variation.
The inverse of $M+\delta M$ up to $\mathcal{O}(\delta M)$ is given by
\begin{equation}\label{inverseexpansion}
\frac{1}{M+\delta M}=\frac{1}{M}(M-\delta M)\frac{1}{M}\pnt
\end{equation} 
The determinant of $M+\delta M$ expanded up to $\mathcal{O}\big((\delta
M)^2\big)$ reads  
\begin{equation}\label{detexpansion}
\det(M+\delta M)=\det M\Big[1+\tr(M^{-1}\delta M)+\frac{1}{2}(\tr(M^{-1}\delta M))^2-\frac{1}{2}\tr(M^{-1}\delta MM^{-1}\delta M)+\dots\Big]\pnt
\end{equation}
With these relations we can now compute the variation of
\eqref{DBIlaginterpol} under $\delta\theta^{\mu\nu}$. 
The variations 
\begin{equation}
\Phi_{\mu\nu}\to\Phi_{\mu\nu}+\delta\Phi_{\mu\nu}\col\qquad G_{\mu\nu}\to G_{\mu\nu}+\delta G_{\mu\nu}\col\qquad\theta^{\mu\nu}\to\theta^{\mu\nu}+\delta\theta^{\mu\nu}
\end{equation}
are not independent due to \eqref{geneffgthetarel} but fulfill
\begin{equation}
(\delta G+\tilde\alpha\delta\Phi)_{\mu\nu}=\Big((G+\tilde\alpha\Phi)\frac{\delta\theta}{\tilde\alpha}(G+\tilde\alpha\Phi)\Big)_{\mu\nu}\col
\end{equation}
as can be seen with the help of \eqref{inverseexpansion} and $\tilde\alpha=2\pi\alpha'$.
Considering the part of $\mathcal{O}(\delta M)$ in \eqref{detexpansion} 
it is then easy to derive the variations
\begin{equation}
\delta\sqrt{\det(G+\tilde\alpha(\Phi+F))}
=\frac{1}{2}\sqrt{\det(G+\tilde\alpha(\Phi+F))}\tr\Big(\frac{1}{G+\tilde\alpha(\Phi+F)}(\delta
G+\tilde\alpha(\delta\Phi+\delta F)\Big)\col
\end{equation}
and thus with one finds in particular from \eqref{genostrincouplingrel} that
\begin{equation}
\delta(G^\Phi_0)^2=\frac{1}{2}(G^\Phi_0)^2\tr\Big((G+\tilde\alpha\Phi)\frac{\delta\theta}{\tilde\alpha}\Big)\pnt
\end{equation}
The variation of the DBI Lagrangian \eqref{DBIlaginterpol} then reads up 
to a factor $g_sT_p$
\begin{equation}
\begin{aligned}\label{DBIlagvar1}
&\delta\Big(\frac{1}{(G^\Phi_0)^2}\sqrt{\det(G+\tilde\alpha(\Phi+F))}\Big)\\
&\qquad=\frac{1}{(2G^\Phi_0)^2}\sqrt{\det(G+\tilde\alpha(\Phi+F))}\tr\Big(\frac{1}{G+\tilde\alpha(\Phi+F)}\Big[-(G+\tilde\alpha\Phi)\delta\theta
F+\tilde\alpha\delta
F\Big]\Big)\pnt
\end{aligned}
\end{equation}
We now insert \eqref{SWdiffeq} which describes how $\delta F_{\mu\nu}$ depends on
$\delta\theta^{\mu\nu}$. For the Abelian case one finds 
\begin{equation}\label{SWFvar}
\begin{aligned}
\delta
F_{\mu\nu}&=\delta\theta^{\alpha\beta}\big(F_{\mu\alpha}F_{\nu\beta}-\frac{1}{2}A_\alpha(\partial_\beta+D_\beta)F_{\mu\nu}+\mathcal{O}(\partial
F\partial F)\big)\\
&=-(F\delta\theta
F)_{\mu\nu}-\frac{1}{2}\vec A\delta\theta(\vec\partial+\vec D)F_{\mu\nu}+\mathcal{O}(\partial
F\partial F)\col
\end{aligned}
\end{equation}
where in the second line we have used matrix notation. Furthermore, it is easy
to see that
\begin{equation}
\begin{aligned}
\partial_l\sqrt{\det(G+\tilde\alpha(\Phi+F))}&=\frac{\tilde\alpha}{2}\sqrt{\det(G+\tilde\alpha(\Phi+F))}\tr\Big(\frac{1}{G+\tilde\alpha(\Phi+F)}\partial_lF\Big)\col\\
D_l\sqrt{\det(G+\tilde\alpha(\Phi+F))}&=\frac{\tilde\alpha}{2}\sqrt{\det(G+\tilde\alpha(\Phi+F))}\tr\Big(\frac{1}{G+\tilde\alpha(\Phi+F)}D_lF\Big)+\mathcal{O}(\partial
FDF)\pnt
\end{aligned}
\end{equation}
One inserts \eqref{SWFvar} into \eqref{DBIlagvar1} and then uses the above 
relations to reexpress the terms where derivatives act on $F_{\mu\nu}$. 
After integrating by parts, one obtains
\begin{equation}
\begin{aligned}\label{DBIlagvar2}
&\delta\Big(\frac{1}{(G^\Phi_0)^2}\sqrt{\det(G+\tilde\alpha(\Phi+F))}\Big)\\
&\qquad=\frac{1}{(2G^\Phi_0)^2}\sqrt{\det(G+\tilde\alpha(\Phi+F))}
\Big[-\tr(\delta\theta
F)+(\partial_\beta+D_\beta)(A\delta\theta)^\beta\Big]+\mathcal{O}(\partial
F)+\text{tot.\ der.}\\
&\qquad=\mathcal{O}(\partial F)+\text{total
  derivatives}
\col
\end{aligned}
\end{equation}
where the last step follows with the identity
\begin{equation}
(\partial_\beta+D_\beta)(A\delta\theta)^\beta=\delta\theta^{\alpha\beta}F_{\beta\alpha}=\tr(\delta\theta F)
\pnt
\end{equation}
Hence, the Seiberg-Witten map translates the field strength
in such a way that \eqref{DBIlagequiv} holds. 
At the end it is important to remark that we had to take into account 
the second term in \eqref{SWFvar} although naively it is $\mathcal{O}(\partial
F)$. The reason is that it contains a factor $A_\mu$ without a derivative 
such that it is not negligible after 
partial integration that shifts the derivative to this `bare' $A_\mu$.
If one works with the known explicit solution of
the differential equation \eqref{SWdiffeq} in the case of an (exactly) constant $F_{\mu\nu}$, the absence of this 
term is responsible for an observed mismatch \cite{Seiberg:1999vs,Yang:2004vd} 
in the relation \eqref{DBIlagequiv}.
An extension of the above given calculation to the non-Abelian case can be
found in \cite{Terashima:2000ej}.

\section{The Weyl operator formalism}
\label{app:NCWeylop}
In a noncommutative spacetime the coordinates no longer commute. Instead one
has the relation
\begin{equation}\label{coordcomm}
\comm{\hat x^\mu}{\hat x^\nu}=i\theta^{\mu\nu}\col
\end{equation}
where $\hat x^\mu$ are Hermitian operators and we assume $\theta^{\mu\nu}$ to
be constant. One can now define the noncommutative counterpart to a
function $f$ in ordinary $d$-dimensional space $\mathds{R}^d$ 
by taking its Fourier transform
 \begin{equation}\label{Fouriertraf}
\tilde f(k)=\int\de^dx\,\e^{-ik_\mu x^\mu}f(x)
\end{equation}
and using the Weyl symbol
\begin{equation}\label{Weyltraf}
\wop[f]=\int\frac{\de^dk}{(2\pi)^D}\tilde f(k)\e^{ik_\mu\hat{x}^\mu}\pnt
\end{equation}
This procedure is described by the Hermitian operator
$\dop(x)=\dop^\dagger(x)$ which is given by
\begin{equation}\label{dopdef}
\dop(x)=\int\frac{\de^dk}{(2\pi)^d}\e^{ik_\mu\hat{x}^\mu}\e^{-ik_\mu x^\mu}\col
\end{equation}
such that
\begin{equation}\label{wopindop}
\wop[f]=\int\de^dx\,f(x)\dop(x)\pnt
\end{equation}
Derivatives can be defined as
\begin{equation}
\comm{\hat\partial_\mu}{\hat{x}^\nu}=\delta_\mu^\nu\col\qquad\comm{\hat\partial_\mu}{\hat\partial_\nu}=0\pnt 
\end{equation}
From this definition it follows immediately that
\begin{equation}
\comm{\hat\partial_\mu}{\dop(x)}=-\partial_\mu\dop(x)\col
\end{equation}
and one thus finds from \eqref{wopindop} after integration by parts
\begin{equation}
\comm{\hat\partial_\mu}{\wop[f]}=\int\de^dx\,\partial_\mu
f(x)\dop(x)=\wop[\partial_\mu f]\pnt
\end{equation}
The translation operator is given by $\e^{v^\mu\hat\partial_\mu}$ and it acts 
as follows
\begin{equation}
\begin{aligned}
\e^{v^\mu\hat\partial_\mu}\dop(x)\e^{-v^\mu\hat\partial_\mu}=\dop(x+v)\pnt
\end{aligned}
\end{equation}
From this it is obvious that a trace defined for the Weyl 
operators is independent of $x\in\mathds{R}^d$ because of its invariance under
cyclic permutations. 
One then finds from
\eqref{wopindop} that the trace `$\hat{\tr}$' corresponds to an integration over
spacetime
\begin{equation}\label{optracedef}
\hat{\tr}\wop[f]=\int\de^dx\,f(x)\col
\end{equation}
where we have normalized $\hat{\tr}\dop(x)=1$.
The products of operators at distinct points can be defined by using the
Baker-Campbell-Hausdorff formula 
\begin{equation}\label{BCHformula}
\e^A\e^B=\e^{A+B+\frac{1}{2}\comm{A}{B}+\frac{1}{12}(\comm{A}{\comm{A}{B}}+\comm{B}{\comm{B}{A}})+\dots}\pnt
\end{equation}
In the special case where the commutator is a c-number as in
\eqref{coordcomm}, one obtains
\begin{equation}\label{BCHccomm}
\e^{ik_\mu\hat{x}^\mu}\e^{ik'_\mu\hat{x}^\mu}=\e^{-\frac{i}{2}\theta^{\mu\nu}k_\mu k'_\nu}e^{i(k+k')_\mu\hat{x}^\mu}\pnt
\end{equation}
Using \eqref{dopdef} and assuming that $\theta^{-1}$ exists (for
which an even spacetime dimension is necessary), the product of two 
$\dop$-operators reads 
\begin{equation}\label{dopprod}
\begin{aligned}
\dop(x)\dop(y)
&=\frac{1}{\pi^d|\det\theta|}\int\de^dz\,\dop(z)e^{-2i(\theta^{-1})_{\mu\nu}(z^\mu-x^\mu)(z^\nu-y^\nu)}\pnt
\end{aligned}
\end{equation}
With the normalization $\hat{\tr}\dop(x)=1$, it follows that the operators $\dop(x)$ and $\dop(y)$ are orthonormal w.\ r.\ t.\ the trace operation 
\begin{equation}
\begin{aligned}
\hat{\tr}\big(\dop(x)\dop(y)\big)
&=\delta^d(x-y)\pnt
\end{aligned}
\end{equation}
Hence, the inverse of the Weyl-operator \eqref{wopindop} is well defined given
by 
\begin{equation}
f(x)=\hat{\tr}\big(\wop[f]\dop(x)\big)
\end{equation}
for a function $f$.
\section{The $\ast$-product}
\label{app:NCastprod}
The operation of multiplication of Weyl operators can be captured by
introducing a noncommutative $\ast$-product in ordinary space that has to 
fulfill the relation
\begin{equation}
\wop[f]\wop[g]=\wop[f\ast g]\pnt
\end{equation}
Using the Fourier transformation \eqref{Fouriertraf}, the Weyl transformation
\eqref{wopindop}, the definition of $\dop$ \eqref{dopdef} and the
Baker-Campbell-Hausdorff relation \eqref{BCHccomm} for the commutator
\eqref{coordcomm}, one finds for the $\ast$-product 
\begin{equation}
(f\ast g)(x)
=\iint\frac{\de^dk}{(2\pi)^d}\frac{\de^dk'}{(2\pi)^d}\tilde f(k)\tilde
g(k')e^{-\frac{i}{2}\theta^{\mu\nu}k_\mu k'_\nu}e^{i(k_\mu+k'_\mu)x^\mu}\pnt
\end{equation}
With the help of the inverse Fourier transformation 
the above expression can be cast into the following form
\begin{equation}\label{astproddefapp}
\begin{aligned}
(f\ast g)(x)
&=\exp\Big\{\frac{i}{2}\theta^{\mu\nu}\parderiv{}{x^\mu}\parderiv{}{y^\nu}\Big\}f(x)g(y)\Big|_{y=x}\\
&=f(x)g(x)+\sum_{n=1}^\infty\Big(\frac{i}{2}\Big)^n\frac{1}{n!}\theta^{i_1j_1}\cdots\theta^{i_nj_n}\partial_{i_1}\cdots\partial_{i_n}f(x)\partial_{j_1}\cdots\partial_{j_n}g(x)\pnt
\end{aligned}
\end{equation}
The second line shows how the exponential function has to be understood. 
One has obtained the Moyal-Weyl $\ast$-product \cite{Moyal:1949sk}. 
From this result it 
is easy to derive expressions for the $\ast$-(anti)commutator  
\begin{equation}\label{astsincos}
\begin{aligned}
\astcomm{f(x)}{g(x)}&=2i\sin\Big\{\frac{i}{2}\theta^{\mu\nu}\parderiv{}{x^\mu}\parderiv{}{y^\nu}\Big\}f(x)g(y)\Big|_{y=x}
=2i f(x)\astsin g(x)\col\\
\astacomm{f(x)}{g(x)}&=2\cos\Big\{\frac{i}{2}\theta^{\mu\nu}\parderiv{}{x^\mu}\parderiv{}{y^\nu}\Big\}f(x)g(y)\Big|_{y=x}=2f(x)\astcos
g(x)\pnt
\end{aligned}
\end{equation}
The $\ast$-product is associative and one finds for the product of $n$
functions $f_a$, $a=1,\dots,n$
\begin{equation}
(f_1\ast \cdots \ast
f_n)(x)=\prod_{a<b}\exp\Big\{\frac{i}{2}\theta^{\mu\nu}\parderiv{}{x_a^\mu}\parderiv{}{x_b^\nu}\Big\}f_1(x_1)\cdots
f_n(x_n)\Big|_{x_1=\cdots=x_n=x}\pnt 
\end{equation}
The cyclicity of the operator trace \eqref{optracedef} 
translates into the invariance of the integral 
\begin{equation}
\hat{\tr}\big(\wop[f_1]\cdots\wop[f_n]\big)=\int\de^dx\,f_1(x)\ast\cdots\ast f_n(x)
\end{equation}
under cyclic permutations of the $f_a$. In particular, a trace over two 
Weyl operators and therefore an integral of two functions multiplied by 
the $\ast$-product reduces to the integral with the two functions being 
multiplied by using the ordinary product
\begin{equation}\label{optracebilin}
\hat{\tr}\big(\wop[f]\wop[g]\big)=\int\de^dx\,f(x)\ast g(x)=\int\de^dx\,f(x)g(x)\pnt
\end{equation}

\section{Noncommutative Yang-Mills theories}
\label{app:NCYM}
In Yang-Mills theories the gauge fields take values in a representation of 
the Lie algebra of the underlying gauge group. One therefore
has to generalize the formalism of appendix \ref{app:NCWeylop} somewhat. 
The Weyl transformation \eqref{wopindop} is redefined as a tensor product of 
the $\dop$-operator and the Lie algebra representation matrices to
\begin{equation}
\wop[A_\mu]=\int\de^dx\,\dop(x)\otimes A_\mu(x)\pnt
\end{equation}
One can then write the action of the noncommutative Yang-Mills theories in the 
operator space as follows
\begin{equation}
S_\text{YM}=-\frac{1}{4g^2}\hat{\tr}\otimes\tr\big(\wop[F_{\mu\nu}]\wop[F^{\mu\nu}]\big)\col
\end{equation}
where `$\hat{\tr}$' and `$\tr$' denote the traces w.\ r.\ t.\
the spacetime part and the gauge group, and the field strength in operator
space reads
\begin{equation}
\begin{aligned}
\wop[F_{\mu\nu}]&=\comm{\hat\partial_\mu}{\wop[A_\nu]}-\comm{\hat\partial_\nu}{\wop[A_\mu]}-i\comm{\wop[A_\mu]}{\wop[A_\nu]}\\
&=\wop\big[\partial_\mu A_\nu-\partial_\nu A_\mu-i\astcomm{A_\mu}{A_\nu}\big]\pnt
\end{aligned}
\end{equation}
Using the definition of the operator trace \eqref{optracedef} one finds
\begin{equation}
S_\text{YM}=-\frac{1}{4g^2}\int\de^dx\,\tr\big(F_{\mu\nu}\ast F^{\mu\nu}\big)\pnt
\end{equation}
Due to the symmetry of the trace and of the integral under cyclic permutations
of the argument, the above action is invariant under the gauge transformation
\begin{equation}\label{NCgaugetrafoWop}
\begin{aligned}
\wop[A_\mu]\to\wop[\tilde A_\mu]&=\wop[U]\wop[A_\mu]\wop[U]^{-1}-i\wop[U]\comm{\partial_\mu}{\wop[U]^{-1}}\\
&=\wop\big[U\ast A_\mu\ast U^{-1_\ast}-iU\ast \partial_\mu U^{-1_\ast}\big]\col
\end{aligned}
\end{equation}
because the field strength transforms according to
\begin{equation}
\wop[F_{\mu\nu}]\to\wop[\tilde F_{\mu\nu}]=\wop[U]\wop[F_{\mu\nu}]\wop[U]^{-1}
=\wop\big[U\ast F_{\mu\nu}\ast U^{-1_\ast}\big]\pnt
\end{equation}
The Weyl transformation of $U^{-1_\ast}$ yields the inverse in the operator
space. Hence, in ordinary space it is the inverse of $U$ w.\ r.\ t.\ the
$\ast$-product. The corresponding relations are given by 
\begin{equation}\label{Uastinversedef}
\wop[U]\wop[U]^{-1}=\wop[U]^{-1}\wop[U]=\hat{\mathds{1}}\otimes\mathds{1}\col\qquad
U\ast U^{-1_\ast}=U^{-1_\ast}\ast U =\mathds{1}\pnt
\end{equation}
They imply for a constant $\theta^{\mu\nu}$ that the following equalities hold
\begin{equation}
\wop[U]\comm{\partial_\mu}{\wop[U]^{-1}}=-\comm{\partial_\mu}{\wop[U]}\wop[U]^{-1}
\col\qquad U\ast\partial_\mu U^{-1_\ast}=-(\partial_\mu U)\ast U^{-1_\ast}\pnt
\end{equation}
It is important to remark that one must not identify $U^{-1_\ast}$ and
$U^{-1}$ because the latter is  
defined as the inverse of $U$ w.\ r.\ t.\ the ordinary (matrix) product 
\begin{equation}
U\,U^{-1}=\mathds{1}\col
\end{equation}
and therefore it is clear that
$\wop[U^{-1}]\neq\wop[U]^{-1}=\wop[U^{-1_\ast}]$. 
A relation between $U^{-1_\ast}$ and $U^{-1}$ can be worked out order by order
in  $\theta^{\mu\nu}$ if one expands the $\ast$-product in 
\eqref{Uastinversedef}.
This leads to (with an invertible $\theta^{\mu\nu}$)
\begin{equation}
U^{-1_\ast}=U^{-1}+\frac{i}{2}\theta^{\mu\nu}U^{-1}(\partial_\mu
U)U^{-1}(\partial_\nu U)U^{-1}+\dots\pnt
\end{equation} 

In the following we will make some general comments on the gauge
transformation \eqref{NCgaugetrafoWop} and especially compare with the 
ordinary case. In the latter we have the ordinary YM gauge field $a_\mu$ and
the finite gauge transformation $u$ which acts like
\begin{equation}\label{ordinarygaugetrafo}
a_\mu\to\tilde a_\mu=u\,a_\mu\, u^{-1}-iu\,\partial_\mu u^{-1}\pnt
\end{equation}
If we require that the gauge field is Hermitian we have to impose the 
unitarity condition on $u$, i.\ e.\
\begin{equation}
u^{-1}=u^\dagger\pnt
\end{equation}
The quantity $u$ is an element of the gauge group and the gauge field
$a_\mu$ takes values in the corresponding Lie algebra. The same is true for the
transformed gauge field  $\tilde a_\mu$ in \eqref{ordinarygaugetrafo}.

The gauge transformation \eqref{NCgaugetrafoWop} preserves the Hermiticity of
a Hermitian gauge field $A_\mu=A_\mu^\dagger$ if one chooses $U$ as
$\ast$-unitary, i.\ e.\  
\begin{equation}
U^{-1_\ast}=U^\dagger\pnt
\end{equation} 
One can therefore define noncommutative gauge theories with $U(N)$ gauge
groups. 
That a naive extension to other gauge groups appears to be difficult can
most easily be seen from the infinitesimal versions of the gauge
transformations. With $\Lambda$ as an infinitesimal gauge transformation 
parameter one finds from \eqref{NCgaugetrafoWop}
\begin{equation}\label{NCgaugetrafoinfinit}
A_\mu\to A_\mu+\delta A_\mu\col\qquad\delta A_\mu=\partial_\mu\Lambda+i\astcomm{\Lambda}{A_\mu}\pnt
\end{equation}
This has to be compared with the infinitesimal gauge transformation with
parameter $\lambda$ in the ordinary case extracted from
\eqref{ordinarygaugetrafo}
\begin{equation}\label{ordinarygaugetrafoinfinit}
a_\mu\to a_\mu+\delta a_\mu\col\qquad\delta a_\mu=\partial_\mu\lambda+i\comm{\lambda}{a_\mu}\pnt
\end{equation}
Since $\lambda$ and $a_\mu$ are Lie algebra valued in the ordinary case, 
the infinitesimal gauge transformation is guaranteed to be Lie algebra 
valued, too. This is because the ordinary commutator of two elements of the
Lie algebra is itself an element. 
In the noncommutative case, however, the situation is different. 
One has to evaluate the $\ast$-commutator that occurs in
\eqref{NCgaugetrafoinfinit} and analyze in which cases it closes on the 
algebra, see section \ref{sec:NCYMgaugegroups}.

\section{The Seiberg-Witten map from the enveloping algebra approach}
\label{SWmapfromenvalg}
In \cite{Jurco:2000ja} the authors deal with the noncommutative coordinates
$\hat x^\mu$ and the enveloping algebra generators
on an equal footing. They replace 
them by ordinary quantities $x^\mu$ and $t^a$ respectively and describe the 
noncommutativity with a $\ast$-product. 
A product of $n$ variables $t^a$ corresponds to the symmetrized product 
of $n$ Lie algebra generators $\mathfrak{t}^a$ given by (see \eqref{envalggen})
\begin{equation}
t^{a_1}\cdots t^{a_n}=\frac{1}{n!}\sum_{\pi\in S_n}\mathfrak{t}^{a_{\pi(1)}}\cdots \mathfrak{t}^{a_{\pi(n)}}\pnt
\end{equation}
These generators span the corresponding enveloping algebra.
The star product that describes the spacetime noncommutativity and the gauge
algebra is defined by 
\begin{equation}\label{astlieproddef}
(f\ast g)(x,t)=e^{\frac{i}{2}(\theta^{\mu\nu}\preparderiv{x^\mu}\preparderiv{x'^\nu}+t^ag_a(i\preparderiv{t},i\preparderiv{t'}))}f(x,t)g(x',t')\big|_{x'=x,t'=t}\pnt
\end{equation}
Here $g_a(u,v)$ follows from the group multiplication
\begin{equation}
\e^{u_a\mathfrak{t}^a}\e^{v_b\mathfrak{t}^b}=\e^{i(u_c+v_c+\frac{1}{2}g_c(u,v))\mathfrak{t}^c}
\end{equation}
and with the Baker-Campbell-Hausdorff formula \eqref{BCHformula} one finds the
expansion 
\begin{equation}\label{lieexpand}
g_c(u,v)=-u_av_b\fuud{a}{b}{c}+\frac{1}{6}u_av_b(v_d-u_d)\fuud{a}{b}{e}\fuud{e}{d}{c}+\dots\pnt
\end{equation}
In \cite{Madore:2000en} infinitesimal gauge transformations on noncommutative
space have been defined for a field $\phi$ as 
\begin{equation}\label{NCgaugetrafoWopphi}
\hat\delta_\Lambda\wop[\phi]=i\wop[\Lambda]\wop[\phi]\pnt
\end{equation} 
Multiplication of a field by a coordinate is not a covariant operation since
\begin{equation}
\hat\delta_\Lambda\big(\wop[x^\mu]\wop[\phi]\big)=i\wop[x^\mu]\wop[\Lambda]\wop[\phi]\neq i\wop[\Lambda]\wop[x^\mu]\wop[\phi]\pnt
\end{equation}
Therefore, one introduces covariant coordinates that commute with the gauge 
transformation  
\begin{equation}
\hat\delta_\Lambda\big(\wop[X^\mu]\wop[\phi]\big)=\wop[X^\mu]\,\hat\delta_\Lambda\wop[\phi]\pnt
\end{equation}
With the ansatz $X^\mu=x^\mu+V^\mu(x)$ one finds for $V^\mu(x)$
\begin{equation}
\hat\delta_\Lambda\wop[V^\mu]=-i\comm{\hat x^\mu}{\wop[\Lambda]}+i\comm{\wop[\Lambda]}{\wop[V^\mu]}\col
\end{equation}
where $\comm{\dummylen}{\dummylen}$ denotes the commutator in the underlying
noncommutative space.
The corresponding expression in ordinary space is found by replacing
all functions by their ordinary counterparts and all products between Weyl
operators by the corresponding $\ast$-product.
The transformation of the gauge connection then becomes
\begin{equation}\label{connectiontraf}
\hat\delta_\Lambda V^\mu=\theta^{\mu\nu}\partial_\nu\Lambda+i\astcomm{\Lambda}{V^\mu}\col
\end{equation}
where we have used the $\ast$-product \eqref{astlieproddef} that leads one to 
the relation
\begin{equation}
-i\astcomm{x^\mu}{\Lambda}=\theta^{\mu\nu}\partial_\nu\Lambda\pnt
\end{equation} 
The second term
in \eqref{connectiontraf} with the $\ast$-product \eqref{astlieproddef}
in the case of a non-vanishing $g_c$ in \eqref{lieexpand} shows, that the
transformation of $V^\mu$ starts either at order $\mathcal{\theta}^0$ 
or linearly in $\theta^{\mu\nu}$ if $V^\mu$ itself is of the order
$\mathcal{\theta}^0$ or of higher order in $\theta^{\mu\nu}$.
It is thus reasonable to assume that $V^\mu$ starts with a
term of at most first order in $\theta^{\mu\nu}$, because with the 
transformation \eqref{connectiontraf} one can always generate a term that is
linear in $\theta^{\mu\nu}$. 
Furthermore, it is important
to stress that the connection $V^\mu$ was introduced to make multiplication in
a noncommutative space a covariant operation. 
It should vanish for $\theta^{\mu\nu}\to0$
as the space becomes commutative. Hence, it should start at linear order in $\theta^{\mu\nu}$ and one can make the ansatz    
\begin{equation}
\begin{aligned}\label{connectiongaugefieldrel}
V^\mu&=\theta^{\mu\nu}A_\nu\col\\
\hat\delta_\Lambda A_\mu&=\partial_\mu\Lambda+i\astcomm{\Lambda}{A_\mu}
\end{aligned}
\end{equation}
for the connection. Its form becomes more clear in a comparison with 
the case in an ordinary space. The gauge connection 
$V^\mu$ itself has no counterpart in ordinary space, where the multiplication
with a coordinate is a covariant operation w.\ r.\ t.\ the ordinary 
counterpart of the gauge transformation \eqref{NCgaugetrafoWopphi}. 
In ordinary space, differentiation becomes covariant under gauge
transformations by introducing a covariant derivative that depends on the 
gauge connection.  
In noncommutative spaces one has to start one step earlier and covariantize
the coordinates themselves. This already includes the covariantization of
ordinary derivatives, because in the  
the $\ast$-product \eqref{astlieproddef} they appear at $\mathcal{O}(\theta)$.
Hence, one should not wonder that a order $\theta^{\mu\nu}$, 
the gauge connection \eqref{connectiongaugefieldrel} coincides with the 
one found in the ordinary case.

We will now analyze \eqref{connectiongaugefieldrel} order by order in
$\theta^{\mu\nu}$. Expansion of \eqref{astlieproddef} gives
\begin{equation}\label{circledastproddef}
\begin{aligned}
(f\ast g)(x,t)&=\Big(1+\frac{i}{2}\theta^{\alpha\beta}\preparderiv{x^\alpha}\preparderiv{x'^\beta}+\dots\Big)f(x,t)\circledast g(x',t')\Big|_{x'= x,t'= t}\col\\
f(x,t)\circledast g(x',t')&=e^{\frac{i}{2}t^a g_a(i\preparderiv{t},i\preparderiv{t'})}f(x,t)g(x',t')\pnt
\end{aligned}
\end{equation}
At $\mathcal{O}(\theta^0)$ one finds that the relation for 
$\hat\delta_\Lambda A_\mu$ 
in \eqref{connectiongaugefieldrel} can be solved with an ansatz where 
$\Lambda$ and $A_\mu$ depend linearly on $t^a$  
\begin{equation}
\Lambda=\lambda_a^1t^a\col\qquad A_\mu=a_{\mu,a}^1t^a\pnt
\end{equation}
That this should work is clear from the previous discussion.
In the $\theta^{\mu\nu}\to0$ limit $A_\mu$ should become the well
known gauge connection on ordinary space which is Lie algebra valued and
which thus depends linearly on $t^a$. Inserting the ansatz into
\eqref{connectiongaugefieldrel} one finds
\begin{equation}\label{theta0gaugetraf}
\hat\delta_\Lambda a_{\mu,a}^1t^a=\partial_\mu\lambda_a^1t^a+i\lambda_a^1a_{\mu,b}^1[t^a\circledast t'^b-t^b\circledast t'^a]\Big|_{t'= t}
=\partial_\mu\lambda_a^1t^a+i\lambda_a^1a_{\mu,b}^1\fuud{a}{b}{c}t^c\col
\end{equation}
where we have used 
\begin{equation}
t^a\circledast t'^b\Big|_{t'=t}=t^at^b+\frac{i}{2}\fuud{a}{b}{c}t^c\pnt
\end{equation}
This follows from the definition of the $\circledast$-product in
\eqref{circledastproddef} with the expansion \eqref{lieexpand}.
At $\mathcal{O}(\theta^1)$ one includes terms which are quadratic in $t^a$ in
the ansatz, such that
\begin{equation}
\Lambda=\lambda_a^1t^a+\lambda_{ab}^2t^at^b\col\qquad A_\mu=a_{\mu,a}^1t^a+a_{\mu,ab}^2t^at^b\pnt
\end{equation}
The reason that this is sufficient follows from the fact that the
$\mathcal{O}(\theta^0)$ terms can contribute at most in second order in
$t^a$. This can be seen from \eqref{circledastproddef}.
Using the result of $\mathcal{O}(\theta^0)$, one finds
\begin{equation}\label{theta1gaugetraf}
\begin{aligned}
\hat\delta_\Lambda
a_{\mu,ab}^2t^at^b&=\partial_\mu\lambda_{ab}^2t^at^b-\frac{1}{2}\theta^{\alpha\beta}\partial_\alpha\lambda_a^1\partial_\beta
a_{\mu,b}^1(t^a\circledast t'^b+t^b\circledast t'^a)\Big|_{t'= t}\\
&\phantom{={}}+i(\lambda_c^1a_{\mu,ab}^2-\lambda_{ab}^2a_{\mu,c}^1)(t^c\circledast t'^at'^b-t^at^b\circledast t'^c)\Big|_{t'= t}\\
&=\partial_\mu\lambda_{ab}^2t^at^b-\theta^{\alpha\beta}\partial_\alpha\lambda_a^1\partial_\beta
a_{\mu,b}^1t^at^b-2(\lambda_a^1a_{\mu,cb}^2+\lambda_{ab}^2a_{\mu,c}^1)\fuud{a}{c}{d}t^bt^d\col
\end{aligned}
\end{equation}
where we have used the symmetry of $\lambda_{ab}^2$ and $a_{\mu,ab}^2$ under
the exchange $a\leftrightarrow b$ and the relations 
\begin{equation}
\begin{aligned}
t^c\circledast t'^at'^b\Big|_{t'= t}
&=t^at^bt^c-\frac{i}{2}t^d\Big[\fuud{a}{c}{d}t^b+\fuud{b}{c}{d}t^a-\frac{i}{6}(\fuud{a}{c}{e}\fuud{e}{b}{d}+\fuud{b}{c}{e}\fuud{e}{a}{d})\Big]\col
\end{aligned}
\end{equation}
\begin{equation}
\begin{aligned}
t^at^b\circledast t'^c\Big|_{t'= t}
&=t^at^bt^c+\frac{i}{2}t^d\Big[\fuud{a}{c}{d}t^b+\fuud{b}{c}{d}t^a+\frac{i}{6}(\fuud{a}{c}{e}\fuud{e}{b}{d}+\fuud{b}{c}{e}\fuud{e}{a}{d})\Big]\pnt
\end{aligned}
\end{equation}
Together, \eqref{theta0gaugetraf} and \eqref{theta1gaugetraf} read
\begin{equation}\label{theta01gaugetraf}
\begin{aligned}
\hat\delta_\Lambda a_{\mu,a}^1&=\partial_\mu\lambda_a^1+i\lambda_b^1a_{\mu,c}^1\fuud{b}{c}{a}\col\\
\hat\delta_\Lambda
a_{\mu,ab}^2&=\partial_\mu\lambda_{ab}^2-\theta^{\alpha\beta}\partial_\alpha\lambda_a^1\partial_\beta
a_{\mu,b}^1-2(\lambda_d^1a_{\mu,cb}^2+\lambda_{db}^2a_{\mu,c}^1)\fuud{d}{c}{a}\pnt
\end{aligned}
\end{equation} 
The second equality can now be reformulated. Remember that the elements 
$t^{a_1}\dots t^{a_n}$ correspond to symmetric products of the Lie algebra
generators $\mathfrak{t}^a$, forming the generators of the enveloping algebra.
If we define the following quantities
\begin{equation}
a_\mu=a_{\mu,a}^1\mathfrak{t}^a\col\qquad A'_\mu=a_{\mu,ab}^2\mathfrak{t}^a\mathfrak{t}^b\col\qquad\lambda=\lambda_a^1\mathfrak{t}^a\col\qquad\Lambda'=\lambda_{ab}^2\mathfrak{t}^a\mathfrak{t}^b\col
\end{equation}
then the second equation of \eqref{theta01gaugetraf} reads
\begin{equation}
\hat\delta_\Lambda
A'_\mu=\partial_\mu\Lambda'-\frac{1}{2}\theta^{\alpha\beta}\acomm{\partial_\alpha\lambda}{\partial_\beta
  a_\mu}+i\comm{\lambda}{A'_\mu}+i\comm{\Lambda}{a_\mu}\pnt
\end{equation}
This is the exactly the $\mathcal{O}(\theta^1)$ expansion \eqref{gaugeorbitmaplintheta} of 
\eqref{gaugeorbitmap} that defines the Seiberg-Witten map. The solution is
given in \eqref{SWmaplin} and its translation to the previously used notation 
is given by
\begin{equation}
\begin{aligned}
a_{\mu,ab}^2t^at^b&=-\frac{1}{2}\theta^{\alpha\beta}a_{\alpha,a}^1(\partial_\beta a_{\mu,b}^1+F_{\beta\mu,b}^1)t^at^b\col\\
\lambda_{ab}^2t^at^b&=\frac{1}{2}\theta^{\alpha\beta}\partial_\alpha\lambda_a^1a_{\beta,b}^2t^at^b\col
\end{aligned}
\end{equation} 
where 
\begin{equation}
F_{\mu\nu,a}^1=\partial_\mu a_{\nu,a}^1-\partial_\nu a_{\mu,a}^1+\fuud{c}{d}{a}a_{\mu,c}^1a_{\nu,d}^1\pnt
\end{equation}
In \cite{Jurco:2001rq} the above summarized formalism is used to work out the
Seiberg-Witten map up to 
$\mathcal{O}(\theta^2)$ .
From the above construction it follows that the gauge connection 
\eqref{connectiongaugefieldrel} and the gauge transformation
\eqref{NCgaugetrafoWopphi} are completely determined by the 
coefficients $a_{\mu,a}^1$ and $\lambda_a^1$ of the Lie algebra valued terms. 
One can write
\begin{equation}\label{NCgaugetrafoliefunc}
\hat\delta_\Lambda\phi(x)=i\Lambda[\lambda^1,a^1]\ast\phi(x)\pnt
\end{equation}
In \cite{Jurco:2000ja} the authors check that the composition property of 
two gauge transformations 
\begin{equation}\label{NCgaugetrafocomp}
(\hat\delta_{\Lambda_1}\hat\delta_{\Lambda_2}-\hat\delta_{\Lambda_2}\hat\delta_{\Lambda_1})=\hat\delta_{i\astcomm{\Lambda_1}{\Lambda_2}}
\end{equation}
holds if the transformations are interpreted in the form
\eqref{NCgaugetrafoliefunc}.

\section{Constraints on the gauge group via anti-automorphisms}
\label{antiautoconstraints}
In \cite{Bonora:2000td,Bars:2001iq} the authors present proposals of how to
construct noncommutative gauge theories with some subgroups of $U(N)$.
The idea is to formulate a constraint for the  
gauge field and the gauge parameter. Here we will present some 
more detail about this construction. 
In \cite{Bonora:2000td} the authors observe that the condition of
(anti)-Hermiticity is preserved by the $\ast$-product \eqref{astproddefapp}, 
i.\ e.\ the relation 
\begin{equation}\label{astprodhermit}
(f\ast g)^\dagger=g^\dagger\ast f^\dagger
\end{equation}
holds for two matrix-valued functions $f$ and $g$. 
It follows that the finite gauge transformation of the 
noncommutative gauge connection $A_\mu$ then preserves (anti)-Hermiticity if
the gauge parameter $\Lambda$ itself is chosen to be (anti)-Hermitian (see the 
discussion in appendix \ref{app:NCYM}). 
One could have the idea to obtain $SO(N)$ or $SP(N)$ gauge group by making the
gauge field and gauge transformations real, and in addition dropping the $i$ in the exponent of the $\ast$-product \eqref{astproddefapp}.
However, this approach fails because then the property
\eqref{astprodhermit} for Hermitian conjugation is no longer valid. 
However it is essential for preserving Hermiticity. 
Instead, the authors of \cite{Bonora:2000td} 
formulate an additional constraint on the gauge field
and the gauge parameter. 
They define an algebra $\mathcal{A}_\theta$ which elements are formal power 
series in $\theta^{\mu\nu}$. That means it is required to 
define the gauge field and gauge transformations as elements of this algebra,
depending explicitly on $\theta^{\mu\nu}$.
Then they define an anti-automorphism $\text{r}$ 
of $\mathcal{A}_\theta$, which acts on $f(x,\theta)\in\mathcal{A}_\theta$
as follows
\begin{equation}
(\dummylen)^\text{r}:f(x,\theta)\mapsto f^\text{r}(x,\theta)=f(x,-\theta)\pnt
\end{equation}
Acting on the coordinates $x^\mu$ themselves, this
map is the identity. It reverses the order of $\ast$-multiplication 
\begin{equation}
(x_1^{\mu_1}\ast\dots\ast x_n^{\mu_n})^\text{r}=(x_n^{\mu_n})^\text{r}\ast\dots\ast(x_1^{\mu_1})^\text{r}\pnt
\end{equation}
The anti-automorphism $\text{r}$ is now combined
with matrix transposition $\text{t}$, acting on the representation matrices of
the gauge Lie-algebra, in a map which is defined as
$(\dummylen)^\text{rt}=((\dummylen)^\text{t})^\text{r}$. 
It has the crucial property that its action on the  $\ast$-product of two
elements $f,g\in\mathcal{A}_\theta$ is given by  
\begin{equation}
(f\ast g)^\text{rt}=g^\text{rt}\ast f^\text{rt}\col
\end{equation}
and hence it provides us with a relation similar to \eqref{astprodhermit}
for Hermitian conjugation.

In addition to the Hermiticity condition\footnote{The authors of
  \cite{Bonora:2000td,Bars:2001iq} work with an anti-Hermitian gauge field and
gauge transformation parameter.} 
\begin{equation}\label{hermitcond}
A_\mu^\dagger(x,\theta)=A_\mu(x,\theta)\col\qquad
\Lambda^\dagger(x,\theta)=\Lambda(x,\theta)
\end{equation}
one can now impose the extra constraint
\begin{equation}
A_\mu^\text{rt}(x,\theta)=-A_\mu(x,\theta)\col\qquad
\Lambda^\text{rt}(x,\theta)=-\Lambda(x,\theta)
\end{equation} 
on the gauge field and gauge transformations. 
Using the definition of the anti-automorphism $\text{r}$, the constraint
becomes 
\begin{equation}\label{thetareflexcond}
A_\mu^\text{t}(x,-\theta)=-A_\mu(x,\theta)\col\qquad
\Lambda^\text{t}(x,-\theta)=-\Lambda(x,\theta)\pnt
\end{equation}
The above relations lead to definite symmetry properties of the matrix valued
expansion coefficients, if one expands the gauge field and gauge
transformation parameter in power series in $\theta^{\mu\nu}$ like
\begin{equation}
A_\mu(x,\theta)=A^0_\mu+\theta^{\alpha\beta}A^1_{\mu,\alpha\beta}+\dots\col\qquad
\Lambda(x,\theta)=\Lambda^0+\theta^{\alpha\beta}\Lambda^1_{\mu,\alpha\beta}+\dots\pnt
\end{equation}
The matrices $A^{2n}$ [$A^{2n+1}$] and $\Lambda^{2n}$ [$\Lambda^{2n+1}$],
$n=0,1\dots$ have to be antisymmetric [symmetric]. 
In combination with the Hermiticity condition
\eqref{hermitcond} this then requires that $A^{2n}$ [$A^{2n+1}$] are purely 
imaginary [real].

One should not interpret the higher order expansion coefficients as new
degrees of freedom. Instead they should be regarded as functions of 
$A^0$ and $\Lambda^0$, as explicitly realized in the Seiberg-Witten
map \eqref{SWmapansatz}.
The latter respects the constraint \eqref{thetareflexcond}. 
For instance, in the explicit expansion of the Seiberg-Witten map
\eqref{SWmaplin} the coefficient at linear order in $\theta$ is symmetric
if the ordinary field $a_\mu$ is antisymmetric under matrix transposition.
An inversion of the Seiberg-Witten map 
gives the constraint \eqref{thetareflexcond} formulated for $A_0$
\begin{equation}
a_\mu[A]=-a_\mu^\text{t}[A]\pnt
\end{equation}
This means that the noncommutative gauge theory with the gauge group 
restricted by \eqref{thetareflexcond} is the image of an ordinary gauge 
theory with gauge group $SO(N)$ under the Seiberg-Witten map. 
  
\section{Proof that \eqref{greenkinexpl} does not vanish}
\label{appnptproof}

To prove that the Green function in \eqref{greenkinexpl} is
non-zero, it is sufficient 
to show that at least one contribution to this quantity with an
\emph{independent} tensor structure is non-vanishing at some configuration of
the external momenta, Lorentz and group indices. Choosing the most symmetric
non-trivial external configuration 
\begin{equation}
p_1=\dots=p_{n-1}=p\col\quad
p_n=-(n-1)p\col\quad\mu_1=\dots=\mu_n=\mu\col\quad m'_1=\dots=m'_n=m'
\label{exconfig}
\end{equation}
simplifies \eqref{greenkinexpl} considerably, e.\ g.\
the summation over permutations of the external
quantities simply lead to a combinatorial factor.   

We first pick out all terms where -- \emph{after} performing the
integral of \eqref{greenkinexpl} -- the tensor structure of the
$\mu_i$ 
is purely constructed with $G_{\mu_i\mu_j}$ such that $\theta^{\alpha\beta}$
does not carry an external
Lorentz index $\mu_i$. To minimize the number of
contributing terms we choose
$\theta^{\alpha\beta}(p_i)_\beta=0$.\footnote{This can be realized for
  the choice \eqref{exconfig}.} In this case the square 
brackets in \eqref{greenkinexpl}  simplify and we use the abbreviations
\begin{equation*}
2\Big[\olett{1}_r+\olett{2}_r+\olett{3}_r\Big]k=
2\Big[-\theta^{\alpha_r\gamma_r}
G_{\mu_{i_r}\alpha_{r+1}}+\theta_{\alpha_{r+1}}^{\phantom{\alpha_{r+1}}
\gamma_r}\delta_{\mu_{i_r}}^{\alpha_r}-\theta^{\alpha_r}_{\phantom{\alpha_r}
\alpha_{r+1}}\delta_{\mu_{i_r}}^{\gamma_r}\Big]k_{\gamma_r}\col
\end{equation*}
where Lorentz indices are not written explicitly. 
For the three terms inside the bracket only the following multiplications
can produce a pure $G_{\mu_i\mu_j}$-structure
\begin{equation*}
\begin{aligned}
\olett{1}_r\olett{2}_{r+1}=&\theta^{\alpha_r\gamma_r}\theta^{\gamma_{r+1}}_{\phantom{\gamma_{r+1}}\alpha_{r+2}}G_{\mu_{i_r}\mu_{i_{r+1}}}\\
\olett{2}_r\olett{1}_{r+1}=&\vphantom{\theta}^{\gamma_r}\theta\theta^{\gamma_{r+1}}\delta_{\mu_{i_r}}^{\alpha_r}G_{\mu_{i_{r+1}}\alpha_{r+2}}\\
\olett{2}_r\olett{3}_{r+1}=&\vphantom{\theta}^{\gamma_r}\theta\theta_{\alpha_{r+2}}\delta_{\mu_{i_r}}^{\alpha_r}\delta_{\mu_{i_{r+1}}}^{\gamma_{r+1}}\\
\olett{3}_r\olett{1}_{r+1}=&\vphantom{\theta}^{\alpha_r}\theta\theta^{\gamma_{r+1}}\delta_{\mu_{i_r}}^{\gamma_r}G_{\mu_{i_{r+1}}\alpha_{r+2}}\\
\olett{3}_r\olett{3}_{r+1}=&\vphantom{\theta}^{\alpha_r}\theta\theta_{\alpha_{r+2}}\delta_{\mu_{i_r}}^{\gamma_r}\delta_{\mu_{i_{r+1}}}^{\gamma_{r+1}}\col
\end{aligned}
\end{equation*}
where we have defined
$\vphantom{\theta}^\alpha\theta\theta^\gamma=\theta^{\alpha\beta}\theta_{\beta}^{\phantom{\beta}\gamma}$.    
These products are the building blocks of the complete terms
with $n$ factors, for instance like
\begin{equation*}
\underbrace{\olett{1}_1\olett{2}_2\dots\olett{1}_{k-1}\olett{2}_k\olett{3}_{k+1}\dots\olett{3}_{j+k}
\dots}_n\col
\end{equation*}
where the $\alpha_1$ index of the first factor is contracted with the
$\alpha_{n+1}$ index of the last. 
 
Further restrictions are imposed on the complete expressions: 
The total number of factors $n$ has to be
even because one cannot construct a pure $G_{\mu_i\mu_j}$ structure
with an odd number of $\mu_i$'s. In addition the number of $\olett{3}$'s in
the complete 
product of $n$ terms has to  be even as otherwise after performing the
integral in \eqref{greenkinexpl} one  $\theta$ would
carry an index $\mu_i$ (see equations below). Then it follows that the
numbers of $\olett{1}$'s and $\olett{2}$'s have to be identical.

Using the configuration \eqref{exconfig}
the contribution of all terms with an even number $j$ of $\olett{3}$'s
and an even number $n-j$ $\olett{1}$'s and $\olett{2}$'s can now be
written as
\begin{equation}
\begin{aligned}
& G_{\text{c}\negphantom{\text{c}{}}\phantom{\text{kin},}\mu\dots\mu}^{\text{kin},m'\dots
  m'}(p,\dots,p,-(n-1)p)\big|_{\propto\theta^ng^{2n}\text{, only }
  G_{\mu\mu}}\\
& \qquad\quad
=\frac{(n-1)!}{2}\frac{g^{2n}}{2^n}\Big[\prod_{r=1}^n\duud{a_r}{m'}{a_{r+1}}\Big]\\
&\phantom{\quad\qquad={}}\times\sum_{j=0,2}^n(\underbrace{\theta\dots\theta}_n)^{\gamma_{j+1}\dots\gamma_n}\underbrace{G_{\mu\mu}\dots
G_{\mu\mu}}_{n-j}\delta_\mu^{\gamma_1}\dots\delta_\mu^{\gamma_j}
\int\frac{\de^dk}{(2\pi)^d}\frac{k_{\gamma_1}\dots
k_{\gamma_n}}{q_1^2\dots q_n^2}\Big|_{\text{only special }G}
\col
\end{aligned}
\label{relstruc}
\end{equation} 
where the factor $\frac{(n-1)!}{2}$ stems from performing the
summation over all proper permutations and $q_r=k+rp$, $r\neq n$, $q_n=k$.  
To make the above expression compact we have used some further abbreviations
which we now explain.

The relevant part of the integral in the above expression is defined
as the tensor component of the integral only made out of the metric where
the metric must not possess a mixed index pair with one index from the
set $\{\gamma_1,\dots,\gamma_j\}$ and one from the set
$\{\gamma_{j+1}\dots\gamma_n\}$. It then reads
\begin{equation}
\int\frac{\de^dk}{(2\pi)^d}\frac{k_{\gamma_1}\dots
  k_{\gamma_n}}{q_1^2\dots q_n^2}\Big|_{\text{only special }G}=I_0\sum_{\scriptsize\begin{array}{c}\text{perm} \\\{i_1,\dots,i_j\}\\
  \{i_{j+1},\dots,i_n\}\end{array}}\prod_{r=1,3}^{n-1}G_{\gamma_{i_r}
\gamma_{i_{r+1}}}\col
\label{relint}
\end{equation}
where $I_0$ denotes a scalar integral which will be discussed later. 

The tensor
$(\theta\dots\theta)^{\gamma_{j+1}\dots\gamma_n}$ in
\eqref{relstruc} is built by summing over all possibilities to
replace $\frac{n-j}{2}$ of the $n$ summation index pairs
$(\alpha_r,\alpha_r)$ in the trace
$\tr\{\theta^n\}=\theta^{\alpha_1}_{\phantom{\alpha_1}\alpha_2}\theta^{\alpha_2}_{\phantom{\alpha_2}\alpha_3}\dots\theta^{\alpha_n}_{\phantom{\alpha_n}   
\alpha_1}$ by the index pairs
$\{(\gamma_{j+1},\gamma_{j+2}),\dots,(\gamma_{n-1},\gamma_n)\}$
keeping the ordering of the $\gamma$-pairs, i.\ e.\  the pair
$(\gamma_{j+1},\gamma_{j+2})$ is inserted at the positions with
smallest index $r$ of all replaced $\alpha_r$ and so on. All indices
$r$ of the replaced $\alpha_r$ either have to be odd or even, since otherwise
at least two substructures
$\vphantom{\theta}^{\gamma_r}\theta\dots\theta^{\gamma_{r+1}}$ would contain
  an odd number of $\theta$'s vanishing when contracted with the
  symmetric $k_{\gamma_r}k_{\gamma_{r+1}}$ in \eqref{relstruc}. Some
examples for illustration: If $j=n$ in \eqref{relstruc} then
$(\theta\dots\theta)=\tr\{\theta^n\}$ and there is only one
contribution. If $j=n-2$ then there are $n$
possibilities\footnote{$\frac{n}{2}$ possibilities to replace
  $\alpha_r$ with odd $r$ and $\frac{n}{2}$ to replace the ones with
  even $r$.} to replace a
pair $\alpha_r$ by the pair $(\gamma_{n-1},\gamma_n)$ such that
$(\theta\dots\theta)^{\gamma_{n-1}\gamma_n}=n\vphantom{\theta}^{\gamma_{n-1}}
\theta\dots\theta^{\gamma_n}$. For general $j\neq n$ there are
$2\binom{n/2}{j/2}$ non-vanishing possibilities to replace summation
indices by the $\gamma$-pairs.   

The contraction of the above defined 
$(\theta\dots\theta)^{\gamma_{j+1}\dots\gamma_n}$ in
\eqref{relstruc} with the tensor
structure of the integral \eqref{relint} leads to a sum over
products of traces  of the form $\prod_i\tr\{\theta^{2k_i}\}$,
$k_i\in\mathbb{N}$ such that 
$\sum_i 2k_i=n$. All these products of traces include the same sign
$(\sgn\tr\{\theta^2\})^\frac{n}{2}$.\footnote{This can be proven by
  using the canonical skew-diagonal form of \cite{Szabo:2001kg}.}
Thus, all summed terms in \eqref{relstruc}  
carry the same sign such that a cancellation
mechanism between different terms cannot be present. Proving
the non-vanishing of \eqref{relstruc} therefore only requires to show
that the group structure factor and the scalar integral $I_0$ in
\eqref{relint} are non-zero. 

For instance, the choice $m'=0$, where the generator $T^0$ is
given by $T^0=\frac{1}{\sqrt{2N}}\mathds{1}$ in an $U(N)$ theory, leads to
$d_{ab0}=\sqrt{\frac{2}{N}}\delta_{ab}$. Hence, with $\dim\mathfrak{g}$ as the
dimension of the Lie algebra $\mathfrak{g}$
\begin{equation*}
\prod_{r=1}^n\duud{a_r}{m'}{a_{r+1}}\Big|_{m'=0}=\Big(\frac{2}{N}\Big)^\frac{n}{2}\dim\mathfrak{g}
\end{equation*}
does not vanish.

In general the integral in \eqref{relstruc} can be decomposed in
scalar integrals like
\begin{equation*}
\int\frac{\de^dk}{(2\pi)^d}\frac{k_{\gamma_1}\dots
  k_{\gamma_n}}{q_1^2\dots
  q_n^2}=I_0\sum_{\scriptsize\begin{array}{c}\text{perm}\\
\{i_1,\dots,i_n\}\end{array}}\prod_{r=1,3}^{n-1}G_{\gamma_{i_r}\gamma_{i_{r+1}}}+\text{
  terms containing }p_{\gamma_i}\col
\end{equation*}
where due to the choice \eqref{exconfig} the $q_i$
\eqref{qdef} now only depend on $p$ such that the above tensor
structure can only be spanned by $G_{\gamma_i\gamma_j}$ and $p_{\gamma_i}$.
Notice that in \eqref{relstruc} only one part of the total symmetric
tensor multiplying $I_0$ given in \eqref{relint} is needed. 
In the above expression we
now choose all indices $\gamma_1=\dots=\gamma_n=\gamma$ and the momentum
$p$ such that it has a vanishing component $p_\gamma$ for the special
choice of $\gamma$. Then one finds for $I_0$
\begin{equation*}
I_0=\frac{1}{n!}\int\frac{\de^dk}{(2\pi)^d}\frac{(k_{\gamma})^n}{q_1^2\dots
  q_n^2}\pnt
\end{equation*}
For even $n$ this is non-vanishing since it is positive definite after
a Wick rotation. 

Thus, the expression \eqref{relstruc} in general does not vanish for
all even $n$ implying that 
at least all connected $n$-point Green functions with an even number
of external points  are
therefore present in the kinetic theory such that it produces infinitely many
building blocks in the $\theta$-summed case.

\section{A counterexample that disproves the $SO(N)$ Feynman rules of \cite{Bonora:2001ny}}
\label{appB}
In this appendix we give an explicit proof for the non-vanishing
of the lowest order contribution in $g^2$ and $\theta$
to $\langle A^{m_1}(x_1)\dots A^{m_8}(x_8)\rangle^{\text{kin}+S'_i}_\text{c}$ 
in the $SO(3)$ case.  

As discussed in the main text, we focus on the $8$-point vertex generated
out of a $4$-point interaction of the non-commutative $A_\mu$ in $S'_i$, 
see Fig.\ \ref{fig:NCUNYMFrules}.
Via the definition of $S'_i$, at least one of the $A_\mu$ has
 to carry a primed group index. Since we look for the lowest order in
 $\theta$ we can replace the $\ast$-product by the usual
 product. The interaction then has the gauge group structure 
$\fddu{N_1}{N_2}{K}f_{N_3n'_4K}$. 
Due to the subgroup property of $G$ this is zero 
if $N_j=n_j,j=1,2,3$. Therefore, we have to start with a 4-point interaction
of the $A$ where two of them carry a primed index. Three interaction terms 
contribute in this case 
\begin{equation}
\begin{aligned}
\frac{i}{g^2}\left(\fddu{m_1}{m_2}{a}f_{n'_3n'_4a}G^{\mu_1\nu_3}G^{\mu_2\nu_4}
+\fddu{n'_4}{m_1}{a'}f_{m_2n'_3a'}G^{\mu_1\nu_3}G^{\mu_2\nu_4}
+\fddu{n'_4}{m_1}{a'}f_{n'_3m_2a'}G^{\mu_1\mu_2}G^{\nu_3\nu_4}\right)\\
\times A^{m_1}_{\mu_1}A^{m_2}_{\mu_2}A^{m'_3}_{\nu_3}A^{m'_4}_{\nu_4}\pnt
\label{cc1}
\end{aligned}
\end{equation}
Now we replace $A^{m_i}_{\mu_i}$ by $a^{m_i}_{\mu_i}$ for $i=1,2$
and $A^{n'_i}_{\nu_i},i=3,4$ by the term with maximum number of $a$ 
within the $\theta^1$ contribution, see \eqref{ALambdaCbarCSWmaplin}, and get
\begin{equation}
\begin{aligned}
\frac{i}{16g^2}\left(\fddu{m_1}{m_2}{a}f_{n'_3n'_4a}G^{\mu_1\mu_5}G^{\mu_2\mu_8}
+\fddu{m_1}{n'_4}{a'}f_{m_2n'_3a'}(G^{\mu_1\mu_2}G^{\mu_5\mu_8}-
G^{\mu_1\mu_5}G^{\mu_2\mu_8})\right)\\
\times \dddu{n'_3}{m_3}{e}f_{em_4m_5}\dddu{n'_4}{m_6}{k}f_{km_7m_8}\theta^{\mu_3\mu_4}
\theta^{\mu_6\mu_7}a^{m_1}_{\mu_1}a^{m_2}_{\mu_2}\dots a^{m_8}_{\mu_8}
\pnt
\label{cc2}
\end{aligned}
\end{equation}
With this interaction the $g^{18}\theta^2$ contribution
to the Fourier transform of\\ $\langle A(x_1)\dots A(x_8)\rangle^{\text{kin}+S'_i}_\text{c}$ becomes up to the momentum
conservation factor equal to
\begin{equation}
\begin{aligned}
M^{\mu_1\dots\mu_8}_{m_1\dots m_8}&=\frac{i}{16}g^{18}
\sum_{\scriptsize\begin{array}{c}\text{perm}\\
\{i_1,\dots,i_8\}\end{array}}\theta^{\mu_{i_3}\mu_{i_4}}\theta^{\mu_{i_6}\mu_{i_7}}
\dddu{n'_3}{m_{i_3}}{e}f_{em_{i_4}m_{i_5}}\dddu{n'_4}{m_{i_6}}{k}f_{km_{i_7}m_{i_8}}\\
&\phantom{=\frac{i}{16}g^{18}
\sum_{\scriptsize\begin{array}{c}\text{perm}\\
\{i_1,\dots,i_8\}\end{array}}}
\times\Big(\frac{1}{2}(G^{\mu_{i_1}\mu_{i_5}}G^{\mu_{i_2}\mu_{i_8}}-
G^{\mu_{i_2}\mu_{i_5}}G^{\mu_{i_1}\mu_{i_8}})\fddu{m_{i_1}}{m_{i_2}}{a}
f_{n'_3n'_4a}\\
&\phantom{=\frac{i}{16}g^{18}
\sum_{\scriptsize\begin{array}{c}\text{perm}\\
\{i_1,\dots,i_8\}\end{array}}\times\Big({}}
+(G^{\mu_{i_1}\mu_{i_2}}G^{\mu_{i_5}\mu_{i_8}}-
G^{\mu_{i_1}\mu_{i_5}}G^{\mu_{i_2}\mu_{i_8}})\fddu{m_{i_1}}{n'_4}{a'}f_{m_{i_2}
n'_3a'}\Big)\pnt
\end{aligned}
\label{cc3}
\end{equation}

We will have reached the goal of this appendix if it can be shown that the 
above quantity is different from zero. Our explicit proof of
$M^{\mu_1\dots\mu_8}_{m_1\dots m_8}\neq 0$ consists in the
numerical calculation for one special choice of 
gauge group and Lorentz indices. To minimize the calculational
effort forced by taking into account all the permutations, we looked
for an index choice with a lot of symmetry with respect to the interchange
of external legs. But we also had to avoid too much symmetry not to
produce a zero result. 

If we use the standard Gell-Mann enumeration of the nine generators
of the $U(3)$ Lie algebra, see e.g. \cite{Itzykson:1980rh}, the generators
of the $SO(3)$ subalgebra carry the indices $2$,$5$,$7$. Then our special
choice for the external legs is 
\begin{equation}
\begin{tabular}{ccccccccc}
        & leg 1 & leg 2 & leg 3 & leg 4 & leg 5 & leg 6 & leg 7 & leg 8\\
$m_i$   & $5$   & $5$   & $2$   & $5$   & $2$   & $2$   & $5$   & $2$\\ 
$\mu_i$ & $\lambda $ & $\lambda $ & $\mu $ & $\nu $ & $\mu $ & $\mu $ & $\nu $
& $\mu$
\end{tabular}\pnt
\end{equation} 
The chosen Lorentz indices are all spacelike and have to fulfill
\begin{equation}\label{cc4}
\begin{tabular}{ccc}
$\mu\neq\nu\col$ & $\mu \neq\lambda\col$ & $\nu\neq\lambda\col$\\
$\theta^{\mu\nu}\neq 0\col$ & $\theta^{\mu\lambda}=0\col$ & $\theta^{\nu\lambda}=0$\pnt
\end{tabular}
\end{equation} 
Taking into account the list of vanishing $d_{ABC}$ and $f_{ABC}$ for
$U(3)$ \cite{Itzykson:1980rh} we find
\begin{equation}
M^{\mu_1\dots\mu_8}_{m_1\dots m_8}\big|_\text{special}
=6ig^{14}\left(\theta^{\mu\nu}\right)^2
f_{257}^2\left[\left(f_{345}d_{247}-f_{123}d_{157}\right)^2+f^2_{458}d^2_{247}\right].
\label{cc5}
\end{equation}
All $f$ and $d$ in \eqref{cc5} are different from zero.

\chapter{Appendix to Part \ref{BMN}}
\label{BMNapp}
\section{The Einstein equations with cosmological constant}
Some spacetimes that we discuss in section \ref{backgrounds} 
are solutions of the Einstein equations or
are direct products of such solutions. 
In the following we will in short present the equations and fix our notations
and conventions.
The Einstein-Hilbert action in $D$ dimensions with a cosmological 
constant is given by
\begin{equation}
S=\frac{1}{2\kappa^2}\int\de^D\sqrt{-g}(\mathcal{R}-2\Lambda)\pnt
\end{equation}
From it one derives the Einstein equations
\begin{equation}\label{Einsteineq}
\mathcal{R}_{\mu\nu}-\frac{1}{2}\mathcal{R}g_{\mu\nu}=-\Lambda g_{\mu\nu}\pnt
\end{equation}
The Ricci tensor $\mathcal{R}_{\mu\nu}$ and the scalar curvature $\mathcal{R}$
are computed from the Riemannian curvature tensor
\begin{equation}\label{curvaturetensor}
\Ruddd{\alpha}{\beta}{\gamma}{\delta}=\partial_\gamma\Gudd{\alpha}{\beta}{\delta}-\partial_\delta\Gudd{\alpha}{\beta}{\gamma}+\Gudd{\alpha}{\rho}{\gamma}\Gudd{\rho}{\beta}{\delta}-\Gudd{\alpha}{\rho}{\delta}\Gudd{\rho}{\beta}{\gamma}
\end{equation}
as follows
\begin{equation}\label{Riccitensor}
\mathcal{R}_{\beta\delta}=\Ruddd{\alpha}{\beta}{\alpha}{\delta}\col\qquad\mathcal{R}=g^{\beta\delta}\mathcal{R}_{\beta\delta}\pnt
\end{equation}
Here $\Gudd{\alpha}{\beta}{\gamma}$ denotes the Christoffel connection
coefficients which are given in terms of the metric as
\begin{equation}\label{connection}
\Gudd{\alpha}{\beta}{\gamma}=\frac{1}{2}g^{\alpha\rho}(\partial_\beta
g_{\rho\gamma}+\partial_\gamma g_{\rho\beta}-\partial_\rho g_{\beta\gamma})\pnt
\end{equation}
\section{Conformal flatness}
\label{app:Weyltens}
One can explicitly check conformal flatness by a computation
of the Weyl tensor. The latter is defined as
\begin{equation}
\begin{aligned}
\mathcal{C}_{ABCD}&=\mathcal{R}_{ABCD}-\frac{1}{D-2}(g_{AC}\mathcal{R}_{BD}-g_{AD}\mathcal{R}_{BC}+g_{BD}\mathcal{R}_{AC}-g_{BC}\mathcal{R}_{AD})\\
&\phantom{{}=\mathcal{R}_{ABCD}}+\frac{\mathcal{R}}{(D-1)(D-2)}(g_{AC}g_{BD}-g_{AD}g_{BC})
\end{aligned}
\end{equation}
for a $D$ dimensional space with coordinate indices $A,B,C,D$.
The Weyl tensor is constructed in such a way that under a conformal
transformation of the metric
\begin{equation}
g_{AB}\to g'_{AB}=\varrho\,g_{AB}
\end{equation}
it transforms homogeneously\footnote{This is equivalent to $\mathcal{C}'^{A}_{\phantom{A}{B}{C}{D}}=\Cuddd{A}{B}{C}{D}$.}\ i.\ e.\ 
\begin{equation}
\mathcal{C}_{ABCD}\to\mathcal{C}'_{ABCD}=\varrho\,\mathcal{C}_{ABCD}\pnt
\end{equation}
Two spaces are called conformal to each other if their Weyl tensors are
related as given in the above equation. In particular a space with
$\mathcal{C}_{ABCD}=0$ can be conformally mapped to flat space and is therefore
called conformally flat. 

It is easy to see that all spaces with a Riemann tensor of the form
\begin{equation}
\mathcal{R}_{ABCD}=\frac{\mathcal{R}}{D(D-1)}(g_{AC}g_{BD}-g_{AD}g_{BC})
\end{equation}
in $D$ dimensions are conformally flat. This is not necessarily true if the
$D$-dimensional space is a direct product of conformally flat spaces. 
In the following we will check that $\AdSS$ with embedding radii $R_1$ and
$R_2$ respectively is conformally flat if and only if $R_1=R_2$ for $d,d'>0$.
Greek indices refer to $\AdS_{d+1}$ and lower case Latin indices refer to
$\text{S}^{d'+1}$ in the following. 
Since we now that $\AdS_{d+1}$ and $\mathds{R}\times\text{S}^{d'+1}$ are
solutions to the Einstein equation with cosmological constant, the Ricci
tensors have to be proportional to the metrics of these spaces. As the 
metric of $\AdSS$ is block diagonal, all contributions to the Ricci tensor
with mixed indices vanish, too.
To find the condition for conformally flatness of $\AdSS$, 
we have to check the vanishing of
\begin{equation}
\mathcal{C}_{\mu\nu\rho\sigma}\col\qquad\mathcal{C}_{mn\rho\sigma}\col\qquad
\mathcal{C}_{m\nu r\sigma}\col\qquad\mathcal{C}_{mnrs}\pnt
\end{equation}
If one inserts the expressions \eqref{AdSScurvature} for the non-vanishing
Riemann and Ricci tensors and the scalar curvature
one finds
\begin{equation}
\begin{aligned}
\mathcal{C}_{\mu\nu\rho\sigma}&=\frac{d'(d'+1)}{(d+d')(d+d'+1)}\Big(-\frac{1}{R_1^2}+\frac{1}{R_2^2}\Big)(g_{\mu\rho}g_{\nu\sigma}-g_{\mu\sigma}g_{\nu\rho})\col\\
\mathcal{C}_{m\nu r\sigma}&=-\frac{dd'}{(d+d')(d+d'+1)}\Big(-\frac{1}{R_1^2}+\frac{1}{R_2^2}\Big)g_{mr}g_{\nu\sigma}\col\\
\mathcal{C}_{mnrs}&=\frac{d(d+1)}{(d+d')(d+d'+1)}\Big(-\frac{1}{R_1^2}+\frac{1}{R_2^2}\Big)(g_{mr}g_{ns}-g_{ms}g_{nr})
\pnt
\end{aligned}
\end{equation}
It is easy to see that these components are zero and thus that $\AdSS$ is
conformally flat if and only if $R_1=R_2$ for $d,d'>0$.
Furthermore, one finds from the above given results that the product of a  
conformally flat higher dimensional space with $\mathds{R}$ or $\text{S}^1$
is always conformally flat.
Hence, the ESU is conformally flat since it is given by the direct 
product of a sphere with $\mathds{R}$.  

\section{Relation of the bulk-to-bulk and the bulk-to-boundary propagator}
\label{app:bulkboundproprel}
We will show for a scalar field in an Euclidean space how the bulk-to-boundary
propagator is related to the bulk-to-bulk propagator, if the boundary has
codimension one w.\ r.\ t.\ the bulk, like in the case of the 
$\AdS/\text{CFT}$ correspondence.
The bulk-to-bulk propagator $G(x,x')$ of a scalar field with mass $m$ is
defined as Green function that fulfills
\begin{equation}\label{Gfuncdiffeq}
(\Box_x -m^2)G(x,x')=-\frac{1}{\sqrt{g}}\delta (x,x')\col
\end{equation} 
with appropriate boundary conditions. Here
$\Box_x$ is the Laplace operator on the $(d+1)$-dimensional Riemannian 
manifold $M$ with a $d$-dimensional boundary which we denote with 
$\partial M$. The
coordinates are $x^i$, the metric is $g_{ij}$ and its determinant is $g$. The
  propagator $G(x,x')$ corresponds to a scalar field $\phi(x)$ which should
  obey 
\begin{equation}\label{phidiffeq}
(\Box_x -m^2)\phi(x)=J(x)\col\qquad\lim_{x_\perp\to 0}\phi(x)x^a_\perp=\bar\phi(\bar x)\col
\end{equation}
where $J(x)$ are sources for the field $\phi$ in the interior. We have split
the coordinates like $x=(x_\perp,\bar x)$ with the boundary at
$x_\perp=0$. Boundary values $\bar\phi$ for the field $\phi$ 
are specified with a
nontrivial scaling with $x^a_\perp$ for later convenience. 
The bulk-to-boundary propagator $K(x,\bar x')$ is defined as the solution of 
the equations
\begin{equation}\label{bulkboundpropdefeq}
(\Box_x -m^2)K(x,\bar x')=0\col\qquad\lim_{x_\perp\to 0}K(x,\bar
x')x^a_\perp=\delta(\bar x,\bar x')\col
\end{equation}
where the second equation implements the necessary singular behaviour at the
boundary. 
A solution of the equations \eqref{phidiffeq} 
with $J(x)=0$ is then given by
\begin{equation}\label{boundvalsol}
\phi(x)=\int_{\partial M}\de^d\bar x' K(x,\bar x')\bar\phi(\bar x')\pnt
\end{equation}
Since we deal with the problem in Euclidean signature
\cite{Witten:1998qj,Freedman:1998tz,D'Hoker:2002aw},
we will denote $K(x,\bar x')$ as the Poisson kernel. It
is not independent from the Green function defined via \eqref{Gfuncdiffeq} as
we will now show. 

With \eqref{Gfuncdiffeq} one can write an identity for the field $\phi$ that
reads 
\begin{equation}
\phi(x)=-\int_M\de^{d+1}x'\sqrt{g}\phi(x')(\Box_{x'}-m^2)G(x,x')\pnt
\end{equation}
After applying partial integration twice and using \eqref{phidiffeq} it
assumes the form
\begin{equation}\label{boundsourcesol}
\phi(x)=\int_{\partial M}\de
A'_\mu\sqrt{g}g^{\mu\nu}\big[(\partial'_\nu\phi(x'))G(x,x')-\phi(x')\partial'_\nu
G(x,x')\big]-\int_M\de^{d+1}x'\sqrt{g}J(x')G(x,x')\col
\end{equation}
where $\de A'_\mu$ denotes the infinitesimal area element on $\partial M$ which
points into the outer normal direction and $\partial'_\mu$ denotes a derivative
w.\ r.\ t.\ $x'_\mu$. If one has the additional restriction that $G(x,x')=0$
for $x'\in\partial M$ ($x'_\perp=0$) the first term in the above boundary
integral is zero.  One then arrives at the `magic rule'  
which for the boundary value problem in presence of a source $J(x)$ in the
interior can be found in \cite{Barton:1989}.  
Here, however, we have to be more
careful. In \eqref{phidiffeq} we have allowed for a scaling of the boundary
value with $x^a_\perp$ as written down in \eqref{bulkboundpropdefeq}. For
$a>0$ the vanishing $G(x,x')$ at the boundary can be compensated and the 
first term in the boundary integral of \eqref{boundsourcesol} then
contributes. 

Considering $\AdS_{d+1}$, this is indeed the case, because the
field $\phi$ with conformal dimension $\Delta$ represents the non-normalizable 
modes $\phi_\Delta$ which scale as given in \eqref{phiboundbehaviour}, i.\ e.\
$a=\Delta-d$.  
Indicating the corresponding propagator with the suffix $\Delta$,
one finds $G(x,x')=0$
at $x'\in\partial M$ but the vanishing is compensated by the singular
behaviour of the non-normalizable modes in the limit $x'_\perp\to 0$.

We now formulate \eqref{boundsourcesol} on Euclidean $\AdS_{d+1}$ in Poincar\'e
coordinates \eqref{AdSmetricPcoord} (with the minus sign in front of $\de x_0^2$ converted to plus) with $J(x')=0$, where one has
\begin{equation}
\de A_\mu=-\de^d\bar
x\delta_\mu^\perp\col\qquad\sqrt{g}=\Big(\frac{R_1}{x_\perp}\Big)^{d+1}\col\qquad g^{\perp\perp}=\Big(\frac{x_\perp}{R_1}\Big)^2\pnt
\end{equation}
The minus sign in the area element stems from the fact that the
$x_\perp$-direction points into the interior of $\AdS_{d+1}$, 
but one has to take
the outer normal vector. Now \eqref{boundsourcesol} reads
\begin{equation}
\phi_\Delta(x)=-R_1^{d-1}\int\de^d\bar
x'x'^{1-d}_\perp\big[(\partial'_\perp\phi(x'))G_\Delta(x,x')-\phi(x')\partial'_\perp G_\Delta(x,x')\big]\pnt
\end{equation}
Using now \eqref{phiboundbehaviour} and comparing with
\eqref{AdSboundvalsol} that replaces \eqref{boundvalsol} in the
$\AdS_{d+1}$-case, one finds that the relation of the bulk-to-bulk and the
bulk-to-boundary propagator is given by 
\begin{equation}\label{bulktoboundarypropagatorrel}
K_\Delta(x,\bar x')=-R_1^{d-1}
\big[(d-\Delta)x'^{-\Delta}_\perp-x'^{1-\Delta}_\perp\partial'_\perp
\big] G_\Delta(x,x')\big|_{x'_\perp=0}\pnt
\end{equation}
If we now insert the explicit expression \eqref{AdSprop} for $G_\Delta(x,x')$, 
we see what we already mentioned: in approaching the boundary  
($x'_\perp\to 0$), $G_\Delta(x,x')$ itself goes to zero 
like $x'^\Delta_\perp$ but this is compensated by the singular behaviour of
the prefactor in the first 
term of \eqref{bulktoboundarypropagatorrel}.
Hence, in contrast to the situation of the 'magic rule' \cite{Barton:1989}, 
it contributes to the bulk-to-boundary propagator. 
One then finds with \eqref{xiinPcoord} and with
$\hypergeometric{a}{b}{c}{0}=1$ that effectively 
\begin{equation}
K_\Delta(x,\bar x')=-R_1^{d-1}(d-2\Delta)x'^{-\Delta}_\perp G_\Delta(x,x')\big|_{x'_\perp=0}\col
\end{equation}
which is in perfect agreement with the explicit expressions
\eqref{bulkboundAdSprop} and \eqref{AdSprop}. 

\section{Geodesics in a warped geometry}
\label{app:geoinwarpedspace}
Consider a metric of the form
\begin{equation}\label{warpedmetric}
\de s_z^2=\de s_x^2+\e^{2w(x)}\de s_y^2\col
\end{equation}
where the subscript indicates which coordinates $z=(x,y)$ the corresponding
part depends on. We use capital Latin indices for the whole space lower case
Greek indices for the first part (that only depends on $x$) and lower case
Latin indices for the second part (that depends on $y$ and on $x$ due to the
warp factor). 
The components of the complete metric are $G_{MN}$ and the two blocks are
$G_{\mu\nu}$ and $G_{mn}=\e^{2w(x)}g_{mn}$. 
The equations for the geodesics, parameterized with an affine
parameter $\tau$ read
\begin{equation}\label{geodesicequations}
\ddot z^R+\Gudd{R}{M}{N}\dot z^M\dot z^N=0\pnt
\end{equation}
Derivatives w.\ r.\ t.\ $\tau$ are denoted with dots.
It is easy to see that the only non-zero components of the connection
$\Gudd{R}{M}{N}$, which has been defined in \eqref{connection} in terms of the
metric, are then given by
\begin{equation}
\begin{aligned}
\Gudd{\rho}{\mu}{\nu}&=\frac{1}{2}g^{\rho\kappa}(\partial_\mu
G_{\kappa\nu}+\partial_\nu G_{\kappa\mu}-\partial_\kappa G_{\mu\nu})\col\\
\Gudd{\rho}{m}{n}&=-G^{\rho\kappa}\e^{2w}(\partial_\kappa w)g_{mn}\col\\
\Gudd{r}{m}{\nu}&=(\partial_\nu w)\delta_m^r\col\\
\Gudd{r}{\mu}{n}&=(\partial_\mu w)\delta_n^r\col\\
\Gudd{r}{m}{n}&=\gudd{r}{m}{n}=\frac{1}{2}g^{rk}(\partial_m
g_{kn}+\partial_n g_{km}-\partial_k g_{mn})\pnt
\end{aligned}
\end{equation}
Here, $\gudd{r}{m}{n}$ is the connection that corresponds to $\de s_y^2$
computed with $g_{mn}$ (without the warp factor).  
The geodesic equations then read
\begin{equation}
\begin{aligned}\label{geoeq}
\ddot x^\rho+\Gudd{\rho}{\mu}{\nu}\dot x^\mu\dot
x^\nu-G^{\rho\kappa}\e^{2w}(\partial_\kappa w)\dot{\vec y}^2&=0\col\\
\ddot y^r+\gudd{r}{m}{n}\dot y^m\dot y^n+2\dot w\dot y^r&=0\col
\end{aligned}
\end{equation}
where we have used $g_{mn}\dot y^m\dot y^n=\dot{\vec y}^2$ and $\dot
x^\mu\partial_\mu w=\dot w$. The effect of the warp factor in
\eqref{warpedmetric} is that the scale in the part of the space with
$y$-coordinates becomes $x$-dependent. Along a geodesic (where $x=x(\tau)$) 
the scale thus depends on $\tau$. This
means that the parameter $\tau$ is not an affine parameter for a geodesic in
the space with metric $\de s_y^2$. However, using a reparameterization,
one can make it affine w.\ r.\ t.\ $\de s_y^2$. We introduce a new
parameter $\sigma=f(\tau)$ and
compute the derivatives of $y(\sigma)=y(f(\tau))$  w.\ r.\ t.\ $\tau$ and
w.\ r.\ t.\ $\sigma$, where we denote the latter with a prime.
One finds
\begin{equation}
\dot y^r=y'^r\dot f\col\qquad\ddot
y^r=y''^r\dot f^2+y'^r\ddot f\pnt 
\end{equation}  
Inserting these results into the second geodesic equations of \eqref{geoeq},
one obtains
\begin{equation}
\dot f^2(y''^r+\Gudd{r}{m}{n}y'^my'^n)+(\ddot f+2\dot
w\dot f)y'^r=0\pnt 
\end{equation}
If the reparameterization $\sigma=f(\tau)$ obeys the differential equation
\begin{equation}\label{sigmawrel}
\ddot f+2\dot w\dot f=0\col
\end{equation}
then $y(\sigma)$ describes the geodesic in the geometry given by $\de s_y^2$
with affine parameter $\sigma$,  and we find the geodesic equations
\begin{equation}
\begin{aligned}
\ddot x^\rho+\Gudd{\rho}{\mu}{\nu}\dot x^\mu\dot
x^\nu-G^{\rho\kappa}\e^{2w}(\partial_\kappa w)c\dot f^2&=0\col\\
y''^r+\gudd{r}{m}{n}y'^my'^n=0\col
\end{aligned}
\end{equation}
where $\sigma=f(\tau)$ has to obey \eqref{sigmawrel}. We have used 
$\vec y'^2=c$, where $c$ is a constant because $\sigma$ is
the affine parameter w.\ r.\ t.\ $\de s_y^2$. It is clear that an affine
parameter remains affine under constant rescalings and henceforth we can 
choose $\sigma=f(\tau)$ in such a way that $c=1$ without 
loss of generality.

The above considerations are especially useful for finding the geodesics of 
$\AdS_{d+1}$ with the metric written down in the first line of
\eqref{AdSmetricglobalcoord}. One can identify 
\begin{equation}
\de s_x^2=-\cosh^2\rho\de t^2+\de\rho^2\col\qquad \de
s_y^2=\de\Omega_{d-1}^2\col\qquad \e^w=\sinh\rho\pnt
\end{equation}
The movement of the geodesics in the $(d-1)$-dimensional spherical part is
clear. They run along a great circle with speed 
$\dot{\vec y}^2=\dot f^2$. We will not discuss its concrete dependence on the
coordinates $y$, but use the
function $f$ to describe the movement in this part.
The only non-vanishing connection coefficient in $(t,\rho)$ are
\begin{equation}
\Gudd{t}{\rho}{t}=\Gudd{t}{t}{\rho}=\tanh\rho\col\qquad\Gudd{\rho}{t}{t}=-\sinh\rho\cosh\rho\pnt
\end{equation}
The differential equations that collectively describe the
movement in the $(d-1)$-dimensional subsphere of $\AdS_{d+1}$ then read
\begin{equation}\label{AdSgeoeq}
\begin{aligned}
\ddot t +2\dot\rho\dot t\tanh\rho&=0\col\\
\ddot\rho+(\dot t^2-\dot f^2)\sinh\rho\cosh\rho&=0\col\\
\ddot f+2\dot\rho\dot f\coth\rho&=0\pnt
\end{aligned}
\end{equation}

\section[Relation of the chordal and the geodesic distance in the 
$10$-dim.\ plane wave]{Relation of the chordal and the geodesic distance in
  the $10$-dimensional plane wave}
\label{app:chordalvsgeodist}
The chordal distance \eqref{pwchordaldist} in the plane wave 
is related to the geodesic distance. This relation will now be determined for
the $10$-dimensional Hpp wave solution, where $H_{ij}=-\delta_{ij}$.
We start from the type B geodesics \eqref{typeBgeo} in the plane wave and 
fix the free parameters in terms of two points which are connected with a 
geodesic segment using the parameter range $0\le\tau\le 1$.
We choose
\begin{equation}\label{initialfinalpoints}
\begin{aligned}
z^+(0)&=z'^+\col&\qquad z^-(0)&=z'^-\col&\qquad z^i(0)&=z'^i\col\\
z^+(1)&=z^+\col&\qquad z^-(1)&=z^-\col&\qquad z^i(1)&=z^i\pnt
\end{aligned}
\end{equation}
Then from \eqref{geo34} it follows immediately
\begin{equation}\label{zplusconstants}
\alpha=z^+-z'^+\col\qquad z_0^+=z'^+\pnt
\end{equation}
After that it is advantageous to use the
theorems for trigonometric functions casting \eqref{typeBgeo} into the
form
\begin{equation}
\begin{aligned}
z^-(\tau)&=\frac{1}{4}\sum_iz^i\beta^i\cos\alpha(\tau+\tau_0^i)+\gamma\tau+z_0^-\\ 
&=\frac{1}{4}\sum_i\big[\beta_1^i\beta_2^i(\cos^2\alpha\tau-\sin^2\alpha\tau)+\big((\beta_1^i)^2-(\beta_2^i)^2\big)\cos\alpha\tau\sin\alpha\tau\big]+\gamma\tau+z_0^-\col\\
z^i(\tau)&=\beta_1^i\sin\alpha\tau+\beta_2^i\cos\alpha\tau\col
\end{aligned}
\end{equation}
where $\beta_1^i$ and $\beta_2^i$ are given by
\begin{equation}\label{betaidef}
\beta_1^i=\beta^i\cos\alpha\tau_0^i\col\qquad\beta_2^i=\beta^i\sin\alpha\tau_0^i\pnt  
\end{equation}
The initial value $z^-(0)=z'^-$ now yields
\begin{equation}
z_0^-=z'^--\frac{1}{4}\sum_i\beta_1^i\beta_2^i\pnt
\end{equation}
Inserting this result into the expression for $z^-$ at the final value $z^-(1)=z^-$
then determines 
\begin{equation}\label{gammainbetai}
\gamma=z^--z'^-+\frac{1}{4}\sum_i\big[2\beta_1^i\beta_2^i\sin^2\alpha-\big((\beta_1^i)^2-(\beta_2^i)^2\big)\cos\alpha\sin\alpha\big]\pnt
\end{equation}
For the parameters $\beta_1^i$ and $\beta_2^i$ one finds from \eqref{betaidef}
at the initial value $z^i(0)=z'^i$
\begin{equation}
\beta_2^i=z'^i\pnt
\end{equation}
Using this the condition to reach the final value $z^i(1)=z^i$ leads to
\begin{equation}
\beta_1^i=\frac{1}{\sin\alpha}(z^i-z'^i\cos\alpha)\pnt
\end{equation}
From this we determine
\begin{equation}
\begin{aligned}
\sum_i\beta_1^i\beta_2^i&=\frac{1}{\sin\alpha}\big(\vec z\vec z'-\vec z'^2\cos\alpha\big)\col\\
\sum_i\big((\beta_1^i)^2-(\beta_2^i)^2\big)&=\frac{1}{\sin^2\alpha}\big(\vec
z^2-2\vec z\vec z'\cos\alpha+\vec z'^2\cos^2\alpha-\vec
z'^2\sin^2\alpha\big)\pnt 
\end{aligned}
\end{equation}
Inserting this into \eqref{gammainbetai} then yields
\begin{equation}\label{gammaexpl}
\begin{aligned}
\gamma&=z^--z'^-+\frac{1}{4\sin\alpha}\big(2\vec z\vec z'-(\vec z^2+\vec
z'^2)\cos\alpha\big)\\
&=z^--z'^-+\frac{1}{4\sin\alpha}\big(2(\vec z^2+\vec
z'^2)\sin^2\tfrac{\alpha}{2}-(\vec z-\vec z')^2\big)\pnt
\end{aligned}
\end{equation}
The constant tangential vector has length $-4\alpha\gamma$. Due to our choice
that $\tau$ runs over an interval of one length unit, this is directly the 
geodesic distance $s(z,z')$ between the two points in 
\eqref{initialfinalpoints}.
From \eqref{zplusconstants} and \eqref{gammaexpl} one finds that it is explicitly
given by
\begin{equation}\label{geoinchordaldist}
\begin{aligned}
s(z,z')
&=-4(z^+-z'^+)(z^--z'^-)
-\frac{z^+-z'^+}{\sin(z^+-z'^+)}\big(2(\vec z^2+\vec
z'^2)\sin^2\tfrac{(z^+-z'^+)}{2}-(\vec z-\vec z')^2\big)\\
&=\frac{z^+-z'^+}{\sin(z^+-z'^+)}\Phi(z,z')\pnt
\end{aligned}
\end{equation}
The last line above shows the
relation between the geodesic and the chordal distance \eqref{pwchordaldist} 
in the case $H_{ij}=-\delta_{ij}$. the chordal distance \eqref{pwchordaldist} 
in the case $H_{ij}=-\delta_{ij}$. It is immediately clear that the above
relation holds for the type A geodesics \eqref{geo35} as well. If one takes
the limit $z'^+\to z^+$ the prefactor becomes one. 
For the two points \eqref{initialfinalpoints} one finds from \eqref{geo35}
\begin{equation}
\beta=z^--z'^-\col\qquad z_0^-=z'^-\col\qquad\gamma^i=z^i-z'^i\col\qquad z_0^i=z'^i\pnt
\end{equation}
The length of the tangent
vector is given by $(\gamma^i)^2$ and therefore the distances are related via
\begin{equation}
s(z,z')=\Phi(z,z')=(\vec z-\vec z')^2\col
\end{equation}
that coincides with \eqref{geoinchordaldist}.

\section{Useful relations for hypergeometric functions}
\label{app:userelHyper}
Most of the relations we present here can be found in
\cite{Bateman:1953,Abramowitz:1972,Gradshteyn:1980} or are derived from there.
An integral representation for the hypergeometric functions is given by
\begin{equation}\label{hypergeometricintrep}
\hypergeometric{a}{b}{c}{z}=\frac{\Gamma(c)}{\Gamma(b)\Gamma(c-b)}\int_0^1\de
u\,u^{b-1}(1-u)^{c-b-1}(1-zu)^{-a}\col\qquad\Re c>\Re b>0\pnt
\end{equation}
In particular, hypergeometric functions with parameters $a$ ,$b$, and $c$ that 
are related to each other are important for us. One finds so called quadratic 
transformation formulas like
\begin{equation}\label{hypergeometricquadratictrafo}
\begin{aligned}
\hypergeometric{a}{a+\tfrac{1}{2}}{c}{z^2}&=(1+z)^{-2a}\hypergeometric{2a}{c-\tfrac{1}{2}}{2c-1}{\tfrac{2z}{1+z}}\col\\
\hypergeometric{a}{b}{a+b-\tfrac{1}{2}}{z}&=\frac{1}{\sqrt{1-z}}\hypergeometric{2a-1}{2b-1}{a+b-\tfrac{1}{2}}{\tfrac{1}{2}(1-\sqrt{1-z})}\col\\
\hypergeometric{a}{b}{a+b+\tfrac{1}{2}}{z}&=\hypergeometric{2a}{2b}{a+b+\tfrac{1}{2}}{\tfrac{1}{2}(1-\sqrt{1-z})}\pnt
\end{aligned}
\end{equation}
The hypergeometric functions in the 
propagators \eqref{AdSprop} with $\Delta_\pm=\frac{d\pm1}{2}$ (at the  mass
value generated by the Weyl invariant coupling) become ordinary analytic expressions 
\begin{equation}\label{hypergeom1}
\begin{aligned}
\hypergeometric{a}{a+\tfrac{1}{2}}{\tfrac{1}{2}}{\xi^2}&=\frac{1}{2}\big[(1+\xi)^{-2a}+(1-\xi)^{2a}\big]\col\\
\hypergeometric{a+\tfrac{1}{2}}{a+1}{\tfrac{3}{2}}{\xi^2}&=-\frac{1}{4a\xi}\big[(1+\xi)^{-2a}-(1-\xi)^{-2a}\big]\pnt
\end{aligned}
\end{equation}
Setting $a=\frac{d-1}{4}$ one finds \eqref{AdSsuperpos}.

To find the hypergeometric functions relevant for the propagators in higher dimensional AdS spaces one can use a recurrence relation (Gau\ss' relation for contiguous functions)  
\begin{equation}\label{recurrencerel}
\hypergeometric{a}{b}{c-1}{z}=\frac{c[c-1-(2c-a-b-1)z]}{c(c-1)(1-z)}\hypergeometric{a}{b}{c}{z}+\frac{(c-a)(c-b)z}{c(c-1)(1-z)}\hypergeometric{a}{b}{c+1}{z}\pnt
\end{equation}
where the hypergeometric functions relevant in lower dimensional AdS spaces enter.

For odd AdS dimensions (even $d$) the relevant hypergeometric functions can be expressed with the above recurrence relation in terms of ordinary analytic functions. This happens because of the explicit expressions
\begin{equation}
\begin{aligned}\label{hypergeomstart}
\hypergeometric{a}{a+\tfrac{1}{2}}{2a}{z}&=\frac{2^{2a-1}}{\sqrt{1-z}}\big[1+\sqrt{1-z}\big]^{1-2a}\col\\
\hypergeometric{a}{a+\tfrac{1}{2}}{2a+1}{z}&=2^{2a}\big[1+\sqrt{1-z}\big]^{-2a}\pnt
\end{aligned}
\end{equation}
One has to apply \eqref{recurrencerel} $n$ times to compute the AdS propagator at generic $\Delta$ in $d+1=3+2n$ dimensions. For $\AdS_3$ one simply uses the first expression in \eqref{hypergeomstart}.
The $\AdS_5$ case is of particular importance and therefore we give the explicit expression for the needed hypergeometric function
\begin{equation}
\hypergeometric{\tfrac{\Delta}{2}}{\tfrac{\Delta}{2}+\tfrac{1}{2}}{\Delta-1}{z}=\frac{1}{2(1-z)^\frac{3}{2}}\Big[\frac{2}{1+\sqrt{1-z}}\Big]^{\Delta-1}\Big[\sqrt{1-z}+\frac{\Delta-1}{\Delta-2}(1-z)+\frac{1}{\Delta-1}\Big]\pnt
\end{equation}

\section{Spheres and spherical harmonics of arbitrary dimensions}
\label{app:SandSH}
Here we have collected some useful facts about spheres and spherical
harmonics, see e.\ g.\ \cite{Bateman:1953} for more details.
A discussion of spherical harmonics require some results for 
Gegenbauer polynomials that can be found for example 
in \cite{Abramowitz:1972}. 

The $(d'+1)$-dimensional sphere can be parameterized in several ways.
One can choose the standard spherical coordinates 
\begin{equation}
0\le\psi_k\le\pi\col\qquad k=1,\dots d'\col\qquad 0\le\psi_{d'+1}<2\pi
\end{equation}
with the embedding
\begin{equation}\label{Scoord}
\begin{aligned}
Y_k&=R_2\prod_{i=1}^{k-1}\sin\psi_i\cos\psi_k\col\qquad k=1,\dots,d'+1\col\\
Y_{d'+2}&=R_2\prod_{i=1}^{d'}\sin\psi_i\sin\psi_{d'+1}
\end{aligned}
\end{equation}
in which the metric reads
\begin{equation}
\de
s^2=R_2^2\big(\de\psi_1^2+\sin^2\psi_1(\de\psi_2^2+\dots+\sin^2\psi_{d'-1}(\de\psi_{d'}^2+\sin^2\psi_{d'}\de\psi_{d'+1}^2\big)\cdots\big)\big)\pnt
\end{equation}
Alternatively it is sometimes advantageous to divide the sphere into two
subspheres of dimensions $\bar d$ and $d'-\bar d$ with the help of the coordinate $\psi$, 
\begin{equation}
0\le\psi\le\frac{\pi}{2}\col
\end{equation}
and the embedding
\begin{equation}
\begin{aligned}
Y_k&=R_2\cos\psi\,\omega_k\col\qquad\sum_{k=1}^{\bar
  d+1}\omega_k^2=1\col\qquad k=1,\dots,\bar d+1\col\\
Y_l&=R_2\sin\psi\,\hat\omega_l\col\qquad\sum_{l=1}^{d'-\bar
  d+1}\hat\omega_l^2\col\qquad l=1,\dots, d'-\bar d+1\col 
\end{aligned}
\end{equation}
such that the metric is then given by
\begin{equation}
\de s^2=R_2^2\big(\de\psi^2+\sin^2\psi\de\Omega_{\bar
  d}^2+\cos^2\psi\de\hat\Omega_{d'-\bar d}^2\big)\pnt
\end{equation}
In the coordinates \eqref{Scoord} the volume form of the unit
$\text{S}^{d'+1}$ reads 
\begin{equation}\label{Svolumeform}
\vol(\Omega_{d'+1})=\prod_{k=1}^{d'}\Big((\sin\psi_k)^{d'-k+1}\de\psi_k\Big)\wedge\de\psi_{d'+1}\col
\end{equation}
where the product has to be understood as the wedge product.
Integrating the volume form leads to the total volume of the unit 
$\text{S}^{d'+1}$
\begin{equation}\label{Svolume}
\Omega_{d'+1}=\frac{2\pi^{\frac{d'}{2}+1}}{\Gamma(\frac{d'}{2}+1)}\pnt
\end{equation}

Spherical harmonics $Y^I(y)$ on $\text{S}^{d'+1}$ are characterized by quantum
numbers 
\begin{equation}\label{multindex}
I = (l,m_1,\dots,m_{d'})\col\qquad l  \ge m_1 \ge \dots \ge m_{d'-1} \ge | m_{d'}| \ge 0
\end{equation}
 and form irreducible representations of $\text{SO}(d'+2)$. They are
 eigenfunctions with respect to the Laplace operator on the sphere  
\begin{equation}\label{SHcasimir}
\Box_y Y^I(y)=-\frac{l(l+d')}{R_2^2} Y^I(y)\pnt
\end{equation}
In the coordinates \eqref{Scoord} they are explicitly given by
\cite{Bateman:1953} 
\begin{equation}\label{SHexpl}
Y^{l,m_1,\dots,m_{d'}}(y)=N(l,m_1,\dots,m_{d'})\e^{im_{d'}\psi_{d'+1}}\prod_{k=1}^{d'}(\sin\psi_k)^{m_k}C_{m_{k-1}-m_k}^{(m_k+\frac{1}{2}(d'-k+1))}(\cos\psi_k)\col
\end{equation}
where $y=(\psi_1,\dots,\psi_{d'+1})$, $m_0=l$ and the $C_l^{(\beta)}$ are the
Gegenbauer Polynomials described below. With their normalization in \eqref{Gnorm} the normalization factor is given by
\begin{equation}\label{SHnorm}
(N(l,m_1,\dots,m_{d'}))^{-2}=2\pi\prod_{k=1}^{d'}N(m_{k-1}-m_k,m_k+\tfrac{d'-k+1}{2})\col
\end{equation}
such that the spherical harmonics are orthonormal w.\ r.\ t.\ integration over the unit $\text{S}^{d'+1}$. 
One very important relation for the spherical harmonics is 
\begin{equation}\label{SHcomplete}
\sum_{\scriptscriptstyle m_1\ge\dots\ge
  m_{d'-1}\ge|m_{d'}|\ge0}^lY^I(y)Y^{\ast I}(y') = \frac{(2l+d')\Gamma (\frac{d'}{2})}{4 \pi^{\frac{d'}{2}+1}}
C_l^{(\frac{d'}{2})}(\cos\Theta)\col\qquad\cos\Theta=\frac{\vec Y\vec Y'}{R_2^2}=1-\frac{v}{2R_2^2}\col
\end{equation}
where $\vec Y$ and $\vec Y'$ are the embedding vectors which coordinates are
given by \eqref{Scoord}. 
The above formula can be easily verified with the help of the homogeneity of
the sphere which allows one to choose $\vec Y'=(1,0,\dots,0)$. At this values 
all angles $y'=(0,\dots,0)$ and therefore 
all spherical harmonics \eqref{SHexpl} except of 
\begin{equation}
Y^{l,0,\dots,0}(y')=N(l,0,\dots,0)C_l^{(\frac{d'}{2})}(1)\prod_{k=1}^{d'-1}C_0^{(\frac{d'-k}{2})}(1)=N(l,0,\dots,0)C_l^{(\frac{d'}{2})}(1)
\end{equation}
are zero. 
The normalization factor reads
\begin{equation}
(N(l,0,\dots,0))^2=\frac{(2l+d')\Gamma(\frac{d'}{2})}{4\pi^{\frac{d'}{2}+1}}\frac{1}{C_0^{(\frac{d'}{2})}(1)}\col
\end{equation}
as can be verified from \eqref{SHnorm} with \eqref{Gstandard} by using the
duplication formula
\begin{equation}\label{Gammaduplication}
2^{1-z}\sqrt{\pi}\Gamma(z)=\Gamma(\tfrac{z}{2})\Gamma(\tfrac{z+1}{2})\pnt
\end{equation}
Only the
first term of the sum \eqref{SHcomplete} survives and yields the R.\ H.\ S. with
$\cos\theta=\cos\psi_1$, and $\psi_1$ being given by the angle between $\vec Y$
and $\vec Y'$ according to \eqref{Scoord}.

The Gegenbauer Polynomials $C_l^{(\beta)}$ can be defined via their
generating function 
\begin{equation}\label{Ggenfunc}
\frac{1}{(1-2q\eta+q^2)^\beta}=\sum_{l=0}^\infty q^lC_l^{(\beta)}(\eta)\pnt
\end{equation} 
Their standardization is given by
\begin{equation}\label{Gstandard}
G_l^{(\beta)}(1)=\frac{\Gamma(l+2\beta)}{\Gamma(l+1)\Gamma(\beta)}
\end{equation}
and they obey the orthogonality relation
\begin{equation}\label{Gorthorel}
\int_{-1}^{1}\de\eta\,
(1-\eta^2)^{\beta-\frac{1}{2}}C_m^{(\beta)}(\eta)C_n^{(\beta)}(\eta)=N(n,\beta)\delta_{mn}\col
\end{equation}
where
\begin{equation}\label{Gnorm}
N(n,\beta)=\frac{2^{1-2\beta}\pi\Gamma(n+2\beta)}{\Gamma(n+1)(n+\beta)\Gamma(\beta)^2}\pnt
\end{equation}
One way to represent the Gegenbauer polynomials is via Rodrigues' formula
\begin{equation}\label{GRodrigues}
C_n^{(\beta)}(\eta)=(-1)^n2^{-n}\frac{\Gamma(\beta+\frac{1}{2})\Gamma(n+2\beta)}{\Gamma(n+1)\Gamma(2\beta)\Gamma(n+\beta+\frac{1}{2})}(1-\eta^2)^{\frac{1}{2}-\beta}\frac{\de^n}{\de\eta^n}(1-\eta^2)^{n+\beta-\frac{1}{2}}\pnt
\end{equation}
An important property is their symmetry under reflection of their arguments.
One can easily read off from \eqref{Ggenfunc} that they obey
\begin{equation}\label{Gsymmetry}
C_l^{(\beta)}(-\eta)=(-1)^lC_l^{(\beta)}(\eta)\pnt
\end{equation}
This symmetry relates the values of the spherical harmonics \eqref{SHexpl} 
at $y=(\psi_1,\dots,\psi_{d'+1})$ to their values at the antipodal point on
the sphere, which is given by
\begin{equation}\label{Santipodal}
\tilde y=(\tilde\psi_1,\dots,\tilde\psi_{d'},\tilde\psi_{d'+1})=(\pi-\psi_1,\dots,\pi-\psi_{d'},\psi_{d'+1}+\pi)\pnt
\end{equation}
Then the spherical harmonics obey
\begin{equation}\label{SHsymmetry}
Y^I(\tilde y)=(-1)^lY^I(y)\pnt
\end{equation}

\section{Proof of the summation theorem}
\label{app:HyperGsummationproof}
Using  \eqref{AdSprop} and \eqref{SHcomplete} for $\Delta_+=l+\frac{d+d'}{2}$ (leading to \eqref{modesumprop}), one finds the solution \eqref{AdSSprop} if the following relation holds for $\alpha\ge\beta>0$,
$2\alpha,2\beta\in\mathds{N}$
\begin{equation}\label{hyperid}
\sum_{l=0}^{\infty}\frac{\Gamma(l+\alpha)}{\Gamma(l+\beta)}\Big(\frac{\xi}{2}\Big)^l\hypergeometric{\tfrac{l}{2}+\tfrac{\alpha}{2}}{\tfrac{l}{2}+\tfrac{\alpha}{2}+\tfrac{1}{2}}{l+\beta+1}{\xi^2}C_l^{(\beta)}(\eta)
=\frac{\Gamma(\alpha)}{\Gamma(\beta)}\frac{1}{(1-\xi\eta)^\alpha}\col
\end{equation} 
with the interpretation $\alpha=\frac{d+d'}{2}$, $\beta=\frac{d'}{2}$.
We could not find the above formula in the literature. It is in fact a
summation rule for a product of a special hypergeometric function for which so called quadratic transformation formulae exist and which can be expressed in terms of a Legendre function \cite{Abramowitz:1972} and of a Gegenbauer polynomial. The identity can therefore be re-expressed in the following way
\begin{equation}\label{polyid}
\Big(\frac{2}{\xi}\Big)^\beta(1-\xi^2)^\frac{\beta-\alpha}{2}\sum_{l=0}^{\infty}\Gamma(l+\alpha)(l+\beta)P_{\alpha-\beta-1}^{-l-\beta}\big(\tfrac{1}{\sqrt{1-\xi^2}}\big)C_l^{(\beta)}(\eta)=\frac{\Gamma(\alpha)}{\Gamma(\beta)}\frac{1}{(1-\xi\eta)^\alpha}\pnt
\end{equation}
The simplest way to prove\footnote{We thank Danilo Diaz for delivering a
  simplification of our original recurrence proof, that allows for an
  extension to more general values of $\alpha$ and $\beta$.} this relation
is to use the orthogonality of the Gegenbauer polynomials \eqref{Gorthorel}
to project out a term with fixed $l$ from the sum in \eqref{hyperid}. One
  finds
\begin{equation}
\frac{\Gamma(l+\alpha)}{\Gamma(l+\beta)}\Big(\frac{\xi}{2}\Big)^l\hypergeometric{\tfrac{l}{2}+\tfrac{\alpha}{2}}{\tfrac{l}{2}+\tfrac{\alpha}{2}+\tfrac{1}{2}}{l+\beta+1}{\xi^2}N(l,\beta)
=\frac{\Gamma(\alpha)}{\Gamma(\beta)}\int_{-1}^{1}\de\eta\,
\frac{(1-\eta^2)^{\beta-\frac{1}{2}}}{(1-\xi\eta)^\alpha}C_l^{(\beta)}(\eta)\pnt 
\end{equation} 
The Gegenbauer polynomials are then expressed with Rodrigues' formula \eqref{GRodrigues}.
Applying partial integration $l$ times and using 
\begin{equation}
\frac{\de^l}{\de\eta^l}(1-\xi\eta)^{-\alpha}=\xi^l\frac{\Gamma(l+\alpha)}{\Gamma(\alpha)}(1-\xi\eta)^{-l-\alpha}
\end{equation}
then yields
\begin{equation}\label{constlrel}
\begin{aligned}
&N(l,\beta)\hypergeometric{\tfrac{l}{2}+\tfrac{\alpha}{2}}{\tfrac{l}{2}+\tfrac{\alpha}{2}+\tfrac{1}{2}}{l+\beta+1}{\xi^2}\\
&\qquad\qquad\qquad\quad=\frac{\Gamma(l+\beta)}{\Gamma(\beta)}\frac{\Gamma(\beta+\frac{1}{2})\Gamma(l+2\beta)}{\Gamma(l+1)\Gamma(2\beta)\Gamma(l+\beta+\frac{1}{2})}
\times\int_{-1}^{1}\de\eta\,
\frac{(1-\eta^2)^{l+\beta-\frac{1}{2}}}{(1-\xi\eta)^{l+\alpha}}\col 
\end{aligned}
\end{equation}
where $N(l,\beta)$ is given by \eqref{Gnorm} 
With the variable substitution $u=\frac{\eta+1}{2}$ the integral on the R.\
H.\ S.\ can be cast into the form
\begin{equation}\label{integral}
\begin{aligned}
\int_{-1}^{1}\de\eta\,\frac{(1-\eta^2)^{l+\beta-\frac{1}{2}}}{(1-\xi\eta)^{l+\alpha}}
&=\frac{4^{l+\beta}}{(1+\xi)^{l+\alpha}}\int_0^{1}\de u\,
u^{l+\beta-\frac{1}{2}}(1-u)^{l+\beta-\frac{1}{2}}\big(1-\tfrac{2\xi}{1+\xi}u\big)^{-l-\alpha}\\
&=\frac{4^{l+\beta}}{(1+\xi)^{l+\alpha}}\frac{(\Gamma(l+\beta+\tfrac{1}{2}))^2}{\Gamma(2l+2\beta+1)}\hypergeometric{l+\alpha}{l+\beta+\tfrac{1}{2}}{2l+2\beta+1}{\tfrac{2\xi}{1+\xi}}\\
&=4^{l+\beta}\frac{\Gamma(l+\beta+\tfrac{1}{2})^2}{\Gamma(2l+2\beta+1)}\hypergeometric{\tfrac{l}{2}+\tfrac{\alpha}{2}}{\tfrac{l}{2}+\tfrac{\alpha}{2}+\tfrac{1}{2}}{l+\beta+1}{\xi^2}
\pnt 
\end{aligned}
\end{equation}
In the second line we have used the integral representation for the
hypergeometric functions \eqref{hypergeometricintrep} 
and the third line follows with the help of
the quadratic transformation formula in the first line of
\eqref{hypergeometricquadratictrafo}.
As the last step, we have to insert \eqref{integral} into \eqref{constlrel}. 
With the duplication formula \eqref{Gammaduplication} one then finds that 
both sides of \eqref{constlrel} match, and the proof of \eqref{hyperid} is
complete. The relation is valid not only for $\alpha\ge\beta>0$,
$2\alpha,2\beta\in\mathds{N}$, but for all $\beta>0$.


\end{appendix}
\clearemptydoublepage
%
\bibliographystyle{../habbrv}
\addcontentsline{toc}{chapter}{Bibliography}
\bibliography{references_history,references_textbooks,references_NC,references_BMN}
\pagebreak

\end{document}